\documentclass[
    reprint,twocolumn,
    aps,prx,10pt,amsmath,amssymb,
    longbibliography,superscriptaddress,
    showpacs,preprintnumbers,
    floatfix,nofootinbib,
]{revtex4-2}

\usepackage{xr-hyper}
\usepackage[colorlinks=true, linkcolor=blue, citecolor=blue, urlcolor=blue, filecolor=blue]{hyperref}
\usepackage{bm,bbm,bbold}
\usepackage{braket}
\usepackage{mathtools}
\usepackage{amsmath,amssymb,amsfonts}
\usepackage[dvipsnames]{xcolor}
\usepackage{graphicx}
\usepackage[section]{placeins}
\usepackage{multirow}
\usepackage{orcidlink}
\usepackage[normalem]{ulem}
\usepackage{xspace}

\newcommand{\mb}[1]{{\bm{\mathrm{#1}}}}
\newcommand{\mc}[1]{{\mathcal{#1}}}

\newcommand{\nnnl}{\nonumber \\}
\newcommand{\veps}[0]{\varepsilon}
\newcommand{\dd}[0]{{\mathrm{d}}}

\renewcommand{\Re}{\operatorname{Re}}
\renewcommand{\Im}{\operatorname{Im}}

\newcommand{\mcT}{{\mathcal{T}}}

\newcommand{\one}[0]{\mathbbm{1}}

\usepackage{cleveref}
\crefname{equation}{Eq.}{Eqs.}
\Crefname{equation}{Equation}{Equations}
\crefname{figure}{Fig.}{Figs.}
\Crefname{figure}{Figure}{Figures}
\crefname{table}{Table}{Tables}
\Crefname{table}{Table}{Tables}
\crefname{section}{Sec.}{Secs.}
\Crefname{section}{Section}{Sections}
\crefname{subsection}{Sec.}{Secs.}
\Crefname{subsection}{Section}{Sections}
\crefname{subsubsection}{Sec.}{Secs.}
\Crefname{subsubsection}{Section}{Sections}
\crefname{appendix}{Appendix}{Appendixes}
\Crefname{appendix}{Appendix}{Appendixes}
\crefname{subappendix}{Appendix}{Appendixes}
\Crefname{subappendix}{Appendix}{Appendixes}
\crefname{subsubappendix}{Appendix}{Appendixes}
\Crefname{subsubappendix}{Appendix}{Appendixes}
\let\MYappendix\appendix
\renewcommand{\appendix}{\MYappendix
    \crefalias{section}{appendix}\crefalias{subsection}{subappendix}%
    \crefalias{subsubsection}{subsubappendix}}

\ExplSyntaxOn
\seq_new:N \l__jml_fig_out_seq
\seq_new:N \l__jml_fig_range_seq
\bool_new:N \l__jml_fig_plural_bool
\cs_new_protected:Npn \__jml_fig_panels:nn #1#2
  {
    \seq_clear:N \l__jml_fig_out_seq
    \bool_set_false:N \l__jml_fig_plural_bool
    \clist_map_inline:nn {#1}
      {
        \seq_set_split:Nnn \l__jml_fig_range_seq { - } {##1}
        \int_compare:nNnTF { \seq_count:N \l__jml_fig_range_seq } > { 1 }
          {
            \bool_set_true:N \l__jml_fig_plural_bool
            \seq_put_right:Nx \l__jml_fig_out_seq
              {
                \exp_not:n { \ref{#2} } ( \seq_item:Nn \l__jml_fig_range_seq { 1 } )
                \exp_not:n { -- \ref{#2} } ( \seq_item:Nn \l__jml_fig_range_seq { 2 } )
              }
          }
          { \seq_put_right:Nn \l__jml_fig_out_seq { \ref{#2} (##1) } }
      }
    \int_compare:nNnT { \seq_count:N \l__jml_fig_out_seq } > { 1 }
      { \bool_set_true:N \l__jml_fig_plural_bool }
  }
\cs_new_protected:Npn \MYFigdo #1#2#3
  {
    \IfNoValueTF {#2}
      { \IfBooleanTF {#1} { Figure } { Fig. } \nobreakspace \ref{#3} }
      {
        \__jml_fig_panels:nn {#2} {#3}
        \IfBooleanTF {#1}
          { \bool_if:NTF \l__jml_fig_plural_bool { Figures } { Figure } }
          { \bool_if:NTF \l__jml_fig_plural_bool { Figs. } { Fig. } }
        \nobreakspace
        \seq_use:Nnnn \l__jml_fig_out_seq { ~and~ } { ,~ } { ,~and~ }
      }
  }
\NewDocumentCommand{\Fig}{ o m } { \MYFigdo {\BooleanFalse} {#1} {#2} }
\NewDocumentCommand{\Figu}{ o m } { \MYFigdo {\BooleanTrue} {#1} {#2} }
\ExplSyntaxOff

\newcommand{\bk}[0]{{\mb{k}}}
\newcommand{\bq}[0]{{\mb{q}}}

\newcommand{\br}[0]{{\mb{r}}}
\newcommand{\bR}[0]{{\mb{R}}}

\newcommand{\brp}[0]{{\mb{r}'}}

\newcommand{\bkq}[0]{{\mb{k}+\mb{q}}}

\newcommand{\nk}[0]{{n\mb{k}}}

\newcommand{\mnk}[0]{{mn\mb{k}}}
\newcommand{\mnkq}[0]{{mn\mb{k}\mb{q}}}
\newcommand{\mkq}[0]{{m\mb{k}+\mb{q}}}

\newcommand{\iR}[0]{{i\mb{R}}}

\newcommand{\ik}[0]{{i\mb{k}}}

\newcommand{\nint}{\!\int\!\!}
\newcommand{\nbint}[2]{\!\int_{#1}^{#2}\!\!}
\newcommand{\nsint}[1]{\!\int_{#1}\!}

\newcommand{\intbk}{{\!\int_\mb{k}}}
\newcommand{\intbq}{{\!\int_\mb{q}}}

\newcommand{\QE}[0]{\textsc{Quantum ESPRESSO}\xspace}

\DeclarePairedDelimiterX\mel[3]{\langle}{\rangle}{#1 \delimsize\vert\mathopen{} #2 \delimsize\vert\mathopen{} #3}
\DeclarePairedDelimiterX\abs[1]{\lvert}{\rvert}{#1}

\makeatletter
\newcommand*{\addFileDependency}[1]{
  \typeout{(#1)}
  \@addtofilelist{#1}
  \IfFileExists{#1}{}{\typeout{No file #1.}}
}
\makeatother

\usepackage{soul}

\makeatletter
\def\maketitle{
\@author@finish
\title@column\titleblock@produce
\suppressfloats[t]}
\makeatother

\makeatletter
\newsavebox{\@brx}
\newcommand{\llangle}[1][]{\savebox{\@brx}{\(\m@th{#1\langle}\)}%
  \mathopen{\copy\@brx\mkern2mu\kern-0.9\wd\@brx\usebox{\@brx}}}
\newcommand{\rrangle}[1][]{\savebox{\@brx}{\(\m@th{#1\rangle}\)}%
  \mathclose{\copy\@brx\mkern2mu\kern-0.9\wd\@brx\usebox{\@brx}}}
\newcommand{\vvert}[1][]{\savebox{\@brx}{\(\m@th{#1\vert}\)}%
  \mathopen{\copy\@brx\mkern2mu\kern-0.9\wd\@brx\usebox{\@brx}}}
\makeatother

\newcommand{\kket}[1]{{\vvert #1 \rrangle}}
\newcommand{\bbra}[1]{{\llangle #1 \vvert}}
\newcommand{\bbrakket}[2]{{\llangle #1 \vvert #2 \rrangle}}

\newcommand{\opone}[0]{\hat{\one}}
\newcommand{\opchiq}[0]{\hat{\chi}_\bq}
\newcommand{\opepsq}[0]{\hat{\epsilon}_\bq}
\newcommand{\opvq}[0]{\hat{v}^{\rm b}_\bq}
\newcommand{\opWq}[0]{\hat{W}_\bq}

\begin{document}

\title{Unifying dynamical electron-electron and electron-phonon interactions\\from first principles}

\author{Jae-Mo Lihm\,\orcidlink{0000-0003-0900-0405}}
\affiliation{Center for Computational Quantum Physics, Flatiron Institute, 162 5th Avenue, New York, NY 10010, USA}
\affiliation{European Theoretical Spectroscopy Facility and Institute of Condensed Matter and Nanosciences, Université catholique de Louvain, Chemin des Étoiles 8, B-1348 Louvain-la-Neuve, Belgium}

\author{Malte R\"osner\,\orcidlink{0000-0002-6199-2176}}
\affiliation{Bielefeld University, Universitätsstr. 25, 33615 Bielefeld, Germany}
\affiliation{Institute for Molecules and Materials, Radboud University, 6525 AJ Nijmegen, The Netherlands}

\author{Cyrus E. Dreyer\,\orcidlink{0000-0003-0035-7520}}
\affiliation{Department of Physics and Astronomy, Stony Brook University, Stony Brook, New York, 11794-3800, USA}
\affiliation{Center for Computational Quantum Physics, Flatiron Institute, 162 5th Avenue, New York, NY 10010, USA}

\author{Massimiliano Stengel\,\orcidlink{0000-0003-4175-3888}}
\affiliation{Institut de Ciencia de Materials de Barcelona (ICMAB-CSIC), Campus UAB, 08193 Bellaterra, Spain}
\affiliation{ICREA-Instituci\'o Catalana de Recerca i Estudis Avançats, 08010 Barcelona, Spain}

\author{Samuel Ponc\'e\,\orcidlink{0000-0003-1159-8389}}
\affiliation{European Theoretical Spectroscopy Facility and Institute of Condensed Matter and Nanosciences, Université catholique de Louvain, Chemin des Étoiles 8, B-1348 Louvain-la-Neuve, Belgium}
\affiliation{WEL Research Institute, avenue Pasteur 6, 1300 Wavre, Belgium}

\date{October 29, 2026}

\begin{abstract}
Electron-phonon interactions in materials with free carriers are dynamically screened by low-energy electronic excitations, yet first-principles calculations almost always treat this screening in the adiabatic approximation.
A fully dynamical treatment of metals and doped semiconductors within time-dependent density functional perturbation theory (TDDFPT) is expensive and numerically delicate.
Here, we introduce DFPT+A downfolding (``A'' for active space), which starts from the static density functional perturbation theory (DFPT) response of a computationally convenient base system and adds the dynamical screening of a low-energy active space, described by Wannier functions.
The required statically screened electron-electron interaction is computed using DFPT directly in the basis of Wannier-function pairs and truncated in a computationally efficient and systematically improvable way.
The method handles entangled bands and accurately reproduces TDDFPT at a fraction of its cost.
We use it to compute the total dynamical electron-electron interaction, including direct Coulomb and phonon-mediated interactions with dynamically screened electron-electron and electron-phonon couplings, all on equal footing.
In hole-doped monolayer hBN, the hybridization of the carrier plasmon with the longitudinal optical phonon makes the low-energy retarded interaction nonmonotonic in the carrier density, an effect missed by the adiabatic approximation.
In metallic monolayer TaS$_2$, nonadiabatic effects are weak and occur only due to the intraband electron-hole continuum and the nesting of nearly parallel Fermi velocities.
We also validate the widely used screened-screened approximation for the phonon spectral function.
This work opens the way to first-principles studies of spectral functions, transport, and superconductivity with dynamically screened interactions.
\end{abstract}

\maketitle

\section{Introduction}

Phonons and electron-phonon (e-ph) interactions play a central role in condensed matter physics, governing phenomena ranging from spectroscopic properties to transport and superconductivity~\cite{Giustino2017}.
Density functional perturbation theory (DFPT) is the standard tool for computing phonon dispersions and e-ph coupling in real materials~\cite{Baroni2001}.
Its combination with Wannier function (WF) interpolation~\cite{Giustino2007, Marzari2012} significantly reduces the cost of evaluating these quantities on dense momentum grids.
This method has enabled the \textit{ab initio} study of e-ph physics and its impact on mobility, superconductivity, and many other properties from first principles~\cite{Ponce2020Review,Franchini2021,Boeri2022,Dai2025}.

Phonon frequencies and e-ph interactions are generally studied within the Born--Oppenheimer approximation, which assumes that electrons respond instantaneously to atomic displacements.
However, in systems with free carriers, such as metals or doped semiconductors, this approximation may break down due to the presence of low-energy electronic excitations with characteristic energy scales comparable to the phonon frequency~\cite{Lazzeri2006, Pisana2007, Bonini2007}.
In this regime, the nonadiabatic~\footnote{In this work, we use the term \textit{``nonadiabatic''} to refer to effects beyond the scalar-potential adiabatic approximation~\cite{Ponce2026}, which contain \textit{dynamical} corrections, with Berry connection and curvature as the low-frequency limit.} nature of phonons and retardation effects in their coupling with electrons become important~\cite{Calandra2010,Berges2023,Stefanucci2025}.
Such nonadiabatic effects manifest as finite phonon linewidths and frequency-dependent screening of the e-ph coupling vertex.
While the nonadiabatic phonon spectral functions and linewidths are being actively studied~\cite{Calandra2010, Berges2023, Guandalini2025b, Park2025PRB, Erhardt2026}, their effect on the e-ph coupling and interaction with electronic excitations has been less explored~\cite{Marchese2024, Fachin2026}.
Moreover, \textit{ab initio} calculations of the nonadiabatic phonon spectral function~\cite{Calandra2010, Berges2023} often use approximations that neglect the dynamical screening of the e-ph coupling.
A reference \textit{ab initio} calculation without this approximation is needed to assess its accuracy in real materials.
While in this work we will exclusively deal with nonmagnetic systems, note that similar issues occur in the context of coupled magnon-phonon dynamics, where spins must be treated as slow variables, on the same footing as nuclei~\cite{Bonini2023,Ren2024,Royo2025}.

A key challenge in computing dynamically screened e-ph interactions from first principles is to solve the self-consistent screening problem with a time-dependent perturbation.
Since nonadiabatic effects arise from electronic screening, one needs to consider both the e-ph interaction and the electron-electron (e-e) interaction on equal footing.
Recent studies of this problem described the electronic screening with simplified models, such as the plasmon-pole or the van der Waals dielectric models~\cite{Macheda2024b, Lihm2024PlPh, Krsnik2024}, rather than computing it fully from first principles.
Such a first-principles treatment is formally provided by a time-dependent generalization of DFPT (TDDFPT), which can address this problem of frequency-dependent screening from free carriers~\cite{Motornyi2018, Motornyi2020}.
However, it is rarely applied to phonons due to both the computational cost and numerical instability under low-frequency resonances, with ongoing research to overcome these challenges~\cite{Royo2025}.

Here, we develop an accurate and efficient framework to compute dynamical, nonadiabatic e-ph interactions from first principles, at the accuracy of TDDFPT but at a fraction of the computational cost.
We use a downfolding approach, which we term \textit{DFPT+A downfolding} (``A'' for active space), to capture the dynamical screening effects from a low-energy active subspace, described by WFs, while treating the high-energy states at the static level using static DFPT.
Since only the active subspace is treated dynamically, we can efficiently calculate the dynamical screening and the resulting nonadiabatic phonon renormalization.
Within the rigid-band approximation, the exact TDDFPT result is recovered by converging the size of the active subspace and the range of the e-e interaction.
We apply this framework to doped hexagonal BN (hBN) and metallic TaS$_2$ monolayers.
In hole-doped hBN, we show that plasmon-phonon hybridization leads to a nonmonotonic change of the low-energy e-e interaction strength with carrier density, an effect driven by the frequency-dependent screening of the e-ph interaction.
In metallic TaS$_2$, we show that the dynamical screening effect is weak, validating the ``screened-screened'' approximation~\cite{Calandra2010, Berges2023} for the phonon spectral function.

Overall, this work makes four key contributions.
First, the DFPT+A downfolding approach provides an affordable framework for calculating nonadiabatic e-ph coupling and phonon spectral functions.
The WF-based representation of the dynamical screening allows for a controlled, rapidly converging truncation of the e-e interaction that works for both isolated and entangled bands.
Second, we compute the total dynamical e-e interaction and find this quantity to be the fundamental building block for understanding the nonadiabatic e-ph coupling and phonon spectral functions.
We show that the e-ph vertex alone is not sufficient to describe the actual interaction, as the e-ph vertex renormalization entails a strong change of the phonon spectral function as well as a sizable direct e-e interaction.
Third, comparing the total interaction with its adiabatic counterpart provides a direct numerical test of the adiabatic approximation for the e-ph interaction in real materials, which is widely used in first-principles calculations~\cite{Giustino2017, Ponce2020Review} without systematic validation.
Fourth, we identify the dominant microscopic origin of the nonadiabatic effects in the two materials: the long-range Fr\"ohlich coupling of the longitudinal optical (LO) phonon to the carrier plasmon in doped hBN, and the coupling to intraband and nested electron-hole excitations in TaS$_2$.
Beyond these examples, the framework opens the way for studying the effect of dynamical screening on electronic spectral functions, superconductivity, and carrier transport in real materials with realistic doping and temperatures.

\section{Dynamical screening from first principles} \label{sec:theory_background}

We begin by introducing the concept of time-dependent responses in the context of dynamical screening, nonadiabatic interatomic force constants (IFCs) and the associated e-ph interactions.
We then review how they are computed with DFPT and its time-dependent extension, and why the latter is costly for metals and doped semiconductors.
While we use the IFC as a concrete example to introduce the key concepts, the formalism itself is general and can be applied to any perturbation, such as external electric and magnetic fields, or a charge density perturbation.

\subsection{Dynamical interatomic force constants}

The dynamical IFCs are the forces induced by time-periodic atomic displacements with frequency $\omega$ and wavevector $\bq$.
We denote the displacement of atom $\kappa$ along the Cartesian direction $\alpha$ with a composite index $a = (\kappa, \alpha)$.
Such a displacement induces a ``bare'' perturbation to the Kohn--Sham Hamiltonian
\begin{equation}
    V^{\rm b}_{a\bq}(\mathbf{r})
    = \sum_{\bR} e^{i\bq \cdot \bR} \, \frac{\partial V^{\rm ion}(\mathbf{r})}{\partial \tau_{a\bR}} \,,
\end{equation}
where $V^{\rm ion}$ is the ionic potential and $\tau_{a\bR}$ is the $\alpha$-component of the position of atom $\kappa$ in unit cell $\bR$.
Equivalently, $V^{\rm b}_{a\bq} = \partial V^{\rm ion} / \partial \tau_{a\bq}$, where $\partial / \partial \tau_{a\bq} \equiv \sum_{\bR} e^{i\bq \cdot \bR} \, \partial / \partial \tau_{a\bR}$ is the derivative with respect to the amplitude $\tau_{a\bq}$ of the periodic displacement pattern $\Delta\tau_{a\bR} = \tau_{a\bq} \, e^{i\bq \cdot \bR}$.
This potential is combined with the self-consistent screening potential to give the screened first-order potential $V^{\rm s}_{a\bq}(\mathbf{r},\omega)$.
To gain an intuitive understanding, we momentarily assume that the perturbation is local and can be represented with a single spatial coordinate $\br$.
Such a potential perturbation induces a ``screened'' charge-density response given by
\begin{multline} \label{eq:rpa_n_s_r}
    n^{\rm s}_{a\bq}(\br, \omega)
    = \intbk \sum_{mn} \frac{f_\nk - f_\mkq}{\omega + \veps_\nk - \veps_\mkq} \, g^{\rm s}_{a\bq,\mnk}(\omega) 
    \\
    \times \psi_\mkq(\br) \, \psi^*_\nk(\br) \,,
\end{multline}
where $f_\nk$ is the Fermi--Dirac occupation factor of band $n$ at wavevector $\bk$, $\veps_\nk$ is the Kohn--Sham eigenvalue, and $g^{\rm s}_{a\bq,\mnk}(\omega) \equiv \mel{\psi_\mkq}{V^{\rm s}_{a\bq}(\omega)}{\psi_\nk}$ is the screened e-ph matrix element involving Kohn--Sham states $\psi_\nk$ and $\psi_\mkq$.
Throughout, we set $\hbar = 1$, so that frequencies are expressed in energy units.
We do not assume spin degeneracy; spinless calculations should have an additional factor of 2.
We denote the Brillouin-zone (BZ) average by
\begin{equation} \label{eq:bz_average}
    \intbk f(\bk)
    \equiv \nint \frac{\dd\bk}{V^{\rm BZ}} f(\bk)
    \approx \frac{1}{N_\bk^{\rm f}} \sum_\bk f(\bk) \,,
\end{equation}
and analogously $\intbq$, where $V^{\rm BZ}$ is the volume of the BZ.
In practice, we evaluate it as a sum over a fine grid of $N_\bk^{\rm f}$ points, as in the second expression.
The induced charge density of \cref{eq:rpa_n_s_r} then gives rise to the local screening potential via the bare e-e interaction kernel $v^{\rm b}_\bq(\br, \br')$ as
\begin{equation} \label{eq:rpa_V_ind_r}
    V^{\rm ind}_{a\bq}(\br, \omega)
    = \nsint{\rm u.c.} \dd\br' \, v^{\rm b}_\bq(\br, \br') \, n^{\rm s}_{a\bq}(\br', \omega) \,.
\end{equation}
Here, u.c.\ denotes the integral over the unit cell.
The kernel $v^{\rm b}_\bq$, which describes the bare e-e interaction within DFPT, is given by the second-order derivative of the Hartree plus exchange-correlation (Hxc) energy at the ground-state density:
\begin{equation}
    v^{\rm b}_\bq(\br, \br')
    \equiv \frac{\partial^2 E^{\rm Hxc}}{\partial n_{-\bq}(\br) \partial n_\bq(\br')}
    \,.
\end{equation}
Here, $n_\bq(\br)$ denotes a modulation of the electron density with crystal momentum $\bq$, $n_\bq(\br + \bR) = e^{i\bq \cdot \bR} n_\bq(\br)$, such as the induced density $n^{\rm s}_{a\bq}(\br, \omega)$ of \cref{eq:rpa_n_s_r}.
If we consider only the Hartree contribution, $v^{\rm b}_\bq$ reduces to the bare Coulomb interaction, and we obtain the random phase approximation (RPA).
The total first-order potential then reads
\begin{equation} \label{eq:rpa_dV_r}
    V^{\rm s}_{a\bq}(\br, \omega)
    = V^{\rm b}_{a\bq}(\br)
    + V^{\rm ind}_{a\bq}(\br, \omega) \,.
\end{equation}
The screened dynamical force constant between two atomic displacements $a$ and $a'$ is given by~\cite{Berges2023}
\begin{multline} \label{eq:dynscreen}
    C^{\rm s}_{aa' \bq}(\omega)
    = C^{\rm b}_{aa' \bq}
    + \intbk \sum_{mn} \frac{f_\nk - f_\mkq}{\omega + \veps_\nk - \veps_\mkq}
    \\
    \times g^{{\rm b}*}_{a\bq,\mnk} \, g^{\rm s}_{a'\bq,\mnk}(\omega) \,,
\end{multline}
where $g^{\rm b}_{a\bq,\mnk} \equiv \mel{\psi_\mkq}{V^{\rm b}_{a\bq}}{\psi_\nk}$.
Here, $C^{\rm b}_{aa' \bq}$ is a frequency-independent bare force constant term that does not include electronic screening:
\begin{equation} \label{eq:Cb_def}
    C^{\rm b}_{aa' \bq}
    = C^{\rm ion}_{aa' \bq}
    + \nsint{\rm u.c.} \dd\br \, n(\br) \, \frac{\partial^2 V^{\rm ion}(\br)}{\partial \tau_{a-\bq} \partial \tau_{a'\bq}} \,,
\end{equation}
where the first term is the second derivative of the ion-ion interaction energy, and $n(\br)$ is the ground-state electron density.

\begin{table}[t]
\begin{tabular}{l|ccc}
    Quantity & Bare & Partially screened & Fully screened
    \\ \hline
    e-ph interaction & $\kket{V^{\rm b}_{a\bq}}$ & $\kket{V^{\rm p}_{a\bq}}$ & $\kket{V^{\rm s}_{a\bq}(\omega)}$
    \\
    e-e interaction & $\hat{v}^{\rm b}_\bq$ & $\opWq^{\rm p}$ & $\opWq^{\rm s}(\omega)$
    \\
    Force constants & $C^{\rm b}_{aa' \bq}$ & $C^{\rm p}_{aa' \bq}$ & $C^{\rm s}_{aa' \bq}(\omega)$
    \\
    Susceptibility & $\hat{\chi}_{\bq}^{\rm b}(\omega)$ & $\hat{\chi}^{\rm p}_\bq$ & $\hat{\chi}^{\rm s}_\bq(\omega)$
    \\
    Dielectric function & $\opone$ & $\opepsq^{\rm p}$ & $\opepsq^{\rm s}(\omega)$
\end{tabular}
\caption{
    Summary of the quantities used in the \textit{ab initio} downfolding framework presented in this work.
    Double brackets denote operators that act on the wavefunctions and a hat denotes two-particle operators.
    In the constrained random phase approximation (cRPA) literature, the partially-screened e-e interaction is usually denoted by $U$~\cite{Aryasetiawan2004}.
    We denote it by $\opWq^{\rm p}$ for consistency with the other partially-screened quantities, and use $U$ only for the local interaction (\cref{sec:theory_Uloc}).
}
\label{tab:notation}
\end{table}

Many \textit{ab initio} calculations use nonlocal pseudopotentials to represent the electron-ion interaction.
In that case, the potential perturbation is nonlocal and is a function of two spatial coordinates $\br$ and $\br'$.
We generalize \crefrange{eq:rpa_n_s_r}{eq:Cb_def} to treat arbitrary perturbations to the single-particle Hamiltonian.
We call the vector space of single-particle operators the two-particle space and denote its elements using double brackets.
Two useful bases to describe the two-particle space are the position and wavefunction-pair bases:
\begin{subequations} \label{eq:rpa_pair_basis}
\begin{align}
    \kket{\br\brp;\bq} &\equiv \sum_\bR e^{i\bq\cdot\bR} \ket{\br+\bR} \bra{\brp+\bR} \,, \\
    \kket{\Psi_\mnkq} &\equiv \ket{\psi_\mkq} \bra{\psi_\nk} \,.
\end{align}
\end{subequations}
The inner product in two-particle space is defined as
\begin{subequations}
\begin{equation}
    \bbrakket{A}{B} \equiv \mathrm{Tr}^{\rm cell} A^\dagger B \,,
\end{equation}
where $\mathrm{Tr}^{\rm cell}$ is the trace per unit cell, and Bloch states are normalized to one per unit cell.
For example, we have
\begin{align}
    \bbrakket{\br\brp;\bq}{\Psi_\mnkq} &= \psi_\mkq(\br) \psi^*_\nk(\brp)
    \,, \\
    \bbrakket{\Psi_\mnkq}{V_{a\bq}} &= \mel{\psi_\mkq}{V_{a\bq}}{\psi_\nk}
    \,.
\end{align}
\end{subequations}
With this inner product, the position pairs are orthonormal, $\bbrakket{\br\brp;\bq}{\br''\br''';\bq} = \delta(\br-\br'')\,\delta(\brp-\br''')$.
They are complete in the space of operators with crystal momentum $\bq$, $\nsint{\rm u.c.} \dd\br \nint \dd\brp \, \kket{\br\brp;\bq} \bbra{\br\brp;\bq} = \opone$, where $\opone$ is the identity operator in the two-particle space.

\Cref{eq:rpa_n_s_r,eq:rpa_V_ind_r} relate the electron density matrix response and the induced potential, both of which can be viewed as single-particle operators, that is, vectors in the two-particle space.
Relations mapping one to the other are linear operators in the two-particle space, which we denote with a hat.
\Cref{eq:rpa_n_s_r} can then be rewritten as
\begin{equation} \label{eq:rpa_n_s_op}
    \kket{n^{\rm s}_{a\bq}(\omega)}
    = \hat{\chi}_{\bq}^{\rm b}(\omega) \, \kket{V^\mathrm{s}_{a\bq}(\omega)} \,,
\end{equation}
where
\begin{equation} \label{eq:rpa_chi_b_def}
    \hat{\chi}_{\bq}^{\rm b}(\omega)
    = \intbk \sum_{mn} \frac{f_\nk - f_\mkq}{\omega + \veps_\nk - \veps_\mkq}
    \kket{\Psi_\mnkq} \bbra{\Psi_\mnkq}
\end{equation}
is the bare susceptibility written in the band pair basis, in which it is diagonal.
\Cref{eq:rpa_V_ind_r} is rewritten as
\begin{equation} \label{eq:rpa_Vind_op}
    \kket{V^{\rm ind}_{a\bq}(\omega)}
    = \hat{v}^{\rm b}_\bq \, \kket{n^{\rm s}_{a\bq}(\omega)} \,,
\end{equation}
where the e-e interaction kernel is written in the position basis as
\begin{equation} \label{eq:rpa_v_def}
    \hat{v}^{\rm b}_\bq
    = \nsint{\rm u.c.} \dd\br \, \dd\brp \, \kket{\br\br;\bq} \, v^{\rm b}_\bq(\br, \brp) \, \bbra{\brp\brp;\bq} \,.
\end{equation}
With these definitions, \cref{eq:rpa_dV_r} reads
\begin{equation} \label{eq:rpa_H_s}
    \kket{V^{\rm s}_{a\bq}(\omega)}
    = \kket{V^{\rm b}_{a\bq}}
    + \hat{v}^{\rm b}_\bq \, \hat{\chi}_{\bq}^{\rm b}(\omega) \, \kket{V^{\rm s}_{a\bq}(\omega)} \,.
\end{equation}
The screened dynamical IFC of \cref{eq:dynscreen} is then expressed as
\begin{align} \label{eq:rpa_C}
    C^{\rm s}_{aa' \bq}(\omega)
    &= C^{\rm b}_{aa' \bq}
    + \bbra{V^{\rm b}_{a\bq}} \, \opchiq^{\rm b}(\omega) \, \kket{V^{\rm s}_{a'\bq}(\omega)}
    \nnnl
    &= C^{\rm b}_{aa' \bq}
    + \bbra{V^{\rm b}_{a\bq}} \, \hat{\chi}^{\rm s}_\bq(\omega) \, \kket{V^{\rm b}_{a'\bq}}
    \,,
\end{align}
where
\begin{equation}
    \hat{\chi}^{\rm s}_\bq(\omega)
    = \opchiq^{\rm b}(\omega) \,
    \bigl[ \opone - \hat{v}^{\rm b}_\bq \, \opchiq^{\rm b}(\omega) \bigr]^{-1}
\end{equation}
is the screened susceptibility.
Correspondingly, the fully-screened dielectric function and e-e interaction are $\opepsq^{\rm s}(\omega) \equiv \opone - \opvq \, \opchiq^{\rm b}(\omega)$ and $\opWq^{\rm s}(\omega) \equiv [\opepsq^{\rm s}(\omega)]^{-1} \opvq$ (\cref{tab:notation}).
Note that the second line of \cref{eq:rpa_C} uses two bare potential perturbations, as the screening is fully contained in the susceptibility.

The presented formalism is beyond the RPA because we include the exchange-correlation kernel in the density response.
It corresponds to linear-response time-dependent density functional theory (TDDFT) with a frequency-independent (adiabatic) exchange-correlation functional.

\subsection{Density functional perturbation theory}

The above formulas cannot be implemented directly in a first-principles calculation because they require a double summation over a complete set of single-particle eigenstates.
This issue is circumvented in DFPT by rewriting the summation as a linear equation called the \textit{Sternheimer equation}~\cite{Sternheimer1954} that can be solved iteratively.
This approach transforms the double summation into a single sum over occupied states only.
This formulation generalizes to time-dependent DFPT (TDDFPT), which allows for an efficient computation of finite-frequency response properties~\cite{Motornyi2018, Motornyi2020}.

For nonmagnetic insulators with time-reversal symmetry, the density matrix response can be rewritten as~\cite{Cao2018, Gorni2018}
\begin{multline} \label{eq:fc_dm}
    \kket{n^{\rm s}_{a\bq}(\omega)}
    = \intbk \sum_{n}^{\rm occ} \Bigl[ \ket{\psi_{\nk,a\bq}(\omega)} \bra{\psi_\nk}
    \\
    +\ket{\mcT^{-1} \psi_\nk} \bra{\mcT^{-1} \psi_{\nk,a\bq}(-\omega)} \Bigr] \,,
\end{multline}
where $\ket{\psi_{\nk,a\bq}(\omega)}$ is the first-order wavefunction response at frequency $\omega$, given below by \cref{eq:fc_sternheimer}, and $\mcT$ is the time-reversal operator (see \cref{app:tddfpt}).
The corresponding charge density is
\begin{equation} \label{eq:fc_density}
    n^{\rm s}_{a\bq}(\br, \omega)
    = \intbk \sum_{n}^{\rm occ} \psi^*_\nk(\br)
    \bigl[ \psi_{\nk,a\bq}(\br, \omega) + \psi_{\nk,a\bq}(\br, -\omega) \bigr] \,.
\end{equation}
The second term, which involves the response at $-\omega$, takes the place of the complex-conjugate term of the static DFPT expression: for $\omega = 0$, \cref{eq:fc_density} reduces to the familiar form $2 \, \intbk \sum_{n}^{\rm occ} \psi^*_\nk(\br) \, \psi_{\nk,a\bq}(\br, 0)$~\cite{Baroni2001}.
The wavefunction response $\ket{\psi_{\nk,a\bq}(\omega)}$ is the solution of the Sternheimer equation
\begin{equation} \label{eq:fc_sternheimer}
    \bigl( H^0_\bkq - \veps_\nk - \omega \bigr) \ket{\psi_{\nk,a\bq}(\omega)}
    = - P^{\rm c}_{\bkq} V_{a\bq}^{\rm s}(\omega) \ket{\psi_\nk} \,,
\end{equation}
where $H^0_\bkq$ is the ground-state Kohn--Sham Hamiltonian at wavevector $\bkq$, and
\begin{equation}
    P^{\rm c}_{\bkq} \equiv \one - \sum_{m}^{\rm occ} \ket{\psi_\mkq} \bra{\psi_\mkq}
\end{equation}
is the projector onto the complementary (unoccupied) states.
Note that for $\omega \neq 0$, we need to solve two Sternheimer equations, one for $\omega$ and one for $-\omega$~\cite{Motornyi2018, Motornyi2020}.
For metals or magnetic systems, the Sternheimer equation is further modified~\cite{Baroni2001, Cao2018, Gorni2018, Urru2019, Ricci2019, Gonze2024, Zabalo2024}; we detail the formalism in \cref{app:tddfpt}.

Finite-frequency TDDFPT is a viable strategy for computing dynamical IFCs and other linear-response properties from first principles.
It has been successfully applied to study finite-frequency dielectric functions, magnons, and nonadiabatic phonons~\cite{Motornyi2018, Cao2018, Gorni2018, Motornyi2020, Royo2025}.
However, there are good reasons to seek further improvements.
The potential $V_{a\bq}^{\rm s}(\omega)$ in the Sternheimer equation, \cref{eq:fc_sternheimer}, is a frequency-dependent object.
Thus, the Sternheimer equation must be solved self-consistently for each frequency.
The situation is further complicated by the fact that for metals, there are low-energy electron-hole pairs in the frequency range of interest for phonons.
Resolving these resonances requires a fine sampling of the Brillouin zone, as well as a large number of iterations to converge, making the self-consistent solution of the Sternheimer equation computationally expensive.

An alternative formulation of TDDFPT, the Liouville--Lanczos approach, computes the response at all frequencies simultaneously, avoiding frequency-by-frequency calculations~\cite{Walker2006, Rocca2008, Timrov2013, Motornyi2018, Motornyi2020}.
However, it still requires a fine sampling of the Brillouin zone.
Also, when low-energy resonances are present, converging the response functions requires a large number of Lanczos iterations, which can be several thousands or more~\cite{Gorni2018}.

\section{DFPT+A downfolding} \label{sec:theory_downfolding}

To circumvent these issues, we use the downfolding technique.
The central idea of downfolding is to partition the electronic susceptibility into an \textit{``active''} contribution that captures the dynamical screening effects from the low-energy electronic states and a high-energy \textit{``rest''} contribution treated at the static level.
Downfolding has been used to compute screened e-e interactions within the constrained RPA (cRPA)~\cite{Aryasetiawan2004} and static and dynamical IFCs with constrained DFPT (cDFPT)~\cite{Nomura2015, Berges2020, vanLoon2021, Berges2023}.

Here, we develop the DFPT+A downfolding approach, a formulation built on standard DFPT.
First, we separate the susceptibility into the static DFPT response and a dynamical correction from the active space (\cref{sec:theory_downfolding_approach}).
This separation is equivalent to cDFPT, but requires only standard DFPT calculations.
Second, we show how the dynamical screening then reduces to a Dyson equation for small matrices in the space of active-space transitions (\cref{sec:theory_active_screening}), which we represent in the basis of WF pairs (\cref{sec:theory_wannier}).
Third, we truncate the partially-screened e-e interaction in this basis (\cref{sec:theory_truncation}) and compute its matrix elements directly with DFPT (\cref{sec:theory_ee_DFPT}).
We then describe the interpolation of the e-e interaction (\cref{sec:theory_itp}) and validate the framework on monolayer hBN and TaS$_2$ (\cref{sec:theory_demo}).

\subsection{The downfolding approach} \label{sec:theory_downfolding_approach}

\begin{figure}[tb]
\centering
\includegraphics[width=0.99\linewidth]{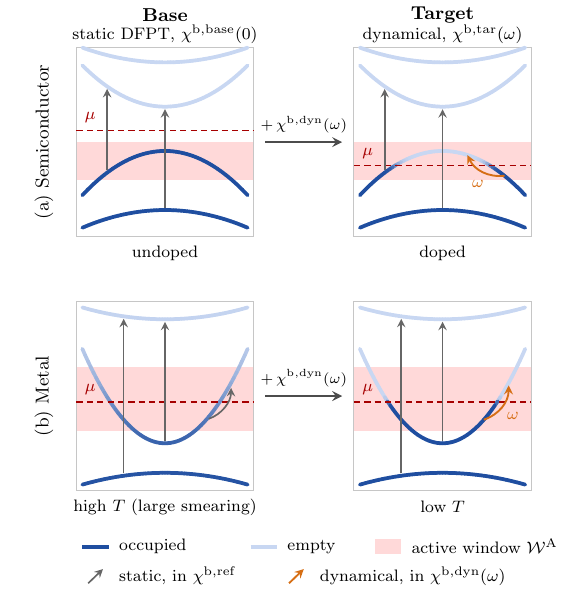}
\caption{
    Schematic of DFPT+A downfolding for (a) a semiconductor, with the insulating case as the base system and the doped system as the target, and (b) a metal, with a large electronic smearing as the base system and a low temperature as the target.
    The color of the bands indicates their occupation, and the shaded region indicates the active window $\mathcal{W}^{\rm A}$.
    The static response of the base system, including all bands, is computed with DFPT.
    Gray arrows show examples of the static transitions it contains, which are kept in the target system.
    Contrary to the cRPA~\cite{Aryasetiawan2004}, it includes all transitions, including those within the active bands, but only the static part of the response is kept.
    The change of the occupations of the active bands and the dynamical response to a perturbation of frequency $\omega$, which arises from low-energy transitions within the active bands (orange arrows), are added through the dynamical correction $\opchiq^{\rm b,dyn}(\omega)$ computed in the WF basis.
}
\label{fig:downfolding_schematic}
\end{figure}

Nonadiabatic effects originate from a small number of low-energy electronic transitions near the Fermi level, while the vast majority of transitions have much higher energies compared to the phonon energy scale.
The idea of downfolding is to treat the former explicitly and dynamically in a reduced basis, and to include the latter through static, partially-screened interactions computed once and for all.
This separation makes the dynamical problem small enough to be solved at many frequencies and on dense $\bk$ grids.

The central quantity of the screening problem is the dynamical susceptibility $\opchiq^{\rm b}(\omega)$ [\cref{eq:rpa_chi_b_def}].
We need it at the physical conditions of interest, namely the temperature, carrier concentration, and electronic smearing (and corresponding $\bk$-point sampling) at which the observables are to be predicted.
We call these the \textit{target} parameters, and all quantities computed at these parameters are target quantities.
We denote the target susceptibility by $\opchiq^{\rm b,tar}(\omega)$; see also the discussion of the choice of parameters at which the screening is evaluated in Ref.~\cite{Berges2023}.
It is costly to compute directly: its frequency dependence requires a TDDFPT calculation at each frequency, and at low temperature or low doping, resolving excitations near the Fermi level requires a fine $\bk$ grid.
We therefore introduce a \textit{base} system, for which a standard DFPT calculation on a coarse $\bk$ grid gives the static susceptibility $\opchiq^{\rm b,base}(0)$.
For the simulation of a doped semiconductor, we choose the base system to be the undoped crystal [\Fig[a]{fig:downfolding_schematic}].
For metals, we choose the base to be the same metal with a large electronic smearing [\Fig[b]{fig:downfolding_schematic}].
We use the rigid-band approximation,
\begin{equation} \label{eq:rigid_band}
    \veps^{\rm tar}_\nk \approx \veps^{\rm base}_\nk \equiv \veps_\nk \,, \quad
    \ket{\psi^{\rm tar}_\nk} \approx \ket{\psi^{\rm base}_\nk} \equiv \ket{\psi_\nk} \,,
\end{equation}
for a change of either the doping or the temperature, so that the target and the base susceptibilities differ only in the occupations, $f^{\rm tar}_\nk$ and $f^{\rm base}_\nk$, and in their frequency dependence.

In both cases, the target has low-energy excitations near the Fermi level that are absent or smeared out in the base system, as shown in the right panels of \cref{fig:downfolding_schematic}.
They change the target susceptibility both statically, through the change of the occupations, and dynamically, through low-energy resonances that are responsible for nonadiabatic effects.
Since both effects arise mainly from transitions between states near the Fermi level, we only treat these transitions explicitly and approximate the remaining ones by their static base values.

The transitions near the Fermi level can be singled out by defining an active space of low-energy states, as commonly done in cRPA~\cite{Aryasetiawan2004}.
We first briefly review the cRPA method, whose approximations our approach shares.
The cRPA method separates the low-energy active states from the high-energy states.
In this work, we define the active space by a fixed energy window around the Fermi level, $\mc{W}^{\rm A} = [\veps^{\rm min}, \veps^{\rm max}]$.
The window must be wide enough that the states outside the window are either empty in both the target and base systems or fully occupied in both.
Restricting the state sum of the bare susceptibility [\cref{eq:rpa_chi_b_def}] to the active states defines the susceptibility within the active window,
\begin{equation} \label{eq:rpa_chib_active}
    \opchiq^{\rm b,A}(\omega)
    = \intbk \sum_{mn}^{\rm A} \kket{\Psi_\mnkq} \frac{f_\nk - f_\mkq}{\omega + \veps_\nk - \veps_\mkq} \bbra{\Psi_\mnkq} \,,
\end{equation}
where $\sum\limits^{\rm A}$ denotes the summation over pairs of states with $\veps_\mkq, \veps_\nk \in \mc{W}^{\rm A}$.
Since the window is defined by energy rather than by band index, a band may enter and leave the active space across the Brillouin zone.
The remaining transitions, which involve at least one state outside the active window, define the rest susceptibility,
\begin{equation} \label{eq:chib_rest_def}
    \opchiq^{\rm b,R}(\omega) \equiv \opchiq^{\rm b}(\omega) - \opchiq^{\rm b,A}(\omega) \,.
\end{equation}
In cRPA and cDFPT, the removal of the active-space transitions amounts to constraining the elements of the density matrix response that correspond to the active-space transitions to zero.

If the window is wide compared with the phonon energies, every transition in $\opchiq^{\rm b,R}(\omega)$ involves a state far from the Fermi level.
The rest susceptibility then has no resonances on the phonon energy scale.
Since phonon energies range from meV to a few hundred meV, a window containing a few bands around the Fermi level is usually sufficient.
We therefore approximate the rest susceptibility by neglecting its frequency dependence,
\begin{equation} \label{eq:chib_approx1}
    \opchiq^{\rm b,tar,R}(\omega) \approx \opchiq^{\rm b,tar,R}(0) \,.
\end{equation}
The rest susceptibility does change between the base and target systems, due to the change of the occupations of the active states, which affects the transitions between active states and states outside the window.
Yet, since the energy of these transitions is at least the distance from the Fermi level to the window edge, screening from such transitions is small in magnitude and decreases as the window is enlarged.
We thus also neglect the dependence of the rest susceptibility on temperature and doping,
\begin{equation} \label{eq:chib_approx2}
    \opchiq^{\rm b,tar,R}(0) \approx \opchiq^{\rm b,base,R}(0) \,,
\end{equation}
so that it can be computed with static calculations on a coarse $\bk$ grid, with a large electronic smearing for metals.
The full target susceptibility then takes the form~\cite{Aryasetiawan2004},
\begin{equation} \label{eq:chib_cRPA}
    \opchiq^{\rm b,tar}(\omega) \approx \opchiq^{\rm b,base,R}(0) + \opchiq^{\rm b,tar,A}(\omega) \,.
\end{equation}
This is the separation used in cRPA~\cite{Aryasetiawan2004} and cDFPT~\cite{Nomura2015}, and we refer to downfolding based on it as cDFPT+A.

However, for our purpose of studying low-energy nonadiabatic effects, the cDFPT+A separation has practical drawbacks.
First, for metals, $\opchiq^{\rm b,base,R}(0)$ excludes all intraband transitions, so that the static screening is insulating~\cite{Berges2023}, and the active-space screening must restore the metallic screening entirely.
This requires a very accurate treatment of the active-space screening, which limits the efficiency of the calculation.
Second, if the window cuts through bands, transitions move abruptly between the active and rest susceptibilities, which makes the screened interaction hard to interpolate.
Third, if bands near the Fermi level are entangled with higher- or lower-lying states without an energy gap, the subtraction in \cref{eq:chib_rest_def} can produce spurious negative screening~\cite{Miyake2009}, although several remedies have been proposed~\cite{Sasioglu2011, Kaltak2015, Kaltak2025}.
Finally, the rest term cannot be obtained from a standard DFPT calculation, and requires a constrained one~\cite{Nomura2015}.

Here, we propose an alternative approach that avoids these problems.
Using $\opchiq^{\rm b,base,R}(0) = \opchiq^{\rm b,base}(0) - \opchiq^{\rm b,base,A}(0)$ [\cref{eq:chib_rest_def}], we rewrite \cref{eq:chib_cRPA} as
\begin{equation} \label{eq:chib_DFPT}
    \opchiq^{\rm b, tar}(\omega)
    \approx \opchiq^{\rm b,base}(0)
    + \bigl[ \opchiq^{\rm b,tar,A}(\omega) - \opchiq^{\rm b,base,A}(0) \bigr] \,.
\end{equation}
This new separation defines our DFPT+A downfolding approach.
The first term on the right-hand side is the full static susceptibility of the base system.
The second term contains only the difference between the active-space susceptibilities of the target and base systems.
Importantly, \cref{eq:chib_DFPT} contains the same terms as \cref{eq:chib_cRPA}, only grouped differently.
To illustrate, we write both separations generically as the sum of a static \textit{reference} susceptibility and a \textit{dynamical} correction:
\begin{subequations} \label{eq:chib_ref_dyn_all}
\begin{alignat}{2}
    &\opchiq^{\rm b,tar}(\omega) &&
    \nnnl
    &\approx \opchiq^{\rm b,ref} && + \opchiq^{\rm b,dyn}(\omega)
    \label{eq:chib_ref_dyn}
    \\
    &= \opchiq^{\rm b,base,R}(0) && + \opchiq^{\rm b,tar,A}(\omega)
    \label{eq:chib_ref_dyn_cDFPTA}
    \\
    &= \opchiq^{\rm b,base}(0) && + \bigl[ \opchiq^{\rm b,tar,A}(\omega) - \opchiq^{\rm b,base,A}(0) \bigr] \,,
    \label{eq:chib_ref_dyn_DFPTA}
\end{alignat}
\end{subequations}
where \cref{eq:chib_ref_dyn_cDFPTA,eq:chib_ref_dyn_DFPTA} are the cDFPT+A and DFPT+A separations, respectively.
The two differ only in whether the static term $\opchiq^{\rm b,base,A}(0)$ is subtracted from the reference susceptibility or from the dynamical correction.
Since the sums are identical, the two separations formally yield the same fully-screened quantities (\cref{app:cDFPT}).

DFPT+A downfolding has several advantages, addressing the drawbacks of the cDFPT+A method.
First, for metals, it is more robust against approximations to the active-space screening, as the reference susceptibility already contains the metallic screening.
Second, since the reference susceptibility is simply the full static susceptibility of the base system, one has more freedom in the choice of the active space, without having to worry about a window cutting through bands, or special treatment of entangled bands.
The window enters only the dynamical correction, where the contributions of states at the window edge do not pose any difficulty.
Third, the reference susceptibility is computed with standard DFPT, whose implementation is widely available.

The cDFPT+A scheme would be preferable for heavily doped semiconductors beyond the rigid-band approximation, where both the base and target systems are doped, but the long-range e-e interaction still plays a significant role.
In this case, using the DFPT reference means that the metallically screened interaction of the base system needs to be unscreened by the active-space screening, which is numerically delicate.
Starting from the cDFPT reference would not have this issue, as the reference is insulating.
When treating doping using the rigid-band approximation, as we do in this work, the DFPT+A scheme also gives an insulating reference.
The DFPT+A and cDFPT+A schemes coincide for an undoped semiconductor base system whose active window contains only valence (or only conduction) states, as for hBN in this work [\Fig[a]{fig:bands}], since then $\opchiq^{\rm b,base,A}(0) = 0$.

\subsection{Separation of reference and active-space screening} \label{sec:theory_active_screening}

With the susceptibility split into the static reference susceptibility and the dynamical correction [\cref{eq:chib_ref_dyn}], we now rewrite the screening equations [\cref{eq:rpa_H_s,eq:rpa_C}] into a two-step procedure, where we first screen with the reference susceptibility and then with the dynamical correction.
This subsection summarizes the standard procedure in cRPA~\cite{Aryasetiawan2004} and cDFPT~\cite{Nomura2015, Berges2020, vanLoon2021, Berges2023}, but we present it in a generic fashion that is applicable to both cDFPT+A and DFPT+A.

In the first step, we screen the bare perturbation with the reference susceptibility only.
We call the resulting quantities \textit{partially-screened} and denote them with a superscript ``p''.
With the partially-screened dielectric function
\begin{equation}
    \opepsq^{\rm p} \equiv \opone - \opvq \, \opchiq^{\rm b,ref} \,,
\end{equation}
we define the partially-screened perturbation and e-e interaction as
\begin{equation} \label{eq:rpa_Wp_def}
    \kket{V_{a\bq}^{\rm p}} \equiv (\opepsq^{\rm p})^{-1} \kket{V_{a\bq}^{\rm b}} \,,\quad
    \opWq^{\rm p} \equiv (\opepsq^{\rm p})^{-1} \opvq \,.
\end{equation}
Analogously to the screened susceptibility of \cref{sec:theory_background}, we also define the partially-screened susceptibility
\begin{equation} \label{eq:rpa_chip_def}
    \opchiq^{\rm p} \equiv \opchiq^{\rm b,ref} \, (\opepsq^{\rm p})^{-1}
    = \opchiq^{\rm b,ref} \bigl[ \opone - \opvq \, \opchiq^{\rm b,ref} \bigr]^{-1} \,,
\end{equation}
so that $(\opepsq^{\rm p})^{-1} = \opone + \opvq \, \opchiq^{\rm p}$ and $\opWq^{\rm p} = \opvq + \opvq \, \opchiq^{\rm p} \, \opvq$.
Since the reference susceptibility in \cref{eq:chib_ref_dyn} is static, these quantities are also static.
The partially-screened perturbation satisfies
\begin{equation} \label{eq:rpa_Hp}
    \kket{V_{a\bq}^{\rm p}} = \kket{V_{a\bq}^{\rm b}} + \opvq \, \opchiq^{\rm b,ref} \, \kket{V_{a\bq}^{\rm p}} \,,
\end{equation}
which is \cref{eq:rpa_H_s} with $\opchiq^{\rm b}(\omega)$ replaced by $\opchiq^{\rm b,ref}$.
Correspondingly, following \cref{eq:rpa_C}, the partially-screened force constants are
\begin{align} \label{eq:rpa_Cp_def}
    C^{\rm p}_{aa' \bq}
    &\equiv C^{\rm b}_{aa' \bq} + \bbra{V^{\rm b}_{a\bq}} \, \opchiq^{\rm b,ref} \, \kket{V^{\rm p}_{a'\bq}}
    \nnnl
    &= C^{\rm b}_{aa' \bq} + \bbra{V^{\rm b}_{a\bq}} \, \opchiq^{\rm p} \, \kket{V^{\rm b}_{a'\bq}} \,.
\end{align}
In DFPT+A downfolding, $\opchiq^{\rm b,ref}$ is the static susceptibility of the base system, so that \cref{eq:rpa_Hp} is the standard static DFPT equation of the base system, and $V^{\rm p}$, $W^{\rm p}$, and $C^{\rm p}$ are simply the static DFPT quantities of the base system.
In practice, however, the full operator $\opWq^{\rm p}$ is never computed: as we show below, only its matrix elements within the small space of active-space transitions are needed.

In the second step, we screen the partially-screened quantities with the dynamical correction $\opchiq^{\rm b,dyn}(\omega)$.
Substituting $\opchiq^{\rm b,tar}(\omega) = \opchiq^{\rm b,ref} + \opchiq^{\rm b,dyn}(\omega)$ into \cref{eq:rpa_H_s,eq:rpa_C} and absorbing the reference screening into the partially-screened quantities, we obtain~\cite{Aryasetiawan2004, Nomura2015}
\begin{equation} \label{eq:rpa_Hs_downfold}
    \kket{V_{a\bq}^{\rm s}(\omega)}
    = \kket{V_{a\bq}^{\rm p}}
    + \opWq^{\rm p} \, \opchiq^{\rm b,dyn}(\omega) \, \kket{V_{a\bq}^{\rm s}(\omega)} \,,
\end{equation}
whose solution is
\begin{equation} \label{eq:rpa_Hs_downfold_inverse}
    \kket{V_{a\bq}^{\rm s}(\omega)}
    = \bigl[ \opone - \opWq^{\rm p} \, \opchiq^{\rm b,dyn}(\omega) \bigr]^{-1} \, \kket{V_{a\bq}^{\rm p}} \,,
\end{equation}
and the dynamical IFCs read
\begin{equation} \label{eq:rpa_C_downfold}
    C^{\rm s}_{aa'\bq}(\omega)
    = C^{\rm p}_{aa'\bq}
    + \bbra{V^{\rm p}_{a\bq}} \, \opchiq^{\rm b,dyn}(\omega) \, \kket{V^{\rm s}_{a'\bq}(\omega)} \,.
\end{equation}
See \cref{app:downfold_derivation} for the derivation.
These downfolded equations have the same form as the original screening equations [\cref{eq:rpa_H_s,eq:rpa_C}], except that the bare quantities are replaced with the partially-screened ones and the susceptibility with its dynamical correction~\cite{Aryasetiawan2004, Nomura2015}.

Unlike $\opchiq^{\rm b,ref}$, the dynamical correction contains only transitions between active states [\cref{eq:chib_ref_dyn_all}], so it acts within a small subspace of the two-particle space.
We expand it in an orthonormal set of two-particle basis functions $\kket{\Phi_{I\bq}}$ that span these transitions,
\begin{equation} \label{eq:rpa_chib_active_W}
    \opchiq^{\rm b,dyn}(\omega)
    = \sum_{II'} \kket{\Phi_{I\bq}} \, X^{\rm b,dyn}_{II'\bq}(\omega) \, \bbra{\Phi_{I'\bq}} \,,
\end{equation}
with the matrix elements
\begin{equation} \label{eq:rpa_X_elel}
    X^{\rm b,dyn}_{II'\bq}(\omega) \equiv \bbra{\Phi_{I\bq}} \opchiq^{\rm b,dyn}(\omega) \kket{\Phi_{I'\bq}} \,.
\end{equation}
Here, $I$ and $I'$ run over this subspace, whose dimension is much smaller than that of the full two-particle space.
Since $\opchiq^{\rm b,dyn}(\omega)$ acts only within this subspace, the dynamical IFCs involve the partially-screened perturbation and e-e interaction only through their projections onto it.
The required quantities are thus the matrix elements of the partially-screened e-e interaction,
\begin{equation} \label{eq:rpa_Wp_WF}
    W_{II'\bq}^{\rm p} \equiv \bbra{\Phi_{I\bq}} \opWq^{\rm p} \kket{\Phi_{I'\bq}} \,,
\end{equation}
and the matrix elements of the dynamical correction involving the partially-screened phonon perturbations,
\begin{subequations} \label{eq:rpa_X_defs}
\begin{align}
    X^{\rm b,dyn}_{aI\bq}(\omega)
    &\equiv \bbra{V^{\rm p}_{a\bq}} \, \opchiq^{\rm b,dyn}(\omega) \, \kket{\Phi_{I\bq}} \,,
    \\
    X^{\rm b,dyn}_{Ia'\bq}(\omega)
    &\equiv \bbra{\Phi_{I\bq}} \, \opchiq^{\rm b,dyn}(\omega) \, \kket{V^{\rm p}_{a'\bq}} \,,
    \\
    X^{\rm b,dyn}_{aa'\bq}(\omega)
    &\equiv \bbra{V^{\rm p}_{a\bq}} \, \opchiq^{\rm b,dyn}(\omega) \, \kket{V^{\rm p}_{a'\bq}} \,.
\end{align}
\end{subequations}
The indices distinguish these matrices: $I$ denotes a basis function of the subspace and $a$ an atomic displacement.
Their explicit expressions as sums over pairs of active states are given in \cref{eq:rpa_X_band}.

Evaluating \cref{eq:rpa_Hs_downfold,eq:rpa_C_downfold} in the subspace (\cref{app:downfold_derivation}), we find that the dynamical IFCs of the target system can be written as~\cite{Nomura2015,Berges2023}
\begin{equation} \label{eq:eph_Cs_downfold}
    C^{\rm s}_{aa'\bq}(\omega)
    = C^{\rm p}_{aa'\bq}
    + \Pi^{\rm A}_{aa'\bq}(\omega) \,,
\end{equation}
where the phonon self-energy from the active space is
\begin{align} \label{eq:eph_phself_matrix}
    \Pi^{\rm A}_{aa'\bq}(\omega)
    &= X^{\rm b,dyn}_{aa'\bq}(\omega)
    \nnnl
    &\quad + \sum_{II'} X^{\rm b,dyn}_{a I\bq}(\omega) \,
    W^{\rm s}_{II'\bq}(\omega) \,
    X^{\rm b,dyn}_{I'a'\bq}(\omega) \,,
\end{align}
and the fully-screened e-e interaction in the subspace is given by the Dyson equation
\begin{equation} \label{eq:eph_Ws_dyson}
    W^{\rm s}_{II'\bq}(\omega)
    = \sum_{I''} \bigl[ \one - W^{\rm p}_{\bq} \, X^{\rm b,dyn}_{\bq}(\omega) \bigr]^{-1}_{II''} \,
    W^{\rm p}_{I''I'\bq} \,.
\end{equation}
The first term of \cref{eq:eph_phself_matrix} is the response of the active space to the partially-screened phonon perturbation, and the second term describes its rescreening by the fully-screened e-e interaction within the subspace.

The phonon self-energy $\Pi^{\rm A}_{aa'\bq}(\omega)$ [\cref{eq:eph_phself_matrix}] gives the dynamical IFCs of the target system [\cref{eq:eph_Cs_downfold}], for both DFPT+A and cDFPT+A downfolding.
This self-energy depends on which method is used to separate the susceptibility into the reference susceptibility and the dynamical correction.
In DFPT+A downfolding [\cref{eq:chib_DFPT}], $\opchiq^{\rm b,dyn}(0)$ and the static self-energy correction $\Pi^{\rm A}_{aa'\bq}(0)$ vanish if the target and base systems are identical, so that $\Pi^{\rm A}_{aa'\bq}(\omega)$ then only contains the nonadiabatic correction, as in Refs.~\cite{Calandra2010, Giustino2017, Berges2023}.
If the parameters differ, $\Pi^{\rm A}_{aa'\bq}(0)$ is nonzero and gives the change of the static IFCs from the base system to the target, e.g., the softening of the phonons upon doping or cooling.
Apart from the numerical approximations, the final dynamical IFC $C^{\rm s}_{aa'\bq}(\omega)$ [\cref{eq:eph_Cs_downfold}] is independent of the choice of the base system.

\Cref{eq:eph_phself_matrix,eq:eph_Ws_dyson} reduce the screening problem in the full two-particle space to a problem in the space of active-space transitions.
We note that reducing the rank of the screening problem is a subject of active research.
In particular, for $GW$ and other many-body perturbation theory calculations, methods based on low-rank approximations of the susceptibility~\cite{Wilson2008, Govoni2015, Shao2016}, localized representations of the screening~\cite{Lu2008}, and interpolative separable density fitting~\cite{Yeh2024} have been developed.
In our DFPT+A downfolding approach, all screening beyond the dynamical correction is contained in the static quantities $C^{\rm p}$, $V^{\rm p}$, and $W^{\rm p}$, which are computed once for the base system, while $X^{\rm b,dyn}(\omega)$ only involves active-space transitions, which makes it cheap to evaluate at many frequencies and on dense $\bk$ grids.
This makes the calculation of the low-energy dynamical IFCs very efficient, at the cost of reduced accuracy at higher frequencies, which are not needed for the study of nonadiabatic phonon effects.
Also, the accuracy and the frequency range can be systematically improved by enlarging the active space.

\subsection{Wannier representation of the active space} \label{sec:theory_wannier}

To lower the cost further, we reduce the size of the active-space matrices.
The efficiency of the DFPT+A downfolding approach is determined by the size of the basis $\kket{\Phi_{I\bq}}$ of active-space transitions, which sets the size of the matrices and the number of elements of $W^{\rm p}$ to be computed.
The natural choice, the band pairs $\kket{\Psi_\mnkq}$ of active states [\cref{eq:rpa_pair_basis}], is not small: their number grows linearly with the number of $\bk$ points on the fine grid.
Instead, we represent the active-space transitions using WFs, localized functions that accurately describe the low-energy electronic structure~\cite{Marzari1997, Souza2001, Marzari2012, Marrazzo2024}.
The number of WF pairs is set by the coarse $\bk$ grid used to construct the WFs, not by the fine grid.
Their spatial localization allows a truncation of the basis: $W^{\rm p}$ can be truncated to a few nearby WF pairs (\cref{sec:theory_truncation}), whose matrix elements we compute directly with DFPT (\cref{sec:theory_ee_DFPT}).
This enables an efficient calculation of the screened interaction and further reduces the matrix sizes in the dynamical screening problem.

The WFs are constructed from the Bloch states on the coarse grid of $N_\bk^{\rm c}$ points,
\begin{equation} \label{eq:wf_construction}
    \ket{w_\iR} = \frac{1}{N_\bk^{\rm c}} \sum_\bk e^{-i\bk\cdot\bR} \sum_n \ket{\psi_\nk} U^{\rm W}_{ni\bk} \,,
\end{equation}
where $U^{\rm W}_{ni\bk}$ is the gauge matrix obtained from the Wannierization, including disentanglement if needed, and $n$ runs over the bands used in the Wannierization.
As usual in Wannier interpolation, we treat the resulting WFs as localized functions in the infinite crystal, with $\bR$ restricted to the Wigner--Seitz supercell of the coarse grid~\cite{Pizzi2020W90}.
Their Bloch sums are
\begin{equation} \label{eq:wf_bloch_sum}
    \ket{w_\ik} \equiv \sum_{\bR} e^{i\bk \cdot \bR} \ket{w_\iR} \,,
\end{equation}
where $\ket{w_\iR}$ is the WF at the unit cell $\bR$ centered at $\br_i + \bR$.
In the WF basis, the energy eigenstates at an arbitrary $\bk$ are
\begin{equation} \label{eq:wf_eigenstates}
    \ket{\psi_\nk}
    = \sum_{i\bR} \ket{w_\iR} U_{in\bk} \, e^{i\bk \cdot \bR}
    = \sum_{i} \ket{w_\ik} U_{in\bk} \,,
\end{equation}
where $U_{in\bk}$ is the eigenvector of the Wannier-interpolated Hamiltonian.
For $\bk$ on the coarse grid and for states inside the inner window, $U_{in\bk} = U^{\rm W*}_{ni\bk}$, up to the arbitrary phase of the eigenvectors (and a unitary rotation within degenerate subspaces), so that \cref{eq:wf_eigenstates} inverts \cref{eq:wf_construction}.
This representation is accurate for states inside the inner window of the Wannierization, which contains the active window $\mc{W}^{\rm A}$.
The WFs are orthonormal: $\braket{w_\ik | w_{i'\bk}} = \delta_{ii'}$.
Analogous to the band pair basis $\kket{\Psi_\mnkq}$ [\cref{eq:rpa_pair_basis}], we define the WF pair basis $\kket{\Phi_{ij\bR\bR'}}$ as the outer product of two WFs,
\begin{equation} \label{eq:rpa_wf_pair_RR}
    \kket{\Phi_{ij\bR\bR'}} \equiv \ket{w_{i\bR'}} \bra{w_{j\bR+\bR'}} \,.
\end{equation}
Fourier transforming over the unit cell coordinate $\bR'$ into the WF pair wavevector $\bq$, we can define
\begin{align} \label{eq:rpa_wf_pair_Rq}
    \kket{\Phi_{ij\bR\bq}}
    &\equiv \sum_{\bR'} e^{i\bq \cdot \bR'} \kket{\Phi_{ij\bR\bR'}}
    \nnnl
    &= \sum_{\bR'} e^{i\bq \cdot \bR'} \ket{w_{i\bR'}} \bra{w_{j\bR+\bR'}}
    \nnnl
    &= \frac{1}{N_\bk^{\rm c}} \sum_{\bk} e^{i\bk \cdot \bR} \ket{w_{i\bkq}} \bra{w_{j\bk}} \,.
\end{align}
This object is a linear combination of WF pairs with fixed relative unit cell distance $\bR$, weighted by the phase factor $e^{i\bq \cdot \bR'}$.
For compactness, we denote $I \equiv (i, j, \bR)$.
The WF pairs are orthonormal, $\bbrakket{\Phi_{I\bq}}{\Phi_{I'\bq}} = \delta_{II'}$.
The band pair basis can be expanded in the WF pair basis as
\begin{equation}
    \kket{\Psi_\mnkq}
    = \sum_{I} \kket{\Phi_{I\bq}} \, U_{im\bkq} \, U^*_{jn\bk} \, e^{-i\bk\cdot\bR} \,,
\end{equation}
which is again accurate if $\veps_\nk$ and $\veps_\mkq$ are inside the inner window of the Wannierization.

We use these WF pairs as the basis $\kket{\Phi_{I\bq}}$ to represent the dynamical correction $\opchiq^{\rm b,dyn}(\omega)$ in \cref{eq:rpa_chib_active_W}.
The explicit expression of the matrix elements $X^{\rm b,dyn}_{II'\bq}(\omega)$ [\cref{eq:rpa_X_elel}] is given in \cref{eq:rpa_X_band}.
The size of the WF pair basis is $N^{\rm pair} = (N^{\rm WF})^2 N_\bk^{\rm c}$, where $N^{\rm WF}$ is the number of WFs and $N_\bk^{\rm c}$ is the number of coarse $\bk$-points used to construct the WFs.

\subsection{Truncation of the screened interaction} \label{sec:theory_truncation}

While the active-space screening problem has a much smaller dimension than the original real-space problem, it still involves matrices of size $N^{\rm pair} \times N^{\rm pair}$, where $N^{\rm pair}$ can be over $10^3$ for realistic materials.
Typically, $N_{\bk}^{\rm c}$ ranges from $5^d$ to $20^d$ in $d$ dimensions.
This is too large to operate on the matrices directly, and it is also too large to compute the partially-screened interaction matrix elements.
Thus, we introduce further approximations, making sure that they are systematically improvable.

The expression of the partially-screened interaction in \cref{eq:rpa_Wp_WF} sets the stage.
Thanks to the spatial localization of the WFs, the WF pair density decays rapidly with increasing WF pair distance $d_I$:
\begin{align} \label{eq:trunc_WF_pair}
    \bbrakket{\br\br;\bq}{\Phi_{I\bq}}
    &= \sum_{\bR'} e^{i\bq \cdot \bR'} w_{i}(\br - \bR') w^*_{j}(\br - \bR' - \bR)
    \nnnl
    &\approx 0 \ \text{if}\ d_I \equiv \abs{\br_i - \br_{j} - \bR}\ \text{is large} \,.
\end{align}
where $w_i(\br) \equiv \braket{\br|w_{i\mb{0}}}$ is the WF in the home unit cell, so that $\braket{\br|w_{i\bR}} = w_i(\br - \bR)$.
Since the interaction kernel $\opvq$ couples only the diagonal elements $\kket{\br\br;\bq}$ of the position basis [\cref{eq:rpa_v_def}], so does the partially-screened interaction $\opWq^{\rm p} = (\opepsq^{\rm p})^{-1} \opvq$; its matrix elements thus involve only the WF pair densities and decay with increasing WF pair distance.
Thus, the matrix elements of $\hat{W}^{\rm p}$ can be truncated to include only WF pairs within a certain distance:
\begin{equation} \label{eq:eph_Wp_trunc}
    \opWq^{\rm p}
    \approx \sum_{II'}^{d^{\rm cut}} \kket{\Phi_{I\bq}} W^{\rm p}_{II'\bq} \bbra{\Phi_{I'\bq}} \,,
\end{equation}
where the summation is truncated to only include WF pairs with $d_I, d_{I'} < d^{\rm cut}$.
This truncation significantly reduces the cost of computing and storing $\hat{W}^{\rm p}$.
A similar idea has been used for the calculation of excitons in the WF basis~\cite{Merkel2024}.

We denote by $N^{\rm pair, trunc}$ the number of truncated WF pairs.
In practice, we find that it suffices to keep only the onsite and a few nearest-neighbor WF pairs, making
\begin{equation} \label{eq:trunc_size}
    N^{\rm pair, trunc} \sim 10 \text{ to } 10^2 \ll N^{\rm pair} \sim 10^3 \text{ to } 10^4 \,.
\end{equation}
This truncation is systematically improvable by increasing the cutoff distance (see \cref{sec:theory_demo,sec:validation} for the convergence).
In this work, we use the truncated set of $N^{\rm pair, trunc}$ WF pairs only for the e-e interaction, where the number of matrix elements scales quadratically with the number of pairs.
The size after truncation is $N^{\rm pair, trunc} \times N^{\rm pair, trunc}$.
For the e-ph interaction, we use the usual Wannier interpolation without truncation, which includes all $N^{\rm pair}$ WF pairs.
Its size is $N^{\rm pair} \times N^{\rm ph}$, where $N^{\rm ph} = 3 N^{\rm at}$ is the number of atomic displacements $a$, with $N^{\rm at}$ the number of atoms in the unit cell.

\subsection{Calculating e-e interactions using DFPT} \label{sec:theory_ee_DFPT}

Next, we discuss how to compute the partially-screened e-e interaction $\opWq^{\rm p}$ in \cref{eq:rpa_Wp_WF}.
Most cRPA implementations~\cite{Nakamura2021, Amadon2014, Kaltak2015, Yeh2024} use the dielectric method~\cite{Adler1962, Wiser1963, Hybertsen1986GW}, which represents the dielectric function in a truncated basis and computes the bare susceptibility with a finite number of bands.
For phonons, the truncation errors of the dielectric method~\cite{Pick1970} converge slowly, which is why DFPT, which avoids both truncations, is the standard method for computing phonons and e-ph interactions~\cite{Baroni2001}.
Thus, we use the DFPT method to compute the partially-screened e-e interactions, which allows for a consistent treatment of e-e and e-ph interactions.

DFPT has been used to compute the onsite density-density interaction of a single WF for Koopmans-compliant functionals~\cite{Colonna2022}, and the density-density interaction between WFs at different sites for real-time simulations of electron dynamics~\cite{Cistaro2026}.
Similarly, the Sternheimer-$GW$ method~\cite{Giustino2010GW} uses the Sternheimer equation to avoid sums over unoccupied states when computing the screened Coulomb interaction in a plane-wave basis.
Here, we generalize the approach of Refs.~\cite{Colonna2022, Cistaro2026} to the full partially-screened interaction between arbitrary WF pairs, including pairs of different WFs and pairs of WFs in different unit cells.

We define $\kket{v^{\rm b} \Phi_{I\bq}} \equiv \opvq \kket{\Phi_{I\bq}}$ and $\kket{W^{\rm p} \Phi_{I\bq}} \equiv \opWq^{\rm p} \kket{\Phi_{I\bq}}$.
These are the bare and screened potentials generated by a test charge density of a WF pair $\kket{\Phi_{I\bq}}$.
From \cref{eq:rpa_Wp_def}, we have
\begin{equation} \label{eq:rpa_partial_Wp}
    \kket{W^{\rm p} \Phi_{I\bq}}
    = \kket{v^{\rm b} \Phi_{I\bq}}
    + \opvq \, \opchiq^{\rm b,ref} \, \kket{W^{\rm p} \Phi_{I\bq}} \,.
\end{equation}
This equation has the structure of a partial screening, \cref{eq:rpa_Hp}, with the bare perturbation being the bare potential perturbation induced by a density perturbation $\kket{\Phi_{I\bq}}$.
In DFPT+A downfolding, where $\opchiq^{\rm b,ref}$ is the static susceptibility of the base system, it is the ordinary static DFPT screening equation of the base system.
Thus, this equation can be solved using a standard DFPT implementation by setting the bare perturbation to be
\begin{align}
    H^{\rm b}_{I\bq}(\br)
    &= \bbrakket{\br\br;\bq}{v^{\rm b} \Phi_{I\bq}}
    \\
    &= \nsint{\rm u.c.} \dd\brp \, v^{\rm b}_\bq(\br, \brp)
    \frac{1}{N_\bk^{\rm c}} \sum_\bk e^{i\bk \cdot \bR} w_{i\bkq}(\brp) w^*_{j\bk}(\brp) \,.
    \nonumber
\end{align}
In the second equality, we used the last line of \cref{eq:rpa_wf_pair_Rq}; recall that $I \equiv (i, j, \bR)$.
The resulting partially-screened potential is
\begin{equation}
    H^{\rm p}_{I\bq}(\br)
    = \bbrakket{\br\br;\bq}{W^{\rm p} \Phi_{I\bq}} \,.
\end{equation}
Then, the matrix elements of the partially-screened e-e interaction are computed as
\begin{align}
    W^{\rm p}_{II'\bq}
    &= \nsint{\rm u.c.} \dd\br \, \bbrakket{\Phi_{I\bq}}{\br\br;\bq} \, H^{\rm p}_{I'\bq}(\br)
    \nnnl
    &= \frac{1}{N_\bk^{\rm c}} \sum_\bk e^{-i\bk \cdot \bR} \mel{w_{i\bkq}}{H^{\rm p}_{I'\bq}}{w_{j\bk}} \,.
\end{align}

Importantly, this approach brings another advantage: the spatial localization of the WFs can be exploited \textit{before} computing the e-e interaction matrix elements, which significantly reduces the computational cost.
In the dielectric method, one first computes the e-e interaction matrix elements in the band pair basis, and then transforms them to the WF pair basis.
However, this approach requires computing and storing a large number of e-e interaction matrix elements, amounting to $(N^{\rm pair})^2 = (N^{\rm WF})^4 (N_\bk^{\rm c})^2$.
Also, a large portion of the matrix elements will be truncated after the transformation.
In contrast, with this DFPT method for computing the e-e interaction, we directly work in the WF pair basis and only compute the necessary elements, significantly reducing the computational cost.
We only compute $(N^{\rm pair, trunc})^2 \ll (N^{\rm pair})^2$ matrix elements per $\bq$ point.

Another distinct advantage of this DFPT approach for the e-e interaction is that it allows one to compute the interactions at an arbitrary $\bq$ point without the need for interpolation.
For an arbitrary $\bq$, the point $\bk+\bq$ in \cref{eq:rpa_wf_pair_Rq} is in general not on the coarse grid for constructing the WFs.
We therefore obtain the WF pair density $\bbrakket{\br\br;\bq}{\Phi_{I\bq}}$ [\cref{eq:trunc_WF_pair}] by Wannier interpolation of the WFs themselves.
At each point $\br$ of the real-space grid, we compute the Bloch sum $w_{i\bk+\bq}(\br)$ [\cref{eq:wf_bloch_sum}] from the WFs $w_{i\bR}(\br)$ with the minimal-distance replica selection~\cite{Pizzi2020W90}.
This capability is analogous to the advantage of DFPT for phonons over the displaced supercell method.
Finally, contrary to the usual cRPA calculations of the e-e interaction, we include the exchange-correlation kernel in the bare interaction $\hat{v}^{\rm b}_\bq$, which gives a unified treatment of the e-e interaction and of the DFPT calculation of phonons.

With these partially-screened e-e interactions, together with the partially-screened IFCs and e-ph matrix elements, all ingredients of the active-space screening problem [\cref{eq:rpa_Hs_downfold}] are available at the $\bq$ points where DFPT calculations are performed.
What remains is to obtain them at the many $\bq$ points needed to compute physical quantities.

\subsection{Interpolation of e-e interactions} \label{sec:theory_itp}

To solve the active-space screening problem [\cref{eq:rpa_Hs_downfold}], we need the partially-screened IFCs, e-ph interaction, and e-e interaction at all $\bq$ points of interest.
To avoid recomputing these quantities for every $\bq$ point, we use interpolation schemes.
For the force constants and e-ph interaction, we use Wannier interpolation~\cite{Giustino2007} with an analytic treatment of the long-range part of the interactions~\cite{Verdi2015, Sjakste2015, Sohier2017, Royo2019, Brunin2020PRL, Brunin2020PRB, Jhalani2020, Park2020, Ponce2023PRL, Ponce2023PRB}, here in its 2D formulation~\cite{Sohier2017, Ponce2023PRB}.

However, Wannier interpolation of the e-e interaction is more challenging.
The Coulomb singularity at $\bq = \mb{0}$ is sharper for the e-e interaction than for the e-ph interaction, as the former has a $1/q^2$ divergence in 3D and a $1/q$ divergence in 2D, while the latter has a $1/q$ divergence in 3D and is finite in 2D.
While the leading $1/q^2$ singularity can be treated using Ewald summation~\cite{Merkel2024}, the subleading terms require an accurate treatment of the anisotropic dielectric screening.
A Wannier interpolation scheme for the e-e interaction including the treatment of the anisotropic Coulomb singularity does not yet exist and is beyond the scope of this work.
Yet, we note that our \textit{ab initio} DFPT approach provides reference data, which would be crucial for developing and testing such a scheme.

Instead, we use the fact that DFPT allows one to compute the e-e interaction at an arbitrary $\bq$ point, not necessarily from a regular grid.
We use a one-dimensional interpolation along the $\bq$ path, similar to that used for the Born effective charge and dielectric functions in Ref.~\cite{Macheda2024a}.
For metals or for a $\bq$ path not including $\bq = \mb{0}$, the partially-screened interaction $\opWq^{\rm p}$ is smooth in $\bq$ and can be directly interpolated.
We compute it on a list of $\bq$ points along the path, $\bq = \bq_0 + t (\bq_1 - \bq_0)$ with $t \in [0, 1]$, and interpolate it with a natural cubic spline.

For insulators, we need to treat the long-wavelength singularity of $\opWq^{\rm p}$ carefully when the $\bq$ path starts from or ends at $\bq = \mb{0}$.
We treat the leading singularity given by the monopole-monopole (m-m) interaction~\cite{Ponce2023PRL, Ponce2023PRB},
\begin{align} \label{eq:itp_Wp_monopole}
    \lim_{\bq \to \mb{0}} W^{\rm p}_{II'\bq} &= W^{\rm p, m\text{-}m}_{\bq} \, \delta_{I0} \, \delta_{I'0}
    + \text{higher order}
    \\
    W^{\rm p, m\text{-}m}_{\bq}
    &= \begin{cases}
        \dfrac{4\pi e^2}{\Omega} \dfrac{1}{\bq \cdot \mb{\epsilon}^{\rm M, p} \cdot \bq}
        & \text{for 3D} \\
        \dfrac{2\pi e^2}{A} \dfrac{1}{\abs{\bq}}
        & \text{for 2D} \,,
    \end{cases}
    \label{eq:itp_Wp_mm}
\end{align}
where $e$ is the elementary charge, $\delta_{I0} = \delta_{ij} \delta_{\bR 0}$ selects the diagonal WF pair, $\Omega$ and $A$ are the volume and area of the unit cell, respectively, and $\mb{\epsilon}^{\rm M,p}$ is the partially-screened macroscopic dielectric tensor computed using the base system.
We then subtract this singularity from $\opWq^{\rm p}$ before performing the interpolation, and add it back after the interpolation.
Since the interaction is discontinuous at $\bq = \mb{0}$, we do not include it in the interpolation knots.
We perform DFPT calculations for $\bq = t_n \bq_1$ with $0 < t_1 < \cdots < t_N = 1$ and interpolate $W^{\rm p}_{II'\bq} - W^{\rm p, m\text{-}m}_{\bq} \delta_{I0} \delta_{I'0}$ with a natural cubic spline through these points.
To obtain the value at $\bq = t' \bq_1$ with $0 < t' < t_1$, we use constant extrapolation of the singularity-subtracted interaction at $t_1$, and add back the singularity at $t'$.
To address the remaining $1/|\bq|$ divergence due to the monopole-dipole term in the 3D case, or the rapid variation of the interaction near $\bq = \mb{0}$ in 2D, we densely sample the small-$\bq$ region using a nonuniform grid of $t_n$ whose spacing increases away from $\bq = \mb{0}$, from 0.002 to 0.1 for hBN (see \cref{app:comp_details}).

\begin{figure}[tb]
\includegraphics[width=0.99\linewidth]{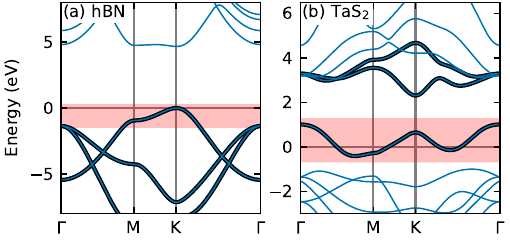}
\caption{
    Electronic band structure of (a) monolayer hBN and (b) monolayer TaS$_2$.
    Thin blue lines show the density functional theory (DFT) bands and thick black lines the Wannier-interpolated bands.
    The shaded region indicates the active window $\mathcal{W}^{\rm A}$.
    For hBN, its lower boundary is $E^{\rm VBM} - 1.5~\mathrm{eV}$, where $E^{\rm VBM}$ is the valence band maximum (VBM) energy.
    For TaS$_2$, it contains the band that crosses the Fermi level.
}
\label{fig:bands}
\end{figure}

\begin{figure}[t]
\includegraphics[width=0.99\linewidth]{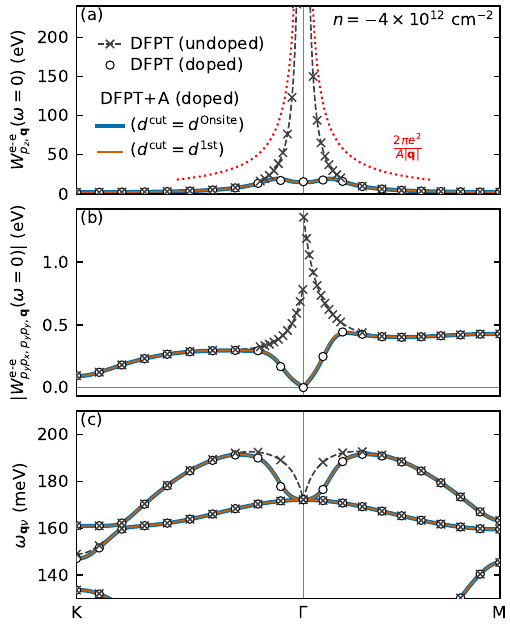}
\caption{
    (a) Static density-density e-e interaction for the N $p_z$ WF of monolayer hBN, for the undoped crystal and upon hole doping with $T=300~\mathrm{K}$ and $n = -4 \times 10^{12}~\mathrm{cm^{-2}}$, where negative carrier concentrations $n$ denote hole doping.
    Crosses show the partially-screened interaction $W^{\rm p}$, computed with direct DFPT for the undoped base system, and dashed gray lines show its interpolation (\cref{sec:theory_itp}).
    Circles show the result of direct DFPT with doping.
    Colored lines show the result of DFPT+A downfolding from the undoped base system to the doped target, obtained with onsite and first-neighbor $d^{\rm cut}$ (\cref{tab:cutoff}).
    The red dotted curve is the 2D monopole-monopole interaction [\cref{eq:itp_Wp_mm}].
    (b) Absolute value of the off-diagonal element of the static e-e interaction between the $p_y p_x$ and $p_y p_y$ WF pairs.
    (c) Dispersion of the optical phonons.
    The dashed gray line shows the dispersion of the undoped base system.
}
\label{fig:hBN_static_ee}
\end{figure}

\begin{figure}[t]
\includegraphics[width=0.99\linewidth]{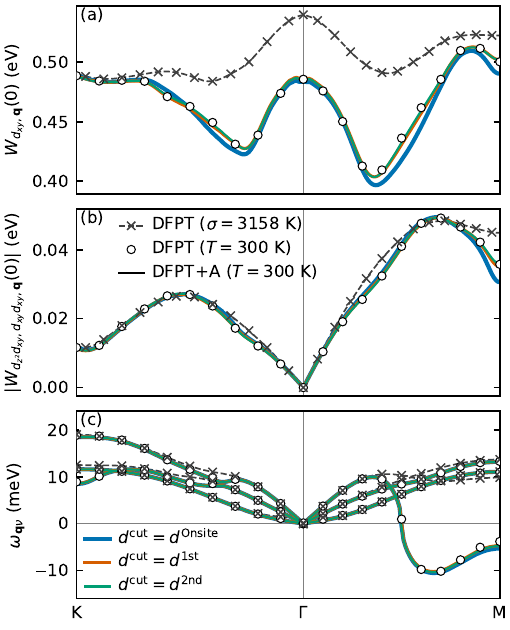}
\caption{
    (a) Static density-density e-e interaction for the $d_{xy}$ WF of monolayer TaS$_2$, for the base system with a Gaussian smearing of $\sigma = 20~\mathrm{mRy}$ ($3158~\mathrm{K}$) and for the target with $T = 300~\mathrm{K}$.
    (b) Absolute value of the off-diagonal element of the static e-e interaction between the $d_{z^2} d_{xy}$ and $d_{xy} d_{xy}$ WF pairs.
    (c) Dispersion of the acoustic phonons.
}
\label{fig:TaS2_static_val}
\end{figure}

\subsection{Computational details} \label{sec:DFTdetails}

We perform DFT, DFPT, and TDDFPT calculations using the \QE package~\cite{Giannozzi2009, Giannozzi2017}.
We significantly refactored and modularized its DFPT implementation to allow the use of arbitrary bare perturbations, which enables the calculation of the partially-screened e-e interaction described in \cref{sec:theory_ee_DFPT}.
We also implemented finite-frequency TDDFPT (\cref{app:tddfpt}) for e-e interaction, and added the cDFPT variant used in \cref{app:cDFPT} using the modified Sternheimer equation adapted from Refs.~\cite{Nomura2015, Berges2023}.
The DFPT and TDDFPT implementations will be released as part of the \texttt{HP} code~\cite{Timrov2022HP}, a code inside the \QE package for calculating Hubbard parameters.
Details of the implementation and of the modularization effort will be presented elsewhere.
We perform the Wannier interpolation and solve the active-space screening equations using the \textsc{ElectronPhonon.jl}~\cite{EPjl} code with GPU acceleration.
The full computational details and parameters are given in \cref{app:comp_details}.

\subsection{Validation on monolayer hBN and TaS$_2$} \label{sec:theory_demo}

\begin{table}[b]
\centering
\caption{
    Cutoff distances and number of truncated WF pairs for different neighbor shells in monolayer hBN (up to the 2nd shell) and TaS$_2$ (up to the 5th shell).
    We denote the cutoffs of the successive columns by $d^{\rm onsite}$, $d^{\rm 1st}$, $d^{\rm 2nd}$, and so on.
}
\label{tab:cutoff}
\small\setlength{\tabcolsep}{2pt}
\begin{tabular}{ll|cccccc}
    & Label & Onsite & 1st & 2nd & 3rd & 4th & 5th \\ \hline
    \multirow{2}{*}{hBN} & $d^{\rm cut}$ (bohr) & 2.34 & 6.41 & 8.75 & -- & -- & -- \\
    & $N^{\rm pair, trunc}$ & 9 & 63 & 117 & -- & -- & -- \\ \hline
    \multirow{2}{*}{TaS$_2$} & $d^{\rm cut}$ (bohr) & 3.16 & 8.62 & 11.78 & 14.66 & 17.82 & 20.40 \\
    & $N^{\rm pair, trunc}$ & 9 & 63 & 117 & 171 & 279 & 333 \\
\end{tabular}
\end{table}

To validate the DFPT+A downfolding scheme, we apply it to the two materials we will study in detail in \cref{sec:applications}: hole-doped monolayer hBN and metallic monolayer TaS$_2$.
They represent the two kinds of base systems of \cref{fig:downfolding_schematic}: an undoped semiconductor, and a metal with a large electronic smearing.
For hBN, the base system is the undoped crystal, and the target is a hole doping of $n = -4 \times 10^{12}~\mathrm{cm^{-2}}$ at $T = 300~\mathrm{K}$.
We construct three $p$-orbital-like WFs centered on the N atom and include in the active space the top of the valence band at K and the parabolic valence band at $\Gamma$ [\Fig[a]{fig:bands}].
For TaS$_2$, the base system uses a Gaussian smearing of $\sigma = 20~\mathrm{mRy}$, so that the base system has no soft phonons, and the target is a realistic temperature of $T = 300~\mathrm{K}$ with the Fermi--Dirac distribution.
We construct three WFs using the $d_{z^2}$, $d_{xy}$, and $d_{x^2-y^2}$ orbitals of Ta as the initial projections~\cite{Berges2020,Berges2023}, and the active window contains only the band that crosses the Fermi level [\Fig[b]{fig:bands}].
A single WF could be constructed for this band, but it would be much less localized than the three-orbital WFs, requiring a larger cutoff distance for the truncation of the e-e interaction.
The Wannierized bands are entangled with the other conduction bands.
This situation poses no difficulty in the DFPT+A scheme, as the base calculation is performed with standard DFPT, not cDFPT, and thus does not depend on the choice of the active space or of the WFs.
In \cref{sec:validation} (\cref{fig:TaS2_dyn_ee_tddfpt_window}), we show that we can enlarge the active space to include the entangled conduction bands.

For both materials, we truncate the partially-screened e-e interaction based on the distance between the centers of the WF pairs, with the cutoffs $d^{\rm cut}$ listed in \cref{tab:cutoff}.
Since the WFs are atom-centered, we use the atomic positions instead of the Wannier centers as the nominal centers.
The number of truncated WF pairs, $N^{\rm pair, trunc}$, is the number of DFPT calculations per $\bq$ point needed to compute the partially-screened e-e interaction.
For real-valued WFs, the densities of the pairs $(i, j, \bR)$ and $(j, i, -\bR)$ differ only by a phase factor, so this number could be reduced, for example from 9 to 6 for the onsite pairs; we do not use this reduction here.
Additional computational details are provided in \cref{app:comp_details}.

\Cref{fig:hBN_static_ee,fig:TaS2_static_val} show the partially-screened e-e interactions of monolayer hBN and TaS$_2$, respectively, computed with DFPT and their interpolations (\cref{sec:theory_itp}).
The density-density interaction [\Fig[a]{fig:hBN_static_ee}] diverges near $\Gamma$ and follows the 2D monopole-monopole term [\cref{eq:itp_Wp_mm}].
The off-diagonal elements [\Fig[b]{fig:hBN_static_ee}] are finite and smooth, and are interpolated directly.
Yet, the values are discontinuous at $\Gamma$ due to higher-order multipole contributions.
Careful treatment of these terms would be required for interpolating the e-e interaction on two- and three-dimensional $\bq$ grids.
For metallic monolayer TaS$_2$, the partially-screened interaction of the high-smearing base system is finite and smooth in $\bq$, and is interpolated directly [\Fig[a,b]{fig:TaS2_static_val}].

To benchmark DFPT+A downfolding, we compare its results against direct DFPT calculations at the target parameters.
For the target, we use a $\bk$ grid of $24 \times 24 \times 1$ for hBN and $36 \times 36 \times 1$ for TaS$_2$, denser than the grids used for the base calculations, but small enough that the direct DFPT calculations are still feasible.
This comparison tests three approximations of DFPT+A downfolding: the rigid-band approximation [\cref{eq:rigid_band}], the replacement of the rest susceptibility of the target with that of the base system [\cref{eq:chib_approx2}], and the truncation of the partially-screened e-e interactions [\cref{eq:eph_Wp_trunc}].

For hBN, doping screens the long-range Coulomb interaction, removing the $\bq \to \mb{0}$ divergence in the density-density interaction [\Fig[a]{fig:hBN_static_ee}] and the discontinuity in the off-diagonal element [\Fig[b]{fig:hBN_static_ee}].
Also, the LO phonon loses its linear dispersion near $\Gamma$ [\Fig[c]{fig:hBN_static_ee}]~\cite{Macheda2022, Macheda2023}.
Downfolding reproduces the direct DFPT results obtained by doping with a uniform background charge.
The onsite and first-neighbor cutoffs give indistinguishable results, indicating that the e-e interaction is well converged with respect to the truncation.

For TaS$_2$, lowering the temperature modifies the density-density and off-diagonal interactions, which are both accurately captured by the downfolding approach [\Fig[a,b]{fig:TaS2_static_val}], with the first-neighbor cutoff being sufficient for convergence.
The emergence of the soft mode at the M point is also quantitatively reproduced, in good agreement with DFPT [\Fig[c]{fig:TaS2_static_val}].
Neglecting $W^{\rm p}$, as in Ref.~\cite{Berges2023}, leaves a small difference [\Fig[b]{fig:TaS2_phonon_static_target}].

In \cref{sec:validation}, we further validate the static approximation for the rest susceptibility [\cref{eq:chib_approx1}] in DFPT+A downfolding by comparing it with explicit TDDFPT calculations.
We find that downfolding accurately reproduces the dynamical e-e interaction of TaS$_2$ up to $\omega \approx 0.5~\mathrm{eV}$, well above the phonon energies.
For the $4000^2$ $\bk$ grid, direct TDDFPT at a single frequency would cost about $10^4$ times more than the full DFPT+A calculation for 101 frequencies (\cref{tab:TaS2_cost}).

\section{Dynamical e-e and e-ph interactions} \label{sec:theory_ee}

Now, using the DFPT+A downfolding formalism, we discuss how to compute the dynamical e-e and e-ph interactions, which are the fundamental quantities for computing properties such as the electron and phonon spectral functions, transport, and superconductivity.
We first evaluate the small-matrix expressions of \cref{sec:theory_active_screening} in the truncated WF pair basis and derive from them the dynamically screened e-ph vertex and the phonon propagator (\cref{sec:theory_eph}).
We then combine them into the total dynamical e-e interaction, the sum of the screened Coulomb and phonon-mediated interactions (\cref{sec:theory_Wtot}).
Finally, we define the local dynamical interaction and its retarded part $\Delta U$, which generalize the Eliashberg function and the e-ph coupling constant to the total interaction (\cref{sec:theory_Uloc}).

\subsection{Calculating dynamical interactions from downfolding} \label{sec:theory_eph}

In this subsection, we evaluate the phonon self-energy [\cref{eq:eph_phself_matrix}] and the fully-screened e-e interaction [\cref{eq:eph_Ws_dyson}] in the truncated WF pair basis, and derive from them the dynamically screened e-ph vertex [\cref{eq:eph_gs_rpa_trunc_2}] and the phonon propagator [\cref{eq:eph_Ds_dyson}].
The inputs are the partially-screened IFCs, e-ph interaction, and e-e interaction, which are computed with DFPT for the base system,
\begin{equation}
    C^{\rm p}_{aa'\bq}\,,\ g^{\rm p}_{a\bq, mn\bk}\,,\ W^{\rm p}_{II'\bq} \,.
\end{equation}
The susceptibilities $X^{\rm b,dyn}$ [\cref{eq:rpa_X_elel,eq:rpa_X_defs}] describe the response of the active space to perturbations in the WF pair density $I$ or the atomic displacement $a$.
In the truncated WF pair basis, $W^{\rm p}_\bq$ and $X^{\rm b,dyn}_{II'\bq}(\omega)$ are $N^{\rm pair, trunc} \times N^{\rm pair, trunc}$ matrices, and $X^{\rm b,dyn}_{Ia\bq}(\omega)$ is an $N^{\rm pair, trunc} \times N^{\rm ph}$ matrix, where $N^{\rm ph}$ is the number of phonon modes.

When the double bracket sandwich is expanded in the band basis, the susceptibilities can be expressed as a summation over band pairs and $\bk$ points, with the static base contribution subtracted [\cref{eq:chib_ref_dyn_DFPTA}]:
\begin{align} \label{eq:rpa_X_band}
    X^{\rm b,dyn}_{YY'\bq}(\omega)
    &= \intbk \sum_{mn}^{\rm A} \bigg[
        \frac{f^{\rm tar}_\nk \!-\! f^{\rm tar}_\mkq}{\omega \!+\! \veps_\nk \!-\! \veps_\mkq}
        - \frac{f^{\rm base}_\nk \!-\! f^{\rm base}_\mkq}{\veps_\nk \!-\! \veps_\mkq}
    \bigg]
    \nnnl
    &\qquad\qquad\times M^*_{\mnkq,Y} M_{\mnkq,Y'}
    \,,
\end{align}
where $Y$ and $Y'$ can be WF pair indices $I$ or atomic displacement indices $a$, and
\begin{align}
    M_{\mnkq,I}
    &\equiv \bbrakket{\Psi_\mnkq}{\Phi_{I\bq}}
    = U^*_{im\bkq} U_{jn\bk} e^{i\bk\cdot\bR}\! \,, \label{eq:rpa_M_I}
    \\
    M_{\mnkq,a}
    &\equiv \bbrakket{\Psi_\mnkq}{V_{a\bq}^{\rm p}}
    = g^{\rm p}_{a\bq,mn\bk} \,. \label{eq:rpa_M_a}
\end{align}

We now derive the dynamically screened e-ph vertex $V^{\rm s}_{a\bq}(\omega)$, which is needed for the total e-e interaction (\cref{sec:theory_Wtot}).
This quantity can be obtained by solving the active-space screening equation [\cref{eq:rpa_Hs_downfold}].
Substituting \cref{eq:eph_Wp_trunc} into it, we have
\begin{multline}
    \kket{V_{a\bq}^{\rm s}(\omega)}
    = \kket{V_{a\bq}^{\rm p}}
    + \sum_{II'}^{d^{\rm cut}} \kket{\Phi_{I\bq}} W^{\rm p}_{II'\bq}
    \\
    \times \bbra{\Phi_{I'\bq}} \, \opchiq^{\rm b,dyn}(\omega) \, \kket{V_{a\bq}^{\rm s}(\omega)} \,.
\end{multline}
The second term is a linear combination of the WF pairs $\kket{\Phi_{I\bq}}$, with coefficients that depend on $V^{\rm s}_{a\bq}(\omega)$ only through its projection onto the subspace.
Solving for this projection (\cref{app:downfold_derivation}), we find
\begin{equation} \label{eq:eph_gs_rpa_trunc_2}
    \kket{V_{a\bq}^{\rm s}(\omega)}
    = \kket{V_{a\bq}^{\rm p}}
    + \sum_{II'} \kket{\Phi_{I\bq}} \, W^{\rm s}_{II'\bq}(\omega) \, X^{\rm b,dyn}_{I'a\bq}(\omega)
    \,,
\end{equation}
where $W^{\rm s}_{II'\bq}(\omega)$ is the fully-screened e-e interaction in the truncated WF pair basis [\cref{eq:eph_Ws_dyson}].
This is the final expression for the dynamically screened e-ph vertex: the active-space electrons respond to the phonon perturbation through the susceptibility $X^{\rm b,dyn}_{I'a\bq}(\omega)$.
The screened e-ph vertex includes renormalization of the e-ph matrix elements between the kept WF pairs, while those between the truncated pairs are not updated.

Keeping only the first term of \cref{eq:eph_phself_matrix} gives the bubble (one-loop) approximation
\begin{equation} \label{eq:phself_1loop}
    \Pi^{\rm A, bubble}_{aa'\bq}(\omega)
    = \bbra{V^{\rm p}_{a\bq}} \, \opchiq^{\rm b,dyn}(\omega) \, \kket{V^{\rm p}_{a'\bq}} \,,
\end{equation}
in which both e-ph vertices are the partially-screened ones, i.e., the screening of the vertex by the active space is neglected.
In DFPT+A downfolding, $V^{\rm p}$ is the static DFPT vertex of the base system, and \cref{eq:phself_1loop} is the ``screened-screened'' approximation used in nonadiabatic phonon spectral function calculations~\cite{Calandra2010, Berges2023}.
In cDFPT+A, $V^{\rm p}$ is the cDFPT vertex, which may significantly differ from the target vertex in metals, as the cDFPT vertex does not include the free-carrier screening of the long-range Coulomb interaction.
We note that the two schemes differ only once approximations, such as the bubble approximation or the truncation of $W^{\rm p}$, are made.
With the full self-energy [\cref{eq:eph_phself_matrix}], the dynamical IFC $C^{\rm s}_{aa'\bq}(\omega) = C^{\rm p}_{aa'\bq} + \Pi^{\rm A}_{aa'\bq}(\omega)$ is independent of the downfolding scheme, as we show in \cref{app:cDFPT}.

Finally, we obtain the phonon propagator and the phonon spectral function from the dynamical IFC.
Defining the partially-screened phonon propagator
\begin{equation} \label{eq:eph_Dp}
    D^{\rm p}_{aa'\bq}(\omega)
    \equiv \bigl( \omega^2 M - C^{\rm p}_{\bq} \bigr)^{-1}_{aa'} \,,
\end{equation}
where $M_{aa'} \equiv M_\kappa \delta_{aa'}$ is the diagonal matrix of the atomic masses, with $a = (\kappa, \alpha)$,
the fully-screened phonon propagator is given by
\begin{align} \label{eq:eph_Ds_dyson}
    D^{\rm s}_{aa'\bq}(\omega)
    &= \bigl[ \omega^2 M - C^{\rm s}_{\bq}(\omega) \bigr]^{-1}_{aa'}
    \nnnl
    &= \bigl( [D^{\rm p}_{\bq}(\omega)]^{-1} - \Pi^{\rm A}_{\bq}(\omega) \bigr)^{-1}_{aa'} \,.
\end{align}
The adiabatically screened phonon frequencies $\omega^{\rm s}_{\nu\bq}$ and eigenvectors $e^{\rm s}_{a,\nu\bq}$ of the target system are obtained from the static IFCs $C^{\rm s}_\bq(0) = C^{\rm p}_\bq + \Pi^{\rm A}_\bq(0)$ by diagonalizing the dynamical matrix,
\begin{equation} \label{eq:ph_eig}
    \sum_{a'} \frac{C^{\rm s}_{aa'\bq}(0)}{\sqrt{M_\kappa M_{\kappa'}}} \, e^{\rm s}_{a',\nu\bq}
    = \bigl( \omega^{\rm s}_{\nu\bq} \bigr)^2 \, e^{\rm s}_{a,\nu\bq} \,.
\end{equation}
The screened e-ph matrix element in the phonon eigenmode basis is
\begin{equation} \label{eq:g_mode}
    g^{\rm s}_{\nu\bq,mn\bk}(\omega)
    \equiv \sqrt{\frac{\hbar}{2 \omega^{\rm s}_{\nu\bq}}} \sum_{a} \frac{e^{\rm s}_{a,\nu\bq}}{\sqrt{M_\kappa}} \, g^{\rm s}_{a\bq,\mnk}(\omega) \,,
\end{equation}
where $g^{\rm s}_{a\bq,\mnk}(\omega) = \bbrakket{\Psi_\mnkq}{V^{\rm s}_{a\bq}(\omega)}$.
The phonon spectral function is computed as
\begin{equation} \label{eq:eph_phspec_def}
    B_{\bq}(\omega)
    = -\frac{2\omega}{\pi} \Im \sum_{a} M_\kappa D^{\rm s}_{aa\bq}(\omega) \,.
\end{equation}
With the factor $2\omega$, each phonon mode contributes a peak of unit weight: for frequency-independent IFCs, $B_\bq(\omega) = \sum_\nu \delta(\omega - \omega^{\rm s}_{\nu\bq})$ for $\omega > 0$.

\subsection{Total dynamical e-e interaction} \label{sec:theory_Wtot}

The previous subsection illustrates a practical recipe to compute the nonadiabatic phonon self-energy using only matrix inversion and multiplications in the truncated WF pair and phonon bases.
Now, we assemble the total dynamical e-e interaction $\opWq^{\rm tot}(\omega)$ from the fully-screened e-e interaction $W^{\rm s}_{II'\bq}(\omega)$, e-ph vertex $V^{\rm s}_{a\bq}(\omega)$, and phonon propagator $D^{\rm s}_{aa'\bq}(\omega)$ [\cref{eq:hyb_W_tot_def}].
This quantity describes the interaction between two electrons, mediated by the screened Coulomb interaction and by the exchange of a phonon, which also screen each other.

The total bare interaction is the sum of the bare Coulomb interaction and the exchange of a bare phonon,
\begin{equation}
    \hat{v}^{\rm tot}_\bq(\omega)
    = \opvq + \sum_{aa'} \kket{V^{\rm b}_{a\bq}} D^{\rm b}_{aa'\bq}(\omega) \bbra{V^{\rm b}_{a'\bq}} \,,
\end{equation}
with $D^{\rm b}_\bq(\omega) = ( \omega^2 M - C^{\rm b}_\bq )^{-1}$.
Screening this interaction by the full susceptibility gives the total screened interaction
\begin{equation} \label{eq:hyb_W_tot_dyson}
    \opWq^{\rm tot}(\omega)
    = \bigl[ \opone - \hat{v}^{\rm tot}_\bq(\omega) \, \opchiq^{\rm b}(\omega) \bigr]^{-1} \hat{v}^{\rm tot}_\bq(\omega) \,.
\end{equation}
The operator to be inverted, $\opone - \hat{v}^{\rm tot}_\bq \opchiq^{\rm b} = \opepsq^{\rm s} - \sum_{aa'} \kket{V^{\rm b}_{a\bq}} D^{\rm b}_{aa'\bq} \bbra{V^{\rm b}_{a'\bq}} \opchiq^{\rm b}$, differs from the dielectric function $\opepsq^{\rm s}$ by a term of rank $N^{\rm ph}$.
We therefore invert it with the Woodbury matrix identity, with which we find
\begin{multline} \label{eq:hyb_woodbury_inv}
    \bigl[ \opone - \hat{v}^{\rm tot}_\bq \opchiq^{\rm b} \bigr]^{-1}
    = (\opepsq^{\rm s})^{-1}
    + \sum_{aa'} \kket{V^{\rm s}_{a\bq}} \, D^{\rm s}_{aa'\bq} \, \bbra{V^{\rm b}_{a'\bq}}
    \\
    \times \opchiq^{\rm b} \, (\opepsq^{\rm s})^{-1} \,,
\end{multline}
where we drop the frequency arguments for brevity.
Multiplying by $\hat{v}^{\rm tot}_\bq$ from the right, and using $(\opepsq^{\rm s})^{-1} \opvq = \opWq^{\rm s}$ and $D^{\rm b} + D^{\rm s} (C^{\rm s} - C^{\rm b}) D^{\rm b} = D^{\rm s}$, gives
\begin{multline} \label{eq:hyb_W_tot_woodbury}
    \opWq^{\rm tot}(\omega)
    = \opWq^{\rm s}(\omega)
    \\
    + \sum_{aa'} \kket{V^{\rm s}_{a\bq}(\omega)} \, D^{\rm s}_{aa'\bq}(\omega) \, \bbra{V^{\rm b}_{a'\bq}} \bigl[ \opone + \opchiq^{\rm b}(\omega) \, \opWq^{\rm s}(\omega) \bigr] \,,
\end{multline}
where the bra in the second term is that of the frequency-dependent vertex,
\begin{align} \label{eq:eph_gs_bra}
    \bbra{V_{a\bq}^{\rm s}(\omega)}
    &\equiv \bbra{V_{a\bq}^{\rm b}} \bigl[ \opone + \opchiq^{\rm b}(\omega) \, \opWq^{\rm s}(\omega) \bigr]
    \nnnl
    &= \bbra{V_{a\bq}^{\rm p}} \bigl[ \opone + \opchiq^{\rm b,dyn}(\omega) \, \opWq^{\rm s}(\omega) \bigr] \,,
\end{align}
which is evaluated at the same frequency as the ket [\cref{eq:eph_gs_rpa_trunc_2}].
The second line, which follows from \cref{eq:rpa_Vp_bra}, is the bra counterpart of \cref{eq:rpa_Hs_downfold_inverse}.
Inserting the truncated $\opWq^{\rm s}$ [\cref{eq:eph_Ws_dyson}] into \cref{eq:hyb_W_tot_woodbury}, we obtain
\begin{align} \label{eq:hyb_W_tot_def}
    \opWq^{\rm tot}(\omega)
    &= \sum_{II'} \kket{\Phi_{I\bq}} \, W^{\rm s}_{II'\bq}(\omega) \, \bbra{\Phi_{I'\bq}}
    \nnnl
    &+ \sum_{aa'} \kket{V^{\rm s}_{a\bq}(\omega)} \, D^{\rm s}_{aa'\bq}(\omega) \, \bbra{V^{\rm s}_{a'\bq}(\omega)}
    \,.
\end{align}
The first term is the screened Coulomb interaction, and the second term is the phonon-mediated interaction.
In the latter, the e-ph vertex $V^{\rm s}_{a\bq}(\omega)$ is dynamically screened by the electrons.
Similar sums of electronic and phonon-mediated interactions have been computed in the WF basis for insulators and dilute doped semiconductors, but with adiabatic phonons and e-ph vertices~\cite{Tubman2025, Coveney2025}, and such a single effective interaction was recently used to unify the $GW$ and Fan--Migdal self-energies of a correlated metal~\cite{Berges2026}.
The interaction between two electronic states $\nk$ and ${m\bkq}$ is a matrix element of this operator,
\begin{equation} \label{eq:W_mels}
    W^{\rm tot}_{\mnkq}(\omega)
    = \bbra{\Psi_\mnkq} \opWq^{\rm tot}(\omega) \kket{\Psi_\mnkq} \,.
\end{equation}

We decompose the total e-e interaction into the electronic, phononic, and e-ph vertex contributions as
\begin{equation}
    \opWq^{\rm tot}(\omega)
    = \opWq^{\rm el}(\omega)
    + \opWq^{\rm ph}(\omega)
    + \opWq^{\rm \text{e-ph}}(\omega)
    \,,
\end{equation}
where
\begin{subequations} \label{eq:W_decomposition}
\begin{align}
    \label{eq:W_decomposition_elel}
    \opWq^{\rm el}(\omega)
    &= \sum_{II'} \kket{\Phi_{I\bq}} W^{\rm s}_{II'\bq}(\omega) \bbra{\Phi_{I'\bq}} \,,
    \\
    \label{eq:W_decomposition_phph}
    \opWq^{\rm ph}(\omega)
    &= \sum_{aa'} \kket{V_{a\bq}^{\rm p}} D^{\rm s}_{aa'\bq}(\omega) \bbra{V_{a'\bq}^{\rm p}} \,,
    \\
    \label{eq:W_decomposition_elph}
    \opWq^{\rm \text{e-ph}}(\omega)
    &= \sum_{aa'} \kket{V^{\rm s}_{a\bq}(\omega)} D^{\rm s}_{aa'\bq}(\omega) \bbra{V^{\rm s}_{a'\bq}(\omega)}
    - \opWq^{\rm ph}(\omega) \,.
\end{align}
\end{subequations}
The first term is a purely electronic contribution, not affected by any e-ph interaction.
The second term is the phonon-mediated interaction without any vertex screening.
From this second term, we can understand the change in the phonon spectral function and its effect on the e-e interaction.
The last term describes the correction due to the frequency-dependent vertex screening.
We compute these decompositions to gain further insight into how the screening from phonons and electrons manifests in the total interaction.

We emphasize that the screening entering $\opWq^{\rm tot}(\omega)$ is the full screening at the level of TDDFPT.
Apart from the approximations of DFPT+A downfolding [\cref{eq:rigid_band,eq:chib_approx1,eq:chib_approx2}] and the truncation of the partially-screened e-e interaction [\cref{eq:eph_Wp_trunc}], the total interaction does not depend on the choice of the base system or of the separation of the susceptibility.

\subsection{Local dynamical interaction and the Eliashberg function} \label{sec:theory_Uloc}

In this subsection, we define the local dynamical interaction and its retarded part $\Delta U$ to quantify the strength of the total interaction and to compare it with the conventional adiabatic description.
The local dynamical interaction reads
\begin{equation} \label{eq:U_loc_def}
    U_i^{\rm tot}(\omega)
    \equiv \intbq W^{\rm tot}_{I_0 I_0 \bq}(\omega) \,,
    \qquad I_0 \equiv (i, i, \mb{0}) \,,
\end{equation}
where $W^{\rm tot}_{II'\bq}(\omega) \equiv \bbra{\Phi_{I\bq}} \opWq^{\rm tot}(\omega) \kket{\Phi_{I'\bq}}$.
It describes the onsite density-density interaction of a single WF $i$.
For the purely electronic part, $U^{\rm el}_i(\omega)$ is the onsite element of the fully-screened e-e interaction $\opWq^{\rm s}(\omega)$.
The total interaction $U_i^{\rm tot}(\omega)$ additionally contains the phonon-mediated contribution~\cite{Nomura2015, Berges2026} and the dynamical e-ph vertex screening effects.
Following \cref{eq:W_decomposition}, and dropping the WF index $i$, we decompose it as
\begin{equation} \label{eq:U_loc_decomposition}
    U^{\rm tot}(\omega)
    = U^{\rm el}(\omega)
    + U^{\rm ph}(\omega)
    + U^{\rm \text{e-ph}}(\omega) \,.
\end{equation}
The frequency dependence of the local interaction is characterized by its spectral function, $-\Im U^{\rm tot}(\omega) / \pi$, which is non-negative for $\omega > 0$, while the individual contributions in \cref{eq:U_loc_decomposition} can have either sign.

To better understand the role of dynamical screening, we compare $U^{\rm tot}(\omega)$ with the \textit{adiabatic approximation}, in which the frequency dependence of the dynamical correction $\opchiq^{\rm b,dyn}(\omega)$ is switched off and approximated by its static value.
The dynamical interaction is then the conventional phonon-mediated one, with the static screened vertex and the adiabatic phonon propagator,
\begin{subequations} \label{eq:W_adiabatic}
\begin{align}
    \hat{W}^{\rm ad}_\bq(\omega)
    &= \sum_{aa'} \kket{V^{\rm s}_{a\bq}(0)} D^{\rm ad}_{aa'\bq}(\omega) \bbra{V^{\rm s}_{a'\bq}(0)} \,,
    \\
    D^{\rm ad}_{aa'\bq}(\omega)
    &= \bigl( \omega^2 M - C^{\rm p}_\bq - \Pi^{\rm A}_\bq(0) \bigr)^{-1}_{aa'} \,.
\end{align}
\end{subequations}
Using downfolding, one can compute the adiabatically screened parameters at a smaller computational cost, still capturing the effect of doping and temperature on the screening.
The momentum average [\cref{eq:U_loc_def}] of $\hat{W}^{\rm ad}_\bq(\omega)$ is called $U^{\rm ad}(\omega)$.
It depends on frequency only through the poles of $D^{\rm ad}_\bq(\omega)$, i.e., the adiabatic phonon frequencies $\omega^{\rm s}_{\nu\bq}$ [\cref{eq:ph_eig}], and vanishes at high frequency.
Its spectral function $-\Im U^{\rm ad}(\omega) / \pi$ is thus a sum of delta peaks weighted by the density-density matrix elements of $V^{\rm s}_{a\bq}(0)$, the local counterpart of the Eliashberg function [\cref{eq:a2F}].
Compared with $U^{\rm tot}(\omega)$, it lacks the dynamically screened Coulomb interaction, the nonadiabatic phonon renormalization, and the frequency-dependent vertex screening.

To relate the local interaction to conventional Eliashberg theory, we recall that the isotropic Eliashberg function is the Fermi-surface average of the squared adiabatic e-ph matrix elements, resolved in phonon frequency~\cite{McMillan1968, Giustino2017},
\begin{align} \label{eq:a2F}
    \alpha^2F(\omega)
    &= \frac{1}{N(E^{\rm F})} \intbk \, \intbq \sum_{mn\nu}
    \abs{g^{\rm s}_{\nu\bq,mn\bk}(0)}^2 \, \delta(\omega - \omega^{\rm s}_{\nu\bq})
    \nnnl
    &\quad \times \delta(\varepsilon_{n\bk} - E^{\rm F}) \,
    \delta(\varepsilon_\mkq - E^{\rm F}) \,,
\end{align}
where $N(E^{\rm F})$ is the density of states per spin at the Fermi energy $E^{\rm F}$.
Its inverse-frequency moment defines the e-ph coupling constant $\lambda = 2 \int_0^\infty \dd\omega \, \alpha^2F(\omega) / \omega$~\cite{McMillan1968}.
The spectral function $-\Im U^{\rm tot}(\omega) / \pi$ is the generalization of $\alpha^2F(\omega)$ to the total interaction: it retains the phonon peaks, adds the electronic spectral weight of the plasmon and of the electron-hole continuum, and corrects both through the e-ph vertex renormalization.

In analogy with $\lambda$, with the spectral function $-\Im U^{\rm tot}(\omega) / \pi$ in place of $\alpha^2F(\omega)$, we quantify the strength of the local interaction by its retarded part
\begin{align} \label{eq:lambda_loc}
    \Delta U
    &\equiv 2 \int_0^\infty \frac{\dd\omega}{\omega} \Bigl[ -\frac{1}{\pi} \Im U^{\rm tot}(\omega) \Bigr]
    \nnnl
    &= U^{\rm tot}(\infty) - \Re U^{\rm tot}(0) \,,
\end{align}
where the second equality follows from the Kramers--Kronig relation.
$\Delta U$ is real, since $U^{\rm tot}(\infty)$ is real, as we discuss below.
The high-frequency limit $U^{\rm tot}(\infty)$ is to be understood within the downfolded model, at frequencies above the active-space excitation and phonon energies, where the phonon propagator and the target active-space susceptibility $\opchiq^{\rm b,tar,A}(\omega)$ vanish.
Taken literally, the $\omega \to \infty$ limit would give the bare interaction, but the static approximation for the rest susceptibility [\cref{eq:chib_approx1}] breaks down before that limit is reached.
Being instantaneous, $U^{\rm tot}(\infty)$ does not contribute to the imaginary part of the interaction, so $\Delta U$ measures only the retarded part of the local interaction, quantifying the strength of the dynamical screening.
We note that, unlike $\lambda$, $\Delta U$ has units of energy and contains no density-of-states factor, which is ill-defined for a nondegenerate semiconductor.

In practice, we evaluate $U^{\rm tot}(\omega)$ at $\omega + i\eta$ with a small positive $\eta$.
The positive sign of $\eta$ selects the retarded interaction, and the finite value of $\eta$ broadens the delta peaks of the phonons and of the electron-hole excitations, which are otherwise not resolved on a discrete $\bk$ grid (see \cref{app:comp_details} for the values of $\eta$).
We also approximate the Brillouin-zone average in \cref{eq:U_loc_def} by the average of one-dimensional integrals along two high-symmetry lines (\cref{app:Uloc_integration}).

\section{Applications} \label{sec:applications}

We apply the framework to two monolayers that realize two limits of the active-space screening.
In hBN, the carriers are extrinsic and their density can be tuned, so that the plasmon can be moved through the phonon spectrum (\cref{sec:result_hBN}).
In metallic TaS$_2$, the carriers are intrinsic, and the low-energy electronic excitations are those of a partially filled band (\cref{sec:result_TaS2}).
The active spaces and WFs are those of \cref{sec:theory_demo}: three N $p$ WFs for hBN, and three Ta $d$ WFs for TaS$_2$.

\subsection{Doped monolayer hBN} \label{sec:result_hBN}

\begin{figure*}[tb]
\includegraphics[width=0.99\linewidth]{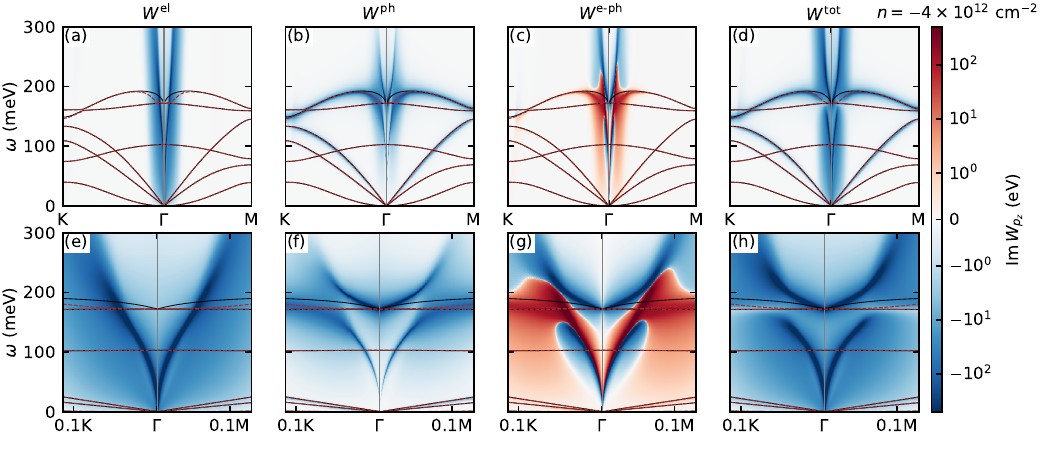}
\caption{
    Decomposition of the total e-e interaction of monolayer hBN at $T=300~\mathrm{K}$ and $n = -4 \times 10^{12}~\mathrm{cm^{-2}}$ into the (a) electronic, (b) phononic, and (c) e-ph vertex screening contributions of \cref{eq:W_decomposition}, and (d) their sum.
    We note that the phononic contribution contains the fully-screened phonon propagator $D^{\rm s}_{aa'\bq}(\omega)$ [\cref{eq:W_decomposition_phph}], which includes the nonadiabatic renormalization of the phonons.
    We show the imaginary part of the density-density element $W_{I_0 I_0 \bq}(\omega)$ for the N $p_z$ WF.
    Black solid and red dashed lines indicate the adiabatic phonon dispersions of the undoped and doped crystal, respectively.
    (e) to (h) Zoom on the region where the plasmon and the LO phonon anticross.
}
\label{fig:hBN_W_decomp}
\end{figure*}

\begin{figure*}[tb]
\includegraphics[width=0.99\linewidth]{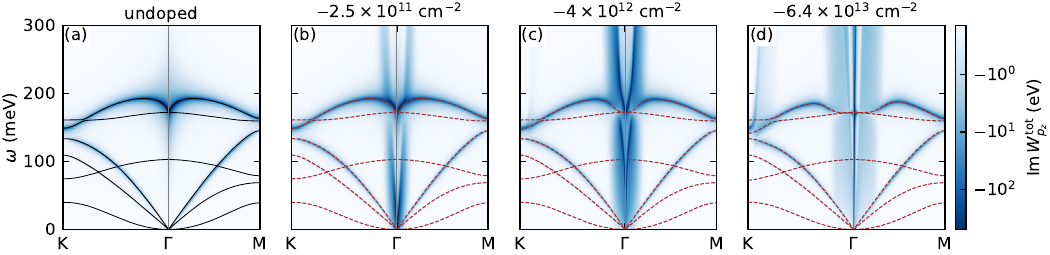}
\caption{
    Doping dependence of the total e-e interaction of monolayer hBN at $T=300~\mathrm{K}$, for (a) the undoped crystal and for carrier concentrations (b) $n = -2.5 \times 10^{11}$, (c) $-4 \times 10^{12}$, and (d) $-6.4 \times 10^{13}~\mathrm{cm^{-2}}$.
    The black solid line in (a) and the red dashed lines in (b) to (d) indicate the adiabatic phonon dispersions of the undoped and doped crystal, respectively.
    See \cref{fig:hBN_W_decomp} for further details.
    The feature near K in (d) originates from K-K$'$ intervalley transitions.
}
\label{fig:hBN_W_doping}
\end{figure*}

Since hBN is a two-dimensional semiconductor, upon doping it develops a plasmon with a $\sqrt{q}$ dispersion.
Depending on the carrier concentration, the wavevector where the plasmon crosses the phonon energies and the strength of the plasmon-phonon hybridization change.

\begin{figure}[tb]
\includegraphics[width=0.99\linewidth]{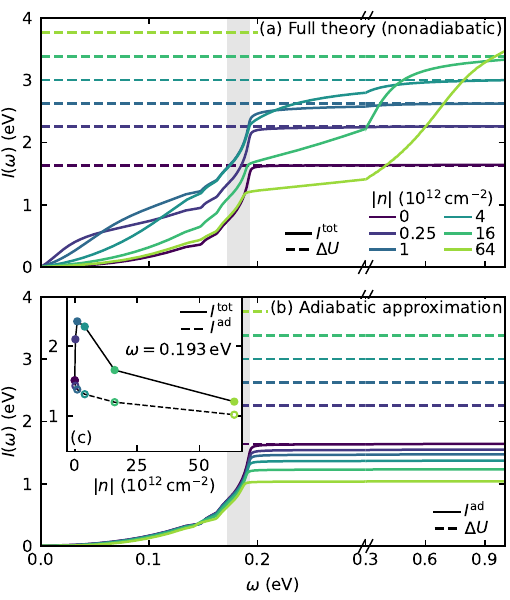}
\caption{
    Cumulative frequency integral [\cref{eq:I_cumulative}] of the local interaction of monolayer hBN at $T=300~\mathrm{K}$ for (a) the full nonadiabatic theory, $I^{\rm tot}(\omega)$, and (b) the adiabatic approximation, $I^{\rm ad}(\omega)$ [\cref{eq:W_adiabatic}].
    The dashed horizontal lines indicate $\Delta U$ [\cref{eq:lambda_loc}], which is the $\omega \to \infty$ limit of the integral in (a).
    The gray region indicates the frequency range of the LO phonon.
    The horizontal frequency axis scale is compressed above 0.3~eV.
    (c) The cumulative integral at the highest LO phonon frequency (0.193~eV) as a function of carrier concentration, for the full theory (filled symbols) and the adiabatic approximation (open symbols).
}
\label{fig:hBN_W_loc_kk}
\end{figure}

We use DFPT+A downfolding [\cref{eq:chib_DFPT}] with the undoped crystal as the base system.
The reference susceptibility is then that of the insulator, and the entire free-carrier response, including the plasmon, is carried by the active space [\Fig[a]{fig:downfolding_schematic}].
Since the doping enters only through the occupations $f^{\rm tar}_\nk$ of the active-space susceptibility, a single base calculation suffices for all doping values.
The partially-screened interaction $W^{\rm p}_\bq$ of the insulating base system retains the $1/q$ singularity of the two-dimensional Coulomb interaction, which we treat with the monopole-monopole term of \cref{eq:itp_Wp_monopole}.
We truncate the short-range part to the onsite N $p$-orbital WF pairs ($d^{\rm cut} = d^{\rm onsite}$, \cref{tab:cutoff}), and set the active window to contain the valence bands within 1.5~eV of the VBM [\Fig[a]{fig:bands}].
This window is much wider than the phonon energies and covers the frequency range considered in this work (up to 1~eV, \cref{fig:hBN_W_loc_kk}), so that all transitions involving the doped holes in this range are treated dynamically.

\Cref{fig:hBN_W_decomp} shows the total e-e interaction of hole-doped hBN at $n = -4 \times 10^{12}~\mathrm{cm^{-2}}$ and $T = 300~\mathrm{K}$ and its decomposition [\cref{eq:W_decomposition}] for the density-density interaction of the N $p_z$ WF, which describes the VBM.
We show the imaginary part, which is directly related to the energy- and momentum-resolved strength of the interaction: it gives the spectrum of the electronic and phonon fluctuations that mediate the interaction between two electrons in the $p_z$ WF.
Its Brillouin-zone average is the spectral function $-\Im U^{\rm tot}(\omega) / \pi$ of the local interaction, so the momentum-resolved map shows from which bosonic excitations, and at which $\bq$ and $\omega$, the retarded local interaction $\Delta U$ [\cref{eq:lambda_loc}] originates.
Causality requires $\Im W^{\rm tot}_{I_0 I_0 \bq}(\omega) \leq 0$ for $\omega > 0$, which holds for the total interaction [\Fig[d,h]{fig:hBN_W_decomp}], while the e-ph vertex screening term alone can take either sign [\Fig[c,g]{fig:hBN_W_decomp}].
\Figu[a,e]{fig:hBN_W_decomp} show the e-e contribution from the plasmon mode, which has a $\sqrt{q}$ dispersion and crosses the undoped optical phonons around $\bq = 0.05\,\bq^{\rm K}$ and $0.05\,\bq^{\rm M}$, where $\bq^{\rm K}$ and $\bq^{\rm M}$ are the wavevectors of the K and M points.
The plasmon-phonon hybridization leads to the splitting of the phonon spectral function into two branches~\cite{Yokota1961, Varga1965, Lihm2024PlPh, Macheda2024b}, reflected in the phonon-mediated interaction shown in \Fig[b,f]{fig:hBN_W_decomp}.
The lower branch starts at zero frequency near $\Gamma$ and approaches the phonon frequency at large $q$, while the upper branch starts at the phonon frequency and further increases following the plasmon dispersion.
The same two branches appear directly in the phonon spectral function [\Fig[a,c]{fig:hBN_ph_spectral}], which we discuss later.

\Figu[c,g]{fig:hBN_W_decomp} show the e-ph vertex term [\cref{eq:W_decomposition_elph}], which accounts for the dynamical screening of the e-ph vertex.
Calculating this term is the key contribution of this work, made possible by combining Wannier-basis screened Coulomb interactions with the joint e-e and e-ph downfolding framework.
We find a nontrivial energy and momentum dependence and a sign structure.
Around the lower and upper plasmon-phonon branches, the e-ph vertex term strengthens the interaction, while in between it weakens the interaction.
This structure reflects the frequency dependence of the screened e-ph vertex $V^{\rm s}_{a\bq}(\omega)$: the LO vertex is resonantly enhanced along the plasmon dispersion, so that the e-ph vertex term is largest where the plasmon-phonon branches lie.
The total interaction is shown in \Fig[d,h]{fig:hBN_W_decomp}, which is the sum of the three contributions.
The e-ph vertex screening term suppresses the direct plasmon contribution, so that the total interaction is dominated by the two plasmon-phonon hybrid branches.

\Cref{fig:hBN_W_doping} shows the doping dependence of the total e-e interaction in monolayer hBN.
For the undoped case, the dynamical e-e interactions are purely phonon-mediated, with the strongest contribution from the LO branch.
In the doped case, the plasmon-phonon hybrid emerges in the small $q$ region, and the interaction shifts to higher energies with increasing carrier concentration.

In two-dimensional materials, the long-range Fr\"ohlich interaction does not split the LO and transverse optical (TO) phonons at $\Gamma$, but gives the LO phonon a steep linear dispersion near $\Gamma$~\cite{Sohier2017} [\Fig[a]{fig:hBN_W_doping}].
The plasmon of the doped carriers changes this dispersion~\cite{Macheda2024b}, in a way that depends on the doping.
At low doping [\Fig[b]{fig:hBN_W_doping}], the plasmon frequency stays below the LO frequency in the momentum range of this linear dispersion, so the carriers cannot follow the LO oscillation, and the dispersion is nearly unchanged.
At intermediate doping [\Fig[c]{fig:hBN_W_doping}], the LO phonon and the plasmon strongly hybridize, and the upper branch is even steeper than the undoped LO phonon, as it starts as the LO phonon at $\Gamma$ but rapidly evolves into a plasmon away from $\Gamma$.
Finally, when the plasmon energy is much higher than the phonon energies, the long-range Fr\"ohlich coupling of the LO phonon is screened by the carriers, which removes its linear dispersion [\Fig[d]{fig:hBN_W_doping}].
Even then, since the plasmon frequency vanishes as $\sqrt{q}$, it falls below the LO frequency at sufficiently small $q$, so the linear LO dispersion should be present at any doping in a region around $\Gamma$ that shrinks with increasing doping.

The feature near the K point in \Fig[d]{fig:hBN_W_doping} is due to K-K$'$ intervalley transitions.
The K and K$'$ valleys are degenerate in energy, and the transition between them carries a crystal momentum close to K.
Similar to the intraband transitions near $\Gamma$, these intervalley transitions form an electron-hole continuum whose energy starts from zero at $\bq = \bq^{\rm K}$ and increases with $\abs{\bq - \bq^{\rm K}}$, and which hybridizes with the optical phonons near K.

\begin{figure}[tb]
\includegraphics[width=0.99\linewidth]{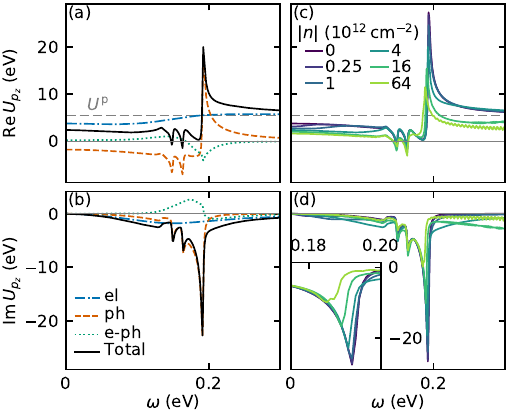}
\caption{
    Local dynamical interaction of monolayer hBN [\cref{eq:U_loc_def,eq:U_loc_decomposition}].
    (a), (b) Real and imaginary parts of the local e-e interaction of hBN at $T=300~\mathrm{K}$ and $n=-4\times10^{12}~\mathrm{cm^{-2}}$, and its decomposition into the electronic, phononic, and e-ph vertex contributions.
    The gray dashed line indicates the local partially-screened interaction $U^{\rm p}$, the $\omega \to \infty$ limit of $\Re U^{\rm tot}(\omega)$.
    (c), (d) The total local e-e interaction of hBN at various carrier concentrations.
    The inset in (d) zooms in on the LO phonon resonance.
    For the doped cases, $\Im U^{\rm tot}(\omega)$ remains finite above the phonon frequencies because of the plasmon and the intraband electron-hole continuum, and it vanishes only at higher frequencies (see \cref{fig:hBN_W_loc_kk}).
}
\label{fig:hBN_W_loc}
\end{figure}

To quantitatively analyze the strength of the total interaction, we compute the total local interaction $U^{\rm tot}(\omega)$ and its spectral function $-\Im U^{\rm tot}(\omega) / \pi$ (\cref{sec:theory_Uloc}).
They are the Brillouin-zone averages of the density-density element shown in \cref{fig:hBN_W_decomp,fig:hBN_W_doping}, approximated by the integral along the $\Gamma$K and $\Gamma$M lines (\cref{app:Uloc_integration}).
\Figu[a,b]{fig:hBN_W_loc} show the real and imaginary parts of this local interaction and its decomposition into the electronic, phononic, and e-ph vertex contributions.
At low frequencies, the real part of the total interaction is smaller than the local partially-screened interaction $U^{\rm p}$, the $\bq$ average of $W^{\rm p}_{I_0 I_0 \bq}$.
Since the undoped base system has no active-space screening, $U^{\rm p}$ is the high-frequency limit $U^{\rm tot}(\infty)$ of \cref{eq:lambda_loc}.
Both electronic screening and lattice screening contribute to the reduction of the interaction.
Around the LO phonon frequency (from 170 to 195~meV), we find a sharp peak in the imaginary part of the total interaction, which is dominated by the phonon-mediated contribution.
The two weaker peaks at about 150 and 165~meV come from the zone-boundary phonons at K and M, respectively, and are likely overemphasized due to the one-dimensional approximation of the Brillouin-zone integral (\cref{app:Uloc_integration}).
The e-ph vertex term has the opposite sign to the phonon-mediated term.

\Figu[c,d]{fig:hBN_W_loc} show the doping dependence of the total local interaction.
As the carrier concentration increases, the sharp peak of the imaginary part around 190~meV [inset of \Fig[d]{fig:hBN_W_loc}] is reduced.
This spectral weight is not lost but redistributed: the plasmon-phonon hybridization transfers it to the lower and upper hybrid branches (\cref{fig:hBN_W_doping}), below and above the LO frequency.
Hence, the real part of the total interaction still monotonically decreases with increasing carrier concentration, even though the phonon-mediated attraction is suppressed.
This result implies that one cannot simply use the phonon-mediated interaction to describe the low-energy physics of doped semiconductors, as one would miss the transfer of spectral weight and thus underestimate the interaction.

The strength of the total interaction can be quantified by the retarded local interaction $\Delta U$ of \cref{eq:lambda_loc}.
To resolve which excitations contribute to it, we define the cumulative integral
\begin{equation} \label{eq:I_cumulative}
    I^{\rm tot}(\omega)
    \equiv 2 \int_0^\omega \frac{\dd\omega'}{\omega'} \Bigl[ -\frac{1}{\pi} \Im U^{\rm tot}(\omega') \Bigr]
    \, \xrightarrow{\omega \to \infty} \,
    \Delta U \,,
\end{equation}
where the limit follows from \cref{eq:lambda_loc}.
We define $I^{\rm ad}(\omega)$ analogously from the adiabatic reference $U^{\rm ad}(\omega)$ of \cref{eq:W_adiabatic}.
\Figu[a]{fig:hBN_W_loc_kk} shows that the $\omega \to \infty$ limit of the integral $I^{\rm tot}(\omega)$ agrees with the direct calculation using the real part of the local interaction at zero frequency, confirming the Kramers--Kronig relation.
We also show in \Fig[a]{fig:hBN_W_loc_kk} that, for the highest carrier concentration, the integral must extend beyond 1~eV for full convergence.
The limit $\Delta U$ increases monotonically with the carrier concentration.
Since $U^{\rm p}$ is fixed by the undoped base system, this growth reflects the stronger screening of the static interaction $\Re U^{\rm tot}(0)$ [\Fig[c]{fig:hBN_W_loc}], i.e., a weaker repulsion at zero frequency.
For the undoped case, there is a dominant contribution in the LO phonon frequency range from 170 to 195~meV.
Upon doping, additional low-energy contributions emerge from the plasmon, which are successively transferred to higher energies as the carrier concentration increases.

We compare this result with the adiabatic approximation [\cref{eq:W_adiabatic}], where the only dynamical interaction comes from the phonons.
Since $U^{\rm ad}(\omega)$ vanishes at high frequency, the adiabatic integral saturates at $I^{\rm ad}(\infty) = -\Re U^{\rm ad}(0)$, the static phonon-mediated attraction, rather than at $\Delta U$.
In other words, the adiabatic approximation misses the retardation of the direct e-e interaction.
In addition, \Fig[b]{fig:hBN_W_loc_kk} shows that this dynamical interaction is monotonically suppressed upon doping, as the carriers statically screen the LO e-ph coupling.
If we focus on the low-energy regime $\omega < 0.2~\mathrm{eV}$, this result agrees with the full calculation only in the two limits of undoped and highly doped cases, where the plasmon is either absent or at a high energy.
In the latter case, the adiabatic approximation cannot capture the high-energy plasmon contribution, but this term is not relevant for low-energy physics.
However, in the intermediate doping regime, the situation is qualitatively different, and the adiabatic approximation fails to describe the enhancement of the low-energy interaction due to the plasmon-phonon hybrid modes.
In particular, the cumulative interaction up to the highest LO phonon frequency (193~meV) is stronger in the intermediately doped case than in the undoped case, which is opposite to the adiabatic result.
In \Fig[c]{fig:hBN_W_loc_kk}, the maximum occurs at $n \sim -1\times10^{12}~\mathrm{cm^{-2}}$.

\begin{figure}[tb]
\includegraphics[width=0.99\linewidth]{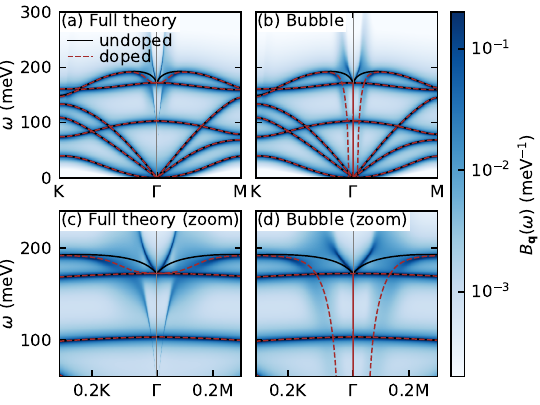}
\caption{
    Phonon spectral function of monolayer hBN at $T=300~\mathrm{K}$ and $n=-4\times10^{12}~\mathrm{cm^{-2}}$.
    (a) Full nonadiabatic theory [\cref{eq:eph_phself_matrix,eq:eph_phspec_def}] and (b) bubble approximation [\cref{eq:phself_1loop}], which neglects the dynamical screening of the e-ph vertex.
    (c), (d) Zoom on the LO phonon near $\Gamma$.
    Black solid and red dashed lines indicate the adiabatic phonon dispersions of the undoped and doped crystal, respectively.
    The doped static phonon dispersions in (b) and (d) are also computed using the bubble approximation.
}
\label{fig:hBN_ph_spectral}
\end{figure}

\begin{figure}[tb]
\includegraphics[width=0.99\linewidth]{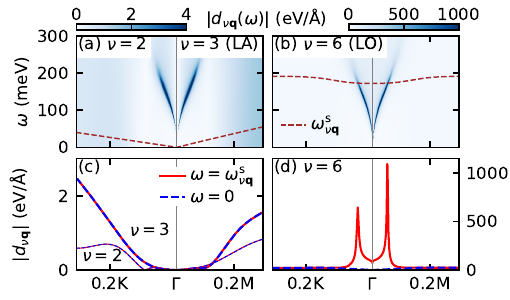}
\caption{
    E-ph vertex of monolayer hBN at $T=300~\mathrm{K}$ and $n=-4\times10^{12}~\mathrm{cm^{-2}}$.
    (a), (b) Frequency- and momentum-resolved deformation potential $\abs{d_{\nu\bq}(\omega)}$ [\cref{eq:eph_D_mode}] for (a) the acoustic mode with the dominant longitudinal character, which is $\nu = 2$ along $\Gamma$K and $\nu = 3$ along $\Gamma$M, and (b) the LO ($\nu = 6$) mode.
    The deformation potential is evaluated for the VBM at $\bk = \mathrm{K}$.
    Solid lines indicate the adiabatic phonon dispersion of the doped system.
    (c), (d) Deformation potential at on-shell ($\omega = \omega^{\rm s}_{\nu\bq}$) and static ($\omega = 0$) frequencies.
}
\label{fig:hBN_eph}
\end{figure}

Finally, to understand the changes in the phonon-mediated interaction, we compute the phonon spectral function and e-ph vertex.
In \cref{fig:hBN_ph_spectral}, we find that the phonon spectral function computed using the full theory displays plasmon-phonon hybridization consistent with \Fig[b]{fig:hBN_W_decomp}.
All other modes are unaffected by the doping.
In contrast, the bubble approximation [\cref{eq:phself_1loop}], which neglects the vertex screening, yields a phonon spectral function that is qualitatively different from the full theory.
Near the $\Gamma$ point, the bubble approximation yields only a single branch, resembling the upper branch of the full theory, while the lower branch is completely absent.
This result occurs because the bubble approximation cannot describe plasmons~\cite{Marini2024}.
The static limit of the bubble approximation breaks down completely, with the LO phonon frequency dropping to zero and becoming imaginary for $\abs{\bq} \lesssim 0.03~\mathrm{bohr}^{-1}$ [\Fig[b,d]{fig:hBN_ph_spectral}].
The over-renormalization happens since the bubble self-energy uses the vertex of the insulating base system, which has the strong long-range Fr\"ohlich coupling, even though the free carriers of the doped target strongly screen the static e-ph vertex.

The failure of the bubble approximation can be further understood by examining the e-ph vertex, which we quantify by the deformation potential in the phonon eigenmode basis [\cref{eq:g_mode}],
\begin{equation} \label{eq:eph_D_mode}
    d_{\nu\bq,mn\bk}(\omega)
    \equiv \sqrt{\frac{2 M^{\rm uc} \omega^{\rm s}_{\nu\bq}}{\hbar}} \, g^{\rm s}_{\nu\bq,mn\bk}(\omega) \,,
\end{equation}
where $M^{\rm uc} = \sum_\kappa M_\kappa$ is the total mass of the unit cell.
For hBN, we take the initial state at the valence band maximum, $n\bk$ with $n$ the topmost valence band and $\bk = \mathrm{K}$, and the final state $m\bkq$ in the same band.
We drop the band indices and write $d_{\nu\bq}(\omega)$ in the following.
\Figu[a,b]{fig:hBN_eph} show the frequency and momentum dependence of the e-ph vertex.
For both the longitudinal acoustic (LA) and LO modes, there is a sharp peak in the vertex along the plasmon dispersion $\omega^{\rm pl}(\bq)$, resulting from the plasmon-phonon resonance; the plasmon reaches the LO phonon frequency, about 175~meV, at $\abs{\bq} \approx 0.05\,\abs{\bq^{\rm K}}$.
While for the LA mode this resonance happens at a frequency much higher than the phonon energy, for the LO mode the resonance is close to the phonon energy.
The on-shell LO vertex is thus enhanced by more than an order of magnitude [\Fig[d]{fig:hBN_eph}], and the static approximation $d_{\nu\bq}(\omega) \approx d_{\nu\bq}(0)$ is no longer valid.
Yet, we also note that the peak in the e-ph vertex does not mean an extremely strong e-ph scattering, since the phonon spectral function at the resonance frequency is suppressed by the plasmon-phonon hybridization.
The resulting total scattering is only moderately altered and simply shifted in frequencies, as seen in \cref{fig:hBN_W_decomp}.

In summary, the dynamical screening is essential in doped hBN because of the long-range Fr\"ohlich coupling between the LO phonon and the carrier plasmon.
The plasmon and the LO phonon hybridize into two branches, the LO e-ph vertex is resonantly enhanced, and the direct e-e interaction contributes substantially to the retarded interaction.
As a result, the low-energy retarded interaction depends nonmonotonically on the carrier concentration, with a maximum at intermediate doping [\Fig[c]{fig:hBN_W_loc_kk}].
The adiabatic approximation misses this behavior, and the bubble phonon self-energy, which neglects the vertex screening, fails even qualitatively for the phonon spectral function.

\subsection{Metallic monolayer TaS$_2$} \label{sec:result_TaS2}

\begin{figure*}[tb]
\includegraphics[width=0.99\linewidth]{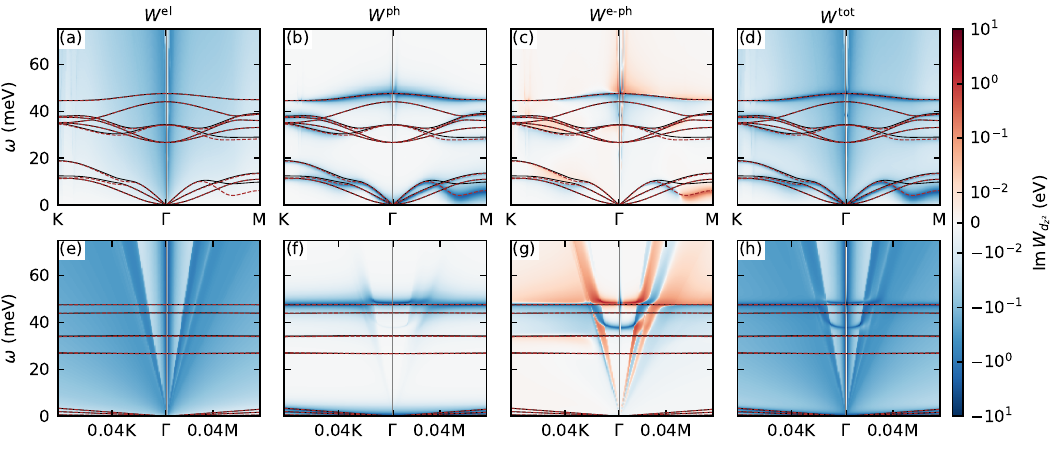}
\caption{
    Frequency- and momentum-resolved total e-e interaction within the active band of monolayer TaS$_2$ at $T = 900~\mathrm{K}$ computed using DFPT+A downfolding, (a) to (d) over the full momentum range and (e) to (h) near the $\Gamma$ point.
    We show the imaginary part of the density-density element for the Ta $d_{z^2}$ WF.
    The total interaction is decomposed into the electronic, phononic, and e-ph vertex contributions, as defined in \cref{eq:W_decomposition}.
    Black solid and red dashed lines indicate the adiabatic phonon dispersions at the base (Gaussian smearing of $20~\mathrm{mRy}$) and target (Fermi--Dirac occupation at $T = 900~\mathrm{K}$) parameters, respectively.
}
\label{fig:TaS2_W}
\end{figure*}

Our second example is metallic monolayer TaS$_2$, which has a single metallic band crossing the Fermi level.
The e-ph-induced phonon softening and nonadiabatic phonon spectral function of this metallic transition-metal dichalcogenide have been studied extensively~\cite{Berges2020, Berges2023}.
We use the Ta $d_{z^2}$, $d_{x^2-y^2}$, and $d_{xy}$ WFs with the e-e interaction truncated to first-nearest-neighbor WF pairs ($N^{\rm pair,trunc} = 63$).
The base system uses a Gaussian smearing of $\sigma = 20~\mathrm{mRy}$, and the target parameter is the Fermi--Dirac occupation at $T = 900~\mathrm{K}$, where the phonon mode near the M point is softened but remains stable~\cite{Berges2020, Berges2023}.

\Figu[a-d]{fig:TaS2_W} show the frequency- and momentum-resolved total e-e interaction and its decomposition [\cref{eq:W_decomposition}] for the density-density interaction of the Ta $d_{z^2}$ WF.
Contrary to the case of hBN, the electronic contribution is finite for a wider frequency-momentum range due to the intraband electron-hole continuum.
Although the plasmon has a $\sqrt{q}$ dispersion, its energy already exceeds 0.5~eV at $\abs{\bq} = 0.01\,\abs{\bq^{\rm M}}$, far above the phonon energies.
The phonon and e-ph vertex contributions are significant throughout the Brillouin zone.
Those contributions are focused around the static phonon dispersion, indicating that the adiabatic approximation is valid for most of the phonon spectrum.

Nonadiabatic features can be seen in the zoom near the $\Gamma$ point, shown in \Fig[e-h]{fig:TaS2_W}.
In the electronic contribution, we observe a linearly dispersing excitation, which is the intraband electron-hole continuum of the metallic band~\cite{Garcia2020, Novko2020}.
This continuum hybridizes with the phonon branches that it crosses, which also renormalizes the e-ph vertex.
These effects are qualitatively similar to the plasmon-phonon hybridization in hBN, but the coupling is weaker as it is driven by the short-range e-ph coupling rather than the long-range Fr\"ohlich coupling.
In the total interaction, a strong contribution from the highest optical mode coexists with a large background from the electronic continuum.

\begin{figure}[tb]
\includegraphics[width=0.99\linewidth]{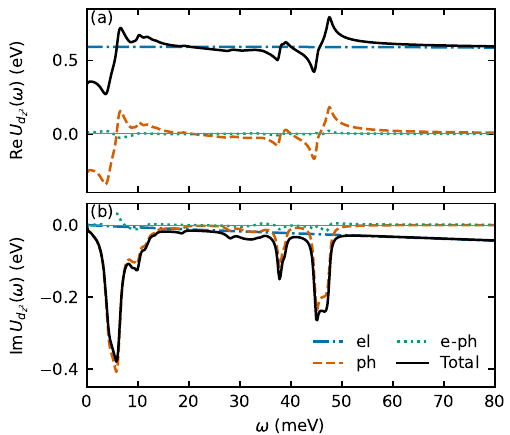}
\caption{
    (a) Real and (b) imaginary parts of the local dynamical interaction [\cref{eq:U_loc_def}] of the Ta $d_{z^2}$ WF of monolayer TaS$_2$ at $T = 900~\mathrm{K}$, computed using DFPT+A downfolding.
    The total interaction is decomposed into the electronic, phononic, and e-ph vertex contributions [\cref{eq:U_loc_decomposition}].
}
\label{fig:TaS2_W_loc}
\end{figure}

\begin{figure}[tb]
\includegraphics[width=0.99\linewidth]{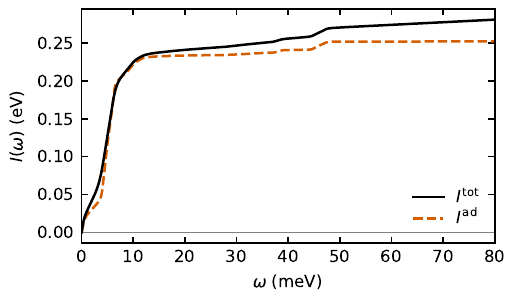}
\caption{
    Cumulative frequency integral [\cref{eq:I_cumulative}] of the local interaction of the Ta $d_{z^2}$ WF of monolayer TaS$_2$ at $T = 900~\mathrm{K}$.
    We show the total integral $I^{\rm tot}(\omega)$ and the adiabatic approximation with purely phononic contribution [\cref{eq:W_adiabatic}].
}
\label{fig:TaS2_W_loc_kk}
\end{figure}

We next turn to the local interaction [\cref{eq:U_loc_def}].
As for hBN, we approximate the Brillouin-zone average by the integral along the $\Gamma$M and $\Gamma$K lines (\cref{app:Uloc_integration}).
\Cref{fig:TaS2_W_loc} shows the local interaction in the phonon frequency range.
The electronic contribution is nearly constant, with a small imaginary part from the electron-hole continuum, so the frequency dependence comes from the phonons.
The soft LA mode near M gives the strongest peak at 6~meV, and the highest optical mode a second peak between 45 and 48~meV.
The e-ph vertex contribution, coming from the dynamical e-ph interaction, is small.

The cumulative integral (\cref{fig:TaS2_W_loc_kk}) shows that the soft mode dominates the retardation effect: $I^{\rm tot}(\omega)$ reaches about 0.22~eV below 10~meV, out of 0.28~eV at 80~meV.
The adiabatic integral, computed from $U^{\rm ad}(\omega)$ of \cref{eq:W_adiabatic}, is very close to the total integral.
Thus, unlike in doped hBN, the low-energy local interaction of TaS$_2$ is well described by the adiabatic approximation.
This is a direct numerical confirmation of the use of the adiabatic approximation for TaS$_2$.
Note that $I^{\rm tot}(\omega)$ does not reach $\Delta U \approx 2.4~\mathrm{eV}$ [\cref{eq:lambda_loc}] at 80~meV, because the electronic dynamical contribution of a metal extends to much higher frequencies.

We now turn to the phonon spectral function, which is the quantity most often reported in the literature.
While the screened-screened approximation [\cref{eq:phself_1loop}] has been applied to TaS$_2$~\cite{Berges2023}, no fully nonadiabatic calculation at the TDDFPT level, which could be used to assess its accuracy, has been done so far.
We provide such a calculation here using DFPT+A downfolding.

\begin{figure}[tb]
\includegraphics[width=0.99\linewidth]{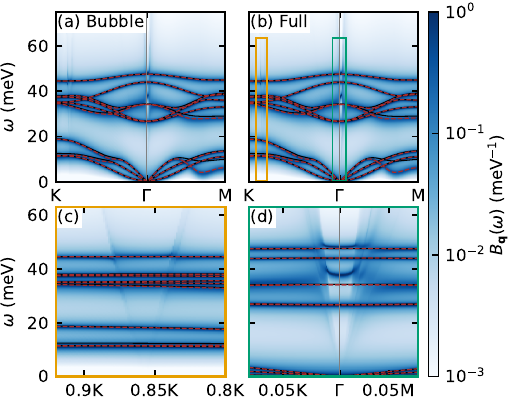}
\caption{
    Phonon spectral function of monolayer TaS$_2$ at $T = 900~\mathrm{K}$ computed using (a) the bubble (screened-screened) approximation [\cref{eq:phself_1loop}] and (b) the full DFPT+A downfolding [\cref{eq:eph_phself_matrix}] with $d^{\rm cut} = d^{\rm 1st}$.
    The solid black and dashed red lines indicate the base (Gaussian smearing of $20~\mathrm{mRy}$) and renormalized target (Fermi--Dirac occupation at $T = 900~\mathrm{K}$) static phonon dispersions, respectively.
    The orange and green boxes in (b) mark the regions with strong nonadiabatic effects, which are shown in (c) and (d).
    (c) Zoom on the $\Gamma$K path near $0.86\,\bq^{\rm K}$ and (d) zoom near $\Gamma$, computed using the full DFPT+A downfolding.
}
\label{fig:TaS2_ph_spectral_DFPT}
\end{figure}

\begin{figure}[tb]
\includegraphics[width=0.99\linewidth]{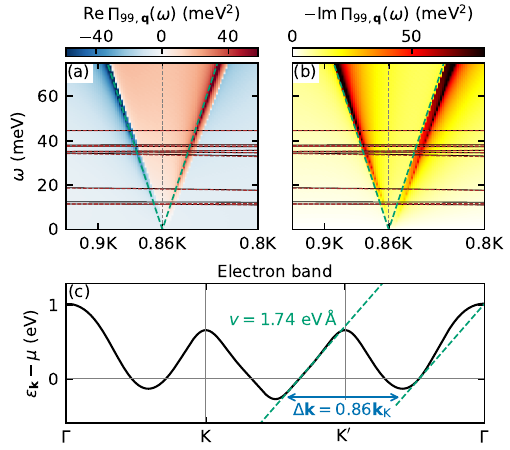}
\caption{
    (a), (b) Real and imaginary parts of the phonon self-energy of monolayer TaS$_2$ at $T = 900~\mathrm{K}$ for the highest phonon branch $\nu=\nu'=9$ in the nonadiabatic region on the $\Gamma$K path [\Fig[c]{fig:TaS2_ph_spectral_DFPT}], computed using DFPT+A downfolding.
    Black solid and red dashed lines indicate the static phonon dispersions at the base and target parameters, respectively.
    (c) Electron band structure along a linear path from $\bk=\mb{0}$ to $3\bq^{\rm K}$.
    The green dashed lines show the Fermi velocity; lines with the same slope are also shown in (a) and (b).
    The nesting vector $\Delta\bk = 0.86\,\bq^{\rm K}$ is also marked with a vertical dashed line in (a) and (b).
}
\label{fig:TaS2_ph_spectral_slice_K}
\end{figure}

\Cref{fig:TaS2_ph_spectral_DFPT} shows the phonon spectral function computed using the bubble approximation and the full DFPT+A downfolding, together with the static phonon dispersions.
The renormalized static phonon dispersions reproduce the softening of the acoustic phonon near M, where the spectral function also shows a large linewidth.
The bubble and full results are nearly indistinguishable on the scale of the figure [\Fig[a,b]{fig:TaS2_ph_spectral_DFPT}].
Unlike for hBN, whose base system is undoped, the static reference susceptibility of TaS$_2$ already contains the metallic intraband screening, so the vertex-screening term [second term of \cref{eq:eph_phself_matrix}] is a small correction.
This finding validates the screened-screened approximation~\cite{Calandra2010, Berges2023} for the nonadiabatic phonon spectral function of TaS$_2$.
In contrast, the bubble approximation of cDFPT+A, whose vertex lacks the intraband screening, fails near $\Gamma$ (see \cref{app:cDFPT}).

Overall, the spectral function in \Fig[b]{fig:TaS2_ph_spectral_DFPT} follows the static dispersion, and nonadiabatic effects are pronounced only in the two regions marked by the orange and green boxes.
In both regions, the spectral function shows a feature with a linear dispersion [\Fig[c,d]{fig:TaS2_ph_spectral_DFPT}].
We note that the spectral functions in these regions obtained from the bubble approximation are nearly indistinguishable from the full DFPT+A results.
Near $\Gamma$, the feature follows the intraband electron-hole continuum of the metallic band.
Similar features have been reported for doped MoS$_2$~\cite{Garcia2020, Novko2020}.
At small $\bq$, where the optical phonon frequency is above the electron-hole continuum, we find anti-adiabatic behavior, where the electrons cannot screen the lattice.

We now examine the $\Gamma$K direction [\Fig[c]{fig:TaS2_ph_spectral_DFPT}].
Here, even though the wavevector is far from $\Gamma$ and thus far from the wavevector of usual intraband electron-hole excitations, we still find a linearly dispersing feature in the phonon spectral function.
Interestingly, this renormalization qualitatively differs from the Kohn anomaly, in that the static phonon dispersion is not renormalized at all.
The linear feature starts at zero frequency at around $0.86\,\bq^{\rm K}$ and has a slope of around 1.74~eV\,\AA.
The corresponding phonon self-energy shown in \Fig[a,b]{fig:TaS2_ph_spectral_slice_K} is also strongly enhanced along the line, with larger magnitudes at higher energies, and vanishing at $\omega = 0$.
We attribute this purely nonadiabatic renormalization to the nearly perfect energy-momentum nesting, i.e., two points on the Fermi surface with a nearly identical Fermi velocity, shown in \Fig[c]{fig:TaS2_ph_spectral_slice_K}.
A one-dimensional model of two linear bands (\cref{app:1d_model}) reproduces the features of the phonon self-energy.
For parallel bands with equal Fermi velocities, the susceptibility displays a linearly dispersing peak, with the static susceptibility at $q = \Delta k$ vanishing by Pauli blocking, consistent with the observations in TaS$_2$ [\Fig[a,b]{fig:TaS2_ph_spectral_slice_K}].
This behavior is in contrast to the usual Kohn anomaly, where the bands have opposite (anti-parallel) Fermi velocities, and the phonon self-energy remains finite at $\omega = 0$.

\begin{figure}[bt]
\includegraphics[width=0.99\linewidth]{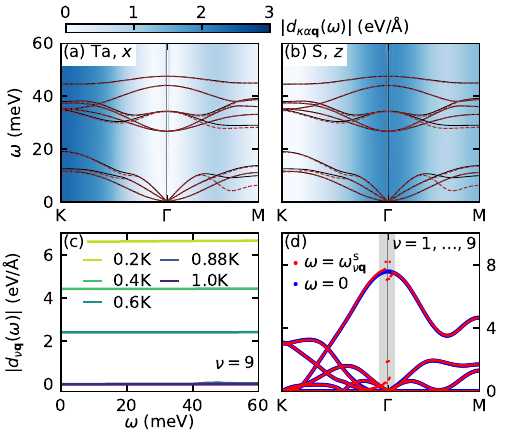}
\caption{
    (a) Frequency- and momentum-resolved deformation potential [\cref{eq:eph_D_cart}] of monolayer TaS$_2$ at $T = 900~\mathrm{K}$ for the Ta atom displaced along $x$, computed using DFPT+A downfolding.
    (b) Same for the S atom displaced along $z$.
    (c) Frequency dependence of the deformation potential in the eigenmode basis [\cref{eq:eph_D_mode}] for the highest optical mode ($\nu = 9$) at selected $\bq$ points along $\Gamma$K.
    (d) On-shell ($\omega = \omega^{\rm s}_{\nu\bq}$, red) and static ($\omega = 0$, blue) deformation potentials for all nine modes.
    The gray region indicates the momentum range where the two differ, which is shown in \cref{fig:TaS2_eph_dyn_narrow}.
    The deformation potential is evaluated for the initial state at $\bk = \Gamma$ in the active band.
}
\label{fig:TaS2_eph_dyn_wide}
\end{figure}

\Figu[a,b]{fig:TaS2_eph_dyn_wide} show the frequency- and momentum-resolved deformation potential for representative atomic displacements,
\begin{equation} \label{eq:eph_D_cart}
    d_{a\bq ,mn\bk}(\omega)
    \equiv \sum_{\nu} e^{{\rm s}*}_{a,\nu\bq} \, d_{\nu\bq,mn\bk}(\omega) \,.
\end{equation}
We use the atomic basis because the phonon eigenmodes cross each other across the Brillouin zone, which makes it difficult to follow a single mode.
Contrary to hBN, the e-e interaction is strongly screened by the free carriers, and thus the e-ph renormalization is very weak.
\Figu[c]{fig:TaS2_eph_dyn_wide} shows that the e-ph coupling indeed has little frequency dependence over the phonon energy range.
This holds even at $\bq = 0.88\,\bq^{\rm K}$, close to the nesting vector $\Delta\bk = 0.86\,\bq^{\rm K}$, where the phonon spectral function is strongly nonadiabatic [\Fig[c]{fig:TaS2_ph_spectral_DFPT}].
Together with the small second term of \cref{eq:eph_phself_matrix}, this weak vertex renormalization explains the agreement between the bubble and full spectral functions [\Fig[a,b]{fig:TaS2_ph_spectral_DFPT}].

\begin{figure}[tb]
\includegraphics[width=0.99\linewidth]{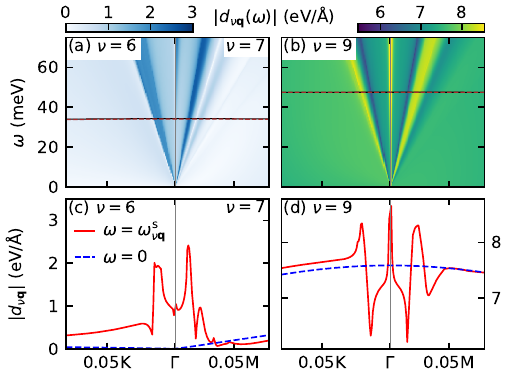}
\caption{
    (a), (b) Frequency- and momentum-resolved deformation potential [\cref{eq:eph_D_mode}] for optical modes (a) $\nu = 6$ along $\Gamma$K and $\nu = 7$ along $\Gamma$M and (b) $\nu = 9$ of monolayer TaS$_2$ at $T = 900~\mathrm{K}$, in the momentum range marked by the gray region in \Fig[d]{fig:TaS2_eph_dyn_wide}.
    (c), (d) On-shell ($\omega = \omega^{\rm s}_{\nu\bq}$, red solid) and static ($\omega = 0$, blue dashed) deformation potentials for the three modes.
}
\label{fig:TaS2_eph_dyn_narrow}
\end{figure}

To quantify the nonadiabatic correction to the e-ph vertex, we compare in \Fig[d]{fig:TaS2_eph_dyn_wide} the on-shell coupling evaluated at the phonon frequency $\omega = \omega^{\rm s}_{\nu\bq}$ with the static limit $\omega = 0$.
The two quantities are nearly indistinguishable throughout the Brillouin zone, indicating that nonadiabatic effects on the e-ph vertex are negligible in monolayer TaS$_2$.
The only exception is the deviation around $\Gamma$.
This region coincides with the large nonadiabatic effects in the phonon spectral function discussed earlier [\Fig[d]{fig:TaS2_ph_spectral_DFPT}].
We zoom in on this region in \cref{fig:TaS2_eph_dyn_narrow}.
We find that there is significant frequency dependence in the deformation potential for optical modes, in the form of a peak or dip with a linear dispersion.
Again, these correspond to the electronic excitations.
If we evaluate the e-ph coupling at the on-shell frequency, large momentum dependence appears whenever the phonon frequency crosses the electronic excitation line.
However, such an on-shell evaluation does not properly capture the nonadiabatic e-ph coupling since the phonon spectral function is also strongly renormalized in this region [\Fig[d]{fig:TaS2_ph_spectral_DFPT}], and thus the phonon is not well described by a single frequency.
Since $g^{\rm s}(\omega)$ inherits its frequency dependence from the screened e-e interaction $W^{\rm s}_\bq(\omega)$ [\cref{eq:eph_gs_rpa_trunc_2}], it is strongly modulated only where the dynamical e-e interactions are also strong.
Thus, one should always use the total interaction $\opWq^{\rm tot}(\omega)$ [\Fig[d,h]{fig:TaS2_W}] when computing physical quantities.

In summary, in metallic TaS$_2$, the static reference susceptibility already contains the metallic intraband screening, so that the dynamical vertex corrections are small.
The adiabatic approximation describes the low-energy local interaction well, and the screened-screened approximation reproduces the full nonadiabatic phonon spectral function.
Nonadiabatic effects are confined to two small regions of the Brillouin zone: near $\Gamma$, where the phonons hybridize with the intraband electron-hole continuum, and near $\bq = 0.86\,\bq^{\rm K}$, where the nesting of nearly parallel Fermi velocities produces a purely nonadiabatic, linearly dispersing feature.

\section{Summary and outlook}

In summary, we developed DFPT+A downfolding to calculate dynamical screening of the phonons, the e-ph vertex, and the Coulomb interaction from first principles.
This method approximates the full TDDFPT response of a target system by a WF-based low-energy Hamiltonian with partially-screened interactions computed using DFPT, offering a systematically improvable and computationally efficient approach.
Using this method, we studied the total dynamical e-e interaction mediated by both electron-hole pairs and phonons in hole-doped monolayer hBN and metallic monolayer TaS$_2$.
In hBN, the hybridization of the carrier plasmon with the LO phonon makes the low-energy retarded interaction nonmonotonic in the carrier density, an effect missed by the adiabatic approximation.
In TaS$_2$, nonadiabatic effects are weak and confined to small regions of the Brillouin zone, and the screened-screened approximation for the phonon spectral function is accurate.

The DFPT+A downfolding approach opens the way to first-principles calculations of electronic properties in which phonons and electronic excitations are dynamically coupled.
The most direct application is the electron self-energy.
In the $G_0 W_0$ approximation, its imaginary part reads~\cite{Lihm2024PlPh}
\begin{align} \label{eq:Sigma_GW}
    \Im \Sigma_\nk(\veps)
    &= \intbq \sum_m \bigl[ f_\mkq + n^{\rm BE}(\veps_\mkq - \veps) \bigr]
    \nnnl
    &\quad \times \Im W^{\rm tot}_{\mnkq}(\veps_\mkq - \veps) \,,
\end{align}
where $n^{\rm BE}(\omega) = 1 / (e^{\omega / k^{\rm B} T} - 1)$ is the Bose--Einstein distribution function, $k^{\rm B}$ is the Boltzmann constant, and $T$ is the temperature.
If one neglects the e-e interaction and the nonadiabatic effects on the phonons, \cref{eq:Sigma_GW} reduces to the standard e-ph self-energy in the Migdal approximation~\cite{Giustino2017}.
When $\hat{W}^{\rm tot}$ is used, \cref{eq:Sigma_GW} describes the scattering of electrons by plasmons, phonons, and their hybrid modes in a single expression.
This self-energy determines the electron spectral function, which can be measured using experiments such as angle-resolved photoemission spectroscopy.
The effects of the electron-plasmon coupling~\cite{Caruso2015PRL, Caruso2021MoS2, Caruso2020Review} and of the screening of the e-ph coupling by free carriers~\cite{Verdi2017, Riley2018} on the spectral function have been studied, but separately and with a model-based description of the screening.
With $\hat{W}^{\rm tot}$, both effects are captured consistently, from first principles.
Also, the e-e interaction in the WF pair basis could provide \textit{ab initio} screened interactions for the Wannier-based Bethe--Salpeter approach to excitons of Ref.~\cite{Merkel2024}.

The imaginary part of the self-energy also gives the quasiparticle lifetimes and electron scattering rates, and hence the conductivity in the self-energy relaxation time approximation~\cite{Ponce2018} or the bubble approximation~\cite{Basov2011RMP}.
However, these approximations are not reliable for e-e scattering, because they do not distinguish normal processes, which conserve the total crystal momentum of the electrons and thus do not relax the current of a single parabolic band, from umklapp processes, which do dissipate the current~\cite{LawrenceWilkins1973, MacDonald1981}.
Beyond these approximations, $\hat{W}^{\rm tot}$ can serve as the scattering kernel of the Boltzmann transport equation~\cite{Sanborn1995, Hauber2017, Gopalan2022, Macheda2026Boltzmann, Dinar2026, Mocatti2026, Paul2026} or of the ladder diagrams~\cite{Holstein1964, PrangeKadanoff1964, Allen1971, Lihm2025, Lihm2026Ladder}, which include the vertex corrections.

The total e-e interaction $\hat{W}^{\rm tot}$ is also the key ingredient for describing superconductivity beyond the Coulomb pseudopotential $\mu^*$~\cite{MorelAnderson1962}.
Recent first-principles studies have included the screened Coulomb interaction, either statically~\cite{Kogler2025} or dynamically~\cite{Davydov2020, Spataru2026}, but the phonons were treated at the adiabatic level.
The interplay of plasmon- and phonon-mediated pairing has been studied in model systems with the full momentum and frequency dependence of the interaction~\cite{inntVeld2023, inntVeld2026, Cui2026}.
DFPT+A downfolding provides a practical way to generalize these model studies to the \textit{ab initio} level.

Finally, the framework can be extended to magnetic systems, where low-energy spin excitations can also cause nonadiabatic effects~\cite{CarrascoAlvarez2026}.
Spin dynamics and magnon spectra can be computed with TDDFPT~\cite{Cao2018, Skovhus2021, Gorni2018, Gorni2022, Binci2025, Skovhus2026}.
Since DFPT+A downfolding is a systematic approximation to TDDFPT, it can significantly accelerate the computation of the spin susceptibility by focusing on the spin responses of the low-energy bands represented in the WF basis, while capturing the screening from the high-energy bands.
It also provides a natural framework to study the nonadiabatic coupling between magnons and phonons~\cite{Bonini2023, Ren2024, Royo2025, Mignolet2026}.

\begin{acknowledgments}
J.-M.L. thanks Iurii Timrov and Nicola Colonna for discussions on the \QE implementation of DFPT and TDDFPT for Coulomb interactions in the Wannier basis.
M.S. and M.R. thank the CCQ at the Flatiron Institute for hospitality during the early stages of this work.
S.P. is a Research Associate of the Fonds de la Recherche Scientifique - FNRS.
This work was supported by the Fonds de la Recherche Scientifique - FNRS under Grants number T.0183.23 (PDR) and T.W011.23 (PDR-WEAVE).
This publication was supported by the Walloon Region in the strategic axe FRFS-WEL-T.
M.S. was supported by the Spanish MCIN/AEI/10.13039/501100011033 through grant PID2023-152710NB-I00 and through a Severo Ochoa Excellence award to ICMAB-CSIC, Grant CEX2023-001263-S.
C.E.D.\ acknowledges support from the National Science Foundation under Grant No.~DMR-2237674.
M.R. acknowledges support from the Vidi ENW research programme of the Dutch Research Council (NWO) [Grant DOI: 10.61686/YDRHT18202] with File No. VI.Vidi.233.077.
Computational resources have been provided
by the EuroHPC JU award granting access to MareNostrum5 at Barcelona Supercomputing Center (BSC), Spain (Project ID: EHPC-EXT-2023E02-050),
by the Consortium des \'Equipements de Calcul Intensif (C\'ECI), funded by the FRS-FNRS under Grant No.~2.5020.11,
and by Lucia, the Tier-1 supercomputer of the Walloon Region with infrastructure funded by the Walloon Region under the grant agreement n$^\circ$~1910247.
The Flatiron Institute is a division of the Simons Foundation.
\end{acknowledgments}

\appendix

\section{Finite-frequency DFPT} \label{app:tddfpt}

In this appendix, we derive the Sternheimer equation for finite-frequency DFPT.
We start from the first-order density matrix response written in terms of the susceptibility operator and derive the first-order wavefunction response computed from the Sternheimer equation.
The Sternheimer equation has many forms, with different numerical stability and efficiency.
We follow the form used in the \QE implementation~\cite{Gironcoli1995, Baroni2001}; in particular, we closely follow Ref.~\cite{Gorni2018}.
Some alternative forms are discussed in Refs.~\cite{Cances2023, Zabalo2024, Gonze2024}.

The first-order density matrix induced by the screened first-order potential $V^{\rm s}_{a\bq}(\omega)$ is [\cref{eq:rpa_n_s_op,eq:rpa_chi_b_def}]
\begin{align} \label{eq:sth_dn_def}
    \kket{n^{\rm s}_{a\bq}(\omega)}
    &= \opchiq^{\rm b}(\omega) \, \kket{V^{\rm s}_{a\bq}(\omega)}
    \nnnl
    &= \intbk \sum_{mn} \frac{f_\nk - f_\mkq}{\omega + \veps_\nk - \veps_\mkq}
    g^{\rm s}_{a\bq,\mnk}(\omega) \, \kket{\Psi_\mnkq} \,.
\end{align}
This equation involves an infinite sum over both bands $m$ and $n$.
The infinite sum can be avoided by rewriting this equation in terms of wavefunction responses and obtaining them by solving the Sternheimer equation.

We first split the density matrix response into two parts by inserting two factors that add up to one,
\begin{equation} \label{eq:sth_dn_split}
    \kket{n^{\rm s}_{a\bq}(\omega)}
    = \kket{n^{\rm s1}_{a\bq}(\omega)} + \kket{n^{\rm s2}_{a\bq}(\omega)} \,,
\end{equation}
where
\begin{subequations}
\begin{align}
    \label{eq:sth_dn_1}
    \kket{n^{\rm s1}_{a\bq}(\omega)}
    &\equiv \intbk \sum_{mn} \kket{\Psi_\mnkq}
    \frac{f_\nk - f_\mkq}{\omega + \veps_\nk - \veps_\mkq}
    \nnnl
    &\qquad\qquad \times \theta_{{m\bkq}, \nk} \,
    g^{\rm s}_{a\bq,\mnk}(\omega) \,,
    \\
    \label{eq:sth_dn_2}
    \kket{n^{\rm s2}_{a\bq}(\omega)}
    &\equiv \intbk \sum_{mn} \kket{\Psi_\mnkq}
    \frac{f_\nk - f_\mkq}{\omega + \veps_\nk - \veps_\mkq}
    \nnnl
    &\qquad\qquad \times \theta_{\nk, {m\bkq}} \,
    g^{\rm s}_{a\bq,\mnk}(\omega) \,,
\end{align}
\end{subequations}
where $\theta$ is a smooth function that adds up to unity:
\begin{equation}
    \theta_{{m\bkq}, \nk} + \theta_{\nk, {m\bkq}} = 1 \,.
\end{equation}
This factor can be any smooth function, and the choice corresponds to choosing a certain gauge for the wavefunction responses~\cite{Cances2023}.
We follow the \QE implementation~\cite{Gorni2018},
\begin{equation}
    \theta_{{m\bkq}, \nk}
    \equiv \frac{1}{2} \, \mathrm{erfc} \Bigl( \frac{\veps_\nk - \veps_\mkq}{\sigma} \Bigr) \,,
\end{equation}
with $\sigma$ being the smearing parameter for the electron occupation.
This factor converges to 0 in the limit $\veps_\nk \gg \veps_\mkq$, and to 1 if $\veps_\nk \ll \veps_\mkq$.

For insulators, $f_\nk - f_\mkq$ is nonzero only if one of the two states is a valence state and the other is a conduction state.
Since there is no smearing, $\theta_{{m\bkq}, \nk}$ is a step function: $\theta_{{m\bkq}, \nk} = 1$ if $\veps_\mkq > \veps_\nk$ and 0 otherwise.
Hence, $\kket{n^{\rm s1}_{a\bq}(\omega)}$ contains the transitions from valence states $n$ to conduction states $m$, and $\kket{n^{\rm s2}_{a\bq}(\omega)}$ those from valence states $m$ to conduction states $n$.
In this case, \cref{eq:sth_unshifted_1} derived below reduces to \cref{eq:fc_sternheimer}.

For metals, to rewrite \cref{eq:sth_dn_1}, we first define the first-order wavefunction
\begin{align} \label{eq:sth_psi_def}
    \ket{\psi_{\nk,a\bq}(\omega)}
    &\equiv \sum_m \ket{\psi_\mkq}
    \frac{f_\nk - f_\mkq}{\omega + \veps_\nk - \veps_\mkq}
    \nnnl
    &\qquad\qquad \times \theta_{{m\bkq}, \nk} \,
    g^{\rm s}_{a\bq,\mnk}(\omega) \,.
\end{align}
Thanks to the $(f_\nk - f_\mkq) \, \theta_{{m\bkq}, \nk}$ factor, this first-order wavefunction vanishes for $\veps_\nk \gg \mu$, where $\mu$ is the chemical potential.
Thus, it suffices to compute it only for states with non-negligible occupation, whose number we denote as $N^{\rm occ}_\bk$,
\begin{equation}
    \kket{n^{\rm s1}_{a\bq}(\omega)}
    = \intbk \sum_{n=1}^{N^{\rm occ}_\bk} \ket{\psi_{\nk,a\bq}(\omega)} \bra{\psi_\nk} \,.
\end{equation}
In practice, we use $\veps_\nk \leq \mu + 3\sigma$ as the criterion for non-negligible occupation.
The first-order wavefunction satisfies a linear equation
\begin{align} \label{eq:sth_unshifted_1}
    &\bigl( H^0_\bkq - \veps_\nk - \omega \bigr) \ket{\psi_{\nk,a\bq}(\omega)}
    \\
    &= - \sum_m \ket{\psi_\mkq} (f_\nk - f_\mkq)
    \theta_{{m\bkq}, \nk} \, g^{\rm s}_{a\bq,\mnk}(\omega) \,.
    \nonumber
\end{align}
We denote the number of bands explicitly calculated as $N^{\rm bnd}$, which is larger than $N^{\rm occ}_\bk$, and large enough such that $f_\mkq \approx 0$ and $\theta_{{m\bkq}, \nk} \approx 1$ for all $m > N^{\rm bnd}$ and $n \leq N^{\rm occ}_\bk$.

For $\omega = 0$, the operator $H^0_\bkq - \veps_\nk - \omega$ becomes singular at degeneracy $\veps_\mkq = \veps_\nk$.
This singularity does not affect $\ket{\psi_{\nk,a\bq}(\omega)}$ because the right-hand side is zero at degeneracy.
Yet, it may cause numerical issues when solving the Sternheimer equation.
To avoid this issue, we shift the linear operator by $\alpha^{\rm PV} P^{\rm occ}_\bkq$, with a positive parameter $\alpha^{\rm PV}$, where $P^{\rm occ}_\bkq \equiv \sum_{m=1}^{N^{\rm occ}_\bkq} \ket{\psi_\mkq} \bra{\psi_\mkq}$ is the projector onto the occupied states~\footnote{Alternatively, one can project the wavefunction response onto the complement of the $N^{\rm bnd}$ bands when solving the Sternheimer equation to solve only for the complement~\cite{Cances2023}.}.
To ensure the solution is unchanged, we add a term to the right-hand side and write~\cite{Baroni2001}
\begin{align} \label{eq:sth_shifted}
    &\bigl( H^0_\bkq - \veps_\nk - \omega + \alpha^{\rm PV} P^{\rm occ}_\bkq \bigr) \ket{\psi_{\nk,a\bq}(\omega)}
    \\
    &= - f_\nk V^{\rm s}_{a\bq}(\omega) \ket{\psi_\nk}
    + \sum_{m=1}^{N^{\rm bnd}} \ket{\psi_\mkq}
    \beta_{\mnk\bq} \, g^{\rm s}_{a\bq,\mnk}(\omega) \,,
    \nonumber
\end{align}
where
\begin{align} \label{eq:sth_beta}
    \beta_{\mnk\bq}
    &= f_\nk (1 - \theta_{\mkq, \nk}) + f_\mkq \theta_{\mkq, \nk}
    \\
    &+ \begin{cases}
        \alpha^{\rm PV} \dfrac{(f_\mkq - f_\nk) \theta_{\mkq, \nk}}{\veps_\mkq - \veps_\nk - \omega}
        & m \leq N^{\rm occ}_\bkq \,,
        \\
        0 & m > N^{\rm occ}_\bkq \,.
    \end{cases}
    \nonumber
\end{align}
These equations correspond to Eqs.~(72) and (73) of Ref.~\cite{Baroni2001} and to Eqs.~(B5) to (B10) of Ref.~\cite{Gorni2018}.
To avoid vanishing denominators at $\omega = 0$ and $\veps_\mkq \approx \veps_\nk$, we replace~\footnote{%
    Using $f'(\tfrac{\veps_\nk + \veps_\mkq}{2})$, or $[f'(\veps_\nk) + f'(\veps_\mkq)] / 2$ as in Eq.~(6) of Ref.~\cite{Stengel2000}, instead of $f'(\veps_\nk)$ will increase the accuracy.
}
\begin{equation} \label{eq:sth_dfde}
    \frac{f_\mkq - f_\nk}{\veps_\mkq - \veps_\nk}
    \to f'(\veps_\nk)
\end{equation}
for $\abs{\veps_\mkq - \veps_\nk} < \Delta \veps^{\rm cutoff}$.
The cutoff $\Delta \veps^{\rm cutoff} = 10^{-5}~\mathrm{Ry}$ is chosen to be large enough to avoid numerical instabilities, but small enough to ensure accurate results.

The parameter $\alpha^{\rm PV}$ should be chosen such that the operator on the left of \cref{eq:sth_shifted} is not singular.
The \QE implementation uses
\begin{equation}
    \alpha^{\rm PV} = \begin{cases}
        2 \bigl( \veps^{\rm occ,max} - \veps^{\rm occ,min} \bigr) & \text{for insulators}\,,
        \\
        \mu + 3\sigma - \veps^{\rm occ,min} & \text{for metals}\,,
    \end{cases}
\end{equation}
where $\veps^{\rm occ,max}$ and $\veps^{\rm occ,min}$ are the maximum and minimum of $\veps_\nk$ and $\veps_\mkq$ over $n \leq N^{\rm occ}_\bk$ and $m \leq N^{\rm occ}_\bkq$.

If $\omega \neq 0$, there can be an actual resonance (singularity) in the response, and $\omega$ must have a finite imaginary part to get a well-defined result.
We use $\omega + i\eta$ with $\eta = 100$~meV in the TDDFPT calculations of \cref{sec:validation}.

We note that in the ensemble DFPT scheme of Ref.~\cite{Zabalo2024}, the Sternheimer equation is solved for all $N^{\rm bnd}$ bands, and it becomes ill-conditioned when the $N^{\rm bnd}$-th and $(N^{\rm bnd}+1)$-th bands are degenerate.
Here, the Sternheimer equation is solved only for the $N^{\rm occ}_\bk$ states with non-negligible occupation, so this issue does not arise.
However, for $\omega = 0$, the operator on the left of \cref{eq:sth_shifted} has eigenvalue $\veps_{N^{\rm occ}_\bkq+1,\bkq} - \veps_\nk$, which becomes small when the $n$-th band at $\bk$ is nearly degenerate with the $(N^{\rm occ}_\bkq+1)$-th band at $\bkq$.
While the right-hand side also vanishes in this limit, such a near-singular operator slows down the convergence of the iterative solution.
This issue can be avoided by using a few buffer bands with negligible occupation and shifting the operator by the projector onto all $N^{\rm bnd}$ bands instead of $P^{\rm occ}_\bkq$~\cite{Cances2023}.

Next, we consider the second term, \cref{eq:sth_dn_2}.
Since the factor $\theta_{\nk, {m\bkq}}$ vanishes for large $\veps_\mkq$, we can only restrict the sum over $m$, but not the sum over $n$, to the occupied states.
This means that the operators must be applied to the bra-states; in other words, we need the Hermitian conjugates of the operators.
While the ground-state Hamiltonian is Hermitian, the first-order potential $V^{\rm s}_{a\bq}(\omega)$ is not Hermitian for $\omega \neq 0$ or $\bq \neq 0$.
Instead of computing the Hermitian conjugate directly, we use the time-reversal operator $\mcT$, which satisfies
\begin{equation}
    \mel{\phi'}{X}{\phi}
    = \mel{\mcT \phi}{\mcT X^\dagger \mcT^{-1}}{\mcT \phi'}
\end{equation}
for any single-particle operator $X$ and states $\ket{\phi}$ and $\ket{\phi'}$.
For spinor wavefunctions, the time-reversal operator acts as $(\mcT \psi)(\br) = -i \sigma_y \psi^*(\br)$, i.e., complex conjugation in real space followed by a spin rotation; without spin-orbit coupling, it reduces to complex conjugation in real space.
We note that this identity does not require the system to be time-reversal symmetric.
We can also show that
\begin{equation}
    \mcT \bigl( \ket{\phi} u \bra{\phi'} \bigr)^\dagger \mcT^{-1}
    = \ket{\mcT \phi'} u \bra{\mcT \phi} \,.
\end{equation}
These identities allow computing $\mcT X^\dagger \mcT^{-1} \ket{\mcT \phi'}$ instead of $\bra{\phi'} X$.
Henceforth, we denote the time-reversed operator as
\begin{equation}
    X^+ \equiv \mcT X^\dagger \mcT^{-1} \,.
\end{equation}

Using these tricks, we write the time-reversed counterpart of \cref{eq:sth_dn_2}, $n^{\rm s2+}_{a\bq}(\omega) \equiv \mcT \bigl[ n^{\rm s2}_{a\bq}(\omega) \bigr]^\dagger \mcT^{-1}$, as
\begin{multline} \label{eq:sth_dn_2_tr1}
    \kket{n^{\rm s2+}_{a\bq}(\omega)}
    = \intbk \sum_{mn} \ket{\mcT \psi_\nk} \bra{\mcT \psi_\mkq}
    \frac{f_\nk - f_\mkq}{\omega + \veps_\nk - \veps_\mkq}
    \\
    \times \theta_{\nk, {m\bkq}}
    \mel{\mcT \psi_\nk}{V^{{\rm s}+}_{a\bq}(\omega)}{\mcT \psi_\mkq} \,.
\end{multline}
Changing dummy variables as $m \leftrightarrow n$ and $\bk \to -\bk-\bq$, we find
\begin{align} \label{eq:sth_dn_2_tr2}
    \kket{n^{\rm s2+}_{a\bq}(\omega)}
    &= \intbk \sum_{mn} \ket{\mcT \psi_{m-\bk-\bq}} \bra{\mcT \psi_{n-\bk}}
    \nnnl
    &\quad \times \frac{f_{m-\bk-\bq} - f_{n-\bk}}{\omega + \veps_{m-\bk-\bq} - \veps_{n-\bk}} \,
    \theta_{m-\bk-\bq, n-\bk}
    \nnnl
    &\quad \times \mel{\mcT \psi_{m-\bk-\bq}}{V^{{\rm s}+}_{a\bq}(\omega)}{\mcT \psi_{n-\bk}} \,.
\end{align}
Defining
\begin{gather}
    \ket{\psi^+_\nk} \equiv \ket{\mcT \psi_{n-\bk}} \,,
    \quad \veps_\nk^+ \equiv \veps_{n-\bk} \,,
    \quad f_\nk^+ \equiv f_{n-\bk} \,,
    \nonumber \\
    \theta^+_{{m\bkq}, \nk} \equiv \theta_{m-\bk-\bq, n-\bk} \,,
\end{gather}
we can rewrite \cref{eq:sth_dn_2_tr2} as
\begin{multline} \label{eq:sth_dn_2_tr3}
    \kket{n^{\rm s2+}_{a\bq}(\omega)}
    = \intbk \sum_{mn} \ket{\psi^+_\mkq} \bra{\psi^+_\nk}
    \frac{f^+_\nk - f^+_\mkq}{-\omega + \veps^+_\nk - \veps^+_\mkq}
    \\
    \times \theta^+_{{m\bkq}, \nk}
    \mel{\psi^+_\mkq}{V^{{\rm s}+}_{a\bq}(\omega)}{\psi^+_\nk} \,.
\end{multline}
Note that this equation has the same form as \cref{eq:sth_dn_1}, with all quantities replaced by their time-reversed counterparts, and $\omega$ replaced by $-\omega$.
Thus, we define the time-reversed first-order wavefunction, analogous to \cref{eq:sth_psi_def},
\begin{multline} \label{eq:sth_psi_tr_def}
    \ket{\psi^+_{\nk,a\bq}(-\omega)}
    \equiv \sum_m \ket{\psi^+_\mkq}
    \frac{f^+_\nk - f^+_\mkq}{-\omega + \veps^+_\nk - \veps^+_\mkq}
    \\
    \times \theta^+_{{m\bkq}, \nk}
    \mel{\psi^+_\mkq}{V^{{\rm s}+}_{a\bq}(\omega)}{\psi^+_\nk} \,.
\end{multline}
It is computed by solving \cref{eq:sth_shifted} with $H^0_\bkq$, $\ket{\psi_\nk}$, $\veps_\nk$, $f_\nk$, $\theta_{{m\bkq}, \nk}$, and $V^{\rm s}_{a\bq}(\omega)$ replaced by their time-reversed counterparts, and $\omega$ replaced by $-\omega$.
As for $\kket{n^{\rm s1}_{a\bq}(\omega)}$, only the states with non-negligible occupation contribute, so that
\begin{equation}
    \kket{n^{\rm s2+}_{a\bq}(\omega)} = \intbk \sum_{n=1}^{N^{\rm occ+}_\bk} \ket{\psi^+_{\nk,a\bq}(-\omega)} \bra{\psi^+_\nk}
\end{equation}
with $N^{\rm occ+}_\bk \equiv N^{\rm occ}_{-\bk}$.
Undoing the time reversal, $\kket{n^{\rm s2}_{a\bq}(\omega)} = \bigl[ \mcT^{-1} n^{\rm s2+}_{a\bq}(\omega) \, \mcT \bigr]^\dagger$, we find
\begin{equation} \label{eq:sth_dn_2_final}
    \kket{n^{\rm s2}_{a\bq}(\omega)}
    = \intbk \sum_{n=1}^{N^{\rm occ+}_\bk}
    \ket{\mcT^{-1} \psi^+_\nk} \bra{\mcT^{-1} \psi^+_{\nk,a\bq}(-\omega)} \,.
\end{equation}

If the ground state is time-reversal symmetric, all symbols with the $+$ superscript reduce to the original ones: $H^{0+} = H^0$, $\ket{\psi^+_\nk} = \ket{\psi_\nk}$, $\veps^+_\nk = \veps_\nk$, $f^+_\nk = f_\nk$, and $\theta^+_{{m\bkq}, \nk} = \theta_{{m\bkq}, \nk}$.
For time-reversal-even perturbations such as atomic displacements, the first-order potential has no magnetic part, so $V^{{\rm s}+}_{a\bq}(\omega) = V^{\rm s}_{a\bq}(\omega)$, which also equals $V^{\rm s}_{a\bq}(-\omega)$ because the density response of a time-reversal symmetric system is even in $\omega$.
Therefore, $\ket{\psi^+_{\nk,a\bq}(-\omega)} = \ket{\psi_{\nk,a\bq}(-\omega)}$.
For a complex frequency $\omega = \omega^{\rm r} + i\eta$, the second Sternheimer equation must be solved at $-\omega = -\omega^{\rm r} - i\eta$, not at $-\omega^{\rm r} + i\eta$.
Thus, the time-reversed quantities reduce to the original ones, and substituting them into \cref{eq:sth_dn_2_final} gives
\begin{equation} \label{eq:sth_dn_2_trs}
    \kket{n^{\rm s2}_{a\bq}(\omega)}
    = \intbk \sum_{n=1}^{N^{\rm occ}_\bk}
    \ket{\mcT^{-1} \psi_\nk} \bra{\mcT^{-1} \psi_{\nk,a\bq}(-\omega)} \,.
\end{equation}
Taking the diagonal part, the first-order density becomes
\begin{equation} \label{eq:sth_density_trs}
    n^{\rm s}_{a\bq}(\br, \omega)
    = \intbk \sum_{n=1}^{N^{\rm occ}_\bk} \psi^\dagger_\nk(\br)
    \bigl[ \psi_{\nk,a\bq}(\br, \omega) + \psi_{\nk,a\bq}(\br, -\omega) \bigr] \,,
\end{equation}
which is \cref{eq:fc_density} of the main text and is the expression implemented in \QE.
Thus, for the time-reversal symmetric case, no calculation at $-\bk$ is needed.
For $\omega = 0$, the first-order wavefunction at $-\omega$ is the same as that at $\omega$, so that only one Sternheimer equation needs to be solved.
For $\omega \neq 0$, we need to solve two Sternheimer equations, one at $\omega$ and one at $-\omega$.

If the system does not have time-reversal symmetry, we also need to solve two Sternheimer equations, one for $\ket{\psi_{\nk,a\bq}(\omega)}$ and one for $\ket{\psi^+_{\nk,a\bq}(-\omega)}$, with the latter using the time-reversed ground-state wavefunctions and Hamiltonian.

\section{Derivation of the downfolded screening equations} \label{app:downfold_derivation}

In this appendix, we derive the downfolded screening equations [\cref{eq:rpa_Hs_downfold,eq:rpa_C_downfold}] and their evaluation in the subspace of active-space transitions [\cref{eq:eph_Cs_downfold,eq:eph_phself_matrix,eq:eph_Ws_dyson}], as well as the screened e-ph vertex with truncated $W^{\rm p}$ [\cref{eq:eph_gs_rpa_trunc_2}].

Substituting $\opchiq^{\rm b,tar}(\omega) = \opchiq^{\rm b,ref} + \opchiq^{\rm b,dyn}(\omega)$ into \cref{eq:rpa_H_s} for the target system, we have
\begin{align} \label{eq:rpa_Hs_split}
    \kket{V_{a\bq}^{\rm s}(\omega)}
    = \kket{V_{a\bq}^{\rm b}}
    &+ \opvq \, \opchiq^{\rm b,ref} \, \kket{V_{a\bq}^{\rm s}(\omega)}
    \nnnl
    &+ \opvq \, \opchiq^{\rm b,dyn}(\omega) \, \kket{V_{a\bq}^{\rm s}(\omega)} \,.
\end{align}
Moving the second term to the left-hand side gives
\begin{equation} \label{eq:rpa_Hs_moved}
    \opepsq^{\rm p} \, \kket{V_{a\bq}^{\rm s}(\omega)}
    = \kket{V_{a\bq}^{\rm b}}
    + \opvq \, \opchiq^{\rm b,dyn}(\omega) \, \kket{V_{a\bq}^{\rm s}(\omega)} \,.
\end{equation}
Multiplying both sides by $(\opepsq^{\rm p})^{-1}$ and using \cref{eq:rpa_Wp_def}, we obtain \cref{eq:rpa_Hs_downfold}.

Similarly, we write the dynamical IFCs in terms of the partially-screened quantities.
We first note that $(\opepsq^{\rm p})^{-1} = \opone + \opvq \, \opchiq^{\rm p}$ and \cref{eq:rpa_Wp_def} give
\begin{equation} \label{eq:rpa_Vp_ket}
    \kket{V^{\rm p}_{a\bq}} = \bigl[ \opone + \opvq \, \opchiq^{\rm p} \bigr] \kket{V^{\rm b}_{a\bq}} \,.
\end{equation}
Since the static operators $\opvq$ and $\opchiq^{\rm p}$ are Hermitian, the adjoint of this relation reads
\begin{equation} \label{eq:rpa_Vp_bra}
    \bbra{V^{\rm p}_{a\bq}}
    = \bbra{V^{\rm b}_{a\bq}} \, \bigl[ \opone + \opchiq^{\rm p} \, \opvq \bigr] \,.
\end{equation}
Substituting the split susceptibility into \cref{eq:rpa_C}, we find
\begin{align}
    & C^{\rm s}_{aa' \bq}(\omega)
    \nnnl
    &= C^{\rm b}_{aa' \bq}
    + \bbra{V^{\rm b}_{a\bq}} \, \bigl[ \opchiq^{\rm b,ref} + \opchiq^{\rm b,dyn}(\omega) \bigr] \, \kket{V^{\rm s}_{a'\bq}(\omega)}
    \nnnl
    &= C^{\rm b}_{aa' \bq}
    + \bbra{V^{\rm b}_{a\bq}} \, \opchiq^{\rm b,ref} \, \kket{V^{\rm p}_{a'\bq}}
    \nnnl
    &\quad + \bbra{V^{\rm b}_{a\bq}} \, \bigl[ \opchiq^{\rm p} \, \opvq \, \opchiq^{\rm b,dyn}(\omega) + \opchiq^{\rm b,dyn}(\omega) \bigr] \, \kket{V^{\rm s}_{a'\bq}(\omega)}
    \nnnl
    &= C^{\rm p}_{aa' \bq}
    + \bbra{V^{\rm b}_{a\bq}} \, \bigl[ \opone + \opchiq^{\rm p} \, \opvq \bigr] \,
    \opchiq^{\rm b,dyn}(\omega) \, \kket{V^{\rm s}_{a'\bq}(\omega)}
    \nnnl
    &= C^{\rm p}_{aa' \bq}
    + \bbra{V^{\rm p}_{a\bq}} \, \opchiq^{\rm b,dyn}(\omega) \, \kket{V^{\rm s}_{a'\bq}(\omega)} \,,
    \label{eq:rpa_C_downfold_deriv}
\end{align}
which is \cref{eq:rpa_C_downfold}.
In the second equality, we substituted \cref{eq:rpa_Hs_downfold} into the reference-screening term and used $\opchiq^{\rm b,ref} \, \opWq^{\rm p} = \opchiq^{\rm b,ref} (\opepsq^{\rm p})^{-1} \opvq = \opchiq^{\rm p} \, \opvq$ [\cref{eq:rpa_Wp_def,eq:rpa_chip_def}].
In the third equality, we used the definition of $C^{\rm p}$ [\cref{eq:rpa_Cp_def}] and collected the terms containing $\opchiq^{\rm b,dyn}(\omega)$.
In the last equality, we used \cref{eq:rpa_Vp_bra}.

Finally, we evaluate \cref{eq:rpa_Hs_downfold,eq:rpa_C_downfold} in a basis for the active subspace.
Since $\opchiq^{\rm b,dyn}(\omega)$ maps any state into the subspace [\cref{eq:rpa_chib_active_W}], \cref{eq:rpa_Hs_downfold} gives
\begin{equation} \label{eq:rpa_Vs_y}
    \kket{V^{\rm s}_{a\bq}(\omega)}
    = \kket{V^{\rm p}_{a\bq}}
    + \sum_{I} \opWq^{\rm p} \kket{\Phi_{I\bq}} \, y_{Ia\bq}(\omega) \,,
\end{equation}
with $y_{Ia\bq}(\omega) \equiv \bbra{\Phi_{I\bq}} \opchiq^{\rm b,dyn}(\omega) \kket{V^{\rm s}_{a\bq}(\omega)}$.
Applying $\bbra{\Phi_{I\bq}} \opchiq^{\rm b,dyn}(\omega)$ to both sides and using \cref{eq:rpa_chib_active_W,eq:rpa_X_defs}, we obtain a closed linear equation
\begin{equation} \label{eq:eph_y_lin}
    y_{Ia\bq}(\omega)
    = X^{\rm b,dyn}_{Ia\bq}(\omega)
    + \sum_{I'I''} X^{\rm b,dyn}_{II'\bq}(\omega) \, W^{\rm p}_{I'I''\bq} \, y_{I''a\bq}(\omega) \,,
\end{equation}
whose solution is
\begin{equation}
    y_{Ia\bq}(\omega)
    = \sum_{I'} \bigl[ \one - X^{\rm b,dyn}_\bq(\omega) W^{\rm p}_\bq \bigr]^{-1}_{II'} X^{\rm b,dyn}_{I'a\bq}(\omega) \,.
\end{equation}
Inserting \cref{eq:rpa_Vs_y} into \cref{eq:rpa_C_downfold} and using $W^{\rm p}_\bq (\one - X^{\rm b,dyn}_\bq W^{\rm p}_\bq)^{-1} = (\one - W^{\rm p}_\bq X^{\rm b,dyn}_\bq)^{-1} W^{\rm p}_\bq$ gives \cref{eq:eph_Cs_downfold,eq:eph_phself_matrix,eq:eph_Ws_dyson}.

If $\opWq^{\rm p}$ is truncated to the subspace [\cref{eq:eph_Wp_trunc}], its action on a basis function stays in the subspace,
\begin{equation} \label{eq:Wp_trunc_action}
    \opWq^{\rm p} \kket{\Phi_{I\bq}} = \sum_{I'} \kket{\Phi_{I'\bq}} W^{\rm p}_{I'I\bq} \,.
\end{equation}
Inserting \cref{eq:Wp_trunc_action} into \cref{eq:rpa_Vs_y} gives
\begin{equation} \label{eq:Vs_trunc_y}
    \kket{V^{\rm s}_{a\bq}(\omega)}
    = \kket{V^{\rm p}_{a\bq}}
    + \sum_{II'} \kket{\Phi_{I\bq}} \, W^{\rm p}_{II'\bq} \, y_{I'a\bq}(\omega) \,.
\end{equation}
With the solution of \cref{eq:eph_y_lin}, the coefficients become, in matrix notation,
\begin{align} \label{eq:Vs_trunc_coef}
    W^{\rm p}_\bq \, y_\bq(\omega)
    &= W^{\rm p}_\bq \bigl[ \one - X^{\rm b,dyn}_\bq(\omega) \, W^{\rm p}_\bq \bigr]^{-1} X^{\rm b,dyn}_\bq(\omega)
    \nnnl
    &= \bigl[ \one - W^{\rm p}_\bq \, X^{\rm b,dyn}_\bq(\omega) \bigr]^{-1} W^{\rm p}_\bq \, X^{\rm b,dyn}_\bq(\omega)
    \nnnl
    &= W^{\rm s}_\bq(\omega) \, X^{\rm b,dyn}_\bq(\omega) \,,
\end{align}
where the last equality uses \cref{eq:eph_Ws_dyson}.
Inserting \cref{eq:Vs_trunc_coef} into \cref{eq:Vs_trunc_y} gives the screened e-ph vertex of \cref{eq:eph_gs_rpa_trunc_2}.

\section{cDFPT+A downfolding} \label{app:cDFPT}

In this appendix, we compare the cDFPT+A and DFPT+A downfolding schemes [\cref{eq:chib_ref_dyn_cDFPTA,eq:chib_ref_dyn_DFPTA}].
We summarize the changes to the equations in the main text for cDFPT+A and show that the two schemes yield the same fully-screened quantities.
Since the downfolded equations are exact rewritings of the screening equations [\cref{eq:rpa_H_s,eq:rpa_C}] for any split of $\opchiq^{\rm b}(\omega)$ into a static reference susceptibility and a dynamical correction, the fully-screened quantities do not depend on the choice of the reference susceptibility, provided that $\opWq^{\rm p}$ is not truncated.
We show this equivalence numerically for monolayer TaS$_2$.
For this system, the cDFPT+A scheme converges more slowly than DFPT+A with respect to the truncation of the WF pairs for the e-e interaction, because the DFPT-screened quantities are metallic whereas the cDFPT-screened quantities are insulating.

Most results in \cref{sec:theory_downfolding,sec:theory_ee} are written in terms of the generic separation of static reference susceptibility and dynamical correction [\cref{eq:chib_ref_dyn}].
Thus, they equally apply to cDFPT+A with $\opchiq^{\rm b,ref} = \opchiq^{\rm b,base,R}(0)$ and $\opchiq^{\rm b,dyn}(\omega) = \opchiq^{\rm b,tar,A}(\omega)$ [\cref{eq:chib_ref_dyn_cDFPTA}].
The only changes are that the static base term is \textit{not} subtracted in the matrix elements of the dynamical correction [\cref{eq:rpa_X_band}],
\begin{multline} \label{eq:X_band_cDFPTA}
    X^{\rm b,dyn}_{YY'\bq}(\omega)
    = \intbk \sum_{mn}^{\rm A} \frac{f^{\rm tar}_\nk - f^{\rm tar}_\mkq}{\omega + \veps_\nk - \veps_\mkq}
    \\
    \times M^*_{\mnkq,Y} M_{\mnkq,Y'} \,,
\end{multline}
with $Y$, $Y'$, and $M_{\mnkq,Y}$ as defined in \cref{eq:rpa_M_I,eq:rpa_M_a}.
Also, the partially-screened quantities should be computed from the cDFPT susceptibility $\opchiq^{\rm b,base,R}(0)$ instead of the DFPT susceptibility $\opchiq^{\rm b,base}(0)$.
We evaluate them by modifying the DFPT implementation to subtract the active-space screening contributions, which corresponds to using a modified Sternheimer equation~\cite{Nomura2015, Berges2023}.

\begin{figure}[tb]
\includegraphics[width=0.99\linewidth]{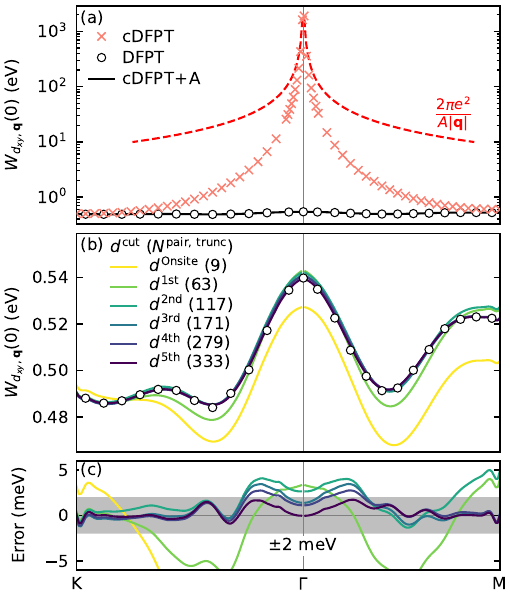}
\caption{
    (a) Static density-density interaction of the $d_{xy}$ WF of monolayer TaS$_2$ using 0.02~Ry Gaussian smearing.
    Crosses show the partially-screened cDFPT interaction, circles the direct DFPT interaction, and the black line the cDFPT+A result with $d^{\rm cut} = d^{\rm 5th}$.
    The red dashed curve denotes the 2D long-range monopole-monopole interaction [\cref{eq:itp_Wp_mm}].
    (b) cDFPT+A results for different cutoffs $d^{\rm cut}$, compared with the direct DFPT result (circles).
    (c) Error of the cDFPT+A results in (b) with respect to direct DFPT.
    The gray band marks $\pm 2$~meV.
}
\label{fig:TaS2_static_ee}
\end{figure}

\begin{figure}[tb]
\includegraphics[width=0.99\linewidth]{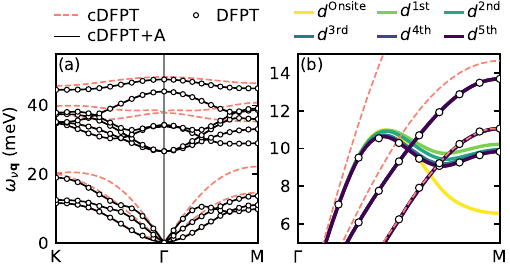}
\caption{
    (a) Phonon dispersion of monolayer TaS$_2$ using 0.02~Ry Gaussian smearing, computed using cDFPT (red dashed lines) and DFPT (circles), and from cDFPT+A downfolding [\cref{eq:eph_Cs_downfold}] using $d^{\rm cut} = d^{\rm 5th}$ (black lines).
    (b) Acoustic phonon dispersion along $\Gamma$M computed using cDFPT+A downfolding with different cutoffs $d^\mathrm{cut}$ for the partially-screened e-e interaction [\cref{eq:eph_Wp_trunc}], compared with the direct DFPT result.
}
\label{fig:TaS2_phonon_static_ref}
\end{figure}

One key practical difference between the two schemes is how they handle entangled bands.
In cDFPT+A, the rest susceptibility is defined by a subtraction [\cref{eq:chib_rest_def}].
When the active states are taken to be the disentangled Wannier-interpolated bands, which differ from the Kohn--Sham states outside the frozen window, the subtracted term contains transitions that are absent in $\opchiq^{\rm b}$, so that $\opchiq^{\rm b,R}$ can produce spurious negative screening~\cite{Miyake2009}.
Defining the active space by an energy window inside the frozen window (\cref{sec:theory_wannier}) avoids this mismatch, but then the window cuts through bands, making the resulting interaction have long-range components.
Several remedies have been proposed in the context of cRPA~\cite{Miyake2009, Sasioglu2011, Kaltak2015, Kaltak2025}.
In DFPT+A, these problems do not arise at all because the reference susceptibility is the full static response of the base system, which does not depend on the active window or on the WFs.
We note, however, that this choice is appropriate here because our aim is to reproduce the TDDFPT result.
When the aim is to embed the downfolded interactions into an impurity problem, or to solve the downfolded Hamiltonian with a higher-level method, the DFPT+A scheme may not be appropriate.

Apart from these practical issues, both the cDFPT+A and DFPT+A schemes rely on the same two approximations introduced in \cref{eq:chib_approx1,eq:chib_approx2}, and yield the same result in the limit of an untruncated partially-screened e-e interaction $W^{\rm p}$.
Differences arise only when the Coulomb interaction is truncated, which is unavoidable in practical calculations.
We now show the equivalence and differences using monolayer TaS$_2$ as an example.

We first consider the static limit with identical target and base systems (0.02~Ry Gaussian smearing).
In the DFPT+A scheme, the dynamical correction [\cref{eq:chib_ref_dyn_DFPTA}] vanishes at $\omega = 0$, and the partially-screened quantities are already the fully-screened ones.
In the cDFPT+A scheme, in contrast, the fully-screened quantities must be recovered by solving the active-space screening equations, which provides a stringent test for the truncation of $W^{\rm p}$ [\cref{eq:eph_Wp_trunc}].
\Figu{fig:TaS2_static_ee} shows that the result of downfolding agrees closely with direct DFPT, even though the starting point is the partially-screened cDFPT result with orders of magnitude larger interaction near $\Gamma$.
Here and in the following figures, ``cDFPT'' and ``DFPT'' denote the partially-screened static quantities, computed with $\opchiq^{\rm b,base,R}(0)$ and $\opchiq^{\rm b,base}(0)$, respectively, and ``cDFPT+A'' and ``DFPT+A'' denote the fully-screened quantities obtained by downfolding with \cref{eq:chib_ref_dyn_cDFPTA,eq:chib_ref_dyn_DFPTA}, respectively.
The fully-screened interaction rapidly converges to the direct DFPT result as the cutoff distance is increased.
With the fifth-neighbor cutoff the error stays within $2$~meV along the whole path.
\Cref{fig:TaS2_phonon_static_ref} shows the corresponding test for the phonon dispersion.
The fully-screened dispersion computed using cDFPT+A downfolding agrees excellently with the direct DFPT result.
These results validate our cDFPT+A downfolding framework and show the practicality of the truncation scheme.

\begin{figure}[t]
\includegraphics[width=0.99\linewidth]{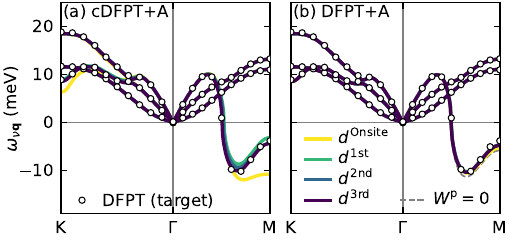}
\caption{
    Acoustic phonon dispersion of monolayer TaS$_2$ at 300~K computed using (a) cDFPT+A and (b) DFPT+A downfolding with different cutoffs $d^{\rm cut}$, compared with the direct DFPT result at the target parameters (circles), namely Fermi--Dirac occupations at $T = 300~\mathrm{K}$ on a $36 \times 36 \times 1$ $\bk$ grid.
    The base system uses a Gaussian smearing of $0.02~\mathrm{Ry}$ on a $12 \times 12 \times 1$ $\bk$ grid.
    The gray dashed line in (b) shows DFPT+A downfolding with $W^{\rm p} = 0$, which is the screened-screened approximation~\cite{Berges2023}.
}
\label{fig:TaS2_phonon_static_target}
\end{figure}

We next downfold from the base system, a 0.02~Ry Gaussian smearing on a $12 \times 12 \times 1$ $\bk$ grid, to the target parameters of \cref{sec:theory_demo}, Fermi--Dirac occupations at $T = 300~\mathrm{K}$ on a $36\times36\times1$ $\bk$ grid, at which TaS$_2$ displays soft phonons around the M point~\cite{Berges2020, Berges2023}.
In \cref{fig:TaS2_phonon_static_target}, we benchmark the cDFPT+A and DFPT+A downfolding methods against a direct DFPT calculation at the target parameters.
For cDFPT+A, the onsite truncation overestimates the softening at M and K points, while for the DFPT+A scheme, the onsite truncation is already almost converged and agrees well with the direct DFPT result.
This slower convergence of cDFPT+A is a consequence of the mismatch between the insulating screening of cDFPT and the metallic full screening.
We also show the dispersion computed using the screened-screened approximation, which means setting $W^{\rm p} = 0$ in the DFPT+A downfolding~\cite{Berges2023}.
It shows a small but nonzero difference with respect to the DFPT+A results.

We note that the small offset at the lowest acoustic mode at M is due to the change, upon changing the occupations, of the transitions between the rest and active bands, which is neglected in both cDFPT+A and DFPT+A downfolding [\cref{eq:chib_approx2}].
Enlarging the active window to include more bands would systematically reduce this offset~\cite{Berges2023}.

\begin{figure*}[tb]
\includegraphics[width=0.99\linewidth]{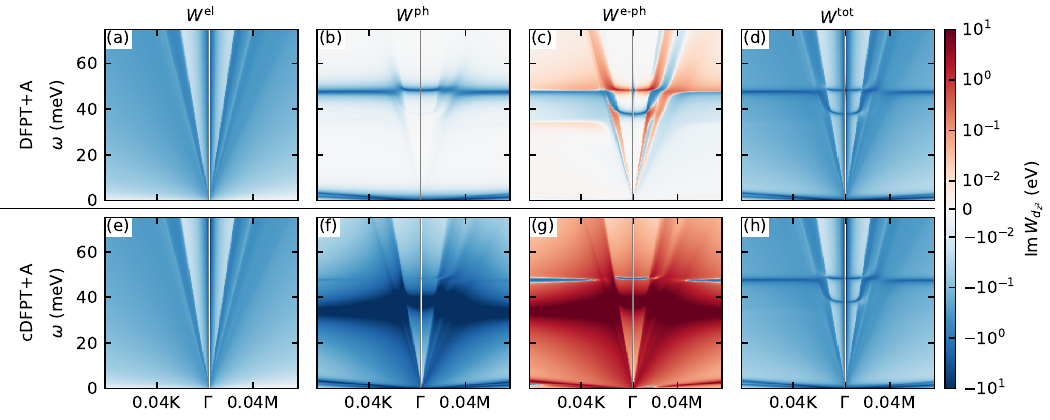}
\caption{
    Frequency- and momentum-resolved e-e interaction for TaS$_2$, same as \Fig[e-h]{fig:TaS2_W}, computed using (a) to (d) DFPT+A and (e) to (h) cDFPT+A downfolding.
    Panels (a) to (d) show the same data as \Fig[e-h]{fig:TaS2_W}.
    For both schemes, the $W^{\rm p}$ matrix is truncated using $d^{\rm cut} = d^{\rm 1st}$.
}
\label{fig:TaS2_W_comparison_zoom}
\end{figure*}

\Cref{fig:TaS2_W_comparison_zoom} compares the total e-e interaction of monolayer TaS$_2$ and its decomposition [\cref{eq:W_decomposition}] computed using the DFPT+A [\Fig[a-d]{fig:TaS2_W_comparison_zoom}] and cDFPT+A [\Fig[e-h]{fig:TaS2_W_comparison_zoom}] downfolding near the $\Gamma$ point.
Since only the reference susceptibility of DFPT+A contains the intraband screening (\cref{sec:theory_downfolding_approach}), the two schemes yield different $W^{\rm p}$, $g^{\rm p}$, and $\Pi^{\rm A}$.
Accordingly, the individual contributions differ.
For the cDFPT+A downfolding, there is a strong phonon contribution in the 30 to 40~meV range [\Fig[f]{fig:TaS2_W_comparison_zoom}], as the partial screening excludes the active-band contributions that strongly screen the e-ph coupling for optical phonon modes.
The e-ph vertex term compensates for this enhancement: its imaginary part has the opposite sign over a broad frequency range [\Fig[g]{fig:TaS2_W_comparison_zoom}] and cancels most of the phonon contribution.
This cancellation simply means that the missing metallic screening of the e-ph vertex is added back.
The total interaction, in contrast, is nearly identical between the two schemes.

Next, we compare the phonon spectral function computed using the two schemes.
We have shown that, while the DFPT+A and cDFPT+A schemes are formally equivalent, they differ once the e-e interaction is approximated.
Here, we consider the extreme case of neglecting the e-e interaction in the active space, $W^{\rm p} = 0$, which gives the bubble approximation of the phonon self-energy [\cref{eq:phself_1loop}].
In this case, the e-ph vertex entering the phonon self-energy is the static partially-screened one of the scheme.
In DFPT+A, this is the static DFPT vertex of the base system, $g^{\rm DFPT}(0)$, and the bubble approximation is the screened-screened approximation~\cite{Calandra2010, Berges2023}, which we denote by $\Pi^{\rm scr\text{-}scr}(\omega)$.
In cDFPT+A, both vertices are the partially-screened cDFPT vertex $g^{\rm cDFPT}(0)$, different from the DFPT vertex.
We denote this approximation by $\Pi^{\rm pscr\text{-}pscr}(\omega)$.

Note that $\Pi^{\rm pscr\text{-}pscr}$ differs from the partially screened-screened approximation~\cite{Berges2023}, which mixes the cDFPT and DFPT vertices.
Since the mixed use of two vertices breaks the symmetry of the two vertices, it gives a non-causal phonon self-energy, leading to a phonon spectral function that can become negative~\cite{Berges2023, Stefanucci2025}.

The full self-energy [\cref{eq:eph_phself_matrix}] adds the e-e interaction term to the bubble one in both schemes (\cref{sec:theory_eph}):
\begin{align} \label{eq:phself_exact}
    C^{\rm s}(\omega)
    &= C^{\rm DFPT}(0) + \Pi^{\rm scr\text{-}scr}(\omega) + (\cdots)
    \nnnl
    &= C^{\rm cDFPT}(0) + \Pi^{\rm pscr\text{-}pscr}(\omega) + (\cdots) \,.
\end{align}
Here, the two parenthetical terms are the terms involving $W^{\rm p}$ in each scheme.
Either approximation becomes exact once its correction term is added.
We note that the correction terms are quadratic in the error made in the e-ph matrix elements~\cite{Calandra2010,Berges2023,Stefanucci2025}.
Therefore, the accuracy of the bubble approximation depends on how well the partially-screened vertex approximates the fully-screened one.

\begin{figure}[tb]
\includegraphics[width=0.99\linewidth]{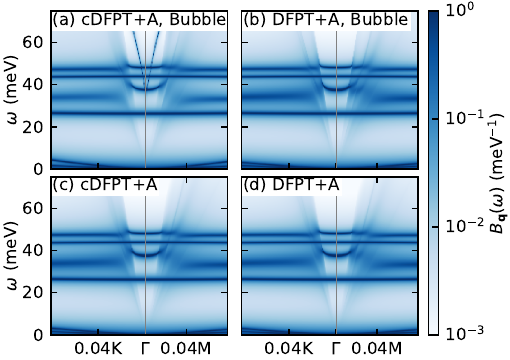}
\caption{
    Phonon spectral function of monolayer TaS$_2$ at $T=900\,\mathrm{K}$ near $\Gamma$, computed using four different approaches:
    (a) bubble approximation [\cref{eq:phself_1loop}] of cDFPT+A,
    (b) bubble approximation of DFPT+A, i.e., the screened-screened approximation,
    (c) full cDFPT+A downfolding [\cref{eq:eph_phself_matrix}], and
    (d) full DFPT+A downfolding.
    For (c) and (d), the $W^{\rm p}$ matrix is truncated with $d^{\rm cut} = d^{\rm 1st}$.
    The steep dispersive feature near $\Gamma$ in (a) is the spurious mode of the bubble approximation of cDFPT+A discussed in the text.
}
\label{fig:TaS2_ph_spectral}
\end{figure}

\Cref{fig:TaS2_ph_spectral} shows the nonadiabatic phonon spectral function near the $\Gamma$ point, computed using the bubble approximations and the full calculation including dynamical screening obtained from both cDFPT+A and DFPT+A downfolding.
The two full results [\Fig[c,d]{fig:TaS2_ph_spectral}] are nearly identical, as expected.
The screened-screened approximation [\Fig[b]{fig:TaS2_ph_spectral}] is close to them, as discussed in \cref{sec:result_TaS2}.
The bubble approximation of cDFPT+A [\Fig[a]{fig:TaS2_ph_spectral}], in contrast, fails qualitatively near $\Gamma$, as it shows a spurious dispersive mode that is not present in the full result.
This failure resembles that of the bubble approximation for doped hBN [\Fig[b,d]{fig:hBN_ph_spectral}].
In both cases, the bubble self-energy uses a partially-screened vertex that lacks the metallic free-carrier screening.
For TaS$_2$, this partially-screened e-ph coupling strongly deviates from the metallic screened vertex, and thus the spectral function deviates from the full result.

We note that the equivalence of the two downfolding schemes can also be generalized to the case of the bare-screened approach, where the active space includes the full set of bands.
In fact, this corresponds to TDDFPT without downfolding.

\let\FloatBarrierSaved\FloatBarrier
\let\FloatBarrier\relax
\section{Numerical validation of the DFPT+A method} \label{sec:validation}
\let\FloatBarrier\FloatBarrierSaved

\begin{figure}[tb]
\includegraphics[width=0.99\linewidth]{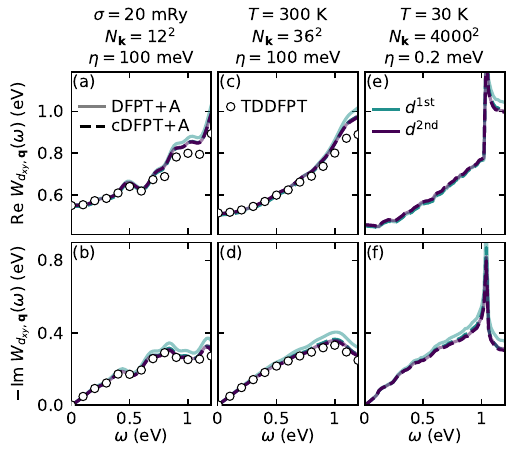}
\caption{
    Dynamical fully-screened density-density e-e interaction of the $d_{xy}$ WF of monolayer TaS$_2$ at $\bq = 0.6\,\bq^{\rm M}$ computed using DFPT+A (solid lines) and cDFPT+A (dashed lines) downfolding with the first- and second-neighbor cutoffs, compared with the direct TDDFPT result (circles).
    The columns correspond to (a), (b) the high-smearing base system (Gaussian smearing of $20~\mathrm{mRy}$, $12^2$ $\bk$ grid), (c), (d) the target parameters of \cref{sec:theory_demo} ($T = 300~\mathrm{K}$, $36^2$ $\bk$ grid), and (e), (f) a low temperature and a dense $\bk$ grid ($T = 30~\mathrm{K}$, $4000^2$ $\bk$ grid), for which only downfolding is feasible.
    We use a complex frequency $\omega + i \eta$, with $\eta = 100~\mathrm{meV}$ in (a) to (d) and $\eta = 0.2~\mathrm{meV}$ in (e) and (f).
}
\label{fig:TaS2_dyn_ee_tddfpt}
\end{figure}

\begin{table}[tb]
\centering
\caption{
    Computational cost, in seconds of wall time, of the dynamical e-e interaction of \cref{fig:TaS2_dyn_ee_tddfpt} for a single $\bq$ point.
    DFPT+A downfolding consists of the DFPT calculation of the partially-screened interaction for the base system and of the active-space calculation.
    The TDDFPT cost is for a single frequency, and that for the $4000^2$ grid, marked with an asterisk, is an estimate based on the $36^2$ grid calculation, assuming a cost linear in the number of $\bk$ points.
    The DFPT and TDDFPT calculations use an AMD Genoa CPU with 96 cores, and the active-space calculations use a single NVIDIA H100 GPU.
    The base DFPT calculation, on a $12^2$ $\bk$ grid, is done only once: its result can be reused for different frequencies, target $\bk$ grids, and temperatures, and can be interpolated to other $\bq$ points.
}
\label{tab:TaS2_cost}
\setlength{\tabcolsep}{5pt}
\begin{tabular}{lccrr}
    & & & \multicolumn{2}{c}{Target $\bk$ grid} \\ \cline{4-5}
    Calculation & $d^{\rm cut}$ & Frequencies & $36^2$ & $4000^2$ \\ \hline
    TDDFPT & -- & 1 & 6\,000 & $7 \times 10^7$* \\ \hline\hline
    \multirow{2}{*}{DFPT (base)} & $d^{\rm 1st}$ & -- & \multicolumn{2}{c}{3\,200} \\
    & $d^{\rm 2nd}$ & -- & \multicolumn{2}{c}{5\,900} \\ \hline
    \multirow{2}{*}{DFPT+A} & $d^{\rm 1st}$ & 101 & 0.03 & 51.7 \\
    & $d^{\rm 2nd}$ & 101 & 0.05 & 121 \\
\end{tabular}
\end{table}

The comparison with direct DFPT in \cref{sec:theory_demo} tests, in the static limit, the rigid-band approximation [\cref{eq:rigid_band}], the replacement of the target rest susceptibility by that of the base system [\cref{eq:chib_approx2}], and the truncation of $W^{\rm p}$ [\cref{eq:eph_Wp_trunc}].
In this appendix, we test the remaining approximation, the static approximation of the rest susceptibility [\cref{eq:chib_approx1}], against direct TDDFPT calculations for monolayer TaS$_2$.
We also test the size of the active space, and assess the effect of the exchange-correlation kernel.
This test required the implementation of TDDFPT in \textsc{Quantum ESPRESSO}, which we describe in \cref{app:tddfpt}.

\begin{figure}[tb]
\includegraphics[width=0.99\linewidth]{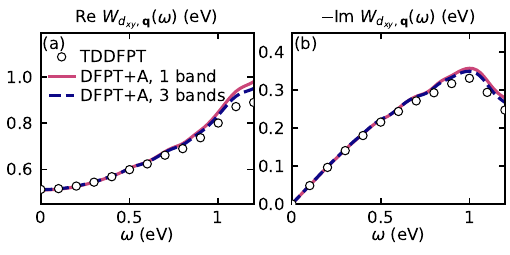}
\caption{
    Dynamical fully-screened e-e interaction as in \Fig[c,d]{fig:TaS2_dyn_ee_tddfpt}, computed using TDDFPT (circles), DFPT+A downfolding with a 1-band active space, and DFPT+A downfolding with a 3-band active space.
    Both use $d^{\rm cut} = d^{\rm 2nd}$.
    We use a complex frequency $\omega + i \eta$, with $\eta = 100~\mathrm{meV}$.
}
\label{fig:TaS2_dyn_ee_tddfpt_window}
\end{figure}

We first validate the static approximation of the rest susceptibility [\cref{eq:chib_approx1}].
We test whether downfolding can reproduce the dynamical fully-screened e-e interaction computed using direct TDDFPT calculations.
\Cref{fig:TaS2_dyn_ee_tddfpt} shows that for both the base and target systems, downfolding does reproduce the TDDFPT result in the low-frequency regime.
The static approximation of the rest susceptibility is valid for $\omega \lesssim 0.5~\mathrm{eV}$.
As we are mainly interested in phonon dynamics, which occur below $0.2~\mathrm{eV}$ in the materials studied here, this approximation is sufficient for our purpose.
If responses at higher frequencies are of interest, one may increase the size of the active space to systematically improve the accuracy, see \cref{fig:TaS2_dyn_ee_tddfpt_window}.
With downfolding, we can reach much lower temperatures and smearing and denser $\bk$ grids that are not accessible to direct TDDFPT, as we show in \Fig[e,f]{fig:TaS2_dyn_ee_tddfpt}.
\Cref{tab:TaS2_cost} compares the computational costs: for the $4000^2$ $\bk$ grid, direct TDDFPT at a single frequency would cost about $10^4$ times more than the full DFPT+A calculation for 101 frequencies, including the base DFPT calculation.
The cDFPT+A and DFPT+A schemes converge to identical results when second-neighbor interactions are included, again confirming the formal equivalence of the two schemes.

Finally, we test whether the DFPT+A downfolding scheme works when the active space includes bands entangled with other bands excluded from Wannierization.
To this end, we include all three Wannier-interpolated bands in the active space, which are entangled with the other conduction bands [\Fig[b]{fig:bands}], instead of only the band crossing the Fermi level.
This poses no difficulty for the DFPT+A scheme, as changing the active space does not affect the partially-screened e-e interactions.
It only affects the dynamical correction that can be easily computed using the same Wannier interpolation.
\Cref{fig:TaS2_dyn_ee_tddfpt_window} shows that the result with an entangled 3-band active space agrees well with that of an isolated 1-band active space in the low-frequency region where 1-band downfolding is valid.
At frequencies above about 0.8~eV, the 3-band active-space result is closer to the TDDFPT result than the 1-band one, as it includes more dynamical screening effects.
The agreement can be systematically improved by further enlarging the active space.

We also examine the effect of the exchange-correlation kernel in the bare interaction $\hat{v}^{\rm b}_\bq$ (\cref{sec:theory_ee_DFPT}).
We repeat the static calculations of \cref{fig:TaS2_static_val} keeping only the Hartree part of the kernel, which corresponds to the RPA.
We label the resulting partially-screened and downfolded quantities by ``RPA'' and ``RPA+A'', respectively.
Without the exchange-correlation kernel, the onsite density-density interaction is larger by about 0.2~eV, for both the base system and the target [\Fig[a]{fig:TaS2_static_val_rpa}], and the off-diagonal element also changes slightly.
In contrast, the phonon dispersions obtained from RPA+A closely match those of DFPT+A, and both agree with the direct DFPT result [\Fig[c]{fig:TaS2_static_val_rpa}].
This result is consistent with the fact that the screened-screened approximation, which does not involve the e-e interaction at all, is already very close to the full result, as shown in \cref{fig:TaS2_ph_spectral}.
Thus, for monolayer TaS$_2$, the exchange-correlation kernel matters for the e-e interaction but has only a minor effect on the phonons.

\begin{figure}[tb]
\includegraphics[width=0.99\linewidth]{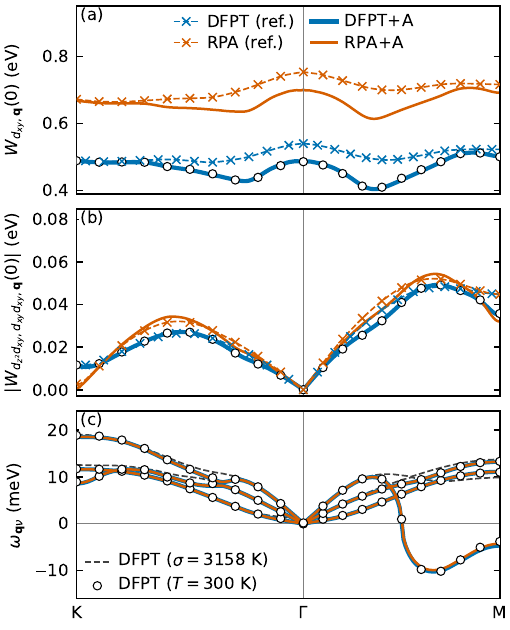}
\caption{
    Effect of the exchange-correlation kernel on the static quantities of monolayer TaS$_2$ shown in \cref{fig:TaS2_static_val}, with the same base system (Gaussian smearing of $20~\mathrm{mRy}$, $12 \times 12 \times 1$ $\bk$ grid) and target (Fermi--Dirac occupations at $T = 300~\mathrm{K}$, $36 \times 36 \times 1$ $\bk$ grid) parameters.
    DFPT and RPA denote calculations with and without the exchange-correlation kernel in $\hat{v}^{\rm b}_\bq$, respectively.
    (a) Static density-density e-e interaction for the $d_{xy}$ WF.
    (b) Absolute value of the off-diagonal element of the static e-e interaction between the $d_{z^2} d_{xy}$ and $d_{xy} d_{xy}$ WF pairs.
    (c) Dispersion of the acoustic phonons.
}
\label{fig:TaS2_static_val_rpa}
\end{figure}

\let\FloatBarrierSaved\FloatBarrier
\let\FloatBarrier\relax
\section{Computational details} \label{app:comp_details}
\let\FloatBarrier\FloatBarrierSaved

\begin{table*}[t]
\centering
\begin{tabular}{c|cccc|cccc|ccc}
\hline
Material & $a$ (bohr) & $L_z$ (bohr) & $E^{\rm cut}$ (Ry) & $N_\bk^{\rm base}$ & $N_{\bk}^{\rm c}$ & Initial guess & $E^{\rm froz}$ (eV) & $N^{\rm band}$ & $N_\bk^{\rm f}$ & $\mc{W}^{\rm A}$ (eV) & $d^{\rm cut}$
\\ \hline
hBN & 4.69 & 40.0 & 160 & $12^2$ & $12^2$ & N: $p$ & & 3 & $2000^2$ & $[-1.5, 0.3]$ & $d^{\rm onsite}$ \\
TaS$_2$ & 6.31 & 28.35 & 100 & $12^2$ & $12^2$ & Ta: $d_{z^2}$, $d_{xy}$, $d_{x^2-y^2}$ & 1.67 & 5 & $4000^2$ & $[-0.7, 1.3]$ & $d^{\rm 1st}$ \\
\hline
\end{tabular}
\caption{
    Computational parameters used in this work:
    in-plane lattice parameter $a$,
    height of the periodic cell $L_z$,
    kinetic energy cutoff $E^{\rm cut}$,
    size of the $\bk$ grid for the DFT and DFPT calculations of the base system $N_\bk^{\rm base}$,
    size of the coarse $\bk$ grid for Wannierization $N_\bk^{\rm c}$,
    initial guesses for constructing the WFs,
    upper bound of the frozen (inner) window $E^{\rm froz}$,
    number of bands included for Wannierization $N^{\rm band}$,
    size of the fine $\bk$ grid for the active-space susceptibility $N_\bk^{\rm f}$,
    active window $\mc{W}^{\rm A}$,
    and the truncation cutoff of $W^{\rm p}$ (\cref{tab:cutoff}) used for the results of \cref{sec:applications}.
    Energies are measured from the valence band maximum for hBN and from the Fermi level for TaS$_2$.
    The base calculation for TaS$_2$ is done with a Gaussian smearing of $20~\mathrm{mRy}$, and that for hBN is the undoped crystal.
}
\label{tab:comp_params}
\end{table*}
Here, we give the details of the calculations summarized in \cref{sec:DFTdetails}.
We perform DFT, DFPT, and TDDFPT calculations using the \QE package~\cite{Giannozzi2009, Giannozzi2017}, with the modifications described in \cref{sec:theory_ee_DFPT}.
We use norm-conserving pseudopotentials generated with \textsc{ONCVPSP}~\cite{Hamann2013ONCVPSP}.
For hBN, we use the local density approximation~\cite{Perdew1992} and the pseudopotentials of Ref.~\cite{Ponce2023PRB}.
For TaS$_2$, we use the Perdew--Burke--Ernzerhof functional~\cite{Perdew1996} and the scalar-relativistic pseudopotentials from \textsc{PseudoDojo} (v0.5)~\cite{Setten2018}.
The long-range Coulomb interaction between periodic images of the monolayers is removed with the 2D Coulomb cutoff~\cite{Sohier2017}.
We then use \textsc{Wannier90}~\cite{Pizzi2020W90} to construct the WFs~\cite{Marzari1997, Souza2001} and \textsc{EPW}~\cite{Giustino2007, Ponce2016EPW, Lee2023EPW} to obtain the real-space e-ph matrix elements and interatomic force constants from DFPT on a $12 \times 12 \times 1$ $\bq$ grid.
For hBN, the long-range dipole contribution to the e-ph matrix elements is treated with the 2D formulation of Refs.~\cite{Sohier2017, Ponce2023PRB}.
Finally, we perform Wannier interpolation and solve the active-space screening equations using the in-house code \textsc{ElectronPhonon.jl}~\cite{EPjl} written in the Julia programming language~\cite{Bezanson2017Julia}.
For the validation calculations in \cref{sec:theory_demo}, we perform direct DFPT calculations at the target parameters using $24^2$ (hBN) and $36^2$ (TaS$_2$) $\bk$ grids.
The DFPT+A calculations compared with them use the same grids.
We summarize the computational parameters in \cref{tab:comp_params}.

In DFT calculations, nonmonotonic occupation functions such as Methfessel--Paxton smearing~\cite{Methfessel1989} or cold smearing~\cite{Marzari1999Smearing} are often used to accelerate the convergence with respect to the $\bk$-point sampling.
These functions exhibit a small region of positive slope $\dd f(\frac{\veps - \mu}{\sigma}) / \dd \veps > 0$ where $\abs{\veps - \mu} \sim 2 \sigma$~\cite{DosSantos2023}.
This nonmonotonicity can lead to positive values of $(f_\nk - f_\mkq)  / (\veps_\nk - \veps_\mkq)$ and thus an unphysical positive contribution to the static susceptibility.
In usual DFPT calculations, such terms are harmless, as they are outweighed by the strong negative contributions from the states near the Fermi level ($\abs{\veps - \mu} \ll 2 \sigma$).
However, in a cDFPT calculation where such states are excluded from the susceptibility, these unphysical contributions can dominate and lead to a negative screening, i.e., screened interactions larger than the bare ones.
To avoid this issue, we use the Gaussian smearing, which yields a monotonic occupation function and decays faster than the Fermi--Dirac distribution.

We compute the partially-screened e-e interaction $W^{\rm p}_\bq$ with DFPT for the base system on $\bq$ points along the $\Gamma$M and $\Gamma$K lines, $\bq = t_n \bq_1$ with $\bq_1 = \bq^{\rm M}$ or $\bq^{\rm K}$, and interpolate it to other $\bq$ points along these lines with the one-dimensional scheme of \cref{sec:theory_itp}.
The two lines are interpolated separately.
For hBN, we use $t_n = 0.002, 0.004, \dots, 0.02$, then $0.03, 0.04, \dots, 0.2$, and then $0.3, 0.4, \dots, 1$ (36 points per line), together with $\bq = \mb{0}$, so the sampling is densest near $\Gamma$, where the long-range part of the interaction varies most rapidly.
For TaS$_2$, we use $t_n = 0.004, 0.008, \dots, 0.08$ and then $0.12, 0.16, \dots, 1$ (43 points per line), together with $\bq = \mb{0}$.
For hBN, and for the cDFPT interaction of TaS$_2$, we subtract the monopole-monopole term [\cref{eq:itp_Wp_mm}] before interpolating, and we do not use the $\bq = \mb{0}$ point as an interpolation knot.
For the DFPT interaction of TaS$_2$, we directly interpolate the interaction, including $\bq = \mb{0}$ as an interpolation knot.

For the dynamical DFPT+A calculations for hBN, we use $\eta = 1$~meV and a frequency grid from 0 to $300$~meV with a spacing of 1~meV.
For TaS$_2$ (\cref{sec:result_TaS2}), we use $\eta = 0.2$~meV and a frequency grid from 0 to 80~meV with a spacing of 0.2~meV.
For the frequency integrals of \cref{fig:hBN_W_loc_kk} for hBN, we extend the frequency range from 300~meV to 1~eV with a separate calculation using $\eta = 5$~meV and a spacing of 5~meV.
The $\bq$ points along $\Gamma$M and $\Gamma$K are sampled uniformly in $\sqrt{\abs{\bq}}$ to resolve the plasmon dispersion $\omega^{\rm pl} \propto \sqrt{q}$ near $\Gamma$.
For the comparison with direct TDDFPT in \cref{sec:validation}, we use $\eta = 100$~meV for the base ($\sigma = 20~\mathrm{mRy}$, $12^2$ $\bk$ grid) and target ($T = 300~\mathrm{K}$, $36^2$ $\bk$ grid) parameters, and $\eta = 0.2$~meV for the downfolding at $T = 30~\mathrm{K}$ on a $4000^2$ $\bk$ grid.

We used Claude Code (Claude Opus and Claude Fable models) and Codex (GPT Sol models) to implement and optimize the GPU offloading of the Julia implementation of Wannier interpolation and downfolding.
We validated the results by comparing them with the reference CPU implementation, and manually inspected and revised the code.
We also used Claude Code (Claude Opus 5.5) to generate the schematic of \cref{fig:downfolding_schematic} from a written description, and to revise the manuscript; all generated content was checked and edited by the authors.
All data, inputs, outputs, pseudopotentials, codes, scripts, and the prompts used are available on the Materials Cloud Archive~\cite{MaterialsCloudArchive}.

\section{One-dimensional model of parallel bands} \label{app:1d_model}

Here we analyze the nonadiabatic renormalization of monolayer TaS$_2$ on the $\Gamma$K path discussed in \cref{sec:result_TaS2}.
We model the band structure of TaS$_2$ with two Fermi-level crossings with a similar Fermi velocity [\Fig[c]{fig:TaS2_ph_spectral_slice_K}] using a simple one-dimensional model with two parallel bands with linear dispersion with the same slope.
We also consider the anti-parallel case of two bands with opposite slopes, as in the usual Kohn anomaly picture.
The dispersions are defined as
\begin{subequations}
\begin{alignat}{2}
    \label{eq:phself_1d_model_energy_1}
    \veps_{1 k} &= + v (k - k_1) \,, &\ \veps_{2 k} &= v (k - k_2) \,, \\
    \label{eq:phself_1d_model_energy_2}
    \veps_{1 k} &= - v (k - k_1) \,, &\ \veps_{2 k} &= v (k - k_2) \,,
\end{alignat}
\end{subequations}
for the parallel and anti-parallel cases, respectively, and are shown in \Fig[a,d]{fig:TaS2_GK_1d_model}.
The susceptibility is computed using
\begin{equation} \label{eq:phself_1d_model_chi}
    \chi^{\rm b}_q(\omega) = \nbint{-\infty}{\infty} \mathrm{d}k \, \frac{f_{1 k} - f_{2 k+q}}{\veps_{1 k} - \veps_{2 k+q} + \omega + i \eta} \,,
\end{equation}
neglecting the matrix element effects for simplicity.

\begin{figure}[tb]
\includegraphics[width=0.99\linewidth]{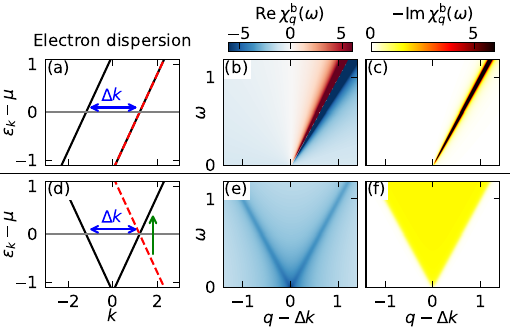}
\caption{
    One-dimensional model with linear dispersion with (a) to (c) parallel and (d) to (f) anti-parallel Fermi velocities.
    (a) The parallel band structure defined in \cref{eq:phself_1d_model_energy_1}.
    The red dashed line indicates the left band shifted by $\Delta k = k_2 - k_1$.
    (b), (c) Real and imaginary parts of the susceptibility computed using \cref{eq:phself_1d_model_chi}.
    (d) to (f) Same as (a) to (c), but with the anti-parallel band structure defined in \cref{eq:phself_1d_model_energy_2}.
    The green arrow in (d) indicates a vertical transition between the right band and the shifted left band, which is not possible in the parallel case.
    We set $v = 1$, $k_1 = -1.2$, $k_2 = 1.2$, and $T = \eta = 0.01$.
}
\label{fig:TaS2_GK_1d_model}
\end{figure}

\Cref{fig:TaS2_GK_1d_model} shows the resulting susceptibility.
First, for the parallel case [\Fig[a-c]{fig:TaS2_GK_1d_model}], we find that the susceptibility has a strong enhancement along the line $\omega = v (q - \Delta k)$, corresponding to excitations between the two bands.
The weight of this peak vanishes at $q = \Delta k$ and increases linearly with $q - \Delta k$, so that it vanishes at $\omega = 0$ and grows with $\omega$.
These behaviors qualitatively agree with the right side of the phonon self-energy of TaS$_2$ shown in \Fig[a,b]{fig:TaS2_ph_spectral_slice_K}.
Note that an additional parallel band pair with velocity $-v$ gives the left-right mirrored version.
This energy and momentum dependence can be understood as the result of Pauli blocking.
If we shift the left band by $\Delta k = k_2 - k_1$ [red dashed line in \Fig[a]{fig:TaS2_GK_1d_model}], the two bands overlap perfectly.
Thus, for $q = \Delta k$, the initial and final states always have the same occupation, yielding a vanishing susceptibility at all frequencies.
For $q \neq \Delta k$, all transitions have the same energy $\veps_{2,k+q} - \veps_{1k} = v (q - \Delta k)$, and the occupation difference is nonzero only on a $k$ interval of length $\abs{q - \Delta k}$.
The situation is analogous to the intraband excitations, except that here the initial and final states are separated in momentum space by $\Delta k$.
Since the interband transition is focused along the line $\omega = v (q - \Delta k)$, there is a strong peak in the susceptibility, which leads to a strong nonadiabatic renormalization of the phonon spectral function.

In contrast, for the anti-parallel case [\Fig[d-f]{fig:TaS2_GK_1d_model}], the susceptibility displays the usual Kohn anomaly behavior, with a logarithmic divergence in the real part at $q = \Delta k$ and a step-like increase in the imaginary part at $\omega = v \abs{q - \Delta k}$.
We can see that the vertical transition between the right band and the shifted left band, which corresponds to a transition with $q = \Delta k$, connects occupied and unoccupied states, yielding a large static susceptibility at $\omega = 0$.
The imaginary part of the susceptibility is nonzero in a wide range of energy-momentum space, in contrast to the parallel case.
Thus, the static phonon dispersion is renormalized, but the nonadiabatic effects are less pronounced.

\section{Brillouin-zone integration of the local dynamical interaction} \label{app:Uloc_integration}

The local dynamical interaction of \cref{eq:U_loc_def} requires a Brillouin-zone average of the interaction over $\bq$.
Calculating the dynamical interaction at each $\bq$ point requires computing the partially-screened e-e interaction $W^{\rm p}_\bq(\omega)$ at that $\bq$ point.
While this interaction can be easily computed for a single $\bq$ point, the Brillouin-zone average requires a dense $\bq$ grid to converge, which is computationally expensive.
Doing a two- and three-dimensional Fourier interpolation of this quantity requires a careful treatment of the long-range Coulomb interaction, which is beyond the scope of this work.
As an alternative, we perform one-dimensional spline interpolation (\cref{sec:theory_itp}), as in Refs.~\cite{Macheda2022, Macheda2023}.

For a two-dimensional hexagonal system, we approximate the Brillouin-zone average of the local interaction by a simple one-dimensional integral along the two high-symmetry lines $\Gamma$M and $\Gamma$K.
For each line, we take the interaction to be isotropic and average it over the disc that reaches the zone boundary along that line,
\begin{equation} \label{eq:U_loc_line}
    U^{\rm M/K}(\omega)
    = \frac{2}{(q^{\rm M/K})^2} \int_0^{q^{\rm M/K}} \dd q \; q \, W^{\rm M/K}(\omega, q) \,,
\end{equation}
where $q^{\rm M/K} \equiv \abs{\bq^{\rm M/K}}$, $\hat{\bq}^{\rm M/K} \equiv \bq^{\rm M/K} / q^{\rm M/K}$, and $W^{\rm M/K}(\omega, q) \equiv W^{\rm tot}_{I_0 I_0, \, q \hat{\bq}^{\rm M/K}}(\omega)$ is the density-density interaction at distance $q$ from $\Gamma$ along the $\Gamma$M or $\Gamma$K direction.
The local interaction is the plain average of the two,
\begin{equation} \label{eq:U_loc_avg}
    U^{\rm tot}(\omega)
    = \frac{1}{2} \bigl[ U^{\rm M}(\omega) + U^{\rm K}(\omega) \bigr] \,.
\end{equation}
Although this approximation neglects the hexagonal shape of the Brillouin zone, it provides a good estimate of the two-dimensional average.
The radial integral uses a dense sampling of the interaction along the two lines with a spline interpolation of the DFPT e-e interaction.

\begin{figure}[tb]
\includegraphics[width=0.99\linewidth]{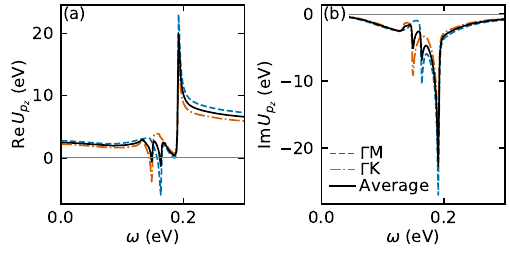}
\caption{
    (a) Real and (b) imaginary parts of the local interaction of monolayer hBN at $T=300~\mathrm{K}$ and $n=-4\times10^{12}~\mathrm{cm^{-2}}$, computed from the $\Gamma$M line alone and from the $\Gamma$K line alone [\cref{eq:U_loc_line}], and their average [\cref{eq:U_loc_avg}].
}
\label{fig:hBN_W_loc_rays}
\end{figure}

\Cref{fig:hBN_W_loc_rays} compares the two single-line integrals with their average for monolayer hBN.
The two lines give similar overall frequency dependence, including the LO phonon resonance.
They differ in the position of additional sharp features between 0.14 and 0.17~eV, which are due to the zone-boundary phonon modes.
The average retains both sets of features with half their weight.
With a full two-dimensional integration, these features would be broadened and their weight would be further reduced.
Yet, the largest peak at 0.19~eV, which is due to the LO phonon near $\Gamma$, would be largely unaffected.

\FloatBarrier
\bibliography{main}

\begin{thebibliography}{139}%
\makeatletter
\providecommand \@ifxundefined [1]{%
 \@ifx{#1\undefined}
}%
\providecommand \@ifnum [1]{%
 \ifnum #1\expandafter \@firstoftwo
 \else \expandafter \@secondoftwo
 \fi
}%
\providecommand \@ifx [1]{%
 \ifx #1\expandafter \@firstoftwo
 \else \expandafter \@secondoftwo
 \fi
}%
\providecommand \natexlab [1]{#1}%
\providecommand \enquote  [1]{``#1''}%
\providecommand \bibnamefont  [1]{#1}%
\providecommand \bibfnamefont [1]{#1}%
\providecommand \citenamefont [1]{#1}%
\providecommand \href@noop [0]{\@secondoftwo}%
\providecommand \href [0]{\begingroup \@sanitize@url \@href}%
\providecommand \@href[1]{\@@startlink{#1}\@@href}%
\providecommand \@@href[1]{\endgroup#1\@@endlink}%
\providecommand \@sanitize@url [0]{\catcode `\\12\catcode `\$12\catcode
  `\&12\catcode `\#12\catcode `\^12\catcode `\_12\catcode `\%12\relax}%
\providecommand \@@startlink[1]{}%
\providecommand \@@endlink[0]{}%
\providecommand \url  [0]{\begingroup\@sanitize@url \@url }%
\providecommand \@url [1]{\endgroup\@href {#1}{\urlprefix }}%
\providecommand \urlprefix  [0]{URL }%
\providecommand \Eprint [0]{\href }%
\providecommand \doibase [0]{https://doi.org/}%
\providecommand \selectlanguage [0]{\@gobble}%
\providecommand \bibinfo  [0]{\@secondoftwo}%
\providecommand \bibfield  [0]{\@secondoftwo}%
\providecommand \translation [1]{[#1]}%
\providecommand \BibitemOpen [0]{}%
\providecommand \bibitemStop [0]{}%
\providecommand \bibitemNoStop [0]{.\EOS\space}%
\providecommand \EOS [0]{\spacefactor3000\relax}%
\providecommand \BibitemShut  [1]{\csname bibitem#1\endcsname}%
\let\auto@bib@innerbib\@empty
\bibitem [{\citenamefont {Giustino}(2017)}]{Giustino2017}%
  \BibitemOpen
  \bibfield  {author} {\bibinfo {author} {\bibfnamefont {F.}~\bibnamefont
  {Giustino}},\ }\bibfield  {title} {\bibinfo {title} {Electron-phonon
  interactions from first principles},\ }\href
  {https://doi.org/10.1103/RevModPhys.89.015003} {\bibfield  {journal}
  {\bibinfo  {journal} {Rev. Mod. Phys.}\ }\textbf {\bibinfo {volume} {89}},\
  \bibinfo {pages} {015003} (\bibinfo {year} {2017})}\BibitemShut {NoStop}%
\bibitem [{\citenamefont {Baroni}\ \emph {et~al.}(2001)\citenamefont {Baroni},
  \citenamefont {de~Gironcoli}, \citenamefont {Dal~Corso},\ and\ \citenamefont
  {Giannozzi}}]{Baroni2001}%
  \BibitemOpen
  \bibfield  {author} {\bibinfo {author} {\bibfnamefont {S.}~\bibnamefont
  {Baroni}}, \bibinfo {author} {\bibfnamefont {S.}~\bibnamefont
  {de~Gironcoli}}, \bibinfo {author} {\bibfnamefont {A.}~\bibnamefont
  {Dal~Corso}},\ and\ \bibinfo {author} {\bibfnamefont {P.}~\bibnamefont
  {Giannozzi}},\ }\bibfield  {title} {\bibinfo {title} {Phonons and related
  crystal properties from density-functional perturbation theory},\ }\href
  {https://doi.org/10.1103/RevModPhys.73.515} {\bibfield  {journal} {\bibinfo
  {journal} {Rev. Mod. Phys.}\ }\textbf {\bibinfo {volume} {73}},\ \bibinfo
  {pages} {515} (\bibinfo {year} {2001})}\BibitemShut {NoStop}%
\bibitem [{\citenamefont {Giustino}\ \emph {et~al.}(2007)\citenamefont
  {Giustino}, \citenamefont {Cohen},\ and\ \citenamefont
  {Louie}}]{Giustino2007}%
  \BibitemOpen
  \bibfield  {author} {\bibinfo {author} {\bibfnamefont {F.}~\bibnamefont
  {Giustino}}, \bibinfo {author} {\bibfnamefont {M.~L.}\ \bibnamefont
  {Cohen}},\ and\ \bibinfo {author} {\bibfnamefont {S.~G.}\ \bibnamefont
  {Louie}},\ }\bibfield  {title} {\bibinfo {title} {Electron-phonon interaction
  using {Wannier} functions},\ }\href
  {https://doi.org/10.1103/PhysRevB.76.165108} {\bibfield  {journal} {\bibinfo
  {journal} {Phys. Rev. B}\ }\textbf {\bibinfo {volume} {76}},\ \bibinfo
  {pages} {165108} (\bibinfo {year} {2007})}\BibitemShut {NoStop}%
\bibitem [{\citenamefont {Marzari}\ \emph {et~al.}(2012)\citenamefont
  {Marzari}, \citenamefont {Mostofi}, \citenamefont {Yates}, \citenamefont
  {Souza},\ and\ \citenamefont {Vanderbilt}}]{Marzari2012}%
  \BibitemOpen
  \bibfield  {author} {\bibinfo {author} {\bibfnamefont {N.}~\bibnamefont
  {Marzari}}, \bibinfo {author} {\bibfnamefont {A.~A.}\ \bibnamefont
  {Mostofi}}, \bibinfo {author} {\bibfnamefont {J.~R.}\ \bibnamefont {Yates}},
  \bibinfo {author} {\bibfnamefont {I.}~\bibnamefont {Souza}},\ and\ \bibinfo
  {author} {\bibfnamefont {D.}~\bibnamefont {Vanderbilt}},\ }\bibfield  {title}
  {\bibinfo {title} {Maximally localized {{Wannier}} functions: {{Theory}} and
  applications},\ }\href {https://doi.org/10.1103/RevModPhys.84.1419}
  {\bibfield  {journal} {\bibinfo  {journal} {Rev. Mod. Phys.}\ }\textbf
  {\bibinfo {volume} {84}},\ \bibinfo {pages} {1419} (\bibinfo {year}
  {2012})}\BibitemShut {NoStop}%
\bibitem [{\citenamefont {Ponc{\'e}}\ \emph {et~al.}(2020)\citenamefont
  {Ponc{\'e}}, \citenamefont {Li}, \citenamefont {Reichardt},\ and\
  \citenamefont {Giustino}}]{Ponce2020Review}%
  \BibitemOpen
  \bibfield  {author} {\bibinfo {author} {\bibfnamefont {S.}~\bibnamefont
  {Ponc{\'e}}}, \bibinfo {author} {\bibfnamefont {W.}~\bibnamefont {Li}},
  \bibinfo {author} {\bibfnamefont {S.}~\bibnamefont {Reichardt}},\ and\
  \bibinfo {author} {\bibfnamefont {F.}~\bibnamefont {Giustino}},\ }\bibfield
  {title} {\bibinfo {title} {First-principles calculations of charge carrier
  mobility and conductivity in bulk semiconductors and two-dimensional
  materials},\ }\href {https://doi.org/10.1088/1361-6633/ab6a43} {\bibfield
  {journal} {\bibinfo  {journal} {Rep. Prog. Phys.}\ }\textbf {\bibinfo
  {volume} {83}},\ \bibinfo {pages} {036501} (\bibinfo {year}
  {2020})}\BibitemShut {NoStop}%
\bibitem [{\citenamefont {Franchini}\ \emph {et~al.}(2021)\citenamefont
  {Franchini}, \citenamefont {Reticcioli}, \citenamefont {Setvin},\ and\
  \citenamefont {Diebold}}]{Franchini2021}%
  \BibitemOpen
  \bibfield  {author} {\bibinfo {author} {\bibfnamefont {C.}~\bibnamefont
  {Franchini}}, \bibinfo {author} {\bibfnamefont {M.}~\bibnamefont
  {Reticcioli}}, \bibinfo {author} {\bibfnamefont {M.}~\bibnamefont {Setvin}},\
  and\ \bibinfo {author} {\bibfnamefont {U.}~\bibnamefont {Diebold}},\
  }\bibfield  {title} {\bibinfo {title} {Polarons in materials},\ }\href
  {https://doi.org/10.1038/s41578-021-00289-w} {\bibfield  {journal} {\bibinfo
  {journal} {Nat. Rev. Mater.}\ }\textbf {\bibinfo {volume} {6}},\ \bibinfo
  {pages} {560} (\bibinfo {year} {2021})}\BibitemShut {NoStop}%
\bibitem [{\citenamefont {Boeri}\ \emph {et~al.}(2022)\citenamefont {Boeri},
  \citenamefont {Hennig}, \citenamefont {Hirschfeld}, \citenamefont {Profeta},
  \citenamefont {Sanna}, \citenamefont {Zurek}, \citenamefont {Pickett},
  \citenamefont {Amsler}, \citenamefont {Dias}, \citenamefont {Eremets},
  \citenamefont {Heil}, \citenamefont {Hemley}, \citenamefont {Liu},
  \citenamefont {Ma}, \citenamefont {Pierleoni}, \citenamefont {Kolmogorov},
  \citenamefont {Rybin}, \citenamefont {Novoselov}, \citenamefont {Anisimov},
  \citenamefont {Oganov}, \citenamefont {Pickard}, \citenamefont {Bi},
  \citenamefont {Arita}, \citenamefont {Errea}, \citenamefont {Pellegrini},
  \citenamefont {Requist}, \citenamefont {Gross}, \citenamefont {Margine},
  \citenamefont {Xie}, \citenamefont {Quan}, \citenamefont {Hire},
  \citenamefont {Fanfarillo}, \citenamefont {Stewart}, \citenamefont {Hamlin},
  \citenamefont {Stanev}, \citenamefont {Gonnelli}, \citenamefont {Piatti},
  \citenamefont {Romanin}, \citenamefont {Daghero},\ and\ \citenamefont
  {Valenti}}]{Boeri2022}%
  \BibitemOpen
  \bibfield  {author} {\bibinfo {author} {\bibfnamefont {L.}~\bibnamefont
  {Boeri}}, \bibinfo {author} {\bibfnamefont {R.}~\bibnamefont {Hennig}},
  \bibinfo {author} {\bibfnamefont {P.}~\bibnamefont {Hirschfeld}}, \bibinfo
  {author} {\bibfnamefont {G.}~\bibnamefont {Profeta}}, \bibinfo {author}
  {\bibfnamefont {A.}~\bibnamefont {Sanna}}, \bibinfo {author} {\bibfnamefont
  {E.}~\bibnamefont {Zurek}}, \bibinfo {author} {\bibfnamefont {W.~E.}\
  \bibnamefont {Pickett}}, \bibinfo {author} {\bibfnamefont {M.}~\bibnamefont
  {Amsler}}, \bibinfo {author} {\bibfnamefont {R.}~\bibnamefont {Dias}},
  \bibinfo {author} {\bibfnamefont {M.~I.}\ \bibnamefont {Eremets}}, \bibinfo
  {author} {\bibfnamefont {C.}~\bibnamefont {Heil}}, \bibinfo {author}
  {\bibfnamefont {R.~J.}\ \bibnamefont {Hemley}}, \bibinfo {author}
  {\bibfnamefont {H.}~\bibnamefont {Liu}}, \bibinfo {author} {\bibfnamefont
  {Y.}~\bibnamefont {Ma}}, \bibinfo {author} {\bibfnamefont {C.}~\bibnamefont
  {Pierleoni}}, \bibinfo {author} {\bibfnamefont {A.~N.}\ \bibnamefont
  {Kolmogorov}}, \bibinfo {author} {\bibfnamefont {N.}~\bibnamefont {Rybin}},
  \bibinfo {author} {\bibfnamefont {D.}~\bibnamefont {Novoselov}}, \bibinfo
  {author} {\bibfnamefont {V.}~\bibnamefont {Anisimov}}, \bibinfo {author}
  {\bibfnamefont {A.~R.}\ \bibnamefont {Oganov}}, \bibinfo {author}
  {\bibfnamefont {C.~J.}\ \bibnamefont {Pickard}}, \bibinfo {author}
  {\bibfnamefont {T.}~\bibnamefont {Bi}}, \bibinfo {author} {\bibfnamefont
  {R.}~\bibnamefont {Arita}}, \bibinfo {author} {\bibfnamefont
  {I.}~\bibnamefont {Errea}}, \bibinfo {author} {\bibfnamefont
  {C.}~\bibnamefont {Pellegrini}}, \bibinfo {author} {\bibfnamefont
  {R.}~\bibnamefont {Requist}}, \bibinfo {author} {\bibfnamefont {E.~K.~U.}\
  \bibnamefont {Gross}}, \bibinfo {author} {\bibfnamefont {E.~R.}\ \bibnamefont
  {Margine}}, \bibinfo {author} {\bibfnamefont {S.~R.}\ \bibnamefont {Xie}},
  \bibinfo {author} {\bibfnamefont {Y.}~\bibnamefont {Quan}}, \bibinfo {author}
  {\bibfnamefont {A.}~\bibnamefont {Hire}}, \bibinfo {author} {\bibfnamefont
  {L.}~\bibnamefont {Fanfarillo}}, \bibinfo {author} {\bibfnamefont {G.~R.}\
  \bibnamefont {Stewart}}, \bibinfo {author} {\bibfnamefont {J.~J.}\
  \bibnamefont {Hamlin}}, \bibinfo {author} {\bibfnamefont {V.}~\bibnamefont
  {Stanev}}, \bibinfo {author} {\bibfnamefont {R.~S.}\ \bibnamefont
  {Gonnelli}}, \bibinfo {author} {\bibfnamefont {E.}~\bibnamefont {Piatti}},
  \bibinfo {author} {\bibfnamefont {D.}~\bibnamefont {Romanin}}, \bibinfo
  {author} {\bibfnamefont {D.}~\bibnamefont {Daghero}},\ and\ \bibinfo {author}
  {\bibfnamefont {R.}~\bibnamefont {Valenti}},\ }\bibfield  {title} {\bibinfo
  {title} {The 2021 room-temperature superconductivity roadmap},\ }\href
  {https://doi.org/10.1088/1361-648x/ac2864} {\bibfield  {journal} {\bibinfo
  {journal} {J. Phys.: Condens. Matter}\ }\textbf {\bibinfo {volume} {34}},\
  \bibinfo {pages} {183002} (\bibinfo {year} {2022})}\BibitemShut {NoStop}%
\bibitem [{\citenamefont {Dai}\ \emph {et~al.}(2025)\citenamefont {Dai},
  \citenamefont {Lafuente-Bartolome},\ and\ \citenamefont
  {Giustino}}]{Dai2025}%
  \BibitemOpen
  \bibfield  {author} {\bibinfo {author} {\bibfnamefont {Z.}~\bibnamefont
  {Dai}}, \bibinfo {author} {\bibfnamefont {J.}~\bibnamefont
  {Lafuente-Bartolome}},\ and\ \bibinfo {author} {\bibfnamefont
  {F.}~\bibnamefont {Giustino}},\ }\href
  {https://doi.org/10.48550/ARXIV.2512.06176} {\bibinfo {title} {Polarons from
  first principles}} (\bibinfo {year} {2025})\BibitemShut {NoStop}%
\bibitem [{\citenamefont {Lazzeri}\ and\ \citenamefont
  {Mauri}(2006)}]{Lazzeri2006}%
  \BibitemOpen
  \bibfield  {author} {\bibinfo {author} {\bibfnamefont {M.}~\bibnamefont
  {Lazzeri}}\ and\ \bibinfo {author} {\bibfnamefont {F.}~\bibnamefont
  {Mauri}},\ }\bibfield  {title} {\bibinfo {title} {Nonadiabatic {Kohn} anomaly
  in a doped graphene monolayer},\ }\href
  {https://doi.org/10.1103/PhysRevLett.97.266407} {\bibfield  {journal}
  {\bibinfo  {journal} {Phys. Rev. Lett.}\ }\textbf {\bibinfo {volume} {97}},\
  \bibinfo {pages} {266407} (\bibinfo {year} {2006})}\BibitemShut {NoStop}%
\bibitem [{\citenamefont {Pisana}\ \emph {et~al.}(2007)\citenamefont {Pisana},
  \citenamefont {Lazzeri}, \citenamefont {Casiraghi}, \citenamefont
  {Novoselov}, \citenamefont {Geim}, \citenamefont {Ferrari},\ and\
  \citenamefont {Mauri}}]{Pisana2007}%
  \BibitemOpen
  \bibfield  {author} {\bibinfo {author} {\bibfnamefont {S.}~\bibnamefont
  {Pisana}}, \bibinfo {author} {\bibfnamefont {M.}~\bibnamefont {Lazzeri}},
  \bibinfo {author} {\bibfnamefont {C.}~\bibnamefont {Casiraghi}}, \bibinfo
  {author} {\bibfnamefont {K.~S.}\ \bibnamefont {Novoselov}}, \bibinfo {author}
  {\bibfnamefont {A.~K.}\ \bibnamefont {Geim}}, \bibinfo {author}
  {\bibfnamefont {A.~C.}\ \bibnamefont {Ferrari}},\ and\ \bibinfo {author}
  {\bibfnamefont {F.}~\bibnamefont {Mauri}},\ }\bibfield  {title} {\bibinfo
  {title} {Breakdown of the adiabatic {Born--Oppenheimer} approximation in
  graphene},\ }\href {https://doi.org/10.1038/nmat1846} {\bibfield  {journal}
  {\bibinfo  {journal} {Nat. Mater.}\ }\textbf {\bibinfo {volume} {6}},\
  \bibinfo {pages} {198} (\bibinfo {year} {2007})}\BibitemShut {NoStop}%
\bibitem [{\citenamefont {Bonini}\ \emph {et~al.}(2007)\citenamefont {Bonini},
  \citenamefont {Lazzeri}, \citenamefont {Marzari},\ and\ \citenamefont
  {Mauri}}]{Bonini2007}%
  \BibitemOpen
  \bibfield  {author} {\bibinfo {author} {\bibfnamefont {N.}~\bibnamefont
  {Bonini}}, \bibinfo {author} {\bibfnamefont {M.}~\bibnamefont {Lazzeri}},
  \bibinfo {author} {\bibfnamefont {N.}~\bibnamefont {Marzari}},\ and\ \bibinfo
  {author} {\bibfnamefont {F.}~\bibnamefont {Mauri}},\ }\bibfield  {title}
  {\bibinfo {title} {Phonon anharmonicities in graphite and graphene},\ }\href
  {https://doi.org/10.1103/PhysRevLett.99.176802} {\bibfield  {journal}
  {\bibinfo  {journal} {Phys. Rev. Lett.}\ }\textbf {\bibinfo {volume} {99}},\
  \bibinfo {pages} {176802} (\bibinfo {year} {2007})}\BibitemShut {NoStop}%
\bibitem [{\citenamefont {Ponc{\'e}}\ and\ \citenamefont
  {Gonze}(2026)}]{Ponce2026}%
  \BibitemOpen
  \bibfield  {author} {\bibinfo {author} {\bibfnamefont {S.}~\bibnamefont
  {Ponc{\'e}}}\ and\ \bibinfo {author} {\bibfnamefont {X.}~\bibnamefont
  {Gonze}},\ }\bibfield  {title} {\bibinfo {title} {In search of the
  electron-phonon contribution to total energy},\ }\href
  {https://doi.org/10.1103/zx5f-cw7t} {\bibfield  {journal} {\bibinfo
  {journal} {PRX Energy}\ }\textbf {\bibinfo {volume} {5}},\ \bibinfo {pages}
  {043004} (\bibinfo {year} {2026})}\BibitemShut {NoStop}%
\bibitem [{\citenamefont {Calandra}\ \emph {et~al.}(2010)\citenamefont
  {Calandra}, \citenamefont {Profeta},\ and\ \citenamefont
  {Mauri}}]{Calandra2010}%
  \BibitemOpen
  \bibfield  {author} {\bibinfo {author} {\bibfnamefont {M.}~\bibnamefont
  {Calandra}}, \bibinfo {author} {\bibfnamefont {G.}~\bibnamefont {Profeta}},\
  and\ \bibinfo {author} {\bibfnamefont {F.}~\bibnamefont {Mauri}},\ }\bibfield
   {title} {\bibinfo {title} {Adiabatic and nonadiabatic phonon dispersion in a
  {Wannier} function approach},\ }\href
  {https://doi.org/10.1103/PhysRevB.82.165111} {\bibfield  {journal} {\bibinfo
  {journal} {Phys. Rev. B}\ }\textbf {\bibinfo {volume} {82}},\ \bibinfo
  {pages} {165111} (\bibinfo {year} {2010})}\BibitemShut {NoStop}%
\bibitem [{\citenamefont {Berges}\ \emph {et~al.}(2023)\citenamefont {Berges},
  \citenamefont {Girotto}, \citenamefont {Wehling}, \citenamefont {Marzari},\
  and\ \citenamefont {Ponc\'e}}]{Berges2023}%
  \BibitemOpen
  \bibfield  {author} {\bibinfo {author} {\bibfnamefont {J.}~\bibnamefont
  {Berges}}, \bibinfo {author} {\bibfnamefont {N.}~\bibnamefont {Girotto}},
  \bibinfo {author} {\bibfnamefont {T.}~\bibnamefont {Wehling}}, \bibinfo
  {author} {\bibfnamefont {N.}~\bibnamefont {Marzari}},\ and\ \bibinfo {author}
  {\bibfnamefont {S.}~\bibnamefont {Ponc\'e}},\ }\bibfield  {title} {\bibinfo
  {title} {Phonon self-energy corrections: To screen, or not to screen},\
  }\href {https://doi.org/10.1103/PhysRevX.13.041009} {\bibfield  {journal}
  {\bibinfo  {journal} {Phys. Rev. X}\ }\textbf {\bibinfo {volume} {13}},\
  \bibinfo {pages} {041009} (\bibinfo {year} {2023})}\BibitemShut {NoStop}%
\bibitem [{\citenamefont {Stefanucci}\ and\ \citenamefont
  {Perfetto}(2025)}]{Stefanucci2025}%
  \BibitemOpen
  \bibfield  {author} {\bibinfo {author} {\bibfnamefont {G.}~\bibnamefont
  {Stefanucci}}\ and\ \bibinfo {author} {\bibfnamefont {E.}~\bibnamefont
  {Perfetto}},\ }\bibfield  {title} {\bibinfo {title} {Exact formula with two
  dynamically screened electron-phonon couplings for positive phonon-linewidths
  approximations},\ }\href {https://doi.org/10.1103/PhysRevB.111.024307}
  {\bibfield  {journal} {\bibinfo  {journal} {Phys. Rev. B}\ }\textbf {\bibinfo
  {volume} {111}},\ \bibinfo {pages} {024307} (\bibinfo {year}
  {2025})}\BibitemShut {NoStop}%
\bibitem [{\citenamefont {Guandalini}\ \emph {et~al.}(2025)\citenamefont
  {Guandalini}, \citenamefont {Macheda}, \citenamefont {Caldarelli},\ and\
  \citenamefont {Mauri}}]{Guandalini2025b}%
  \BibitemOpen
  \bibfield  {author} {\bibinfo {author} {\bibfnamefont {A.}~\bibnamefont
  {Guandalini}}, \bibinfo {author} {\bibfnamefont {F.}~\bibnamefont {Macheda}},
  \bibinfo {author} {\bibfnamefont {G.}~\bibnamefont {Caldarelli}},\ and\
  \bibinfo {author} {\bibfnamefont {F.}~\bibnamefont {Mauri}},\ }\bibfield
  {title} {\bibinfo {title} {Excitonic effects in phonons: Reshaping the
  graphene {Kohn} anomalies and lifetimes},\ }\href
  {https://doi.org/10.1103/y1dn-m6pc} {\bibfield  {journal} {\bibinfo
  {journal} {Phys. Rev. Lett.}\ }\textbf {\bibinfo {volume} {135}},\ \bibinfo
  {pages} {076401} (\bibinfo {year} {2025})}\BibitemShut {NoStop}%
\bibitem [{\citenamefont {Park}(2025)}]{Park2025PRB}%
  \BibitemOpen
  \bibfield  {author} {\bibinfo {author} {\bibfnamefont {C.-H.}\ \bibnamefont
  {Park}},\ }\bibfield  {title} {\bibinfo {title} {Nonadiabatic phonon
  self-energy due to electrons with finite linewidths},\ }\href
  {https://doi.org/10.1103/6jq1-cbwq} {\bibfield  {journal} {\bibinfo
  {journal} {Phys. Rev. B}\ }\textbf {\bibinfo {volume} {112}},\ \bibinfo
  {pages} {104314} (\bibinfo {year} {2025})}\BibitemShut {NoStop}%
\bibitem [{\citenamefont {Erhardt}\ and\ \citenamefont
  {Ponc\'e}(2026)}]{Erhardt2026}%
  \BibitemOpen
  \bibfield  {author} {\bibinfo {author} {\bibfnamefont {N.~G.}\ \bibnamefont
  {Erhardt}}\ and\ \bibinfo {author} {\bibfnamefont {S.}~\bibnamefont
  {Ponc\'e}},\ }\href {https://arxiv.org/abs/2608.14882} {\bibinfo {title}
  {Higher-order nonadiabaticity governs the temperature dependence of the
  phonon spectrum}},\ \bibinfo {howpublished} {arXiv:2608.14882} (\bibinfo
  {year} {2026}),\ \Eprint {https://arxiv.org/abs/2608.14882}
  {arXiv:2608.14882} \BibitemShut {NoStop}%
\bibitem [{\citenamefont {Marchese}\ \emph {et~al.}(2024)\citenamefont
  {Marchese}, \citenamefont {Macheda}, \citenamefont {Binci}, \citenamefont
  {Calandra}, \citenamefont {Barone},\ and\ \citenamefont
  {Mauri}}]{Marchese2024}%
  \BibitemOpen
  \bibfield  {author} {\bibinfo {author} {\bibfnamefont {G.}~\bibnamefont
  {Marchese}}, \bibinfo {author} {\bibfnamefont {F.}~\bibnamefont {Macheda}},
  \bibinfo {author} {\bibfnamefont {L.}~\bibnamefont {Binci}}, \bibinfo
  {author} {\bibfnamefont {M.}~\bibnamefont {Calandra}}, \bibinfo {author}
  {\bibfnamefont {P.}~\bibnamefont {Barone}},\ and\ \bibinfo {author}
  {\bibfnamefont {F.}~\bibnamefont {Mauri}},\ }\bibfield  {title} {\bibinfo
  {title} {Born effective charges and vibrational spectra in superconducting
  and bad conducting metals},\ }\href
  {https://doi.org/10.1038/s41567-023-02203-3} {\bibfield  {journal} {\bibinfo
  {journal} {Nat. Phys.}\ }\textbf {\bibinfo {volume} {20}},\ \bibinfo {pages}
  {88} (\bibinfo {year} {2024})}\BibitemShut {NoStop}%
\bibitem [{\citenamefont {Fachin}\ \emph {et~al.}(2026)\citenamefont {Fachin},
  \citenamefont {Macheda}, \citenamefont {Barone},\ and\ \citenamefont
  {Mauri}}]{Fachin2026}%
  \BibitemOpen
  \bibfield  {author} {\bibinfo {author} {\bibfnamefont {P.}~\bibnamefont
  {Fachin}}, \bibinfo {author} {\bibfnamefont {F.}~\bibnamefont {Macheda}},
  \bibinfo {author} {\bibfnamefont {P.}~\bibnamefont {Barone}},\ and\ \bibinfo
  {author} {\bibfnamefont {F.}~\bibnamefont {Mauri}},\ }\href
  {https://arxiv.org/abs/2607.20117} {\bibinfo {title} {Frequency-dependent
  electron-phonon coupling and vibrational responses in tight-binding and
  continuous {Dirac} models with nuclear velocity correction}},\ \bibinfo
  {howpublished} {arXiv:2607.20117} (\bibinfo {year} {2026}),\ \Eprint
  {https://arxiv.org/abs/2607.20117} {arXiv:2607.20117} \BibitemShut {NoStop}%
\bibitem [{\citenamefont {Bonini}\ \emph {et~al.}(2023)\citenamefont {Bonini},
  \citenamefont {Ren}, \citenamefont {Vanderbilt}, \citenamefont {Stengel},
  \citenamefont {Dreyer},\ and\ \citenamefont {Coh}}]{Bonini2023}%
  \BibitemOpen
  \bibfield  {author} {\bibinfo {author} {\bibfnamefont {J.}~\bibnamefont
  {Bonini}}, \bibinfo {author} {\bibfnamefont {S.}~\bibnamefont {Ren}},
  \bibinfo {author} {\bibfnamefont {D.}~\bibnamefont {Vanderbilt}}, \bibinfo
  {author} {\bibfnamefont {M.}~\bibnamefont {Stengel}}, \bibinfo {author}
  {\bibfnamefont {C.~E.}\ \bibnamefont {Dreyer}},\ and\ \bibinfo {author}
  {\bibfnamefont {S.}~\bibnamefont {Coh}},\ }\bibfield  {title} {\bibinfo
  {title} {Frequency splitting of chiral phonons from broken time-reversal
  symmetry in {CrI$_3$}},\ }\href
  {https://doi.org/10.1103/PhysRevLett.130.086701} {\bibfield  {journal}
  {\bibinfo  {journal} {Phys. Rev. Lett.}\ }\textbf {\bibinfo {volume} {130}},\
  \bibinfo {pages} {086701} (\bibinfo {year} {2023})}\BibitemShut {NoStop}%
\bibitem [{\citenamefont {Ren}\ \emph {et~al.}(2024)\citenamefont {Ren},
  \citenamefont {Bonini}, \citenamefont {Stengel}, \citenamefont {Dreyer},\
  and\ \citenamefont {Vanderbilt}}]{Ren2024}%
  \BibitemOpen
  \bibfield  {author} {\bibinfo {author} {\bibfnamefont {S.}~\bibnamefont
  {Ren}}, \bibinfo {author} {\bibfnamefont {J.}~\bibnamefont {Bonini}},
  \bibinfo {author} {\bibfnamefont {M.}~\bibnamefont {Stengel}}, \bibinfo
  {author} {\bibfnamefont {C.~E.}\ \bibnamefont {Dreyer}},\ and\ \bibinfo
  {author} {\bibfnamefont {D.}~\bibnamefont {Vanderbilt}},\ }\bibfield  {title}
  {\bibinfo {title} {Adiabatic dynamics of coupled spins and phonons in
  magnetic insulators},\ }\href {https://doi.org/10.1103/PhysRevX.14.011041}
  {\bibfield  {journal} {\bibinfo  {journal} {Phys. Rev. X}\ }\textbf {\bibinfo
  {volume} {14}},\ \bibinfo {pages} {011041} (\bibinfo {year}
  {2024})}\BibitemShut {NoStop}%
\bibitem [{\citenamefont {Royo}\ and\ \citenamefont
  {Stengel}(2026)}]{Royo2025}%
  \BibitemOpen
  \bibfield  {author} {\bibinfo {author} {\bibfnamefont {M.}~\bibnamefont
  {Royo}}\ and\ \bibinfo {author} {\bibfnamefont {M.}~\bibnamefont {Stengel}},\
  }\bibfield  {title} {\bibinfo {title} {Dynamical response of noncollinear
  spin systems at constrained magnetic moments},\ }\href
  {https://doi.org/10.1103/c68v-8tkb} {\bibfield  {journal} {\bibinfo
  {journal} {Phys. Rev. X}\ }\textbf {\bibinfo {volume} {16}},\ \bibinfo
  {pages} {011049} (\bibinfo {year} {2026})}\BibitemShut {NoStop}%
\bibitem [{\citenamefont {Macheda}\ \emph
  {et~al.}(2024{\natexlab{a}})\citenamefont {Macheda}, \citenamefont {Mauri},\
  and\ \citenamefont {Sohier}}]{Macheda2024b}%
  \BibitemOpen
  \bibfield  {author} {\bibinfo {author} {\bibfnamefont {F.}~\bibnamefont
  {Macheda}}, \bibinfo {author} {\bibfnamefont {F.}~\bibnamefont {Mauri}},\
  and\ \bibinfo {author} {\bibfnamefont {T.}~\bibnamefont {Sohier}},\
  }\bibfield  {title} {\bibinfo {title} {{\textit Ab initio} {Van der Waals}
  electrodynamics: Polaritons and electron scattering from plasmons and phonons
  in {BN}-capped graphene},\ }\href
  {https://doi.org/10.1103/PhysRevB.110.115407} {\bibfield  {journal} {\bibinfo
   {journal} {Phys. Rev. B}\ }\textbf {\bibinfo {volume} {110}},\ \bibinfo
  {pages} {115407} (\bibinfo {year} {2024}{\natexlab{a}})}\BibitemShut
  {NoStop}%
\bibitem [{\citenamefont {Lihm}\ and\ \citenamefont
  {Park}(2024)}]{Lihm2024PlPh}%
  \BibitemOpen
  \bibfield  {author} {\bibinfo {author} {\bibfnamefont {J.-M.}\ \bibnamefont
  {Lihm}}\ and\ \bibinfo {author} {\bibfnamefont {C.-H.}\ \bibnamefont
  {Park}},\ }\bibfield  {title} {\bibinfo {title} {Plasmon-phonon hybridization
  in doped semiconductors from first principles},\ }\href
  {https://doi.org/10.1103/PhysRevLett.133.116402} {\bibfield  {journal}
  {\bibinfo  {journal} {Phys. Rev. Lett.}\ }\textbf {\bibinfo {volume} {133}},\
  \bibinfo {pages} {116402} (\bibinfo {year} {2024})}\BibitemShut {NoStop}%
\bibitem [{\citenamefont {Krsnik}\ \emph {et~al.}(2024)\citenamefont {Krsnik},
  \citenamefont {Novko},\ and\ \citenamefont {Bari\ifmmode \check{s}\else
  \v{s}\fi{}i\ifmmode~\acute{c}\else \'{c}\fi{}}}]{Krsnik2024}%
  \BibitemOpen
  \bibfield  {author} {\bibinfo {author} {\bibfnamefont {J.}~\bibnamefont
  {Krsnik}}, \bibinfo {author} {\bibfnamefont {D.}~\bibnamefont {Novko}},\ and\
  \bibinfo {author} {\bibfnamefont {O.~S.}\ \bibnamefont {Bari\ifmmode
  \check{s}\else \v{s}\fi{}i\ifmmode~\acute{c}\else \'{c}\fi{}}},\ }\bibfield
  {title} {\bibinfo {title} {Superconductivity in two-dimensional systems
  enhanced by nonadiabatic phonon-production effects},\ }\href
  {https://doi.org/10.1103/PhysRevB.110.L180505} {\bibfield  {journal}
  {\bibinfo  {journal} {Phys. Rev. B}\ }\textbf {\bibinfo {volume} {110}},\
  \bibinfo {pages} {L180505} (\bibinfo {year} {2024})}\BibitemShut {NoStop}%
\bibitem [{\citenamefont {Motornyi}\ \emph {et~al.}(2018)\citenamefont
  {Motornyi}, \citenamefont {Raynaud}, \citenamefont {Dal~Corso},\ and\
  \citenamefont {Vast}}]{Motornyi2018}%
  \BibitemOpen
  \bibfield  {author} {\bibinfo {author} {\bibfnamefont {O.}~\bibnamefont
  {Motornyi}}, \bibinfo {author} {\bibfnamefont {M.}~\bibnamefont {Raynaud}},
  \bibinfo {author} {\bibfnamefont {A.}~\bibnamefont {Dal~Corso}},\ and\
  \bibinfo {author} {\bibfnamefont {N.}~\bibnamefont {Vast}},\ }\bibfield
  {title} {\bibinfo {title} {Simulation of electron energy loss spectra with
  the {turboEELS} and {thermo\_pw} codes},\ }\href
  {https://doi.org/10.1088/1742-6596/1136/1/012008} {\bibfield  {journal}
  {\bibinfo  {journal} {J. Phys.: Conf. Ser.}\ }\textbf {\bibinfo {volume}
  {1136}},\ \bibinfo {pages} {012008} (\bibinfo {year} {2018})}\BibitemShut
  {NoStop}%
\bibitem [{\citenamefont {Motornyi}\ \emph {et~al.}(2020)\citenamefont
  {Motornyi}, \citenamefont {Vast}, \citenamefont {Timrov}, \citenamefont
  {Baseggio}, \citenamefont {Baroni},\ and\ \citenamefont
  {Dal~Corso}}]{Motornyi2020}%
  \BibitemOpen
  \bibfield  {author} {\bibinfo {author} {\bibfnamefont {O.}~\bibnamefont
  {Motornyi}}, \bibinfo {author} {\bibfnamefont {N.}~\bibnamefont {Vast}},
  \bibinfo {author} {\bibfnamefont {I.}~\bibnamefont {Timrov}}, \bibinfo
  {author} {\bibfnamefont {O.}~\bibnamefont {Baseggio}}, \bibinfo {author}
  {\bibfnamefont {S.}~\bibnamefont {Baroni}},\ and\ \bibinfo {author}
  {\bibfnamefont {A.}~\bibnamefont {Dal~Corso}},\ }\bibfield  {title} {\bibinfo
  {title} {Electron energy loss spectroscopy of bulk gold with ultrasoft
  pseudopotentials and the {Liouville-Lanczos} method},\ }\href
  {https://doi.org/10.1103/PhysRevB.102.035156} {\bibfield  {journal} {\bibinfo
   {journal} {Phys. Rev. B}\ }\textbf {\bibinfo {volume} {102}},\ \bibinfo
  {pages} {035156} (\bibinfo {year} {2020})}\BibitemShut {NoStop}%
\bibitem [{\citenamefont {Aryasetiawan}\ \emph {et~al.}(2004)\citenamefont
  {Aryasetiawan}, \citenamefont {Imada}, \citenamefont {Georges}, \citenamefont
  {Kotliar}, \citenamefont {Biermann},\ and\ \citenamefont
  {Lichtenstein}}]{Aryasetiawan2004}%
  \BibitemOpen
  \bibfield  {author} {\bibinfo {author} {\bibfnamefont {F.}~\bibnamefont
  {Aryasetiawan}}, \bibinfo {author} {\bibfnamefont {M.}~\bibnamefont {Imada}},
  \bibinfo {author} {\bibfnamefont {A.}~\bibnamefont {Georges}}, \bibinfo
  {author} {\bibfnamefont {G.}~\bibnamefont {Kotliar}}, \bibinfo {author}
  {\bibfnamefont {S.}~\bibnamefont {Biermann}},\ and\ \bibinfo {author}
  {\bibfnamefont {A.~I.}\ \bibnamefont {Lichtenstein}},\ }\bibfield  {title}
  {\bibinfo {title} {Frequency-dependent local interactions and low-energy
  effective models from electronic structure calculations},\ }\href
  {https://doi.org/10.1103/PhysRevB.70.195104} {\bibfield  {journal} {\bibinfo
  {journal} {Phys. Rev. B}\ }\textbf {\bibinfo {volume} {70}},\ \bibinfo
  {pages} {195104} (\bibinfo {year} {2004})}\BibitemShut {NoStop}%
\bibitem [{\citenamefont {Sternheimer}(1954)}]{Sternheimer1954}%
  \BibitemOpen
  \bibfield  {author} {\bibinfo {author} {\bibfnamefont {R.~M.}\ \bibnamefont
  {Sternheimer}},\ }\bibfield  {title} {\bibinfo {title} {Electronic
  polarizabilities of ions from the {Hartree-Fock} wave functions},\ }\href
  {https://doi.org/10.1103/PhysRev.96.951} {\bibfield  {journal} {\bibinfo
  {journal} {Phys. Rev.}\ }\textbf {\bibinfo {volume} {96}},\ \bibinfo {pages}
  {951} (\bibinfo {year} {1954})}\BibitemShut {NoStop}%
\bibitem [{\citenamefont {Cao}\ \emph {et~al.}(2018)\citenamefont {Cao},
  \citenamefont {Lambert}, \citenamefont {Radaelli},\ and\ \citenamefont
  {Giustino}}]{Cao2018}%
  \BibitemOpen
  \bibfield  {author} {\bibinfo {author} {\bibfnamefont {K.}~\bibnamefont
  {Cao}}, \bibinfo {author} {\bibfnamefont {H.}~\bibnamefont {Lambert}},
  \bibinfo {author} {\bibfnamefont {P.~G.}\ \bibnamefont {Radaelli}},\ and\
  \bibinfo {author} {\bibfnamefont {F.}~\bibnamefont {Giustino}},\ }\bibfield
  {title} {\bibinfo {title} {Ab initio calculation of spin fluctuation spectra
  using time-dependent density functional perturbation theory, plane waves, and
  pseudopotentials},\ }\href {https://doi.org/10.1103/PhysRevB.97.024420}
  {\bibfield  {journal} {\bibinfo  {journal} {Phys. Rev. B}\ }\textbf {\bibinfo
  {volume} {97}},\ \bibinfo {pages} {024420} (\bibinfo {year}
  {2018})}\BibitemShut {NoStop}%
\bibitem [{\citenamefont {Gorni}\ \emph {et~al.}(2018)\citenamefont {Gorni},
  \citenamefont {Timrov},\ and\ \citenamefont {Baroni}}]{Gorni2018}%
  \BibitemOpen
  \bibfield  {author} {\bibinfo {author} {\bibfnamefont {T.}~\bibnamefont
  {Gorni}}, \bibinfo {author} {\bibfnamefont {I.}~\bibnamefont {Timrov}},\ and\
  \bibinfo {author} {\bibfnamefont {S.}~\bibnamefont {Baroni}},\ }\bibfield
  {title} {\bibinfo {title} {Spin dynamics from time-dependent density
  functional perturbation theory},\ }\href
  {https://doi.org/10.1140/epjb/e2018-90247-9} {\bibfield  {journal} {\bibinfo
  {journal} {Eur. Phys. J. B}\ }\textbf {\bibinfo {volume} {91}},\ \bibinfo
  {pages} {249} (\bibinfo {year} {2018})}\BibitemShut {NoStop}%
\bibitem [{\citenamefont {Urru}\ and\ \citenamefont
  {Dal~Corso}(2019)}]{Urru2019}%
  \BibitemOpen
  \bibfield  {author} {\bibinfo {author} {\bibfnamefont {A.}~\bibnamefont
  {Urru}}\ and\ \bibinfo {author} {\bibfnamefont {A.}~\bibnamefont
  {Dal~Corso}},\ }\bibfield  {title} {\bibinfo {title} {Density functional
  perturbation theory for lattice dynamics with fully relativistic ultrasoft
  pseudopotentials: The magnetic case},\ }\href
  {https://doi.org/10.1103/PhysRevB.100.045115} {\bibfield  {journal} {\bibinfo
   {journal} {Phys. Rev. B}\ }\textbf {\bibinfo {volume} {100}},\ \bibinfo
  {pages} {045115} (\bibinfo {year} {2019})}\BibitemShut {NoStop}%
\bibitem [{\citenamefont {Ricci}\ \emph {et~al.}(2019)\citenamefont {Ricci},
  \citenamefont {Prokhorenko}, \citenamefont {Torrent}, \citenamefont
  {Verstraete},\ and\ \citenamefont {Bousquet}}]{Ricci2019}%
  \BibitemOpen
  \bibfield  {author} {\bibinfo {author} {\bibfnamefont {F.}~\bibnamefont
  {Ricci}}, \bibinfo {author} {\bibfnamefont {S.}~\bibnamefont {Prokhorenko}},
  \bibinfo {author} {\bibfnamefont {M.}~\bibnamefont {Torrent}}, \bibinfo
  {author} {\bibfnamefont {M.~J.}\ \bibnamefont {Verstraete}},\ and\ \bibinfo
  {author} {\bibfnamefont {E.}~\bibnamefont {Bousquet}},\ }\bibfield  {title}
  {\bibinfo {title} {Density functional perturbation theory within noncollinear
  magnetism},\ }\href {https://doi.org/10.1103/PhysRevB.99.184404} {\bibfield
  {journal} {\bibinfo  {journal} {Phys. Rev. B}\ }\textbf {\bibinfo {volume}
  {99}},\ \bibinfo {pages} {184404} (\bibinfo {year} {2019})}\BibitemShut
  {NoStop}%
\bibitem [{\citenamefont {Gonze}\ \emph {et~al.}(2024)\citenamefont {Gonze},
  \citenamefont {Rostami},\ and\ \citenamefont {Tantardini}}]{Gonze2024}%
  \BibitemOpen
  \bibfield  {author} {\bibinfo {author} {\bibfnamefont {X.}~\bibnamefont
  {Gonze}}, \bibinfo {author} {\bibfnamefont {S.}~\bibnamefont {Rostami}},\
  and\ \bibinfo {author} {\bibfnamefont {C.}~\bibnamefont {Tantardini}},\
  }\bibfield  {title} {\bibinfo {title} {Variational density functional
  perturbation theory for metals},\ }\href
  {https://doi.org/10.1103/PhysRevB.109.014317} {\bibfield  {journal} {\bibinfo
   {journal} {Phys. Rev. B}\ }\textbf {\bibinfo {volume} {109}},\ \bibinfo
  {pages} {014317} (\bibinfo {year} {2024})}\BibitemShut {NoStop}%
\bibitem [{\citenamefont {Zabalo}\ and\ \citenamefont
  {Stengel}(2024)}]{Zabalo2024}%
  \BibitemOpen
  \bibfield  {author} {\bibinfo {author} {\bibfnamefont {A.}~\bibnamefont
  {Zabalo}}\ and\ \bibinfo {author} {\bibfnamefont {M.}~\bibnamefont
  {Stengel}},\ }\bibfield  {title} {\bibinfo {title} {Ensemble density
  functional perturbation theory: Spatial dispersion in metals},\ }\href
  {https://doi.org/10.1103/PhysRevB.109.245116} {\bibfield  {journal} {\bibinfo
   {journal} {Phys. Rev. B}\ }\textbf {\bibinfo {volume} {109}},\ \bibinfo
  {pages} {245116} (\bibinfo {year} {2024})}\BibitemShut {NoStop}%
\bibitem [{\citenamefont {Walker}\ \emph {et~al.}(2006)\citenamefont {Walker},
  \citenamefont {Saitta}, \citenamefont {Gebauer},\ and\ \citenamefont
  {Baroni}}]{Walker2006}%
  \BibitemOpen
  \bibfield  {author} {\bibinfo {author} {\bibfnamefont {B.}~\bibnamefont
  {Walker}}, \bibinfo {author} {\bibfnamefont {A.~M.}\ \bibnamefont {Saitta}},
  \bibinfo {author} {\bibfnamefont {R.}~\bibnamefont {Gebauer}},\ and\ \bibinfo
  {author} {\bibfnamefont {S.}~\bibnamefont {Baroni}},\ }\bibfield  {title}
  {\bibinfo {title} {Efficient approach to time-dependent density-functional
  perturbation theory for optical spectroscopy},\ }\href
  {https://doi.org/10.1103/PhysRevLett.96.113001} {\bibfield  {journal}
  {\bibinfo  {journal} {Phys. Rev. Lett.}\ }\textbf {\bibinfo {volume} {96}},\
  \bibinfo {pages} {113001} (\bibinfo {year} {2006})}\BibitemShut {NoStop}%
\bibitem [{\citenamefont {Rocca}\ \emph {et~al.}(2008)\citenamefont {Rocca},
  \citenamefont {Gebauer}, \citenamefont {Saad},\ and\ \citenamefont
  {Baroni}}]{Rocca2008}%
  \BibitemOpen
  \bibfield  {author} {\bibinfo {author} {\bibfnamefont {D.}~\bibnamefont
  {Rocca}}, \bibinfo {author} {\bibfnamefont {R.}~\bibnamefont {Gebauer}},
  \bibinfo {author} {\bibfnamefont {Y.}~\bibnamefont {Saad}},\ and\ \bibinfo
  {author} {\bibfnamefont {S.}~\bibnamefont {Baroni}},\ }\bibfield  {title}
  {\bibinfo {title} {Turbo charging time-dependent density-functional theory
  with {Lanczos} chains},\ }\href {https://doi.org/10.1063/1.2899649}
  {\bibfield  {journal} {\bibinfo  {journal} {J. Chem. Phys.}\ }\textbf
  {\bibinfo {volume} {128}},\ \bibinfo {pages} {154105} (\bibinfo {year}
  {2008})}\BibitemShut {NoStop}%
\bibitem [{\citenamefont {Timrov}\ \emph {et~al.}(2013)\citenamefont {Timrov},
  \citenamefont {Vast}, \citenamefont {Gebauer},\ and\ \citenamefont
  {Baroni}}]{Timrov2013}%
  \BibitemOpen
  \bibfield  {author} {\bibinfo {author} {\bibfnamefont {I.}~\bibnamefont
  {Timrov}}, \bibinfo {author} {\bibfnamefont {N.}~\bibnamefont {Vast}},
  \bibinfo {author} {\bibfnamefont {R.}~\bibnamefont {Gebauer}},\ and\ \bibinfo
  {author} {\bibfnamefont {S.}~\bibnamefont {Baroni}},\ }\bibfield  {title}
  {\bibinfo {title} {Electron energy loss and inelastic x-ray scattering cross
  sections from time-dependent density-functional perturbation theory},\ }\href
  {https://doi.org/10.1103/PhysRevB.88.064301} {\bibfield  {journal} {\bibinfo
  {journal} {Phys. Rev. B}\ }\textbf {\bibinfo {volume} {88}},\ \bibinfo
  {pages} {064301} (\bibinfo {year} {2013})}\BibitemShut {NoStop}%
\bibitem [{\citenamefont {Nomura}\ and\ \citenamefont
  {Arita}(2015)}]{Nomura2015}%
  \BibitemOpen
  \bibfield  {author} {\bibinfo {author} {\bibfnamefont {Y.}~\bibnamefont
  {Nomura}}\ and\ \bibinfo {author} {\bibfnamefont {R.}~\bibnamefont {Arita}},\
  }\bibfield  {title} {\bibinfo {title} {Ab initio downfolding for
  electron-phonon-coupled systems: Constrained density-functional perturbation
  theory},\ }\href {https://doi.org/10.1103/PhysRevB.92.245108} {\bibfield
  {journal} {\bibinfo  {journal} {Phys. Rev. B}\ }\textbf {\bibinfo {volume}
  {92}},\ \bibinfo {pages} {245108} (\bibinfo {year} {2015})}\BibitemShut
  {NoStop}%
\bibitem [{\citenamefont {Berges}\ \emph {et~al.}(2020)\citenamefont {Berges},
  \citenamefont {van Loon}, \citenamefont {Schobert}, \citenamefont
  {R\"osner},\ and\ \citenamefont {Wehling}}]{Berges2020}%
  \BibitemOpen
  \bibfield  {author} {\bibinfo {author} {\bibfnamefont {J.}~\bibnamefont
  {Berges}}, \bibinfo {author} {\bibfnamefont {E.~G. C.~P.}\ \bibnamefont {van
  Loon}}, \bibinfo {author} {\bibfnamefont {A.}~\bibnamefont {Schobert}},
  \bibinfo {author} {\bibfnamefont {M.}~\bibnamefont {R\"osner}},\ and\
  \bibinfo {author} {\bibfnamefont {T.~O.}\ \bibnamefont {Wehling}},\
  }\bibfield  {title} {\bibinfo {title} {Ab initio phonon self-energies and
  fluctuation diagnostics of phonon anomalies: Lattice instabilities from
  {Dirac} pseudospin physics in transition metal dichalcogenides},\ }\href
  {https://doi.org/10.1103/PhysRevB.101.155107} {\bibfield  {journal} {\bibinfo
   {journal} {Phys. Rev. B}\ }\textbf {\bibinfo {volume} {101}},\ \bibinfo
  {pages} {155107} (\bibinfo {year} {2020})}\BibitemShut {NoStop}%
\bibitem [{\citenamefont {van Loon}\ \emph {et~al.}(2021)\citenamefont {van
  Loon}, \citenamefont {Berges},\ and\ \citenamefont {Wehling}}]{vanLoon2021}%
  \BibitemOpen
  \bibfield  {author} {\bibinfo {author} {\bibfnamefont {E.~G. C.~P.}\
  \bibnamefont {van Loon}}, \bibinfo {author} {\bibfnamefont {J.}~\bibnamefont
  {Berges}},\ and\ \bibinfo {author} {\bibfnamefont {T.~O.}\ \bibnamefont
  {Wehling}},\ }\bibfield  {title} {\bibinfo {title} {Downfolding approaches to
  electron-ion coupling: Constrained density-functional perturbation theory for
  molecules},\ }\href {https://doi.org/10.1103/PhysRevB.103.205103} {\bibfield
  {journal} {\bibinfo  {journal} {Phys. Rev. B}\ }\textbf {\bibinfo {volume}
  {103}},\ \bibinfo {pages} {205103} (\bibinfo {year} {2021})}\BibitemShut
  {NoStop}%
\bibitem [{\citenamefont {Miyake}\ \emph {et~al.}(2009)\citenamefont {Miyake},
  \citenamefont {Aryasetiawan},\ and\ \citenamefont {Imada}}]{Miyake2009}%
  \BibitemOpen
  \bibfield  {author} {\bibinfo {author} {\bibfnamefont {T.}~\bibnamefont
  {Miyake}}, \bibinfo {author} {\bibfnamefont {F.}~\bibnamefont
  {Aryasetiawan}},\ and\ \bibinfo {author} {\bibfnamefont {M.}~\bibnamefont
  {Imada}},\ }\bibfield  {title} {\bibinfo {title} {Ab initio procedure for
  constructing effective models of correlated materials with entangled band
  structure},\ }\href {https://doi.org/10.1103/PhysRevB.80.155134} {\bibfield
  {journal} {\bibinfo  {journal} {Phys. Rev. B}\ }\textbf {\bibinfo {volume}
  {80}},\ \bibinfo {pages} {155134} (\bibinfo {year} {2009})}\BibitemShut
  {NoStop}%
\bibitem [{\citenamefont {\ifmmode \mbox{\c{S}}\else \c{S}\fi{}a\ifmmode
  \mbox{\c{s}}\else \c{s}\fi{}\ifmmode \imath \else \i
  \fi{}o\ifmmode~\breve{g}\else \u{g}\fi{}lu}\ \emph
  {et~al.}(2011)\citenamefont {\ifmmode \mbox{\c{S}}\else \c{S}\fi{}a\ifmmode
  \mbox{\c{s}}\else \c{s}\fi{}\ifmmode \imath \else \i
  \fi{}o\ifmmode~\breve{g}\else \u{g}\fi{}lu}, \citenamefont {Friedrich},\ and\
  \citenamefont {Bl\"ugel}}]{Sasioglu2011}%
  \BibitemOpen
  \bibfield  {author} {\bibinfo {author} {\bibfnamefont {E.}~\bibnamefont
  {\ifmmode \mbox{\c{S}}\else \c{S}\fi{}a\ifmmode \mbox{\c{s}}\else
  \c{s}\fi{}\ifmmode \imath \else \i \fi{}o\ifmmode~\breve{g}\else
  \u{g}\fi{}lu}}, \bibinfo {author} {\bibfnamefont {C.}~\bibnamefont
  {Friedrich}},\ and\ \bibinfo {author} {\bibfnamefont {S.}~\bibnamefont
  {Bl\"ugel}},\ }\bibfield  {title} {\bibinfo {title} {Effective {Coulomb}
  interaction in transition metals from constrained random-phase
  approximation},\ }\href {https://doi.org/10.1103/PhysRevB.83.121101}
  {\bibfield  {journal} {\bibinfo  {journal} {Phys. Rev. B}\ }\textbf {\bibinfo
  {volume} {83}},\ \bibinfo {pages} {121101} (\bibinfo {year}
  {2011})}\BibitemShut {NoStop}%
\bibitem [{\citenamefont {Kaltak}(2015)}]{Kaltak2015}%
  \BibitemOpen
  \bibfield  {author} {\bibinfo {author} {\bibfnamefont {M.}~\bibnamefont
  {Kaltak}},\ }\emph {\bibinfo {title} {Merging GW with DMFT}},\ \href
  {https://doi.org/10.25365/thesis.38099} {\bibinfo {type} {Dissertation}},\
  \bibinfo  {school} {Universität Wien}, \bibinfo {address} {Wien, Austria}
  (\bibinfo {year} {2015})\BibitemShut {NoStop}%
\bibitem [{\citenamefont {Kaltak}\ \emph {et~al.}(2025)\citenamefont {Kaltak},
  \citenamefont {Hampel}, \citenamefont {Schlipf}, \citenamefont {Reddy},
  \citenamefont {Kim},\ and\ \citenamefont {Kresse}}]{Kaltak2025}%
  \BibitemOpen
  \bibfield  {author} {\bibinfo {author} {\bibfnamefont {M.}~\bibnamefont
  {Kaltak}}, \bibinfo {author} {\bibfnamefont {A.}~\bibnamefont {Hampel}},
  \bibinfo {author} {\bibfnamefont {M.}~\bibnamefont {Schlipf}}, \bibinfo
  {author} {\bibfnamefont {I.~R.}\ \bibnamefont {Reddy}}, \bibinfo {author}
  {\bibfnamefont {B.}~\bibnamefont {Kim}},\ and\ \bibinfo {author}
  {\bibfnamefont {G.}~\bibnamefont {Kresse}},\ }\bibfield  {title} {\bibinfo
  {title} {Constrained random phase approximation: The spectral method},\
  }\href {https://doi.org/10.1103/m3gh-g6r6} {\bibfield  {journal} {\bibinfo
  {journal} {Phys. Rev. B}\ }\textbf {\bibinfo {volume} {112}},\ \bibinfo
  {pages} {245102} (\bibinfo {year} {2025})}\BibitemShut {NoStop}%
\bibitem [{\citenamefont {Wilson}\ \emph {et~al.}(2008)\citenamefont {Wilson},
  \citenamefont {Gygi},\ and\ \citenamefont {Galli}}]{Wilson2008}%
  \BibitemOpen
  \bibfield  {author} {\bibinfo {author} {\bibfnamefont {H.~F.}\ \bibnamefont
  {Wilson}}, \bibinfo {author} {\bibfnamefont {F.}~\bibnamefont {Gygi}},\ and\
  \bibinfo {author} {\bibfnamefont {G.}~\bibnamefont {Galli}},\ }\bibfield
  {title} {\bibinfo {title} {Efficient iterative method for calculations of
  dielectric matrices},\ }\href {https://doi.org/10.1103/PhysRevB.78.113303}
  {\bibfield  {journal} {\bibinfo  {journal} {Phys. Rev. B}\ }\textbf {\bibinfo
  {volume} {78}},\ \bibinfo {pages} {113303} (\bibinfo {year}
  {2008})}\BibitemShut {NoStop}%
\bibitem [{\citenamefont {Govoni}\ and\ \citenamefont
  {Galli}(2015)}]{Govoni2015}%
  \BibitemOpen
  \bibfield  {author} {\bibinfo {author} {\bibfnamefont {M.}~\bibnamefont
  {Govoni}}\ and\ \bibinfo {author} {\bibfnamefont {G.}~\bibnamefont {Galli}},\
  }\bibfield  {title} {\bibinfo {title} {Large scale {GW} calculations},\
  }\href {https://doi.org/10.1021/ct500958p} {\bibfield  {journal} {\bibinfo
  {journal} {J. Chem. Theory Comput.}\ }\textbf {\bibinfo {volume} {11}},\
  \bibinfo {pages} {2680} (\bibinfo {year} {2015})}\BibitemShut {NoStop}%
\bibitem [{\citenamefont {Shao}\ \emph {et~al.}(2016)\citenamefont {Shao},
  \citenamefont {Lin}, \citenamefont {Yang}, \citenamefont {Liu}, \citenamefont
  {{da Jornada}}, \citenamefont {Deslippe},\ and\ \citenamefont
  {Louie}}]{Shao2016}%
  \BibitemOpen
  \bibfield  {author} {\bibinfo {author} {\bibfnamefont {M.}~\bibnamefont
  {Shao}}, \bibinfo {author} {\bibfnamefont {L.}~\bibnamefont {Lin}}, \bibinfo
  {author} {\bibfnamefont {C.}~\bibnamefont {Yang}}, \bibinfo {author}
  {\bibfnamefont {F.}~\bibnamefont {Liu}}, \bibinfo {author} {\bibfnamefont
  {F.~H.}\ \bibnamefont {{da Jornada}}}, \bibinfo {author} {\bibfnamefont
  {J.}~\bibnamefont {Deslippe}},\ and\ \bibinfo {author} {\bibfnamefont
  {S.~G.}\ \bibnamefont {Louie}},\ }\bibfield  {title} {\bibinfo {title} {Low
  rank approximation in {$G_0W_0$} calculations},\ }\href
  {https://doi.org/10.1007/s11425-016-0296-x} {\bibfield  {journal} {\bibinfo
  {journal} {Sci. China Math.}\ }\textbf {\bibinfo {volume} {59}},\ \bibinfo
  {pages} {1593} (\bibinfo {year} {2016})}\BibitemShut {NoStop}%
\bibitem [{\citenamefont {Lu}\ \emph {et~al.}(2008)\citenamefont {Lu},
  \citenamefont {Gygi},\ and\ \citenamefont {Galli}}]{Lu2008}%
  \BibitemOpen
  \bibfield  {author} {\bibinfo {author} {\bibfnamefont {D.}~\bibnamefont
  {Lu}}, \bibinfo {author} {\bibfnamefont {F.}~\bibnamefont {Gygi}},\ and\
  \bibinfo {author} {\bibfnamefont {G.}~\bibnamefont {Galli}},\ }\bibfield
  {title} {\bibinfo {title} {Dielectric properties of ice and liquid water from
  first-principles calculations},\ }\href
  {https://doi.org/10.1103/PhysRevLett.100.147601} {\bibfield  {journal}
  {\bibinfo  {journal} {Phys. Rev. Lett.}\ }\textbf {\bibinfo {volume} {100}},\
  \bibinfo {pages} {147601} (\bibinfo {year} {2008})}\BibitemShut {NoStop}%
\bibitem [{\citenamefont {Yeh}\ and\ \citenamefont {Morales}(2024)}]{Yeh2024}%
  \BibitemOpen
  \bibfield  {author} {\bibinfo {author} {\bibfnamefont {C.-N.}\ \bibnamefont
  {Yeh}}\ and\ \bibinfo {author} {\bibfnamefont {M.~A.}\ \bibnamefont
  {Morales}},\ }\bibfield  {title} {\bibinfo {title} {Low-scaling algorithms
  for {GW} and constrained random phase approximation using symmetry-adapted
  interpolative separable density fitting},\ }\href
  {https://doi.org/10.1021/acs.jctc.4c00085} {\bibfield  {journal} {\bibinfo
  {journal} {J. Chem. Theory Comput.}\ }\textbf {\bibinfo {volume} {20}},\
  \bibinfo {pages} {3184} (\bibinfo {year} {2024})}\BibitemShut {NoStop}%
\bibitem [{\citenamefont {Marzari}\ and\ \citenamefont
  {Vanderbilt}(1997)}]{Marzari1997}%
  \BibitemOpen
  \bibfield  {author} {\bibinfo {author} {\bibfnamefont {N.}~\bibnamefont
  {Marzari}}\ and\ \bibinfo {author} {\bibfnamefont {D.}~\bibnamefont
  {Vanderbilt}},\ }\bibfield  {title} {\bibinfo {title} {Maximally-localized
  generalized {{Wannier}} functions for composite energy bands},\ }\href
  {https://doi.org/10.1103/PhysRevB.56.12847} {\bibfield  {journal} {\bibinfo
  {journal} {Phys. Rev. B}\ }\textbf {\bibinfo {volume} {56}},\ \bibinfo
  {pages} {12847} (\bibinfo {year} {1997})}\BibitemShut {NoStop}%
\bibitem [{\citenamefont {Souza}\ \emph {et~al.}(2001)\citenamefont {Souza},
  \citenamefont {Marzari},\ and\ \citenamefont {Vanderbilt}}]{Souza2001}%
  \BibitemOpen
  \bibfield  {author} {\bibinfo {author} {\bibfnamefont {I.}~\bibnamefont
  {Souza}}, \bibinfo {author} {\bibfnamefont {N.}~\bibnamefont {Marzari}},\
  and\ \bibinfo {author} {\bibfnamefont {D.}~\bibnamefont {Vanderbilt}},\
  }\bibfield  {title} {\bibinfo {title} {Maximally localized {{Wannier}}
  functions for entangled energy bands},\ }\href
  {https://doi.org/10.1103/PhysRevB.65.035109} {\bibfield  {journal} {\bibinfo
  {journal} {Phys. Rev. B}\ }\textbf {\bibinfo {volume} {65}},\ \bibinfo
  {pages} {035109} (\bibinfo {year} {2001})}\BibitemShut {NoStop}%
\bibitem [{\citenamefont {Marrazzo}\ \emph {et~al.}(2024)\citenamefont
  {Marrazzo}, \citenamefont {Beck}, \citenamefont {Margine}, \citenamefont
  {Marzari}, \citenamefont {Mostofi}, \citenamefont {Qiao}, \citenamefont
  {Souza}, \citenamefont {Tsirkin}, \citenamefont {Yates},\ and\ \citenamefont
  {Pizzi}}]{Marrazzo2024}%
  \BibitemOpen
  \bibfield  {author} {\bibinfo {author} {\bibfnamefont {A.}~\bibnamefont
  {Marrazzo}}, \bibinfo {author} {\bibfnamefont {S.}~\bibnamefont {Beck}},
  \bibinfo {author} {\bibfnamefont {E.~R.}\ \bibnamefont {Margine}}, \bibinfo
  {author} {\bibfnamefont {N.}~\bibnamefont {Marzari}}, \bibinfo {author}
  {\bibfnamefont {A.~A.}\ \bibnamefont {Mostofi}}, \bibinfo {author}
  {\bibfnamefont {J.}~\bibnamefont {Qiao}}, \bibinfo {author} {\bibfnamefont
  {I.}~\bibnamefont {Souza}}, \bibinfo {author} {\bibfnamefont {S.~S.}\
  \bibnamefont {Tsirkin}}, \bibinfo {author} {\bibfnamefont {J.~R.}\
  \bibnamefont {Yates}},\ and\ \bibinfo {author} {\bibfnamefont
  {G.}~\bibnamefont {Pizzi}},\ }\bibfield  {title} {\bibinfo {title}
  {Wannier-function software ecosystem for materials simulations},\ }\href
  {https://doi.org/10.1103/RevModPhys.96.045008} {\bibfield  {journal}
  {\bibinfo  {journal} {Rev. Mod. Phys.}\ }\textbf {\bibinfo {volume} {96}},\
  \bibinfo {pages} {045008} (\bibinfo {year} {2024})}\BibitemShut {NoStop}%
\bibitem [{\citenamefont {Pizzi}\ \emph {et~al.}(2020)\citenamefont {Pizzi}
  \emph {et~al.}}]{Pizzi2020W90}%
  \BibitemOpen
  \bibfield  {author} {\bibinfo {author} {\bibfnamefont {G.}~\bibnamefont
  {Pizzi}} \emph {et~al.},\ }\bibfield  {title} {\bibinfo {title} {Wannier90 as
  a community code: New features and applications},\ }\href
  {https://doi.org/10.1088/1361-648X/ab51ff} {\bibfield  {journal} {\bibinfo
  {journal} {J. Condens. Matter Phys.}\ }\textbf {\bibinfo {volume} {32}},\
  \bibinfo {pages} {165902} (\bibinfo {year} {2020})}\BibitemShut {NoStop}%
\bibitem [{\citenamefont {Merkel}\ and\ \citenamefont
  {Ortmann}(2023)}]{Merkel2024}%
  \BibitemOpen
  \bibfield  {author} {\bibinfo {author} {\bibfnamefont {K.}~\bibnamefont
  {Merkel}}\ and\ \bibinfo {author} {\bibfnamefont {F.}~\bibnamefont
  {Ortmann}},\ }\bibfield  {title} {\bibinfo {title} {Linear scaling approach
  for optical excitations using maximally localized {Wannier} functions},\
  }\href {https://doi.org/10.1088/2515-7639/ad06cd} {\bibfield  {journal}
  {\bibinfo  {journal} {J. Phys. Mater.}\ }\textbf {\bibinfo {volume} {7}},\
  \bibinfo {pages} {015001} (\bibinfo {year} {2023})}\BibitemShut {NoStop}%
\bibitem [{\citenamefont {Nakamura}\ \emph {et~al.}(2021)\citenamefont
  {Nakamura}, \citenamefont {Yoshimoto}, \citenamefont {Nomura}, \citenamefont
  {Tadano}, \citenamefont {Kawamura}, \citenamefont {Kosugi}, \citenamefont
  {Yoshimi}, \citenamefont {Misawa},\ and\ \citenamefont
  {Motoyama}}]{Nakamura2021}%
  \BibitemOpen
  \bibfield  {author} {\bibinfo {author} {\bibfnamefont {K.}~\bibnamefont
  {Nakamura}}, \bibinfo {author} {\bibfnamefont {Y.}~\bibnamefont {Yoshimoto}},
  \bibinfo {author} {\bibfnamefont {Y.}~\bibnamefont {Nomura}}, \bibinfo
  {author} {\bibfnamefont {T.}~\bibnamefont {Tadano}}, \bibinfo {author}
  {\bibfnamefont {M.}~\bibnamefont {Kawamura}}, \bibinfo {author}
  {\bibfnamefont {T.}~\bibnamefont {Kosugi}}, \bibinfo {author} {\bibfnamefont
  {K.}~\bibnamefont {Yoshimi}}, \bibinfo {author} {\bibfnamefont
  {T.}~\bibnamefont {Misawa}},\ and\ \bibinfo {author} {\bibfnamefont
  {Y.}~\bibnamefont {Motoyama}},\ }\bibfield  {title} {\bibinfo {title}
  {{RESPACK}: An ab initio tool for derivation of effective low-energy model of
  material},\ }\href {https://doi.org/10.1016/j.cpc.2020.107781} {\bibfield
  {journal} {\bibinfo  {journal} {Comput. Phys. Commun.}\ }\textbf {\bibinfo
  {volume} {261}},\ \bibinfo {pages} {107781} (\bibinfo {year}
  {2021})}\BibitemShut {NoStop}%
\bibitem [{\citenamefont {Amadon}\ \emph {et~al.}(2014)\citenamefont {Amadon},
  \citenamefont {Applencourt},\ and\ \citenamefont {Bruneval}}]{Amadon2014}%
  \BibitemOpen
  \bibfield  {author} {\bibinfo {author} {\bibfnamefont {B.}~\bibnamefont
  {Amadon}}, \bibinfo {author} {\bibfnamefont {T.}~\bibnamefont
  {Applencourt}},\ and\ \bibinfo {author} {\bibfnamefont {F.}~\bibnamefont
  {Bruneval}},\ }\bibfield  {title} {\bibinfo {title} {Screened {Coulomb}
  interaction calculations: {cRPA} implementation and applications to dynamical
  screening and self-consistency in uranium dioxide and cerium},\ }\href
  {https://doi.org/10.1103/PhysRevB.89.125110} {\bibfield  {journal} {\bibinfo
  {journal} {Phys. Rev. B}\ }\textbf {\bibinfo {volume} {89}},\ \bibinfo
  {pages} {125110} (\bibinfo {year} {2014})}\BibitemShut {NoStop}%
\bibitem [{\citenamefont {Adler}(1962)}]{Adler1962}%
  \BibitemOpen
  \bibfield  {author} {\bibinfo {author} {\bibfnamefont {S.~L.}\ \bibnamefont
  {Adler}},\ }\bibfield  {title} {\bibinfo {title} {Quantum theory of the
  dielectric constant in real solids},\ }\href
  {https://doi.org/10.1103/PhysRev.126.413} {\bibfield  {journal} {\bibinfo
  {journal} {Phys. Rev.}\ }\textbf {\bibinfo {volume} {126}},\ \bibinfo {pages}
  {413} (\bibinfo {year} {1962})}\BibitemShut {NoStop}%
\bibitem [{\citenamefont {Wiser}(1963)}]{Wiser1963}%
  \BibitemOpen
  \bibfield  {author} {\bibinfo {author} {\bibfnamefont {N.}~\bibnamefont
  {Wiser}},\ }\bibfield  {title} {\bibinfo {title} {Dielectric constant with
  local field effects included},\ }\href
  {https://doi.org/10.1103/PhysRev.129.62} {\bibfield  {journal} {\bibinfo
  {journal} {Phys. Rev.}\ }\textbf {\bibinfo {volume} {129}},\ \bibinfo {pages}
  {62} (\bibinfo {year} {1963})}\BibitemShut {NoStop}%
\bibitem [{\citenamefont {Hybertsen}\ and\ \citenamefont
  {Louie}(1986)}]{Hybertsen1986GW}%
  \BibitemOpen
  \bibfield  {author} {\bibinfo {author} {\bibfnamefont {M.~S.}\ \bibnamefont
  {Hybertsen}}\ and\ \bibinfo {author} {\bibfnamefont {S.~G.}\ \bibnamefont
  {Louie}},\ }\bibfield  {title} {\bibinfo {title} {Electron correlation in
  semiconductors and insulators: Band gaps and quasiparticle energies},\ }\href
  {https://doi.org/10.1103/PhysRevB.34.5390} {\bibfield  {journal} {\bibinfo
  {journal} {Phys. Rev. B}\ }\textbf {\bibinfo {volume} {34}},\ \bibinfo
  {pages} {5390} (\bibinfo {year} {1986})}\BibitemShut {NoStop}%
\bibitem [{\citenamefont {Pick}\ \emph {et~al.}(1970)\citenamefont {Pick},
  \citenamefont {Cohen},\ and\ \citenamefont {Martin}}]{Pick1970}%
  \BibitemOpen
  \bibfield  {author} {\bibinfo {author} {\bibfnamefont {R.~M.}\ \bibnamefont
  {Pick}}, \bibinfo {author} {\bibfnamefont {M.~H.}\ \bibnamefont {Cohen}},\
  and\ \bibinfo {author} {\bibfnamefont {R.~M.}\ \bibnamefont {Martin}},\
  }\bibfield  {title} {\bibinfo {title} {Microscopic theory of force constants
  in the adiabatic approximation},\ }\href
  {https://doi.org/10.1103/PhysRevB.1.910} {\bibfield  {journal} {\bibinfo
  {journal} {Phys. Rev. B}\ }\textbf {\bibinfo {volume} {1}},\ \bibinfo {pages}
  {910} (\bibinfo {year} {1970})}\BibitemShut {NoStop}%
\bibitem [{\citenamefont {Colonna}\ \emph {et~al.}(2022)\citenamefont
  {Colonna}, \citenamefont {De~Gennaro}, \citenamefont {Linscott},\ and\
  \citenamefont {Marzari}}]{Colonna2022}%
  \BibitemOpen
  \bibfield  {author} {\bibinfo {author} {\bibfnamefont {N.}~\bibnamefont
  {Colonna}}, \bibinfo {author} {\bibfnamefont {R.}~\bibnamefont {De~Gennaro}},
  \bibinfo {author} {\bibfnamefont {E.}~\bibnamefont {Linscott}},\ and\
  \bibinfo {author} {\bibfnamefont {N.}~\bibnamefont {Marzari}},\ }\bibfield
  {title} {\bibinfo {title} {{Koopmans} spectral functionals in periodic
  boundary conditions},\ }\href {https://doi.org/10.1021/acs.jctc.2c00161}
  {\bibfield  {journal} {\bibinfo  {journal} {J. Chem. Theory Comput.}\
  }\textbf {\bibinfo {volume} {18}},\ \bibinfo {pages} {5435} (\bibinfo {year}
  {2022})}\BibitemShut {NoStop}%
\bibitem [{\citenamefont {Cistaro}\ \emph {et~al.}(2026)\citenamefont
  {Cistaro}, \citenamefont {S\'a}, \citenamefont {Sangalli}, \citenamefont
  {Pic\'on},\ and\ \citenamefont {Colonna}}]{Cistaro2026}%
  \BibitemOpen
  \bibfield  {author} {\bibinfo {author} {\bibfnamefont {G.}~\bibnamefont
  {Cistaro}}, \bibinfo {author} {\bibfnamefont {M.}~\bibnamefont {S\'a}},
  \bibinfo {author} {\bibfnamefont {D.}~\bibnamefont {Sangalli}}, \bibinfo
  {author} {\bibfnamefont {A.}~\bibnamefont {Pic\'on}},\ and\ \bibinfo {author}
  {\bibfnamefont {N.}~\bibnamefont {Colonna}},\ }\href
  {https://arxiv.org/abs/2608.16226} {\bibinfo {title} {Efficient
  nonequilibrium electron dynamics from first-principles: leveraging {Koopmans}
  spectral functionals and {Wannier} localization}} (\bibinfo {year} {2026}),\
  \Eprint {https://arxiv.org/abs/2608.16226} {arXiv:2608.16226
  [cond-mat.mtrl-sci]} \BibitemShut {NoStop}%
\bibitem [{\citenamefont {Giustino}\ \emph {et~al.}(2010)\citenamefont
  {Giustino}, \citenamefont {Cohen},\ and\ \citenamefont
  {Louie}}]{Giustino2010GW}%
  \BibitemOpen
  \bibfield  {author} {\bibinfo {author} {\bibfnamefont {F.}~\bibnamefont
  {Giustino}}, \bibinfo {author} {\bibfnamefont {M.~L.}\ \bibnamefont
  {Cohen}},\ and\ \bibinfo {author} {\bibfnamefont {S.~G.}\ \bibnamefont
  {Louie}},\ }\bibfield  {title} {\bibinfo {title} {{GW} method with the
  self-consistent {Sternheimer} equation},\ }\href
  {https://doi.org/10.1103/PhysRevB.81.115105} {\bibfield  {journal} {\bibinfo
  {journal} {Phys. Rev. B}\ }\textbf {\bibinfo {volume} {81}},\ \bibinfo
  {pages} {115105} (\bibinfo {year} {2010})}\BibitemShut {NoStop}%
\bibitem [{\citenamefont {Verdi}\ and\ \citenamefont
  {Giustino}(2015)}]{Verdi2015}%
  \BibitemOpen
  \bibfield  {author} {\bibinfo {author} {\bibfnamefont {C.}~\bibnamefont
  {Verdi}}\ and\ \bibinfo {author} {\bibfnamefont {F.}~\bibnamefont
  {Giustino}},\ }\bibfield  {title} {\bibinfo {title} {Fr\"ohlich
  electron-phonon vertex from first principles},\ }\href
  {https://doi.org/10.1103/PhysRevLett.115.176401} {\bibfield  {journal}
  {\bibinfo  {journal} {Phys. Rev. Lett.}\ }\textbf {\bibinfo {volume} {115}},\
  \bibinfo {pages} {176401} (\bibinfo {year} {2015})}\BibitemShut {NoStop}%
\bibitem [{\citenamefont {Sjakste}\ \emph {et~al.}(2015)\citenamefont
  {Sjakste}, \citenamefont {Vast}, \citenamefont {Calandra},\ and\
  \citenamefont {Mauri}}]{Sjakste2015}%
  \BibitemOpen
  \bibfield  {author} {\bibinfo {author} {\bibfnamefont {J.}~\bibnamefont
  {Sjakste}}, \bibinfo {author} {\bibfnamefont {N.}~\bibnamefont {Vast}},
  \bibinfo {author} {\bibfnamefont {M.}~\bibnamefont {Calandra}},\ and\
  \bibinfo {author} {\bibfnamefont {F.}~\bibnamefont {Mauri}},\ }\bibfield
  {title} {\bibinfo {title} {Wannier interpolation of the electron-phonon
  matrix elements in polar semiconductors: Polar-optical coupling in {GaAs}},\
  }\href {https://doi.org/10.1103/PhysRevB.92.054307} {\bibfield  {journal}
  {\bibinfo  {journal} {Phys. Rev. B}\ }\textbf {\bibinfo {volume} {92}},\
  \bibinfo {pages} {054307} (\bibinfo {year} {2015})}\BibitemShut {NoStop}%
\bibitem [{\citenamefont {Sohier}\ \emph {et~al.}(2017)\citenamefont {Sohier},
  \citenamefont {Gibertini}, \citenamefont {Calandra}, \citenamefont {Mauri},\
  and\ \citenamefont {Marzari}}]{Sohier2017}%
  \BibitemOpen
  \bibfield  {author} {\bibinfo {author} {\bibfnamefont {T.}~\bibnamefont
  {Sohier}}, \bibinfo {author} {\bibfnamefont {M.}~\bibnamefont {Gibertini}},
  \bibinfo {author} {\bibfnamefont {M.}~\bibnamefont {Calandra}}, \bibinfo
  {author} {\bibfnamefont {F.}~\bibnamefont {Mauri}},\ and\ \bibinfo {author}
  {\bibfnamefont {N.}~\bibnamefont {Marzari}},\ }\bibfield  {title} {\bibinfo
  {title} {Breakdown of optical phonons' splitting in two-dimensional
  materials},\ }\href {https://doi.org/10.1021/acs.nanolett.7b01090} {\bibfield
   {journal} {\bibinfo  {journal} {Nano Lett.}\ }\textbf {\bibinfo {volume}
  {17}},\ \bibinfo {pages} {3758} (\bibinfo {year} {2017})}\BibitemShut
  {NoStop}%
\bibitem [{\citenamefont {Royo}\ and\ \citenamefont
  {Stengel}(2019)}]{Royo2019}%
  \BibitemOpen
  \bibfield  {author} {\bibinfo {author} {\bibfnamefont {M.}~\bibnamefont
  {Royo}}\ and\ \bibinfo {author} {\bibfnamefont {M.}~\bibnamefont {Stengel}},\
  }\bibfield  {title} {\bibinfo {title} {First-principles theory of spatial
  dispersion: Dynamical quadrupoles and flexoelectricity},\ }\href
  {https://doi.org/10.1103/PhysRevX.9.021050} {\bibfield  {journal} {\bibinfo
  {journal} {Phys. Rev. X}\ }\textbf {\bibinfo {volume} {9}},\ \bibinfo {pages}
  {021050} (\bibinfo {year} {2019})}\BibitemShut {NoStop}%
\bibitem [{\citenamefont {Brunin}\ \emph
  {et~al.}(2020{\natexlab{a}})\citenamefont {Brunin}, \citenamefont {Miranda},
  \citenamefont {Giantomassi}, \citenamefont {Royo}, \citenamefont {Stengel},
  \citenamefont {Verstraete}, \citenamefont {Gonze}, \citenamefont
  {Rignanese},\ and\ \citenamefont {Hautier}}]{Brunin2020PRL}%
  \BibitemOpen
  \bibfield  {author} {\bibinfo {author} {\bibfnamefont {G.}~\bibnamefont
  {Brunin}}, \bibinfo {author} {\bibfnamefont {H.~P.~C.}\ \bibnamefont
  {Miranda}}, \bibinfo {author} {\bibfnamefont {M.}~\bibnamefont
  {Giantomassi}}, \bibinfo {author} {\bibfnamefont {M.}~\bibnamefont {Royo}},
  \bibinfo {author} {\bibfnamefont {M.}~\bibnamefont {Stengel}}, \bibinfo
  {author} {\bibfnamefont {M.~J.}\ \bibnamefont {Verstraete}}, \bibinfo
  {author} {\bibfnamefont {X.}~\bibnamefont {Gonze}}, \bibinfo {author}
  {\bibfnamefont {G.-M.}\ \bibnamefont {Rignanese}},\ and\ \bibinfo {author}
  {\bibfnamefont {G.}~\bibnamefont {Hautier}},\ }\bibfield  {title} {\bibinfo
  {title} {Electron-phonon beyond {Fr\"ohlich}: Dynamical quadrupoles in polar
  and covalent solids},\ }\href
  {https://doi.org/10.1103/PhysRevLett.125.136601} {\bibfield  {journal}
  {\bibinfo  {journal} {Phys. Rev. Lett.}\ }\textbf {\bibinfo {volume} {125}},\
  \bibinfo {pages} {136601} (\bibinfo {year} {2020}{\natexlab{a}})}\BibitemShut
  {NoStop}%
\bibitem [{\citenamefont {Brunin}\ \emph
  {et~al.}(2020{\natexlab{b}})\citenamefont {Brunin}, \citenamefont {Miranda},
  \citenamefont {Giantomassi}, \citenamefont {Royo}, \citenamefont {Stengel},
  \citenamefont {Verstraete}, \citenamefont {Gonze}, \citenamefont
  {Rignanese},\ and\ \citenamefont {Hautier}}]{Brunin2020PRB}%
  \BibitemOpen
  \bibfield  {author} {\bibinfo {author} {\bibfnamefont {G.}~\bibnamefont
  {Brunin}}, \bibinfo {author} {\bibfnamefont {H.~P.~C.}\ \bibnamefont
  {Miranda}}, \bibinfo {author} {\bibfnamefont {M.}~\bibnamefont
  {Giantomassi}}, \bibinfo {author} {\bibfnamefont {M.}~\bibnamefont {Royo}},
  \bibinfo {author} {\bibfnamefont {M.}~\bibnamefont {Stengel}}, \bibinfo
  {author} {\bibfnamefont {M.~J.}\ \bibnamefont {Verstraete}}, \bibinfo
  {author} {\bibfnamefont {X.}~\bibnamefont {Gonze}}, \bibinfo {author}
  {\bibfnamefont {G.-M.}\ \bibnamefont {Rignanese}},\ and\ \bibinfo {author}
  {\bibfnamefont {G.}~\bibnamefont {Hautier}},\ }\bibfield  {title} {\bibinfo
  {title} {Phonon-limited electron mobility in {Si}, {GaAs}, and {GaP} with
  exact treatment of dynamical quadrupoles},\ }\href
  {https://doi.org/10.1103/PhysRevB.102.094308} {\bibfield  {journal} {\bibinfo
   {journal} {Phys. Rev. B}\ }\textbf {\bibinfo {volume} {102}},\ \bibinfo
  {pages} {094308} (\bibinfo {year} {2020}{\natexlab{b}})}\BibitemShut
  {NoStop}%
\bibitem [{\citenamefont {Jhalani}\ \emph {et~al.}(2020)\citenamefont
  {Jhalani}, \citenamefont {Zhou}, \citenamefont {Park}, \citenamefont
  {Dreyer},\ and\ \citenamefont {Bernardi}}]{Jhalani2020}%
  \BibitemOpen
  \bibfield  {author} {\bibinfo {author} {\bibfnamefont {V.~A.}\ \bibnamefont
  {Jhalani}}, \bibinfo {author} {\bibfnamefont {J.-J.}\ \bibnamefont {Zhou}},
  \bibinfo {author} {\bibfnamefont {J.}~\bibnamefont {Park}}, \bibinfo {author}
  {\bibfnamefont {C.~E.}\ \bibnamefont {Dreyer}},\ and\ \bibinfo {author}
  {\bibfnamefont {M.}~\bibnamefont {Bernardi}},\ }\bibfield  {title} {\bibinfo
  {title} {Piezoelectric electron-phonon interaction from ab initio dynamical
  quadrupoles: Impact on charge transport in wurtzite {GaN}},\ }\href
  {https://doi.org/10.1103/PhysRevLett.125.136602} {\bibfield  {journal}
  {\bibinfo  {journal} {Phys. Rev. Lett.}\ }\textbf {\bibinfo {volume} {125}},\
  \bibinfo {pages} {136602} (\bibinfo {year} {2020})}\BibitemShut {NoStop}%
\bibitem [{\citenamefont {Park}\ \emph {et~al.}(2020)\citenamefont {Park},
  \citenamefont {Zhou}, \citenamefont {Jhalani}, \citenamefont {Dreyer},\ and\
  \citenamefont {Bernardi}}]{Park2020}%
  \BibitemOpen
  \bibfield  {author} {\bibinfo {author} {\bibfnamefont {J.}~\bibnamefont
  {Park}}, \bibinfo {author} {\bibfnamefont {J.-J.}\ \bibnamefont {Zhou}},
  \bibinfo {author} {\bibfnamefont {V.~A.}\ \bibnamefont {Jhalani}}, \bibinfo
  {author} {\bibfnamefont {C.~E.}\ \bibnamefont {Dreyer}},\ and\ \bibinfo
  {author} {\bibfnamefont {M.}~\bibnamefont {Bernardi}},\ }\bibfield  {title}
  {\bibinfo {title} {Long-range quadrupole electron-phonon interaction from
  first principles},\ }\href {https://doi.org/10.1103/PhysRevB.102.125203}
  {\bibfield  {journal} {\bibinfo  {journal} {Phys. Rev. B}\ }\textbf {\bibinfo
  {volume} {102}},\ \bibinfo {pages} {125203} (\bibinfo {year}
  {2020})}\BibitemShut {NoStop}%
\bibitem [{\citenamefont {Ponc\'e}\ \emph
  {et~al.}(2023{\natexlab{a}})\citenamefont {Ponc\'e}, \citenamefont {Royo},
  \citenamefont {Gibertini}, \citenamefont {Marzari},\ and\ \citenamefont
  {Stengel}}]{Ponce2023PRL}%
  \BibitemOpen
  \bibfield  {author} {\bibinfo {author} {\bibfnamefont {S.}~\bibnamefont
  {Ponc\'e}}, \bibinfo {author} {\bibfnamefont {M.}~\bibnamefont {Royo}},
  \bibinfo {author} {\bibfnamefont {M.}~\bibnamefont {Gibertini}}, \bibinfo
  {author} {\bibfnamefont {N.}~\bibnamefont {Marzari}},\ and\ \bibinfo {author}
  {\bibfnamefont {M.}~\bibnamefont {Stengel}},\ }\bibfield  {title} {\bibinfo
  {title} {Accurate prediction of {Hall} mobilities in two-dimensional
  materials through gauge-covariant quadrupolar contributions},\ }\href
  {https://doi.org/10.1103/PhysRevLett.130.166301} {\bibfield  {journal}
  {\bibinfo  {journal} {Phys. Rev. Lett.}\ }\textbf {\bibinfo {volume} {130}},\
  \bibinfo {pages} {166301} (\bibinfo {year} {2023}{\natexlab{a}})}\BibitemShut
  {NoStop}%
\bibitem [{\citenamefont {Ponc\'e}\ \emph
  {et~al.}(2023{\natexlab{b}})\citenamefont {Ponc\'e}, \citenamefont {Royo},
  \citenamefont {Stengel}, \citenamefont {Marzari},\ and\ \citenamefont
  {Gibertini}}]{Ponce2023PRB}%
  \BibitemOpen
  \bibfield  {author} {\bibinfo {author} {\bibfnamefont {S.}~\bibnamefont
  {Ponc\'e}}, \bibinfo {author} {\bibfnamefont {M.}~\bibnamefont {Royo}},
  \bibinfo {author} {\bibfnamefont {M.}~\bibnamefont {Stengel}}, \bibinfo
  {author} {\bibfnamefont {N.}~\bibnamefont {Marzari}},\ and\ \bibinfo {author}
  {\bibfnamefont {M.}~\bibnamefont {Gibertini}},\ }\bibfield  {title} {\bibinfo
  {title} {Long-range electrostatic contribution to electron-phonon couplings
  and mobilities of two-dimensional and bulk materials},\ }\href
  {https://doi.org/10.1103/PhysRevB.107.155424} {\bibfield  {journal} {\bibinfo
   {journal} {Phys. Rev. B}\ }\textbf {\bibinfo {volume} {107}},\ \bibinfo
  {pages} {155424} (\bibinfo {year} {2023}{\natexlab{b}})}\BibitemShut
  {NoStop}%
\bibitem [{\citenamefont {Macheda}\ \emph
  {et~al.}(2024{\natexlab{b}})\citenamefont {Macheda}, \citenamefont {Barone},\
  and\ \citenamefont {Mauri}}]{Macheda2024a}%
  \BibitemOpen
  \bibfield  {author} {\bibinfo {author} {\bibfnamefont {F.}~\bibnamefont
  {Macheda}}, \bibinfo {author} {\bibfnamefont {P.}~\bibnamefont {Barone}},\
  and\ \bibinfo {author} {\bibfnamefont {F.}~\bibnamefont {Mauri}},\ }\bibfield
   {title} {\bibinfo {title} {First-principles calculations of dynamical {Born}
  effective charges, quadrupoles, and higher order terms from the charge
  response in large semiconducting and metallic systems},\ }\href
  {https://doi.org/10.1103/PhysRevB.110.094306} {\bibfield  {journal} {\bibinfo
   {journal} {Phys. Rev. B}\ }\textbf {\bibinfo {volume} {110}},\ \bibinfo
  {pages} {094306} (\bibinfo {year} {2024}{\natexlab{b}})}\BibitemShut
  {NoStop}%
\bibitem [{\citenamefont {Giannozzi}\ \emph {et~al.}(2009)\citenamefont
  {Giannozzi} \emph {et~al.}}]{Giannozzi2009}%
  \BibitemOpen
  \bibfield  {author} {\bibinfo {author} {\bibfnamefont {P.}~\bibnamefont
  {Giannozzi}} \emph {et~al.},\ }\bibfield  {title} {\bibinfo {title}
  {{QUANTUM} {ESPRESSO}: a modular and open-source software project for quantum
  simulations of materials},\ }\href
  {https://doi.org/10.1088/0953-8984/21/39/395502} {\bibfield  {journal}
  {\bibinfo  {journal} {J. Phys.: Condens. Matter}\ }\textbf {\bibinfo {volume}
  {21}},\ \bibinfo {pages} {395502} (\bibinfo {year} {2009})}\BibitemShut
  {NoStop}%
\bibitem [{\citenamefont {Giannozzi}\ \emph {et~al.}(2017)\citenamefont
  {Giannozzi} \emph {et~al.}}]{Giannozzi2017}%
  \BibitemOpen
  \bibfield  {author} {\bibinfo {author} {\bibfnamefont {P.}~\bibnamefont
  {Giannozzi}} \emph {et~al.},\ }\bibfield  {title} {\bibinfo {title} {Advanced
  capabilities for materials modelling with {Quantum} {ESPRESSO}},\ }\href
  {https://doi.org/10.1088/1361-648x/aa8f79} {\bibfield  {journal} {\bibinfo
  {journal} {J. Condens. Matter Phys.}\ }\textbf {\bibinfo {volume} {29}},\
  \bibinfo {pages} {465901} (\bibinfo {year} {2017})}\BibitemShut {NoStop}%
\bibitem [{\citenamefont {Timrov}\ \emph {et~al.}(2022)\citenamefont {Timrov},
  \citenamefont {Marzari},\ and\ \citenamefont {Cococcioni}}]{Timrov2022HP}%
  \BibitemOpen
  \bibfield  {author} {\bibinfo {author} {\bibfnamefont {I.}~\bibnamefont
  {Timrov}}, \bibinfo {author} {\bibfnamefont {N.}~\bibnamefont {Marzari}},\
  and\ \bibinfo {author} {\bibfnamefont {M.}~\bibnamefont {Cococcioni}},\
  }\bibfield  {title} {\bibinfo {title} {{HP} -- a code for the calculation of
  {Hubbard} parameters using density-functional perturbation theory},\ }\href
  {https://doi.org/10.1016/j.cpc.2022.108455} {\bibfield  {journal} {\bibinfo
  {journal} {Comput. Phys. Commun.}\ }\textbf {\bibinfo {volume} {279}},\
  \bibinfo {pages} {108455} (\bibinfo {year} {2022})}\BibitemShut {NoStop}%
\bibitem [{\citenamefont {Lihm}(2026)}]{EPjl}%
  \BibitemOpen
  \bibfield  {author} {\bibinfo {author} {\bibfnamefont {J.-M.}\ \bibnamefont
  {Lihm}},\ }\href {https://github.com/jaemolihm/ElectronPhonon.jl} {\bibinfo
  {title} {{ElectronPhonon.jl}}} (\bibinfo {year} {2026})\BibitemShut {NoStop}%
\bibitem [{\citenamefont {Macheda}\ \emph {et~al.}(2022)\citenamefont
  {Macheda}, \citenamefont {Barone},\ and\ \citenamefont
  {Mauri}}]{Macheda2022}%
  \BibitemOpen
  \bibfield  {author} {\bibinfo {author} {\bibfnamefont {F.}~\bibnamefont
  {Macheda}}, \bibinfo {author} {\bibfnamefont {P.}~\bibnamefont {Barone}},\
  and\ \bibinfo {author} {\bibfnamefont {F.}~\bibnamefont {Mauri}},\ }\bibfield
   {title} {\bibinfo {title} {Electron-phonon interaction and
  longitudinal-transverse phonon splitting in doped semiconductors},\ }\href
  {https://doi.org/10.1103/PhysRevLett.129.185902} {\bibfield  {journal}
  {\bibinfo  {journal} {Phys. Rev. Lett.}\ }\textbf {\bibinfo {volume} {129}},\
  \bibinfo {pages} {185902} (\bibinfo {year} {2022})}\BibitemShut {NoStop}%
\bibitem [{\citenamefont {Macheda}\ \emph {et~al.}(2023)\citenamefont
  {Macheda}, \citenamefont {Sohier}, \citenamefont {Barone},\ and\
  \citenamefont {Mauri}}]{Macheda2023}%
  \BibitemOpen
  \bibfield  {author} {\bibinfo {author} {\bibfnamefont {F.}~\bibnamefont
  {Macheda}}, \bibinfo {author} {\bibfnamefont {T.}~\bibnamefont {Sohier}},
  \bibinfo {author} {\bibfnamefont {P.}~\bibnamefont {Barone}},\ and\ \bibinfo
  {author} {\bibfnamefont {F.}~\bibnamefont {Mauri}},\ }\bibfield  {title}
  {\bibinfo {title} {Electron-phonon interaction and phonon frequencies in
  two-dimensional doped semiconductors},\ }\href
  {https://doi.org/10.1103/PhysRevB.107.094308} {\bibfield  {journal} {\bibinfo
   {journal} {Phys. Rev. B}\ }\textbf {\bibinfo {volume} {107}},\ \bibinfo
  {pages} {094308} (\bibinfo {year} {2023})}\BibitemShut {NoStop}%
\bibitem [{\citenamefont {Tubman}\ \emph {et~al.}(2026)\citenamefont {Tubman},
  \citenamefont {Coveney}, \citenamefont {Hsu}, \citenamefont
  {Montoya-Castillo}, \citenamefont {Filip}, \citenamefont {Neaton},
  \citenamefont {Li}, \citenamefont {Vlcek},\ and\ \citenamefont
  {Alvertis}}]{Tubman2025}%
  \BibitemOpen
  \bibfield  {author} {\bibinfo {author} {\bibfnamefont {N.~M.}\ \bibnamefont
  {Tubman}}, \bibinfo {author} {\bibfnamefont {C.~J.~N.}\ \bibnamefont
  {Coveney}}, \bibinfo {author} {\bibfnamefont {C.-E.}\ \bibnamefont {Hsu}},
  \bibinfo {author} {\bibfnamefont {A.}~\bibnamefont {Montoya-Castillo}},
  \bibinfo {author} {\bibfnamefont {M.~R.}\ \bibnamefont {Filip}}, \bibinfo
  {author} {\bibfnamefont {J.~B.}\ \bibnamefont {Neaton}}, \bibinfo {author}
  {\bibfnamefont {Z.}~\bibnamefont {Li}}, \bibinfo {author} {\bibfnamefont
  {V.}~\bibnamefont {Vlcek}},\ and\ \bibinfo {author} {\bibfnamefont {A.~M.}\
  \bibnamefont {Alvertis}},\ }\bibfield  {title} {\bibinfo {title} {Theory of
  ab initio downfolding with arbitrary range electron-phonon coupling},\ }\href
  {https://doi.org/10.1103/7hqv-hn2v} {\bibfield  {journal} {\bibinfo
  {journal} {Phys. Rev. B}\ }\textbf {\bibinfo {volume} {113}},\ \bibinfo
  {pages} {245144} (\bibinfo {year} {2026})}\BibitemShut {NoStop}%
\bibitem [{\citenamefont {Coveney}\ \emph {et~al.}(2025)\citenamefont
  {Coveney}, \citenamefont {Tubman}, \citenamefont {Hsu}, \citenamefont
  {Montoya-Castillo}, \citenamefont {Filip}, \citenamefont {Neaton},
  \citenamefont {Li}, \citenamefont {Vlcek},\ and\ \citenamefont
  {Alvertis}}]{Coveney2025}%
  \BibitemOpen
  \bibfield  {author} {\bibinfo {author} {\bibfnamefont {C.~J.~N.}\
  \bibnamefont {Coveney}}, \bibinfo {author} {\bibfnamefont {N.~M.}\
  \bibnamefont {Tubman}}, \bibinfo {author} {\bibfnamefont {C.-E.}\
  \bibnamefont {Hsu}}, \bibinfo {author} {\bibfnamefont {A.}~\bibnamefont
  {Montoya-Castillo}}, \bibinfo {author} {\bibfnamefont {M.~R.}\ \bibnamefont
  {Filip}}, \bibinfo {author} {\bibfnamefont {J.~B.}\ \bibnamefont {Neaton}},
  \bibinfo {author} {\bibfnamefont {Z.}~\bibnamefont {Li}}, \bibinfo {author}
  {\bibfnamefont {V.}~\bibnamefont {Vlcek}},\ and\ \bibinfo {author}
  {\bibfnamefont {A.~M.}\ \bibnamefont {Alvertis}},\ }\href
  {https://arxiv.org/abs/2501.17230} {\bibinfo {title} {Phonon-mediated
  electron attraction in {SrTiO$_3$} via the generalized {Fr\"ohlich} and
  deformation potential mechanisms}} (\bibinfo {year} {2025}),\ \Eprint
  {https://arxiv.org/abs/2501.17230} {arXiv:2501.17230 [cond-mat.supr-con]}
  \BibitemShut {NoStop}%
\bibitem [{\citenamefont {Berges}\ \emph {et~al.}(2026)\citenamefont {Berges},
  \citenamefont {Ponc\'e}, \citenamefont {Caserta}, \citenamefont {Marzari},\
  and\ \citenamefont {Chiarotti}}]{Berges2026}%
  \BibitemOpen
  \bibfield  {author} {\bibinfo {author} {\bibfnamefont {J.}~\bibnamefont
  {Berges}}, \bibinfo {author} {\bibfnamefont {S.}~\bibnamefont {Ponc\'e}},
  \bibinfo {author} {\bibfnamefont {M.}~\bibnamefont {Caserta}}, \bibinfo
  {author} {\bibfnamefont {N.}~\bibnamefont {Marzari}},\ and\ \bibinfo {author}
  {\bibfnamefont {T.}~\bibnamefont {Chiarotti}},\ }\href
  {https://arxiv.org/abs/2610.09656} {\bibinfo {title} {Unified treatment of
  local dynamical interactions in correlated metals using {Eliashberg} theory}}
  (\bibinfo {year} {2026}),\ \Eprint {https://arxiv.org/abs/2610.09656}
  {arXiv:2610.09656 [cond-mat.str-el]} \BibitemShut {NoStop}%
\bibitem [{\citenamefont {McMillan}(1968)}]{McMillan1968}%
  \BibitemOpen
  \bibfield  {author} {\bibinfo {author} {\bibfnamefont {W.~L.}\ \bibnamefont
  {McMillan}},\ }\bibfield  {title} {\bibinfo {title} {Transition temperature
  of strong-coupled superconductors},\ }\href
  {https://doi.org/10.1103/PhysRev.167.331} {\bibfield  {journal} {\bibinfo
  {journal} {Phys. Rev.}\ }\textbf {\bibinfo {volume} {167}},\ \bibinfo {pages}
  {331} (\bibinfo {year} {1968})}\BibitemShut {NoStop}%
\bibitem [{\citenamefont {Yokota}(1961)}]{Yokota1961}%
  \BibitemOpen
  \bibfield  {author} {\bibinfo {author} {\bibfnamefont {I.}~\bibnamefont
  {Yokota}},\ }\bibfield  {title} {\bibinfo {title} {On the coupling between
  optical lattice vibrations and carrier plasma oscillations in polar
  semiconductors},\ }\href {https://doi.org/10.1143/JPSJ.16.2075} {\bibfield
  {journal} {\bibinfo  {journal} {J. Phys. Soc. Jpn.}\ }\textbf {\bibinfo
  {volume} {16}},\ \bibinfo {pages} {2075} (\bibinfo {year}
  {1961})}\BibitemShut {NoStop}%
\bibitem [{\citenamefont {Varga}(1965)}]{Varga1965}%
  \BibitemOpen
  \bibfield  {author} {\bibinfo {author} {\bibfnamefont {B.~B.}\ \bibnamefont
  {Varga}},\ }\bibfield  {title} {\bibinfo {title} {Coupling of plasmons to
  polar phonons in degenerate semiconductors},\ }\href
  {https://doi.org/10.1103/PhysRev.137.A1896} {\bibfield  {journal} {\bibinfo
  {journal} {Phys. Rev.}\ }\textbf {\bibinfo {volume} {137}},\ \bibinfo {pages}
  {A1896} (\bibinfo {year} {1965})}\BibitemShut {NoStop}%
\bibitem [{\citenamefont {Marini}(2025)}]{Marini2024}%
  \BibitemOpen
  \bibfield  {author} {\bibinfo {author} {\bibfnamefont {A.}~\bibnamefont
  {Marini}},\ }\href {https://doi.org/10.48550/arXiv.2501.01866} {\bibinfo
  {title} {Dynamical electron-phonon vertex correction}} (\bibinfo {year}
  {2025}),\ \Eprint {https://arxiv.org/abs/2501.01866} {arXiv:2501.01866
  [cond-mat.mtrl-sci]} \BibitemShut {NoStop}%
\bibitem [{\citenamefont {Garcia-Goiricelaya}\ \emph
  {et~al.}(2020)\citenamefont {Garcia-Goiricelaya}, \citenamefont
  {Lafuente-Bartolome}, \citenamefont {Gurtubay},\ and\ \citenamefont
  {Eiguren}}]{Garcia2020}%
  \BibitemOpen
  \bibfield  {author} {\bibinfo {author} {\bibfnamefont {P.}~\bibnamefont
  {Garcia-Goiricelaya}}, \bibinfo {author} {\bibfnamefont {J.}~\bibnamefont
  {Lafuente-Bartolome}}, \bibinfo {author} {\bibfnamefont {I.~G.}\ \bibnamefont
  {Gurtubay}},\ and\ \bibinfo {author} {\bibfnamefont {A.}~\bibnamefont
  {Eiguren}},\ }\bibfield  {title} {\bibinfo {title} {Emergence of large
  nonadiabatic effects induced by the electron-phonon interaction on the
  complex vibrational quasiparticle spectrum of doped monolayer
  {${\mathrm{MoS}}_{2}$}},\ }\href
  {https://doi.org/10.1103/PhysRevB.101.054304} {\bibfield  {journal} {\bibinfo
   {journal} {Phys. Rev. B}\ }\textbf {\bibinfo {volume} {101}},\ \bibinfo
  {pages} {054304} (\bibinfo {year} {2020})}\BibitemShut {NoStop}%
\bibitem [{\citenamefont {Novko}(2020)}]{Novko2020}%
  \BibitemOpen
  \bibfield  {author} {\bibinfo {author} {\bibfnamefont {D.}~\bibnamefont
  {Novko}},\ }\bibfield  {title} {\bibinfo {title} {Broken adiabaticity induced
  by {Lifshitz} transition in {MoS$_2$} and {WS$_2$} single layers},\ }\href
  {https://doi.org/10.1038/s42005-020-0299-1} {\bibfield  {journal} {\bibinfo
  {journal} {Commun. Phys.}\ }\textbf {\bibinfo {volume} {3}},\ \bibinfo
  {pages} {30} (\bibinfo {year} {2020})}\BibitemShut {NoStop}%
\bibitem [{\citenamefont {Caruso}\ \emph {et~al.}(2015)\citenamefont {Caruso},
  \citenamefont {Lambert},\ and\ \citenamefont {Giustino}}]{Caruso2015PRL}%
  \BibitemOpen
  \bibfield  {author} {\bibinfo {author} {\bibfnamefont {F.}~\bibnamefont
  {Caruso}}, \bibinfo {author} {\bibfnamefont {H.}~\bibnamefont {Lambert}},\
  and\ \bibinfo {author} {\bibfnamefont {F.}~\bibnamefont {Giustino}},\
  }\bibfield  {title} {\bibinfo {title} {Band structures of plasmonic
  polarons},\ }\href {https://doi.org/10.1103/PhysRevLett.114.146404}
  {\bibfield  {journal} {\bibinfo  {journal} {Phys. Rev. Lett.}\ }\textbf
  {\bibinfo {volume} {114}},\ \bibinfo {pages} {146404} (\bibinfo {year}
  {2015})}\BibitemShut {NoStop}%
\bibitem [{\citenamefont {Caruso}\ \emph {et~al.}(2021)\citenamefont {Caruso},
  \citenamefont {Amsalem}, \citenamefont {Ma}, \citenamefont {Aljarb},
  \citenamefont {Schultz}, \citenamefont {Zacharias}, \citenamefont {Tung},
  \citenamefont {Koch},\ and\ \citenamefont {Draxl}}]{Caruso2021MoS2}%
  \BibitemOpen
  \bibfield  {author} {\bibinfo {author} {\bibfnamefont {F.}~\bibnamefont
  {Caruso}}, \bibinfo {author} {\bibfnamefont {P.}~\bibnamefont {Amsalem}},
  \bibinfo {author} {\bibfnamefont {J.}~\bibnamefont {Ma}}, \bibinfo {author}
  {\bibfnamefont {A.}~\bibnamefont {Aljarb}}, \bibinfo {author} {\bibfnamefont
  {T.}~\bibnamefont {Schultz}}, \bibinfo {author} {\bibfnamefont
  {M.}~\bibnamefont {Zacharias}}, \bibinfo {author} {\bibfnamefont
  {V.}~\bibnamefont {Tung}}, \bibinfo {author} {\bibfnamefont {N.}~\bibnamefont
  {Koch}},\ and\ \bibinfo {author} {\bibfnamefont {C.}~\bibnamefont {Draxl}},\
  }\bibfield  {title} {\bibinfo {title} {Two-dimensional plasmonic polarons in
  $n$-doped monolayer {MoS$_2$}},\ }\href
  {https://doi.org/10.1103/PhysRevB.103.205152} {\bibfield  {journal} {\bibinfo
   {journal} {Phys. Rev. B}\ }\textbf {\bibinfo {volume} {103}},\ \bibinfo
  {pages} {205152} (\bibinfo {year} {2021})}\BibitemShut {NoStop}%
\bibitem [{\citenamefont {Caruso}\ \emph {et~al.}(2020)\citenamefont {Caruso},
  \citenamefont {Verdi},\ and\ \citenamefont {Giustino}}]{Caruso2020Review}%
  \BibitemOpen
  \bibfield  {author} {\bibinfo {author} {\bibfnamefont {F.}~\bibnamefont
  {Caruso}}, \bibinfo {author} {\bibfnamefont {C.}~\bibnamefont {Verdi}},\ and\
  \bibinfo {author} {\bibfnamefont {F.}~\bibnamefont {Giustino}},\ }\bibinfo
  {title} {Many-body calculations of plasmon and phonon satellites in
  angle-resolved photoelectron spectra using the cumulant expansion approach},\
  in\ \href {https://doi.org/10.1007/978-3-319-44677-6_2} {\emph {\bibinfo
  {booktitle} {Handbook of Materials Modeling: Methods: Theory and
  Modeling}}},\ \bibinfo {editor} {edited by\ \bibinfo {editor} {\bibfnamefont
  {W.}~\bibnamefont {Andreoni}}\ and\ \bibinfo {editor} {\bibfnamefont
  {S.}~\bibnamefont {Yip}}}\ (\bibinfo  {publisher} {Springer International
  Publishing},\ \bibinfo {address} {Cham},\ \bibinfo {year} {2020})\ pp.\
  \bibinfo {pages} {341--365}\BibitemShut {NoStop}%
\bibitem [{\citenamefont {Verdi}\ \emph {et~al.}(2017)\citenamefont {Verdi},
  \citenamefont {Caruso},\ and\ \citenamefont {Giustino}}]{Verdi2017}%
  \BibitemOpen
  \bibfield  {author} {\bibinfo {author} {\bibfnamefont {C.}~\bibnamefont
  {Verdi}}, \bibinfo {author} {\bibfnamefont {F.}~\bibnamefont {Caruso}},\ and\
  \bibinfo {author} {\bibfnamefont {F.}~\bibnamefont {Giustino}},\ }\bibfield
  {title} {\bibinfo {title} {Origin of the crossover from polarons to {{Fermi}}
  liquids in transition metal oxides},\ }\href
  {https://doi.org/10.1038/ncomms15769} {\bibfield  {journal} {\bibinfo
  {journal} {Nat. Commun.}\ }\textbf {\bibinfo {volume} {8}},\ \bibinfo {pages}
  {15769} (\bibinfo {year} {2017})}\BibitemShut {NoStop}%
\bibitem [{\citenamefont {Riley}\ \emph {et~al.}(2018)\citenamefont {Riley},
  \citenamefont {Caruso}, \citenamefont {Verdi}, \citenamefont {Duffy},
  \citenamefont {Watson}, \citenamefont {Bawden}, \citenamefont {Volckaert},
  \citenamefont {Van Der~Laan}, \citenamefont {Hesjedal}, \citenamefont
  {Hoesch}, \citenamefont {Giustino},\ and\ \citenamefont {King}}]{Riley2018}%
  \BibitemOpen
  \bibfield  {author} {\bibinfo {author} {\bibfnamefont {J.~M.}\ \bibnamefont
  {Riley}}, \bibinfo {author} {\bibfnamefont {F.}~\bibnamefont {Caruso}},
  \bibinfo {author} {\bibfnamefont {C.}~\bibnamefont {Verdi}}, \bibinfo
  {author} {\bibfnamefont {L.~B.}\ \bibnamefont {Duffy}}, \bibinfo {author}
  {\bibfnamefont {M.~D.}\ \bibnamefont {Watson}}, \bibinfo {author}
  {\bibfnamefont {L.}~\bibnamefont {Bawden}}, \bibinfo {author} {\bibfnamefont
  {K.}~\bibnamefont {Volckaert}}, \bibinfo {author} {\bibfnamefont
  {G.}~\bibnamefont {Van Der~Laan}}, \bibinfo {author} {\bibfnamefont
  {T.}~\bibnamefont {Hesjedal}}, \bibinfo {author} {\bibfnamefont
  {M.}~\bibnamefont {Hoesch}}, \bibinfo {author} {\bibfnamefont
  {F.}~\bibnamefont {Giustino}},\ and\ \bibinfo {author} {\bibfnamefont
  {P.~D.~C.}\ \bibnamefont {King}},\ }\bibfield  {title} {\bibinfo {title}
  {Crossover from lattice to plasmonic polarons of a spin-polarised electron
  gas in ferromagnetic {EuO}},\ }\href
  {https://doi.org/10.1038/s41467-018-04749-w} {\bibfield  {journal} {\bibinfo
  {journal} {Nat. Commun.}\ }\textbf {\bibinfo {volume} {9}},\ \bibinfo {pages}
  {2305} (\bibinfo {year} {2018})}\BibitemShut {NoStop}%
\bibitem [{\citenamefont {Ponc{\'e}}\ \emph {et~al.}(2018)\citenamefont
  {Ponc{\'e}}, \citenamefont {Margine},\ and\ \citenamefont
  {Giustino}}]{Ponce2018}%
  \BibitemOpen
  \bibfield  {author} {\bibinfo {author} {\bibfnamefont {S.}~\bibnamefont
  {Ponc{\'e}}}, \bibinfo {author} {\bibfnamefont {E.~R.}\ \bibnamefont
  {Margine}},\ and\ \bibinfo {author} {\bibfnamefont {F.}~\bibnamefont
  {Giustino}},\ }\bibfield  {title} {\bibinfo {title} {Towards predictive
  many-body calculations of phonon-limited carrier mobilities in
  semiconductors},\ }\href {https://doi.org/10.1103/PhysRevB.97.121201}
  {\bibfield  {journal} {\bibinfo  {journal} {Phys. Rev. B}\ }\textbf {\bibinfo
  {volume} {97}},\ \bibinfo {pages} {121201} (\bibinfo {year}
  {2018})}\BibitemShut {NoStop}%
\bibitem [{\citenamefont {Basov}\ \emph {et~al.}(2011)\citenamefont {Basov},
  \citenamefont {Averitt}, \citenamefont {van~der Marel}, \citenamefont
  {Dressel},\ and\ \citenamefont {Haule}}]{Basov2011RMP}%
  \BibitemOpen
  \bibfield  {author} {\bibinfo {author} {\bibfnamefont {D.~N.}\ \bibnamefont
  {Basov}}, \bibinfo {author} {\bibfnamefont {R.~D.}\ \bibnamefont {Averitt}},
  \bibinfo {author} {\bibfnamefont {D.}~\bibnamefont {van~der Marel}}, \bibinfo
  {author} {\bibfnamefont {M.}~\bibnamefont {Dressel}},\ and\ \bibinfo {author}
  {\bibfnamefont {K.}~\bibnamefont {Haule}},\ }\bibfield  {title} {\bibinfo
  {title} {Electrodynamics of correlated electron materials},\ }\href
  {https://doi.org/10.1103/RevModPhys.83.471} {\bibfield  {journal} {\bibinfo
  {journal} {Rev. Mod. Phys.}\ }\textbf {\bibinfo {volume} {83}},\ \bibinfo
  {pages} {471} (\bibinfo {year} {2011})}\BibitemShut {NoStop}%
\bibitem [{\citenamefont {Lawrence}\ and\ \citenamefont
  {Wilkins}(1973)}]{LawrenceWilkins1973}%
  \BibitemOpen
  \bibfield  {author} {\bibinfo {author} {\bibfnamefont {W.~E.}\ \bibnamefont
  {Lawrence}}\ and\ \bibinfo {author} {\bibfnamefont {J.~W.}\ \bibnamefont
  {Wilkins}},\ }\bibfield  {title} {\bibinfo {title} {Electron-electron
  scattering in the transport coefficients of simple metals},\ }\href
  {https://doi.org/10.1103/PhysRevB.7.2317} {\bibfield  {journal} {\bibinfo
  {journal} {Phys. Rev. B}\ }\textbf {\bibinfo {volume} {7}},\ \bibinfo {pages}
  {2317} (\bibinfo {year} {1973})}\BibitemShut {NoStop}%
\bibitem [{\citenamefont {MacDonald}\ \emph {et~al.}(1981)\citenamefont
  {MacDonald}, \citenamefont {Taylor},\ and\ \citenamefont
  {Geldart}}]{MacDonald1981}%
  \BibitemOpen
  \bibfield  {author} {\bibinfo {author} {\bibfnamefont {A.~H.}\ \bibnamefont
  {MacDonald}}, \bibinfo {author} {\bibfnamefont {R.}~\bibnamefont {Taylor}},\
  and\ \bibinfo {author} {\bibfnamefont {D.~J.~W.}\ \bibnamefont {Geldart}},\
  }\bibfield  {title} {\bibinfo {title} {Umklapp electron-electron scattering
  and the low-temperature electrical resistivity of the alkali metals},\ }\href
  {https://doi.org/10.1103/PhysRevB.23.2718} {\bibfield  {journal} {\bibinfo
  {journal} {Phys. Rev. B}\ }\textbf {\bibinfo {volume} {23}},\ \bibinfo
  {pages} {2718} (\bibinfo {year} {1981})}\BibitemShut {NoStop}%
\bibitem [{\citenamefont {Sanborn}(1995)}]{Sanborn1995}%
  \BibitemOpen
  \bibfield  {author} {\bibinfo {author} {\bibfnamefont {B.~A.}\ \bibnamefont
  {Sanborn}},\ }\bibfield  {title} {\bibinfo {title} {Nonequilibrium
  total-dielectric-function approach to the electron {Boltzmann} equation for
  inelastic scattering in doped polar semiconductors},\ }\href
  {https://doi.org/10.1103/PhysRevB.51.14247} {\bibfield  {journal} {\bibinfo
  {journal} {Phys. Rev. B}\ }\textbf {\bibinfo {volume} {51}},\ \bibinfo
  {pages} {14247} (\bibinfo {year} {1995})}\BibitemShut {NoStop}%
\bibitem [{\citenamefont {Hauber}\ and\ \citenamefont
  {Fahy}(2017)}]{Hauber2017}%
  \BibitemOpen
  \bibfield  {author} {\bibinfo {author} {\bibfnamefont {A.}~\bibnamefont
  {Hauber}}\ and\ \bibinfo {author} {\bibfnamefont {S.}~\bibnamefont {Fahy}},\
  }\bibfield  {title} {\bibinfo {title} {Scattering of carriers by coupled
  plasmon-phonon modes in bulk polar semiconductors and polar semiconductor
  heterostructures},\ }\href {https://doi.org/10.1103/PhysRevB.95.045210}
  {\bibfield  {journal} {\bibinfo  {journal} {Phys. Rev. B}\ }\textbf {\bibinfo
  {volume} {95}},\ \bibinfo {pages} {045210} (\bibinfo {year}
  {2017})}\BibitemShut {NoStop}%
\bibitem [{\citenamefont {Gopalan}\ \emph {et~al.}(2022)\citenamefont
  {Gopalan}, \citenamefont {Van~de Put}, \citenamefont {Gaddemane},\ and\
  \citenamefont {Fischetti}}]{Gopalan2022}%
  \BibitemOpen
  \bibfield  {author} {\bibinfo {author} {\bibfnamefont {S.}~\bibnamefont
  {Gopalan}}, \bibinfo {author} {\bibfnamefont {M.~L.}\ \bibnamefont {Van~de
  Put}}, \bibinfo {author} {\bibfnamefont {G.}~\bibnamefont {Gaddemane}},\ and\
  \bibinfo {author} {\bibfnamefont {M.~V.}\ \bibnamefont {Fischetti}},\
  }\bibfield  {title} {\bibinfo {title} {Theoretical study of electronic
  transport in two-dimensional transition metal dichalcogenides: Effects of the
  dielectric environment},\ }\href
  {https://doi.org/10.1103/PhysRevApplied.18.054062} {\bibfield  {journal}
  {\bibinfo  {journal} {Phys. Rev. Appl.}\ }\textbf {\bibinfo {volume} {18}},\
  \bibinfo {pages} {054062} (\bibinfo {year} {2022})}\BibitemShut {NoStop}%
\bibitem [{\citenamefont {Macheda}\ and\ \citenamefont
  {Sohier}(2026)}]{Macheda2026Boltzmann}%
  \BibitemOpen
  \bibfield  {author} {\bibinfo {author} {\bibfnamefont {F.}~\bibnamefont
  {Macheda}}\ and\ \bibinfo {author} {\bibfnamefont {T.}~\bibnamefont
  {Sohier}},\ }\bibfield  {title} {\bibinfo {title} {Coupled dynamical
  {Boltzmann} transport equations with long-range electron-phonon and
  electron-electron interactions in two-dimensional materials},\ }\href
  {https://doi.org/10.1103/9dt4-gzhn} {\bibfield  {journal} {\bibinfo
  {journal} {Phys. Rev. B}\ }\textbf {\bibinfo {volume} {114}},\ \bibinfo
  {pages} {245419} (\bibinfo {year} {2026})}\BibitemShut {NoStop}%
\bibitem [{\citenamefont {Dinar}\ \emph {et~al.}(2026)\citenamefont {Dinar},
  \citenamefont {Macheda}, \citenamefont {Guandalini}, \citenamefont {Paillet},
  \citenamefont {Consejo}, \citenamefont {Teppe}, \citenamefont {Jouault},
  \citenamefont {Sohier},\ and\ \citenamefont {Nanot}}]{Dinar2026}%
  \BibitemOpen
  \bibfield  {author} {\bibinfo {author} {\bibfnamefont {K.}~\bibnamefont
  {Dinar}}, \bibinfo {author} {\bibfnamefont {F.}~\bibnamefont {Macheda}},
  \bibinfo {author} {\bibfnamefont {A.}~\bibnamefont {Guandalini}}, \bibinfo
  {author} {\bibfnamefont {M.}~\bibnamefont {Paillet}}, \bibinfo {author}
  {\bibfnamefont {C.}~\bibnamefont {Consejo}}, \bibinfo {author} {\bibfnamefont
  {F.}~\bibnamefont {Teppe}}, \bibinfo {author} {\bibfnamefont
  {B.}~\bibnamefont {Jouault}}, \bibinfo {author} {\bibfnamefont
  {T.}~\bibnamefont {Sohier}},\ and\ \bibinfo {author} {\bibfnamefont
  {S.}~\bibnamefont {Nanot}},\ }\href {https://arxiv.org/abs/2604.00678}
  {\bibinfo {title} {Electronic transport in {BN}-encapsulated graphene limited
  by remote phonon scattering}},\ \bibinfo {howpublished} {arXiv:2604.00678}
  (\bibinfo {year} {2026}),\ \Eprint {https://arxiv.org/abs/2604.00678}
  {arXiv:2604.00678} \BibitemShut {NoStop}%
\bibitem [{\citenamefont {Mocatti}\ \emph {et~al.}(2026)\citenamefont
  {Mocatti}, \citenamefont {Marini}, \citenamefont {Volpato}, \citenamefont
  {Cudazzo},\ and\ \citenamefont {Calandra}}]{Mocatti2026}%
  \BibitemOpen
  \bibfield  {author} {\bibinfo {author} {\bibfnamefont {S.}~\bibnamefont
  {Mocatti}}, \bibinfo {author} {\bibfnamefont {G.}~\bibnamefont {Marini}},
  \bibinfo {author} {\bibfnamefont {G.}~\bibnamefont {Volpato}}, \bibinfo
  {author} {\bibfnamefont {P.}~\bibnamefont {Cudazzo}},\ and\ \bibinfo {author}
  {\bibfnamefont {M.}~\bibnamefont {Calandra}},\ }\bibfield  {title} {\bibinfo
  {title} {Nonequilibrium photocarrier and phonon dynamics from first
  principles: a unified treatment of carrier-carrier, carrier-phonon, and
  phonon-phonon scattering},\ }\href
  {https://doi.org/10.1038/s41524-026-02104-y} {\bibfield  {journal} {\bibinfo
  {journal} {npj Comput. Mater.}\ }\textbf {\bibinfo {volume} {12}},\ \bibinfo
  {pages} {260} (\bibinfo {year} {2026})}\BibitemShut {NoStop}%
\bibitem [{\citenamefont {Paul}\ \emph {et~al.}(2026)\citenamefont {Paul},
  \citenamefont {Trukhan},\ and\ \citenamefont {Protik}}]{Paul2026}%
  \BibitemOpen
  \bibfield  {author} {\bibinfo {author} {\bibfnamefont {D.}~\bibnamefont
  {Paul}}, \bibinfo {author} {\bibfnamefont {E.}~\bibnamefont {Trukhan}},\ and\
  \bibinfo {author} {\bibfnamefont {N.~H.}\ \bibnamefont {Protik}},\ }\bibfield
   {title} {\bibinfo {title} {Ab initio study of {Coulomb} drag driven
  electron-hole bifluidity in doped graphene},\ }\href
  {https://doi.org/10.1103/q4mx-9tmn} {\bibfield  {journal} {\bibinfo
  {journal} {Phys. Rev. B}\ }\textbf {\bibinfo {volume} {114}},\ \bibinfo
  {pages} {L051405} (\bibinfo {year} {2026})}\BibitemShut {NoStop}%
\bibitem [{\citenamefont {Holstein}(1964)}]{Holstein1964}%
  \BibitemOpen
  \bibfield  {author} {\bibinfo {author} {\bibfnamefont {T.}~\bibnamefont
  {Holstein}},\ }\bibfield  {title} {\bibinfo {title} {Theory of transport
  phenomena in an electron-phonon gas},\ }\href
  {https://doi.org/https://doi.org/10.1016/0003-4916(64)90008-9} {\bibfield
  {journal} {\bibinfo  {journal} {Ann. Phys.}\ }\textbf {\bibinfo {volume}
  {29}},\ \bibinfo {pages} {410} (\bibinfo {year} {1964})}\BibitemShut
  {NoStop}%
\bibitem [{\citenamefont {Prange}\ and\ \citenamefont
  {Kadanoff}(1964)}]{PrangeKadanoff1964}%
  \BibitemOpen
  \bibfield  {author} {\bibinfo {author} {\bibfnamefont {R.~E.}\ \bibnamefont
  {Prange}}\ and\ \bibinfo {author} {\bibfnamefont {L.~P.}\ \bibnamefont
  {Kadanoff}},\ }\bibfield  {title} {\bibinfo {title} {Transport theory for
  electron-phonon interactions in metals},\ }\href
  {https://doi.org/10.1103/PhysRev.134.A566} {\bibfield  {journal} {\bibinfo
  {journal} {Phys. Rev.}\ }\textbf {\bibinfo {volume} {134}},\ \bibinfo {pages}
  {A566} (\bibinfo {year} {1964})}\BibitemShut {NoStop}%
\bibitem [{\citenamefont {Allen}(1971)}]{Allen1971}%
  \BibitemOpen
  \bibfield  {author} {\bibinfo {author} {\bibfnamefont {P.~B.}\ \bibnamefont
  {Allen}},\ }\bibfield  {title} {\bibinfo {title} {Electron-phonon effects in
  the infrared properties of metals},\ }\href
  {https://doi.org/10.1103/PhysRevB.3.305} {\bibfield  {journal} {\bibinfo
  {journal} {Phys. Rev. B}\ }\textbf {\bibinfo {volume} {3}},\ \bibinfo {pages}
  {305} (\bibinfo {year} {1971})}\BibitemShut {NoStop}%
\bibitem [{\citenamefont {Lihm}\ and\ \citenamefont
  {Ponc\'e}(2025)}]{Lihm2025}%
  \BibitemOpen
  \bibfield  {author} {\bibinfo {author} {\bibfnamefont {J.-M.}\ \bibnamefont
  {Lihm}}\ and\ \bibinfo {author} {\bibfnamefont {S.}~\bibnamefont {Ponc\'e}},\
  }\bibfield  {title} {\bibinfo {title} {Nonperturbative self-consistent
  electron-phonon spectral functions and transport},\ }\href
  {https://doi.org/10.1103/PhysRevLett.134.186401} {\bibfield  {journal}
  {\bibinfo  {journal} {Phys. Rev. Lett.}\ }\textbf {\bibinfo {volume} {134}},\
  \bibinfo {pages} {186401} (\bibinfo {year} {2025})}\BibitemShut {NoStop}%
\bibitem [{\citenamefont {Lihm}\ and\ \citenamefont
  {Ponc\'e}(2026)}]{Lihm2026Ladder}%
  \BibitemOpen
  \bibfield  {author} {\bibinfo {author} {\bibfnamefont {J.-M.}\ \bibnamefont
  {Lihm}}\ and\ \bibinfo {author} {\bibfnamefont {S.}~\bibnamefont {Ponc\'e}},\
  }\bibfield  {title} {\bibinfo {title} {Beyond-quasiparticle transport with
  vertex correction: Self-consistent ladder formalism for electron-phonon
  interactions},\ }\href {https://doi.org/10.1103/mtjf-ylf7} {\bibfield
  {journal} {\bibinfo  {journal} {Phys. Rev. X}\ }\textbf {\bibinfo {volume}
  {16}},\ \bibinfo {pages} {011008} (\bibinfo {year} {2026})}\BibitemShut
  {NoStop}%
\bibitem [{\citenamefont {Morel}\ and\ \citenamefont
  {Anderson}(1962)}]{MorelAnderson1962}%
  \BibitemOpen
  \bibfield  {author} {\bibinfo {author} {\bibfnamefont {P.}~\bibnamefont
  {Morel}}\ and\ \bibinfo {author} {\bibfnamefont {P.~W.}\ \bibnamefont
  {Anderson}},\ }\bibfield  {title} {\bibinfo {title} {Calculation of the
  superconducting state parameters with retarded electron-phonon interaction},\
  }\href {https://doi.org/10.1103/PhysRev.125.1263} {\bibfield  {journal}
  {\bibinfo  {journal} {Phys. Rev.}\ }\textbf {\bibinfo {volume} {125}},\
  \bibinfo {pages} {1263} (\bibinfo {year} {1962})}\BibitemShut {NoStop}%
\bibitem [{\citenamefont {Kogler}\ \emph {et~al.}(2025)\citenamefont {Kogler},
  \citenamefont {Spath}, \citenamefont {Lucrezi}, \citenamefont {Mori},
  \citenamefont {Zhu}, \citenamefont {Li}, \citenamefont {Margine},\ and\
  \citenamefont {Heil}}]{Kogler2025}%
  \BibitemOpen
  \bibfield  {author} {\bibinfo {author} {\bibfnamefont {E.}~\bibnamefont
  {Kogler}}, \bibinfo {author} {\bibfnamefont {D.}~\bibnamefont {Spath}},
  \bibinfo {author} {\bibfnamefont {R.}~\bibnamefont {Lucrezi}}, \bibinfo
  {author} {\bibfnamefont {H.}~\bibnamefont {Mori}}, \bibinfo {author}
  {\bibfnamefont {Z.}~\bibnamefont {Zhu}}, \bibinfo {author} {\bibfnamefont
  {Z.}~\bibnamefont {Li}}, \bibinfo {author} {\bibfnamefont {E.~R.}\
  \bibnamefont {Margine}},\ and\ \bibinfo {author} {\bibfnamefont
  {C.}~\bibnamefont {Heil}},\ }\bibfield  {title} {\bibinfo {title} {{IsoME}:
  Streamlining high-precision {Eliashberg} calculations},\ }\href
  {https://doi.org/10.1016/j.cpc.2025.109720} {\bibfield  {journal} {\bibinfo
  {journal} {Comput. Phys. Commun.}\ }\textbf {\bibinfo {volume} {315}},\
  \bibinfo {pages} {109720} (\bibinfo {year} {2025})}\BibitemShut {NoStop}%
\bibitem [{\citenamefont {Davydov}\ \emph {et~al.}(2020)\citenamefont
  {Davydov}, \citenamefont {Sanna}, \citenamefont {Pellegrini}, \citenamefont
  {Dewhurst}, \citenamefont {Sharma},\ and\ \citenamefont
  {Gross}}]{Davydov2020}%
  \BibitemOpen
  \bibfield  {author} {\bibinfo {author} {\bibfnamefont {A.}~\bibnamefont
  {Davydov}}, \bibinfo {author} {\bibfnamefont {A.}~\bibnamefont {Sanna}},
  \bibinfo {author} {\bibfnamefont {C.}~\bibnamefont {Pellegrini}}, \bibinfo
  {author} {\bibfnamefont {J.~K.}\ \bibnamefont {Dewhurst}}, \bibinfo {author}
  {\bibfnamefont {S.}~\bibnamefont {Sharma}},\ and\ \bibinfo {author}
  {\bibfnamefont {E.~K.~U.}\ \bibnamefont {Gross}},\ }\bibfield  {title}
  {\bibinfo {title} {Ab initio theory of plasmonic superconductivity within the
  {Eliashberg} and density-functional formalisms},\ }\href
  {https://doi.org/10.1103/PhysRevB.102.214508} {\bibfield  {journal} {\bibinfo
   {journal} {Phys. Rev. B}\ }\textbf {\bibinfo {volume} {102}},\ \bibinfo
  {pages} {214508} (\bibinfo {year} {2020})}\BibitemShut {NoStop}%
\bibitem [{\citenamefont {Spataru}\ \emph {et~al.}(2026)\citenamefont
  {Spataru}, \citenamefont {Renskers},\ and\ \citenamefont
  {Margine}}]{Spataru2026}%
  \BibitemOpen
  \bibfield  {author} {\bibinfo {author} {\bibfnamefont {C.~D.}\ \bibnamefont
  {Spataru}}, \bibinfo {author} {\bibfnamefont {C.}~\bibnamefont {Renskers}},\
  and\ \bibinfo {author} {\bibfnamefont {E.~R.}\ \bibnamefont {Margine}},\
  }\bibfield  {title} {\bibinfo {title} {Quasiparticle {GW} for
  superconductors: Toward a unified treatment of electron-phonon and
  electron-plasmon couplings},\ }\href {https://doi.org/10.1103/f8qb-2l59}
  {\bibfield  {journal} {\bibinfo  {journal} {Phys. Rev. B}\ }\textbf {\bibinfo
  {volume} {114}},\ \bibinfo {pages} {094501} (\bibinfo {year}
  {2026})}\BibitemShut {NoStop}%
\bibitem [{\citenamefont {{in 't Veld}}\ \emph {et~al.}(2023)\citenamefont {{in
  't Veld}}, \citenamefont {Katsnelson}, \citenamefont {Millis},\ and\
  \citenamefont {R\"osner}}]{inntVeld2023}%
  \BibitemOpen
  \bibfield  {author} {\bibinfo {author} {\bibfnamefont {Y.}~\bibnamefont {{in
  't Veld}}}, \bibinfo {author} {\bibfnamefont {M.~I.}\ \bibnamefont
  {Katsnelson}}, \bibinfo {author} {\bibfnamefont {A.~J.}\ \bibnamefont
  {Millis}},\ and\ \bibinfo {author} {\bibfnamefont {M.}~\bibnamefont
  {R\"osner}},\ }\bibfield  {title} {\bibinfo {title} {Screening induced
  crossover between phonon- and plasmon-mediated pairing in layered
  superconductors},\ }\href {https://doi.org/10.1088/2053-1583/acf944}
  {\bibfield  {journal} {\bibinfo  {journal} {2D Mater.}\ }\textbf {\bibinfo
  {volume} {10}},\ \bibinfo {pages} {045031} (\bibinfo {year}
  {2023})}\BibitemShut {NoStop}%
\bibitem [{\citenamefont {{in 't Veld}}\ \emph {et~al.}(2026)\citenamefont {{in
  't Veld}}, \citenamefont {Katsnelson}, \citenamefont {Millis},\ and\
  \citenamefont {R\"osner}}]{inntVeld2026}%
  \BibitemOpen
  \bibfield  {author} {\bibinfo {author} {\bibfnamefont {Y.}~\bibnamefont {{in
  't Veld}}}, \bibinfo {author} {\bibfnamefont {M.~I.}\ \bibnamefont
  {Katsnelson}}, \bibinfo {author} {\bibfnamefont {A.~J.}\ \bibnamefont
  {Millis}},\ and\ \bibinfo {author} {\bibfnamefont {M.}~\bibnamefont
  {R\"osner}},\ }\bibfield  {title} {\bibinfo {title} {Enhancing plasmonic
  superconductivity in layered materials via dynamical {Coulomb} engineering},\
  }\href {https://doi.org/10.1038/s41699-026-00668-3} {\bibfield  {journal}
  {\bibinfo  {journal} {npj 2D Mater. Appl.}\ }\textbf {\bibinfo {volume}
  {10}},\ \bibinfo {pages} {30} (\bibinfo {year} {2026})}\BibitemShut {NoStop}%
\bibitem [{\citenamefont {Cui}\ \emph {et~al.}(2026)\citenamefont {Cui},
  \citenamefont {Sous}, \citenamefont {Millis},\ and\ \citenamefont
  {Reichman}}]{Cui2026}%
  \BibitemOpen
  \bibfield  {author} {\bibinfo {author} {\bibfnamefont {Z.-H.}\ \bibnamefont
  {Cui}}, \bibinfo {author} {\bibfnamefont {J.}~\bibnamefont {Sous}}, \bibinfo
  {author} {\bibfnamefont {A.~J.}\ \bibnamefont {Millis}},\ and\ \bibinfo
  {author} {\bibfnamefont {D.~R.}\ \bibnamefont {Reichman}},\ }\href
  {https://arxiv.org/abs/2607.18757} {\bibinfo {title} {Self-consistent {GW}
  theory for superconductivity in {SrTiO$_3$} models}},\ \bibinfo
  {howpublished} {arXiv:2607.18757} (\bibinfo {year} {2026}),\ \Eprint
  {https://arxiv.org/abs/2607.18757} {arXiv:2607.18757} \BibitemShut {NoStop}%
\bibitem [{\citenamefont {Carrasco~{\'A}lvarez}\ \emph
  {et~al.}(2026)\citenamefont {Carrasco~{\'A}lvarez}, \citenamefont
  {Giantomassi}, \citenamefont {Lihm}, \citenamefont {Allemand}, \citenamefont
  {Mignolet}, \citenamefont {Verstraete},\ and\ \citenamefont
  {Ponc{\'e}}}]{CarrascoAlvarez2026}%
  \BibitemOpen
  \bibfield  {author} {\bibinfo {author} {\bibfnamefont {{\'A}.~A.}\
  \bibnamefont {Carrasco~{\'A}lvarez}}, \bibinfo {author} {\bibfnamefont
  {M.}~\bibnamefont {Giantomassi}}, \bibinfo {author} {\bibfnamefont {J.-M.}\
  \bibnamefont {Lihm}}, \bibinfo {author} {\bibfnamefont {G.~E.}\ \bibnamefont
  {Allemand}}, \bibinfo {author} {\bibfnamefont {M.}~\bibnamefont {Mignolet}},
  \bibinfo {author} {\bibfnamefont {M.}~\bibnamefont {Verstraete}},\ and\
  \bibinfo {author} {\bibfnamefont {S.}~\bibnamefont {Ponc{\'e}}},\ }\bibfield
  {title} {\bibinfo {title} {Electron-phonon coupling in magnetic materials
  using the local spin density approximation},\ }\href
  {https://doi.org/10.1103/gpf5-p4sn} {\bibfield  {journal} {\bibinfo
  {journal} {Phys. Rev. B}\ }\textbf {\bibinfo {volume} {114}},\ \bibinfo
  {pages} {074411} (\bibinfo {year} {2026})}\BibitemShut {NoStop}%
\bibitem [{\citenamefont {Skovhus}\ and\ \citenamefont
  {Olsen}(2021)}]{Skovhus2021}%
  \BibitemOpen
  \bibfield  {author} {\bibinfo {author} {\bibfnamefont {T.}~\bibnamefont
  {Skovhus}}\ and\ \bibinfo {author} {\bibfnamefont {T.}~\bibnamefont
  {Olsen}},\ }\bibfield  {title} {\bibinfo {title} {Dynamic transverse magnetic
  susceptibility in the projector augmented-wave method: Application to {Fe},
  {Ni}, and {Co}},\ }\href {https://doi.org/10.1103/PhysRevB.103.245110}
  {\bibfield  {journal} {\bibinfo  {journal} {Phys. Rev. B}\ }\textbf {\bibinfo
  {volume} {103}},\ \bibinfo {pages} {245110} (\bibinfo {year}
  {2021})}\BibitemShut {NoStop}%
\bibitem [{\citenamefont {Gorni}\ \emph {et~al.}(2022)\citenamefont {Gorni},
  \citenamefont {Baseggio}, \citenamefont {Delugas}, \citenamefont {Baroni},\
  and\ \citenamefont {Timrov}}]{Gorni2022}%
  \BibitemOpen
  \bibfield  {author} {\bibinfo {author} {\bibfnamefont {T.}~\bibnamefont
  {Gorni}}, \bibinfo {author} {\bibfnamefont {O.}~\bibnamefont {Baseggio}},
  \bibinfo {author} {\bibfnamefont {P.}~\bibnamefont {Delugas}}, \bibinfo
  {author} {\bibfnamefont {S.}~\bibnamefont {Baroni}},\ and\ \bibinfo {author}
  {\bibfnamefont {I.}~\bibnamefont {Timrov}},\ }\bibfield  {title} {\bibinfo
  {title} {{turboMagnon} -- a code for the simulation of spin-wave spectra
  using the {Liouville-Lanczos} approach to time-dependent density-functional
  perturbation theory},\ }\href {https://doi.org/10.1016/j.cpc.2022.108500}
  {\bibfield  {journal} {\bibinfo  {journal} {Comput. Phys. Commun.}\ }\textbf
  {\bibinfo {volume} {280}},\ \bibinfo {pages} {108500} (\bibinfo {year}
  {2022})}\BibitemShut {NoStop}%
\bibitem [{\citenamefont {Binci}\ \emph {et~al.}(2025)\citenamefont {Binci},
  \citenamefont {Marzari},\ and\ \citenamefont {Timrov}}]{Binci2025}%
  \BibitemOpen
  \bibfield  {author} {\bibinfo {author} {\bibfnamefont {L.}~\bibnamefont
  {Binci}}, \bibinfo {author} {\bibfnamefont {N.}~\bibnamefont {Marzari}},\
  and\ \bibinfo {author} {\bibfnamefont {I.}~\bibnamefont {Timrov}},\
  }\bibfield  {title} {\bibinfo {title} {Magnons from time-dependent
  density-functional perturbation theory and nonempirical {Hubbard}
  functionals},\ }\href {https://doi.org/10.1038/s41524-025-01570-0} {\bibfield
   {journal} {\bibinfo  {journal} {npj Comput. Mater.}\ }\textbf {\bibinfo
  {volume} {11}},\ \bibinfo {pages} {100} (\bibinfo {year} {2025})}\BibitemShut
  {NoStop}%
\bibitem [{\citenamefont {Skovhus}\ and\ \citenamefont
  {Olsen}(2026)}]{Skovhus2026}%
  \BibitemOpen
  \bibfield  {author} {\bibinfo {author} {\bibfnamefont {T.}~\bibnamefont
  {Skovhus}}\ and\ \bibinfo {author} {\bibfnamefont {T.}~\bibnamefont
  {Olsen}},\ }\bibfield  {title} {\bibinfo {title} {Classifying magnons in
  itinerant ferromagnets from linear-response time-dependent {DFT}: {Fe}, {Ni},
  and {Co} revisited},\ }\href {https://doi.org/10.1103/z3hk-wrgl} {\bibfield
  {journal} {\bibinfo  {journal} {Phys. Rev. B}\ }\textbf {\bibinfo {volume}
  {114}},\ \bibinfo {pages} {175124} (\bibinfo {year} {2026})}\BibitemShut
  {NoStop}%
\bibitem [{\citenamefont {Mignolet}\ \emph {et~al.}(2026)\citenamefont
  {Mignolet}, \citenamefont {Royo}, \citenamefont {Stengel},\ and\
  \citenamefont {Verstraete}}]{Mignolet2026}%
  \BibitemOpen
  \bibfield  {author} {\bibinfo {author} {\bibfnamefont {M.}~\bibnamefont
  {Mignolet}}, \bibinfo {author} {\bibfnamefont {M.}~\bibnamefont {Royo}},
  \bibinfo {author} {\bibfnamefont {M.}~\bibnamefont {Stengel}},\ and\ \bibinfo
  {author} {\bibfnamefont {M.~J.}\ \bibnamefont {Verstraete}},\ }\href
  {https://arxiv.org/abs/2607.26986} {\bibinfo {title} {Theory of phonon-magnon
  hybridization and angular momentum in {CrI$_3$} and {CrBr$_3$}}},\ \bibinfo
  {howpublished} {arXiv:2607.26986} (\bibinfo {year} {2026}),\ \Eprint
  {https://arxiv.org/abs/2607.26986} {arXiv:2607.26986} \BibitemShut {NoStop}%
\bibitem [{\citenamefont {de~Gironcoli}(1995)}]{Gironcoli1995}%
  \BibitemOpen
  \bibfield  {author} {\bibinfo {author} {\bibfnamefont {S.}~\bibnamefont
  {de~Gironcoli}},\ }\bibfield  {title} {\bibinfo {title} {Lattice dynamics of
  metals from density-functional perturbation theory},\ }\href
  {https://doi.org/10.1103/PhysRevB.51.6773} {\bibfield  {journal} {\bibinfo
  {journal} {Phys. Rev. B}\ }\textbf {\bibinfo {volume} {51}},\ \bibinfo
  {pages} {6773} (\bibinfo {year} {1995})}\BibitemShut {NoStop}%
\bibitem [{\citenamefont {Canc{\`e}s}\ \emph {et~al.}(2023)\citenamefont
  {Canc{\`e}s}, \citenamefont {Herbst}, \citenamefont {Kemlin}, \citenamefont
  {Levitt},\ and\ \citenamefont {Stamm}}]{Cances2023}%
  \BibitemOpen
  \bibfield  {author} {\bibinfo {author} {\bibfnamefont {E.}~\bibnamefont
  {Canc{\`e}s}}, \bibinfo {author} {\bibfnamefont {M.~F.}\ \bibnamefont
  {Herbst}}, \bibinfo {author} {\bibfnamefont {G.}~\bibnamefont {Kemlin}},
  \bibinfo {author} {\bibfnamefont {A.}~\bibnamefont {Levitt}},\ and\ \bibinfo
  {author} {\bibfnamefont {B.}~\bibnamefont {Stamm}},\ }\bibfield  {title}
  {\bibinfo {title} {Numerical stability and efficiency of response property
  calculations in density functional theory},\ }\href
  {https://doi.org/10.1007/s11005-023-01645-3} {\bibfield  {journal} {\bibinfo
  {journal} {Lett. Math. Phys.}\ }\textbf {\bibinfo {volume} {113}},\ \bibinfo
  {pages} {21} (\bibinfo {year} {2023})}\BibitemShut {NoStop}%
\bibitem [{\citenamefont {Stengel}\ and\ \citenamefont
  {De~Vita}(2000)}]{Stengel2000}%
  \BibitemOpen
  \bibfield  {author} {\bibinfo {author} {\bibfnamefont {M.}~\bibnamefont
  {Stengel}}\ and\ \bibinfo {author} {\bibfnamefont {A.}~\bibnamefont
  {De~Vita}},\ }\bibfield  {title} {\bibinfo {title} {First-principles
  molecular dynamics of metals: A {Lagrangian} formulation},\ }\href
  {https://doi.org/10.1103/PhysRevB.62.15283} {\bibfield  {journal} {\bibinfo
  {journal} {Phys. Rev. B}\ }\textbf {\bibinfo {volume} {62}},\ \bibinfo
  {pages} {15283} (\bibinfo {year} {2000})}\BibitemShut {NoStop}%
\bibitem [{\citenamefont {Hamann}(2013)}]{Hamann2013ONCVPSP}%
  \BibitemOpen
  \bibfield  {author} {\bibinfo {author} {\bibfnamefont {D.~R.}\ \bibnamefont
  {Hamann}},\ }\bibfield  {title} {\bibinfo {title} {Optimized norm-conserving
  {Vanderbilt} pseudopotentials},\ }\href
  {https://doi.org/10.1103/PhysRevB.88.085117} {\bibfield  {journal} {\bibinfo
  {journal} {Phys. Rev. B}\ }\textbf {\bibinfo {volume} {88}},\ \bibinfo
  {pages} {085117} (\bibinfo {year} {2013})}\BibitemShut {NoStop}%
\bibitem [{\citenamefont {Perdew}\ and\ \citenamefont
  {Wang}(1992)}]{Perdew1992}%
  \BibitemOpen
  \bibfield  {author} {\bibinfo {author} {\bibfnamefont {J.~P.}\ \bibnamefont
  {Perdew}}\ and\ \bibinfo {author} {\bibfnamefont {Y.}~\bibnamefont {Wang}},\
  }\bibfield  {title} {\bibinfo {title} {Accurate and simple analytic
  representation of the electron-gas correlation energy},\ }\href
  {https://doi.org/10.1103/PhysRevB.45.13244} {\bibfield  {journal} {\bibinfo
  {journal} {Phys. Rev. B}\ }\textbf {\bibinfo {volume} {45}},\ \bibinfo
  {pages} {13244} (\bibinfo {year} {1992})}\BibitemShut {NoStop}%
\bibitem [{\citenamefont {Perdew}\ \emph {et~al.}(1996)\citenamefont {Perdew},
  \citenamefont {Burke},\ and\ \citenamefont {Ernzerhof}}]{Perdew1996}%
  \BibitemOpen
  \bibfield  {author} {\bibinfo {author} {\bibfnamefont {J.~P.}\ \bibnamefont
  {Perdew}}, \bibinfo {author} {\bibfnamefont {K.}~\bibnamefont {Burke}},\ and\
  \bibinfo {author} {\bibfnamefont {M.}~\bibnamefont {Ernzerhof}},\ }\bibfield
  {title} {\bibinfo {title} {Generalized gradient approximation made simple},\
  }\href {https://doi.org/10.1103/PhysRevLett.77.3865} {\bibfield  {journal}
  {\bibinfo  {journal} {Phys. Rev. Lett.}\ }\textbf {\bibinfo {volume} {77}},\
  \bibinfo {pages} {3865} (\bibinfo {year} {1996})}\BibitemShut {NoStop}%
\bibitem [{\citenamefont {van Setten}\ \emph {et~al.}(2018)\citenamefont {van
  Setten}, \citenamefont {Giantomassi}, \citenamefont {Bousquet}, \citenamefont
  {Verstraete}, \citenamefont {Hamann}, \citenamefont {Gonze},\ and\
  \citenamefont {Rignanese}}]{Setten2018}%
  \BibitemOpen
  \bibfield  {author} {\bibinfo {author} {\bibfnamefont {M.}~\bibnamefont {van
  Setten}}, \bibinfo {author} {\bibfnamefont {M.}~\bibnamefont {Giantomassi}},
  \bibinfo {author} {\bibfnamefont {E.}~\bibnamefont {Bousquet}}, \bibinfo
  {author} {\bibfnamefont {M.}~\bibnamefont {Verstraete}}, \bibinfo {author}
  {\bibfnamefont {D.}~\bibnamefont {Hamann}}, \bibinfo {author} {\bibfnamefont
  {X.}~\bibnamefont {Gonze}},\ and\ \bibinfo {author} {\bibfnamefont {G.-M.}\
  \bibnamefont {Rignanese}},\ }\bibfield  {title} {\bibinfo {title} {The
  {PseudoDojo}: Training and grading a 85 element optimized norm-conserving
  pseudopotential table},\ }\href
  {https://doi.org/https://doi.org/10.1016/j.cpc.2018.01.012} {\bibfield
  {journal} {\bibinfo  {journal} {Comput. Phys. Commun.}\ }\textbf {\bibinfo
  {volume} {226}},\ \bibinfo {pages} {39 } (\bibinfo {year}
  {2018})}\BibitemShut {NoStop}%
\bibitem [{\citenamefont {Poncé}\ \emph {et~al.}(2016)\citenamefont {Poncé},
  \citenamefont {Margine}, \citenamefont {Verdi},\ and\ \citenamefont
  {Giustino}}]{Ponce2016EPW}%
  \BibitemOpen
  \bibfield  {author} {\bibinfo {author} {\bibfnamefont {S.}~\bibnamefont
  {Poncé}}, \bibinfo {author} {\bibfnamefont {E.}~\bibnamefont {Margine}},
  \bibinfo {author} {\bibfnamefont {C.}~\bibnamefont {Verdi}},\ and\ \bibinfo
  {author} {\bibfnamefont {F.}~\bibnamefont {Giustino}},\ }\bibfield  {title}
  {\bibinfo {title} {{EPW}: Electron-phonon coupling, transport and
  superconducting properties using maximally localized {Wannier} functions},\
  }\href {https://doi.org/http://dx.doi.org/10.1016/j.cpc.2016.07.028}
  {\bibfield  {journal} {\bibinfo  {journal} {Comput. Phys. Commun.}\ }\textbf
  {\bibinfo {volume} {209}},\ \bibinfo {pages} {116 } (\bibinfo {year}
  {2016})}\BibitemShut {NoStop}%
\bibitem [{\citenamefont {Lee}\ \emph {et~al.}(2023)\citenamefont {Lee},
  \citenamefont {Ponc{\'e}}, \citenamefont {Bushick}, \citenamefont
  {Hajinazar}, \citenamefont {{Lafuente-Bartolome}}, \citenamefont {Leveillee},
  \citenamefont {Lian}, \citenamefont {Lihm}, \citenamefont {Macheda},
  \citenamefont {Mori}, \citenamefont {Paudyal}, \citenamefont {Sio},
  \citenamefont {Tiwari}, \citenamefont {Zacharias}, \citenamefont {Zhang},
  \citenamefont {Bonini}, \citenamefont {Kioupakis}, \citenamefont {Margine},\
  and\ \citenamefont {Giustino}}]{Lee2023EPW}%
  \BibitemOpen
  \bibfield  {author} {\bibinfo {author} {\bibfnamefont {H.}~\bibnamefont
  {Lee}}, \bibinfo {author} {\bibfnamefont {S.}~\bibnamefont {Ponc{\'e}}},
  \bibinfo {author} {\bibfnamefont {K.}~\bibnamefont {Bushick}}, \bibinfo
  {author} {\bibfnamefont {S.}~\bibnamefont {Hajinazar}}, \bibinfo {author}
  {\bibfnamefont {J.}~\bibnamefont {{Lafuente-Bartolome}}}, \bibinfo {author}
  {\bibfnamefont {J.}~\bibnamefont {Leveillee}}, \bibinfo {author}
  {\bibfnamefont {C.}~\bibnamefont {Lian}}, \bibinfo {author} {\bibfnamefont
  {J.-M.}\ \bibnamefont {Lihm}}, \bibinfo {author} {\bibfnamefont
  {F.}~\bibnamefont {Macheda}}, \bibinfo {author} {\bibfnamefont
  {H.}~\bibnamefont {Mori}}, \bibinfo {author} {\bibfnamefont {H.}~\bibnamefont
  {Paudyal}}, \bibinfo {author} {\bibfnamefont {W.~H.}\ \bibnamefont {Sio}},
  \bibinfo {author} {\bibfnamefont {S.}~\bibnamefont {Tiwari}}, \bibinfo
  {author} {\bibfnamefont {M.}~\bibnamefont {Zacharias}}, \bibinfo {author}
  {\bibfnamefont {X.}~\bibnamefont {Zhang}}, \bibinfo {author} {\bibfnamefont
  {N.}~\bibnamefont {Bonini}}, \bibinfo {author} {\bibfnamefont
  {E.}~\bibnamefont {Kioupakis}}, \bibinfo {author} {\bibfnamefont {E.~R.}\
  \bibnamefont {Margine}},\ and\ \bibinfo {author} {\bibfnamefont
  {F.}~\bibnamefont {Giustino}},\ }\bibfield  {title} {\bibinfo {title}
  {Electron{\textendash}phonon physics from first principles using the {{EPW}}
  code},\ }\href {https://doi.org/10.1038/s41524-023-01107-3} {\bibfield
  {journal} {\bibinfo  {journal} {npj Comput. Mater.}\ }\textbf {\bibinfo
  {volume} {9}},\ \bibinfo {pages} {156} (\bibinfo {year} {2023})}\BibitemShut
  {NoStop}%
\bibitem [{\citenamefont {Bezanson}\ \emph {et~al.}(2017)\citenamefont
  {Bezanson}, \citenamefont {Edelman}, \citenamefont {Karpinski},\ and\
  \citenamefont {Shah}}]{Bezanson2017Julia}%
  \BibitemOpen
  \bibfield  {author} {\bibinfo {author} {\bibfnamefont {J.}~\bibnamefont
  {Bezanson}}, \bibinfo {author} {\bibfnamefont {A.}~\bibnamefont {Edelman}},
  \bibinfo {author} {\bibfnamefont {S.}~\bibnamefont {Karpinski}},\ and\
  \bibinfo {author} {\bibfnamefont {V.~B.}\ \bibnamefont {Shah}},\ }\bibfield
  {title} {\bibinfo {title} {Julia: A fresh approach to numerical computing},\
  }\href {https://doi.org/10.1137/141000671} {\bibfield  {journal} {\bibinfo
  {journal} {SIAM Rev.}\ }\textbf {\bibinfo {volume} {59}},\ \bibinfo {pages}
  {65} (\bibinfo {year} {2017})}\BibitemShut {NoStop}%
\bibitem [{\citenamefont {Methfessel}\ and\ \citenamefont
  {Paxton}(1989)}]{Methfessel1989}%
  \BibitemOpen
  \bibfield  {author} {\bibinfo {author} {\bibfnamefont {M.}~\bibnamefont
  {Methfessel}}\ and\ \bibinfo {author} {\bibfnamefont {A.~T.}\ \bibnamefont
  {Paxton}},\ }\bibfield  {title} {\bibinfo {title} {High-precision sampling
  for {Brillouin}-zone integration in metals},\ }\href
  {https://doi.org/10.1103/PhysRevB.40.3616} {\bibfield  {journal} {\bibinfo
  {journal} {Phys. Rev. B}\ }\textbf {\bibinfo {volume} {40}},\ \bibinfo
  {pages} {3616} (\bibinfo {year} {1989})}\BibitemShut {NoStop}%
\bibitem [{\citenamefont {Marzari}\ \emph {et~al.}(1999)\citenamefont
  {Marzari}, \citenamefont {Vanderbilt}, \citenamefont {De~Vita},\ and\
  \citenamefont {Payne}}]{Marzari1999Smearing}%
  \BibitemOpen
  \bibfield  {author} {\bibinfo {author} {\bibfnamefont {N.}~\bibnamefont
  {Marzari}}, \bibinfo {author} {\bibfnamefont {D.}~\bibnamefont {Vanderbilt}},
  \bibinfo {author} {\bibfnamefont {A.}~\bibnamefont {De~Vita}},\ and\ \bibinfo
  {author} {\bibfnamefont {M.~C.}\ \bibnamefont {Payne}},\ }\bibfield  {title}
  {\bibinfo {title} {Thermal contraction and disordering of the {Al}(110)
  surface},\ }\href {https://doi.org/10.1103/PhysRevLett.82.3296} {\bibfield
  {journal} {\bibinfo  {journal} {Phys. Rev. Lett.}\ }\textbf {\bibinfo
  {volume} {82}},\ \bibinfo {pages} {3296} (\bibinfo {year}
  {1999})}\BibitemShut {NoStop}%
\bibitem [{\citenamefont {dos Santos}\ and\ \citenamefont
  {Marzari}(2023)}]{DosSantos2023}%
  \BibitemOpen
  \bibfield  {author} {\bibinfo {author} {\bibfnamefont {F.~J.}\ \bibnamefont
  {dos Santos}}\ and\ \bibinfo {author} {\bibfnamefont {N.}~\bibnamefont
  {Marzari}},\ }\bibfield  {title} {\bibinfo {title} {Fermi energy
  determination for advanced smearing techniques},\ }\href
  {https://doi.org/10.1103/PhysRevB.107.195122} {\bibfield  {journal} {\bibinfo
   {journal} {Phys. Rev. B}\ }\textbf {\bibinfo {volume} {107}},\ \bibinfo
  {pages} {195122} (\bibinfo {year} {2023})}\BibitemShut {NoStop}%
\bibitem [{\citenamefont {Lihm}\ \emph {et~al.}(2026)\citenamefont {Lihm},
  \citenamefont {R\"osner}, \citenamefont {Dreyer}, \citenamefont {Stengel},\
  and\ \citenamefont {Ponc\'e}}]{MaterialsCloudArchive}%
  \BibitemOpen
  \bibfield  {author} {\bibinfo {author} {\bibfnamefont {J.-M.}\ \bibnamefont
  {Lihm}}, \bibinfo {author} {\bibfnamefont {M.}~\bibnamefont {R\"osner}},
  \bibinfo {author} {\bibfnamefont {C.~E.}\ \bibnamefont {Dreyer}}, \bibinfo
  {author} {\bibfnamefont {M.}~\bibnamefont {Stengel}},\ and\ \bibinfo {author}
  {\bibfnamefont {S.}~\bibnamefont {Ponc\'e}},\ }\bibfield  {title} {\bibinfo
  {title} {Unifying dynamical electron-electron and electron-phonon
  interactions from first principles},\ }\bibfield  {journal} {\bibinfo
  {journal} {Materials Cloud Archive}\ }\href
  {https://doi.org/10.24435/materialscloud:y2-7s}
  {10.24435/materialscloud:y2-7s} (\bibinfo {year} {2026}),\ \bibinfo {note}
  {(reference will be updated upon acceptance)}\BibitemShut {NoStop}%
\end{thebibliography}%

\end{document}